 \documentclass[journal,12pt, draftclsnofoot, onecolumn, twoside]{IEEEtran}

\usepackage{fancyhdr}
\usepackage{graphicx}
\usepackage{cite}
\usepackage{amssymb,amsmath}
\usepackage{algorithm}
\usepackage{color,soul,colortbl}
\usepackage{cleveref}
\usepackage{subcaption}
\usepackage{booktabs}
\usepackage{xcolor}
\usepackage{caption}
\usepackage{multirow}
\usepackage{hhline}
\usepackage[acronym,shortcuts]{glossaries}
\usepackage{amsthm}

\newtheorem{corollary}{Corollary}
\newtheorem{definitioninner}{Definition}

\newenvironment{definition}[1]{%
  \renewcommand\thedefinitioninner{#1}
  \definitioninner
}{\enddefinitioninner}

\theoremstyle{plain}

\theoremstyle{plain}

\theoremstyle{plain}

\providecommand{\lemmaname}{Lemma}
\providecommand{\propositionname}{Proposition}
\providecommand{\theoremname}{Theorem}
\providecommand{\lemmaname}{Lemma}
\providecommand{\propositionname}{Proposition}
\providecommand{\theoremname}{Theorem}
\makeatother
\providecommand{\lemmaname}{Lemma}
\providecommand{\propositionname}{Proposition}
\providecommand{\theoremname}{Theorem}

\definecolor{G}{gray}{0.9}
\definecolor{LC}{rgb}{0.88,1,1}

\newcommand{\notimplies}{\mathrel{{\ooalign{\hidewidth$\not\phantom{=}$\hidewidth\cr$\Longleftarrow$}}}}
\newcommand{\V}[1]{\mathbf{\lowercase{#1}}}                  
\newcommand{\M}[1]{\mathbf{\uppercase{#1}}}                  
\newcommand{\RM}[1]{{\mathrm{#1}}}  
 \newenvironment{psmallmatrix}
  {\left(\begin{smallmatrix}}
  {\end{smallmatrix}\right)}

\newacronym{5G}{5G}{fifth generation}
\newacronym{AWGN}{AWGN}{additive white Gaussian noise}
\newacronym{CDF}{CDF}{cumulative distribution function}
\newacronym{CSI}{CSI}{channel state information}
\newacronym{FDR}{FDR}{full-duplex relaying}
\newacronym{HDR}{HDR}{half-duplex relaying}
\newacronym{IC}{IC}{interference channel}
\newacronym{IGS}{IGS}{improper Gaussian signaling}
\newacronym{MHDF}{MHDF}{multi-hop decode-and-forward}
\newacronym{MIMO}{MIMO}{multiple-input multiple-output}
\newacronym{MISO}{MISO}{multiple-input single-output}
\newacronym{MRC}{MRC}{maximum ratio combining}
\newacronym{PDF}{PDF}{probability density function}
\newacronym{PGS}{PGS}{proper Gaussian signaling}
\newacronym{RSI}{RSI}{residual self-interference}
\newacronym{RV}{RV}{random vector}
\newacronym{r.v.}{r.v.}{random variable}
\newacronym{SISO}{SISO}{single-input single-output}

\newacronym{CEMSE}{CEMSE}{Computer, Electrical, and Mathematical Sciences and Engineering}
\newacronym{KAUST}{KAUST}{King Abdullah University of Science and Technology}

\begin{document}
\title{\LARGE{A Journey from Improper Gaussian Signaling to Asymmetric Signaling}}
\author{{ Sidrah Javed, Osama Amin, Basem Shihada and Mohamed-Slim Alouini }   

\thanks{S. Javed, O. Amin, B. Shihada and M.-S. Alouini are with CEMSE Division,  King Abdullah University of Science and Technology (KAUST),  Thuwal, Makkah Province, Saudi Arabia.   E-mail: \{sidrah.javed, osama.amin, basem.shihada, slim.alouini\}@kaust.edu.sa } 
}

  \maketitle
\thispagestyle{fancy}
\fancyhf{}
\rhead{ \footnotesize{DOI: 10.1109/COMST.2020.2989626}}
 \lhead{\footnotesize{ \copyright~2020 IEEE.}}

\begin{abstract}
The deviation of continuous and discrete complex random variables from the traditional proper and symmetric assumption to a generalized improper and asymmetric characterization (accounting correlation between a random entity and its complex conjugate), respectively, introduces new design freedom and various potential merits. As such, the theory of impropriety has vast applications in medicine, geology, acoustics, optics, image and pattern recognition, computer vision, and other numerous research fields with our main focus on the communication systems. The journey begins from the design of improper Gaussian signaling in the interference-limited communications and leads to a more elaborate and practically feasible asymmetric discrete modulation design. Such asymmetric shaping bridges the gap between theoretically and practically achievable limits with sophisticated transceiver and detection schemes in both coded/uncoded wireless/optical communication systems. Interestingly, introducing asymmetry and adjusting the transmission parameters according to some design criterion render optimal performance without affecting the bandwidth or power requirements of the systems. This dual-flavored article  initially presents the tutorial base content covering the interplay of reality/complexity, propriety/impropriety and circularity/non-circularity and then surveys majority of the contributions in this enormous journey.
\end{abstract}
\begin{IEEEkeywords}
Achievable rate, asymmetric signaling, circularity, data analysis, detection, DoF, equalization, error probability, estimation, filtering, IGS,  Impropriety, outage probability, power efficiency, separation, signal processing, widely linear processing, wireless communications
\end{IEEEkeywords}
\section{Introduction}
The ever growing demands of energy and bandwidth efficient systems in diverse fields of science and technology have compelled researchers to think beyond the traditional approaches and techniques. Divergence from norms/conventions can sometimes beat the traditional and long-practiced assumptions. Same is the case with long assumed circularly symmetric complex (CSC) signal conjecture followed before the advent of last three decades. The advancement from the real stochastic domain to the complex stochastic domain came with the naive assumption of equal energy and uncorrelated real and imaginary components of a complex random entity, later named as proper complex random variables (r.v.). A proper complex r.v. is thus uncorrelated with its complex conjugate whereas any correlation between the two results in improper complex r.v. The investigation of the presence and absence of this property is coined as propriety and impropriety, respectively.
Another related yet distinct phenomenon is the concept of CSC or circular r.v., demonstrating rotationally invariant probability distribution in the complex plane \cite{neeser1993proper,picinbono1994circularity,picinbono1997second,schreier2003second,ollila2008circularity,schreier2010statistical,
adali2011complex}. The absence of this property renders non-circular (NC) complex r.v. and the evaluation study to determine the circular and NC nature of complex entities is termed as circularity and non-circularity, respectively.

Majority of the contributions assumed proper signal model for the underlying complex phenomenon rendering simplified computations, which is in contrast with most of the real-world scenarios. For instance, improper nature of complex modelled entities has been proven for the accumulative additive thermal noise model in communication systems \cite{alsmadi2018ssk,javed2018multiple},  complex envelop of the scalar optical fields \cite{erkmen2006optical},  empirical speech model  \cite{rivet2007log}, complex traces of the seismic signals \cite{taner1979complex}, ocean-current spectra \cite{calman1978interpretation}, wind fields \cite{khalili2014collaborative,knight2018long,burt1974mesoscale} and fluid dynamics \cite{sykulski2017frequency}, complex valued model of unbalanced three-phase voltage in power systems \cite{xia2012widely,dini2013widely} and neural activity in brain/spinal cord as measured by functional magnetic resonance imaging (fMRI)~\cite{schreier2010statistical,rowe2005modeling, rodriguez2012noising}. 

Besides covering the broader aspects of various disciplines, this survey is focused on the communication systems with intentional/unintentional improper signatures. Various discrete modulation schemes that realize the asymmetric complex characteristics include Binary Phase Shift Keying (BPSK) \cite{schreier2003second,buzzi2001new}, Pulse Amplitude Modulation (PAM) \cite{ollila2008circularity}, Gaussian Minimum Shift Keying (GMSK) \cite{hellings2015block,chevalier2006new}, Offset Quadrature Phase Shift Keying (OQPSK) or staggered QPSK \cite{mirbagheri2006enhanced}, single sideband (SSB), vestigial sideband (VSB) \cite{yeo2011optimal} and baseband (but not passband) Orthogonal Frequency Division Multiplexing (OFDM) or discrete multitone (DMT) modulation \cite{taubock2007complex}. On the other hand, coding schemes like space-time block coding (STBC) also results in asymmetric signaling \cite{alamouti1998simple}. Apart from these naturally occurring improper signals, some phenomenon like transceiver in-phase and quadrature-phase (I/Q) imbalance (IQI) transforms proper signal to improper signal as well as proper thermal and distortion noise to improper accumulative noise in the wireless communication systems \cite{javed2018asymmetric}. 

\subsection{Background}
Complete second-order statistical (SOS) characterization of the random signal vector $\V{s}$ involves the covariance matrix ${\rm E} \left[ \V{s}\V{s}^{\RM H}  \right]$ as well as pseudo-covariance matrix ${\rm E} \left[ \V{s}\V{s}^{\RM T}  \right]$ analysis besides statistical mean ${\rm E} \left[ \V{s}  \right]$.  
The traditional complex-valued
signal processing assumes a vanishing pseudo-covariance,
which is inadequate for many real problems where the
signals involved are frequently improper \cite{xia2010augmented}. 
This superficial assumption not only leads to misleading analysis and incorrect results but also deprives us from seeking the maximum benefit of the additional design freedom offered by improper/asymmetric signaling. Thus, much can be gained if the information contained in the complementary/pseudo
covariance is also exploited \cite{adali2011complex,mandic2009complex}.

Surprisingly, the propriety concept is not unheard-of and vast majority of contributions were put forward in one domain or the other. For instance, improper processes and models have
seen growing interest in the statistics community \cite{schreier2005detection,walden2009testing,
mohammadi2015improper}. Similarly, the pioneering works in signal processing community which emphasized the significance of detailed propriety characterization include  \cite{picinbono1994circularity,picinbono1996second,
schreier2003second,ollila2004generalized,ollila2008circularity,
adali2011complex}. 
Moreover, the significant contributions from the information theory group comprises of \cite{dunbridge1967asymmetric,kassam1982robust,neeser1993proper,van1995multivariate,eriksson2006complex,navarro2009estimation,taubock2012complex,via2010properness}. Furthermore, few early contributions from the communication circle include  \cite{divsalar1987trellis,yoon1997maximizing,gelli2000blind,isaka2000multilevel,buzzi2001new,gerstacker2003receivers}. 

The naturally existing improper phenomenon and advanced statistical tools supported various studies to exploit the complete SOS characterization of the involved signals.
Paradoxically, researchers resort to the more tedious two-dimensional real modeling of the complex variables for their statistical investigation \cite{picinbono1996second} and the complex theory lost most of its
beauty, elegance, and interest \cite{picinbono1994circularity}. 

\subsection{Motivation}
The convenience of modeling multivariate data as a complex entity owing to its compact form, simple notation and fast computation has endorsed it's utility in various applications. Nonetheless, the complete statistical  characterization is inevitable to not only exploit the information contained in complementary covariance matrix but also utilize the additional design freedom. Unlike the conventional assumption, complementary covariance function has generally no reason to be equal to zero \cite{picinbono1997second}. This is especially significant when dealing with the concepts of distribution function and entropy quantification. Circularity is an assumption that was originally introduced for the definition of the probability distribution function \cite{picinbono1994circularity}. 
For instance, the \ac{PDF} of a complex
Gaussian random vector (RV) assumes the anticipated and familiar ‘natural’ form only for proper RVs \cite{neeser1993proper,schreier2003second}. Moreover, 
the definitions of independence and/or correlation are inadequate without complete SOS characterization. For example, it is shown that contrary to the real case, uncorrelated complex normal r.vs. are not generally independent \cite{picinbono1996second}.
Similarly, the propriety characterization is inevitable for the appropriate treatment and accurate entropy quantification of the complex signals and complex impulse responses of the equivalent base-band channels.

The conceptual analysis of properness in various signal processing and information theoretic studies can render significant benefits when extended to communication applications. Initially, people argued the superiority of the proper complex signaling as it maximizes the entropy of a given system  \cite{neeser1993proper}. However, the reduced-entropy improper Gaussian signaling (IGS) can be beneficial in unlicensed spectrum-sharing techniques with minimal interference to the legitimate users in  underlay \cite{gaafar2017underlay,lameiro2018improper,oliveira2018physical,lameiro2015benefits,amin2016underlay}, overlay \cite{amin2017overlay} and interweave  \cite{hedhly2020benefits} cognitive radio setups. Followed by the opinion of dominant improper/asymmetric signaling when the system is contaminated by the improper complex interference \cite{
kim2016asymptotically,lameiro2016maximally}, self-interference \cite{javed2017full,korpi2014widely,sornalatha2017modeling}, improper noise \cite{alsmadi2018ssk,
hellings2017reduced,tugnait2017multisensor,yoon1997maximizing}, asymmetric noise/distortions \cite{javed2017asymmetric,kassam1982robust,javed2018improper} and NC hardware imperfections \cite{zarei2016q,
zhang2017widely,
anttila2008circularity,hakkarainen2013widely,
li2017noncircular,javed2018multiple}. 

Recent studies have demonstrated the perks of improper/asymmetric signaling in the general interference-limited scenarios even in the absence of improper contamination. Notable edge attained by improper transmission over proper transmission in various interference-limited scenarios include interference broadcast channel (IBC) \cite{shin2011achievable,lin2018multi}, broadcast channel (BC) \cite{bai2018optimal,hellings2013qos,hellings2015iterative}, multiple access channel (MAC) \cite{kariminezhad2016improper,kariminezhad2017interference},
 cross-interference channel (X-IC or IC) \cite{cadambe2010interference,
 yang2014interference,lameiro2013degrees,kim2013potential,
 zeng2013transmit,ho2012improper,zeng2013miso,lagen2016coexisting},
 one-sided interference channel (Z-IC) \cite{lagen2014improper,kurniawan2015improper,lagen2016superiority,lameiro2017rate}, 
 relaying systems \cite{gaafar2018full,kim2012asymmetric, ho2013optimal,zhang2013widely,kariminezhad2017power}
multi-antenna systems \cite{tugnait2016multiantenna, lameiro2018performance}, 
multi-cell systems \cite{park2013sinr}, and multi-tier networks \cite{kariminezhad2016heterogeneous}, etc.

Besides communication systems, general estimation, detection and filtering techniques of circular cases differ from the NC cases. 
Thus, significant performance gains can be reaped by exploiting circularity/non-circularity in, e.g., wireless transceivers with beamformers or DoA algorithms \cite{haardt2004enhancements,charge2001non,abeida2008statistical},
source identification \cite{napolitano2004doppler,clark2012existence,jelfs2012adaptive,yeo2011optimal}, blind source separation 
\cite{eriksson2006complex,li2008class,douglas2007fixed,schreier2003second,novey2008extending}, estimation \cite{navarro2009estimation,mohammadi2015distributed,
xia2017augmented,trampitsch2013complex,picinbono1995widely,schreier2005detection,lang2017classical}, detection \cite{novey2009circularity,matalkah2008generalized,aghaei2008maximum}, linear and nonlinear filtering \cite{witzke2005linear,yoon1995matched,mohammadi2015complex},
and Cramer-Rao performance bounds \cite{jagannatham2004cramer}, etc. 

\begin{table*}[t]	
\renewcommand{\arraystretch}{1.25}
\caption{Summary of Relevant Contributions Surveying Impropriety Concepts}
\begin{center}
  \begin{tabular}{||p{1.5cm} ||p{3.3cm} ||p{1cm}||p{7.5cm} ||p{1cm}||}
 \toprule
\textbf{Focal Area} & \textbf{Description} & \textbf{Year} & \textbf{Contributions} & \textbf{Ref}\\
\midrule
Filtering & Nonlinear adaptive filtering & 2009 & {{Focused on the complex nonlinear adaptive filters incorporating non-circularity, widely linear processing and neural models.}} & \cite{mandic2009complex} \\
\hline
 Theoretical Fundamentals & Theory of improper and NC signals & 2010 & Basics of improper and
NC complex-valued signals with comprehensive illustration of the main applications, covering detection, estimation,
and signal analysis. & \cite{schreier2010statistical} \\
  \hline  
 Interference mitigation & Interference alignment in communication networks & 2011 & Provides a tutorial while surveying the state-of-the-art interference alignment schemes with a special emphasis on asymmetric complex signal alignment & \cite{jafar2011interference}\\    
   \hline  
 Gaussianity Deviations & Complex elliptically symmetric distributions & 2012 & Comprehensive survey and overview of applications involving complex elliptically symmetric distributions & \cite{ollila2012complex}\\    
   \hline  
\multirow{4}{*}{Estimation} & Statistical Robust Estimation& 2012-2018 &    A tutorial-style treatment of fundamental concepts for  robust statistical estimation in signal processing integrating deviations from the distributional assumptions & \cite{zoubir2012robust,zoubir2018robust}\\    
  \cline{2-5}  
 & Source identification and separation & 2013 & Blind identification  and separation of general complex-valued signals &\cite{moreau2013blind} \\    
 \cline{2-5}  
& Optimization and Estimation & 2014 & Thorough review of the necessary tools in filtering  and blind  source separation emphasizing the use of the full statistical information in complex-valued optimization and estimation.  & \cite{adali2014optimization}\\    
 \cline{2-5}  
& Parameter Estimation & 2015 & Presents a meticulous statistical  study of complex-valued parameter estimators  with a special emphasis on properness &\cite{delmas2015survey}   \\
    \hline 
Detection & MIMO Detection & 2015 & Exhaustive survey on the fifty years of MIMO detection entailing  emerging massive/large-scale MIMO detection schemes and briefly highlighting the importance of appropriate treatment of improper complex-valued systems& \cite{yang2015fifty}   \\
\bottomrule
  \end{tabular}
\end{center}
\label{tab:RelatedContributions}
\end{table*}
Conclusively, all of these and many other relevant contributions motivate the incorporation of propriety characterization for accurate system analysis and improved system design. The significant performance gains marked by various applications in numerous disciplines helped as a precursor to put forward this extensive survey.
\subsection{Contributions}
Various studies discussed the significance of a complete characterization and appropriate treatment of the systems involving improper and NC signals, as summarized in Table \ref{tab:RelatedContributions}. These studies deal with impropriety concepts that are focused on Gaussianity deviants \cite{ollila2012complex}, {interference mitigation} \cite{jafar2011interference}, {filtering} \cite{mandic2009complex}, {detection} \cite{yang2015fifty}, {estimation} \cite{zoubir2012robust,zoubir2018robust,delmas2015survey}, {source identification} \cite{moreau2013blind} {and separation} 
\cite{adali2014optimization}.  Furthermore, a comprehensive treatment of theoretical fundamentals of improper and NC signals along with their diverse applications is carried out in \cite{schreier2010statistical}. 
Despite of all these contributions, an exhaustive yet comprehensive survey is required which not only encompasses all these domains but also furnishes complete evolution from the theoretical aspects to the realization ones. Furthermore, this article bridges the interdisciplinary gap between the fields of information theory and signal processing for the wireless communication applications pertaining to signal characterization. The main contributions of this paper in a chronological order include:
\begin{itemize}
\item {Emphasizing the significance of propriety characterization with three different data representations and their inter-relations.} 
\item {Highlighting main differences between the intermingled terms of impropriety and circularity along with the corresponding measures to determine their extent.}
\item  {Presenting the generalized entropy and probability distribution functions of the complex Gaussian distributions followed by various tests for impropriety.}
\item {Taxonomy of literature to feature naturally occurring sources of impropriety and their consequences.}
\item {Performance comparison of the theoretical limits achieved by IGS as compared to the conventional PGS e.g., achievable rate, outage probability, power/energy efficiency and DoF.} 
\item {Various design guidelines covering suitable optimization tools for the IGS design in addition to the relevant impropriety detection, estimation, filtering and separation procedures.} 
\item {Encompassing the journey from theoretical IGS to practical asymmetric discrete signaling and corresponding asymmetric signal recovery methodologies namely equalization, estimation, filtering and detection.} 
\item  {Error probability (EP) analysis to figure out the maximum reported percentage decrease and the corresponding SNR gains to attain a certain error rate with asymmetric characterization relative to symmetric characterization in various system configurations.} 
\item {Comprehensive survey of the applications, in data analysis, signal processing and communication theory domains, reaping benefits by exploiting or incorporating impropriety concepts.}
\item {Summarizing the lesson learned throughout this study while pointing out the main challenges and way forward.}
\end{itemize}
The proposed study not only encloses the existing contributions but also serves as an introductory and motivational guideline for the beginners in this domain. It further elaborates various tools and techniques for appropriate improper/asymmetric signaling to reap the maximum benefits.
 \subsection{Organization and Road map}
\begin{figure}[t]
\begin{minipage}[b]{1.0\linewidth}
  \centering
  \centerline{\includegraphics[width=16cm]{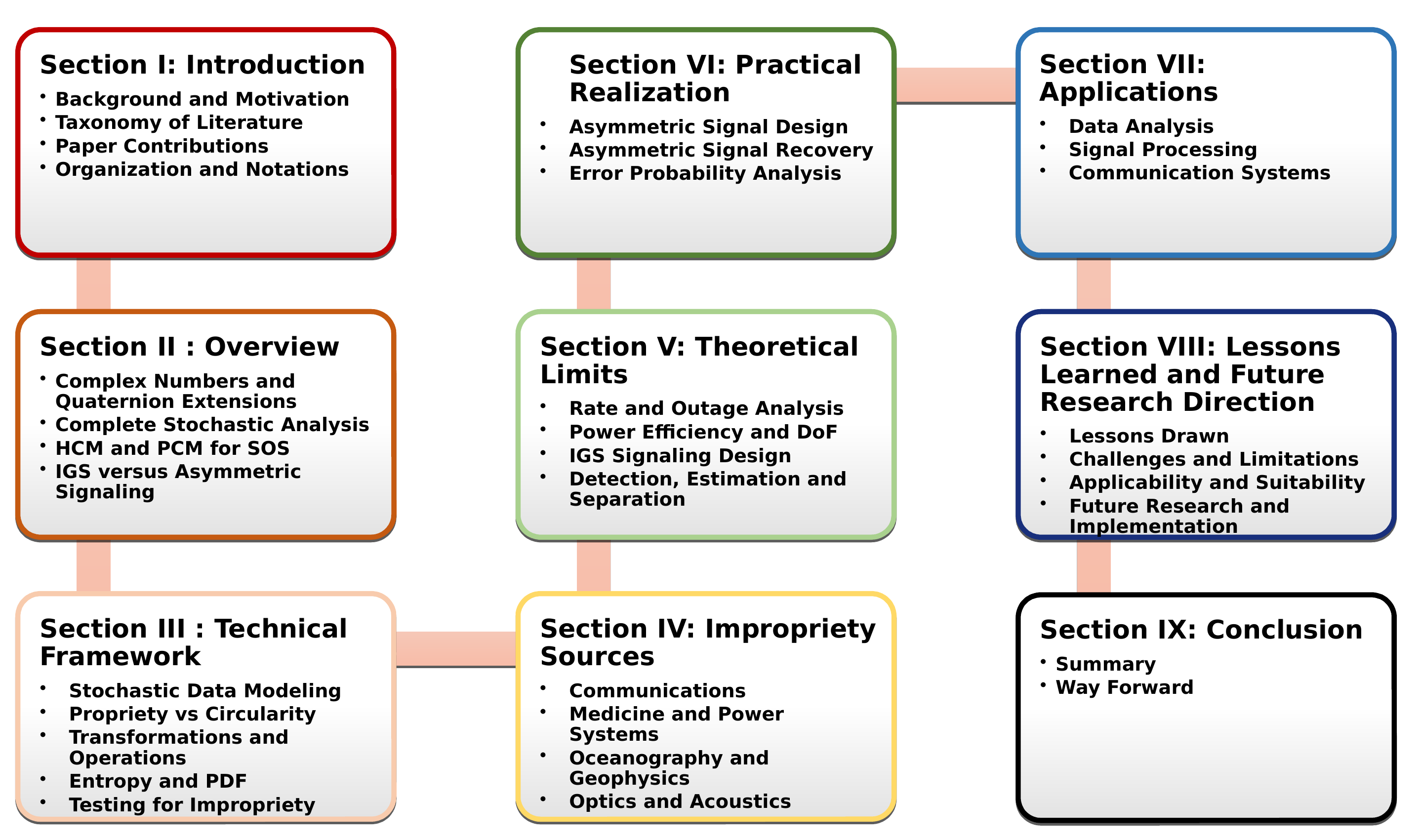}}
\end{minipage}
\caption{{Paper Organization and Contents}}
\label{fig:Outline}
\end{figure}

{The paper covers the journey of improper/asymmetric signaling from theory to practical design by first encompassing the preliminaries and overview of the propriety concepts in Section II.
Section III encompasses the detailed technical framework for the theoretical analysis.  Followed by the brief review of naturally occurring sources of impropriety in various diverse fields in Section IV. Next, Section V covers the theoretical analysis and performance bounds achieved by the IGS in terms of achievable rates, outage probability, power efficiency and degrees of freedom (DoF) in numerous systems.  Furthermore, it elaborates the general guidelines for appropriate IGS design algorithms and recovery procedures. The motivation drawn from the theoretical analysis is used as a precursor for practical realization and design of asymmetric discrete constellation in Section VI. This further includes the more revealing metric i.e., error rate analysis to emphasize the superiority of the adopted/advertised signaling or widely linear (WL) processing in several contributions. Later, Section VII classifies numerous applications utilizing the impropriety concepts to further
enhance their performance. Subsequently, Section VIII and IX 
furnish the lessons learned, challenges/limitations and critical future research directions followed by the concluding remarks. The comprehensive outline of the paper organization is detailed in Fig. \ref{fig:Outline}.}
 \subsection{Notation}
  Throughout the paper, lower-case letter $x$ denotes variables, while boldface lower-case letter  $\V{x}$ and boldface upper-case letter $\M{x}$  denote vectors and matrices, respectively.
$\M{0}_N$ and $\M{I}_N$ depict the $N{\rm x}N$ square null and identity matrix, respectively. The complex-conjugate, inverse conjugate, transpose and conjugate-transpose operators are denoted as ${ \left\{ {\RM{.}} \right\}^{\rm{*}}}$, ${ \left\{ {\RM{.}} \right\}^{\rm{-*}}}$ , $  \left\{ {\RM{.}} \right\}^{\RM{T}}  $ and   $   \left\{ {\RM{.}} \right\} ^{\RM{H}}  $, respectively.
 The complex augmented \ac{RV} $\underline{\V{x}}$ is defined as $\underline{\V{x}} = [{\V{x}}^{\RM{T}} \;  {\V{x}}^{\RM{H}}]^{\RM{T}}$. 
The L2-norm and absolute value of a scalar $x$ are represented as ${\left\| \V{x} \right\|_{\rm{2}}}$ and $\left| x \right|$, respectively.  The notation $\M{x} \in \mathbb{F}^{{{M \times N}}}$ describes that the  $\M{X}$ is a $M \times N$ matrix with entries in $\mathbb{F}$, where $\mathbb{F}$ can be the field of real numbers $\mathbb{R}$, the field of complex numbers $\mathbb{C}$ , or the skew-field of quaternion numbers $\mathbb{H}$. Moreover, the trace, determinant, Frobenius norm, and pseudo-inverse of ${\M{X}}$ are given as ${\RM{Tr}}\left(  \M{X}\right)$,  $ \left|  \M{X} \right| $, $||{\M{X}}||$ and  $\M{x}^\dagger$, respectively. Similarly, for a square invertible matrix $\M{X}$, the  inverse and square-root matrix are represented by ${\M{X}}^{-1}$ and ${\M{X}}^{1/2}$, respectively. Additionally,  $\M{X} \, \underline{\succ} \, 0$ and $\M{X} \succ 0$ denote that $\M{X}$ is a positive semi-definite (PSD) matrix and positive definite (PD) matrix, respectively. The square block diagonal matrix $\M{J}=\textbf{Block-Diag} \left(\M{J}_1,\M{J}_2,\ldots,\M{J}_3\right)$ comprises of square matrices 
 $\M{J}_1,\M{J}_2,\ldots,\M{J}_3$ of any size as the diagonal elements and the off-diagonal elements are all zero.  The subscripts $\left\{ . \right\}_{+}$ and $\left\{ . \right\}_{++}$ are used to define non-negative and strictly positive real numbers of a set, respectively. The expected value operator is expressed as ${\rm E}\left[ . \right]$ and the probability of occurrence of an event $\bf \Omega$ is given by $\Pr \left\{ \bf \Omega  \right\}$. The operation $\odot$ represents a point-wise product. The pure imaginary unit $i$ and the Euler's number $e$ are defined by $i^2 = 1$ and the limit of $(1 + 1/n)^n$ as $n$ approaches infinity, respectively. The term `iff' is used for the condition `if and only if' and the notations $ = ,\triangleq \text{and} \equiv$  describe the general mathematical equality, equality in distribution and equivalence, respectively. 
\section{Overview}
\label{sec:overview}
Many applications in applied sciences employ complex analysis to model the real-world data in complex domain. The complex analysis brings advantages like comprehensiveness, computational economy, extra dimension, elegant analysis and much more. For instance, multi-component alternating current circuits assume complex impedance, in place of real resistance and reactance, for refined circuit analysis. The resistance and reactance are treated as the real and imaginary components, respectively,  of a complex impedance \cite{hayt1986engineering}. Analogously, in digital communications, complex numbers enabled us to deal with quadrature amplitude modulation (QAM) in a tractable mathematical way. Moreover, one square-QAM signal compactly carries the information of two PAM signals in its inphase and quadrature phase components \cite{john2008digital}. 
The data from some physical systems should be analyzed as complex-valued signals because the data represent motion on the complex plane (e.g., tidal analysis in
oceanography and two-component observations in meteorology \cite{rubin2008kinematics}). Furthermore, directional processes (radar, sonar, Doppler ultrasound, vector fields, bearings only estimation), where both the ``intensity''
(amplitude) and ``direction'' (phase) components carry
the information, are also most conveniently analyzed as
complex valued processes \cite{mandic2007complex,olhede2003noise,walden2009testing}.

 But why stop at complex numbers? In fact, there are several applications which require higher dimensional representation such as quaternions, which is convenient to represent the rotations of three-dimensional space. Quaternions are used to characterize data of several systems/applications including aerospace \cite{fortuna2001comparison}, computer graphics \cite{hanson498visualizing}, signal array processing \cite{miron2006quaternion}, Fourier transforms of images \cite{bulow2001hypercomplex}, design of orthogonal polarized STBC \cite{seberry2008theory}, wave separation \cite{buchholz2008polarized}, wind forecasting  \cite{took2009quaternion}, nonlinear estimation \cite{navarro2014quaternion}, adaptive filtering \cite{xiang2018performance,krupinski2018generating}, and vector sensors \cite{le2004singular}. One compelling application is the unified treatment of the relative position and orientation in hand-eye calibration of a robot \cite{daniilidis1999hand}.
 
In a nutshell, complex numbers and their extensions are immensely used in countless real-world applications.
Some of these applications treat variables, signals or images as deterministic quantities. However, many applications require a stochastic modeling of the underlying phenomena such as electromagnetic waves carrying random codes, polarized magnetic disturbances,  and noise in image processing etc. \cite{amblard2004properness}. Therefore, complex and quaternion r.v. require appropriate and complete characterization to be fully understood and applied. 

{At the outset of complex theory, the complex stochastic analysis was straightforwardly derived from the real stochastic analysis. First-order statistics were given by the statistical average/mean, whereas SOS were provided by the Hermitian covariance matrix (HCM) in complex analysis as opposed to the symmetric covariance matrix in real analysis. Interestingly, this completes SOS for real analysis but soon realized to be highly insufficient for the complex analysis. 
HCM accounts for the correlation between different entries of a \ac{RV} but not the correlation between the two quadrature components within a complex r.v. Moreover, the diagonal elements of HCM render the cumulative variance of the real and imaginary components of the corresponding r.vs but fail to provide the difference of energy between them. Furthermore, the non-diagonal elements of HCM render 
\begin{itemize}
\item The aggregate covariance between reals and imaginaries of two different r.vs, respectively but does not show the difference between them
\item The difference in the cross-covariances which are evaluated between the real of one r.v. and the imaginary of the other r.v. and vice versa. However, it is unsuccessful in declaring the accumulative cross-covariances values.
\end{itemize}
The lackings mentioned above necessitated  adopting another matrix, to complement the missing information in HCM, later named as \textit{pseudo-covariance matrix} (PCM) \cite{neeser1993proper,ollila2008circularity}, \textit{conjugate covariance matrix} \cite{schreier2010statistical}, \textit{complementary covariance matrix} \cite{schreier2003second}, \textit{relation matrix} \cite{picinbono1996second} or \textit{pseudo-scatter matrix} \cite{ollila2004generalized}.}

{The incorporation of PCM for the complete characterization is essential to account for the cases when real and imaginary compositions of a r.v. exhibit different variances/powers and/or correlated with each other.
Additionally, a null PCM not only refers to uncorrelated components of r.vs but also dictates the correlation between the reals and the imaginaries of two different r.vs in a RV. 
Moreover, the cross-PCM inclusion is motivated by the fact that the regular definition of two uncorrelated Gaussian RVs  requires cross-HCM to be zero. However, two uncorrelated Gaussian RVs may not be independent pertaining to non-zero cross-PCM, which is counterintuitive \cite{schreier2010statistical}. Therefore, non-zero PCM must be considered for consonance. Remarkably, a Gaussian RV can be completely described using mean vector, HCM, and PCM, where a null PCM designates proper Gaussian RV and a non-zero PCM dictates an improper Gaussian R.V.   
A stronger version of propriety is termed as circularity and it accounts for the rotationally invariant PDF and all existing moments. Thus, circularity implies zero-mean and propriety but not vice versa \cite{adali2011complex}. However, they are exactly same for a zero-mean Gaussian RV.}    

{This work highlights the importance of SOS characterization for complex analysis of Gaussian RVs and higher order statistics for other RVs. Many real-world problems render improper random signals in diverse fields requiring appropriate analysis and treatment e.g., in 
medicine \cite{zarzoso2010robust,adali2014optimization,li2011application,schreier2010statistical,rowe2005modeling, rodriguez2012noising,javidi2010complex,clark2012existence}, 
oceanography \cite{gonella1972rotary,mooers1973technique,calman1978interpretation},
geology \cite{taner1979complex,khalili2014collaborative,knight2018long,sykulski2016widely,
burt1974mesoscale,jelfs2012adaptive,
adali2010complex,xia2011augmented}, 
optics \cite{erkmen2006optical,yuen1976two,erkmen2006phase}, 
acoustics \cite{looney2011augmented,layton2006augmented,rivet2007log}, 
power systems \cite{xia2012widely,dini2013widely,xia2012adaptive,arablouei2013adaptive}.
 and communication systems \cite{schreier2003second,buzzi2001new,ollila2008circularity,
 hellings2015block,chevalier2006new,mirbagheri2006enhanced,
 yeo2011optimal,alamouti1998simple,schreier2005detection,adali2011complex,roemer2006efficient,
 gerstacker2003receivers,mohammadi2015distributed,
 xia2017augmented,javed2018multiple,anttila2008circularity,
 hakkarainen2013widely,li2017noncircular,schreier2010statistical,mokhtar2013ofdm,
li2014q}. Focusing on communication systems, IGS with appropriate design and treatment is proven to improve the theoretical bounds of various system configurations.  The journey begins with the theoretical IGS analysis and leads to the practical asymmetric signaling scheme for the modern communication systems. Asymmetric signaling is the discrete counterpart of IGS as symmetric signaling is of PGS. This expedition encompasses various forms of asymmetric signaling along with their vast applications in numerous fields. }

{
\section{Technical Framework}
This section explains the basic technical framework to understand the interplay of proper/improper and circular/non-circular RVs at length. It begins with the introduction of various stochastic data representations and distinguishes intermingled terms of propriety and circularity.  It further elaborates the appropriate transformations, operations, expressions and testing for the improper characterizations. }

\subsection{Stochastic Data Modeling}
 Various data presentation techniques are proposed in order to connect  data with its modeling domain. In the remaining section, we focus on complex and quaternion forms pertaining to their vast applications \cite{adali2011complex,mandic2007complex,olhede2003noise,walden2009testing,fortuna2001comparison,hanson498visualizing,miron2006quaternion}.

\subsubsection{Complex Random Vectors}
Complex analysis of data can be carried out based on three data representations known as complex, real composite and complex augmented representation of the complex RVs \cite{schreier2010statistical,adali2011complex}. Based on these representations, we will introduce the first-order and SOS characteristics of the RVs and their intuitive meanings.  

\paragraph{Complex Representation}
Consider an $N$-dimensional complex RV $ {\V{z}} = {\V{x}} + i {\V{y}}$, where ${\V{x}}, {\V{y}} \in {\mathbb{R}}^{N}$ and ${\V{z}}  \in {\mathbb{C}}^{N}$. The first order statistic given by the statistical mean) of ${\V{z}}$ is    
 \begin{equation} \label{eq1}
{\V{\mu}}_{\V{z}} = {\rm E}\{{\V{z}}\}= {\rm E}\{{\V{x}}\}+ i { \rm E}\{{\V{y}}\} = {\V{\mu}}_{\V{x}}+i{\V{\mu}}_{\V{y}},
  \end{equation}
  where, $\V{\mu}_{\V{x}},\V{\mu}_{\V{y}} \in {\mathbb{R}}^{N}$ and $\V{\mu}_{\V{z}} \in {\mathbb{C}}^{N}$. The conventional SOS characterization of ${\V{z}}$ is perceived by the covariance matrix, which is defined as
   \begin{align}  \label{eq2}
   {\M{R}_{\V{zz}}} & = {\rm E} \{(\V{z}-\V{\mu}_{\V{z}})(\V{z}-\V{\mu}_{\V{z}})^{\RM{H}}\} \nonumber \\ 
& = {\M{R}_{\V{xx}}} + {\M{R}_{\V{yy}}} + i ({\M{R}_{\V{xy}}^\RM{T}}-{\M{R}_{\V{xy}}}),
\end{align}   
where ${\M{R}_{\V{xx}}}$ and ${\M{R}_{\V{yy}}}$ are the auto covariance matrices of the real and imaginary components, respectively and ${\M{R}_{\V{xy}}}$ is the cross covariance matrix between them.  Furthermore, ${\M{R}_{\V{xx}}}, {\M{R}_{\V{yy}}},{\M{R}_{\V{xy}}} \in \mathbb{R}^{N{\rm x}N}$ and  ${{\M{R}}_{\V{zz}}}\in \mathbb{C}^{N{\rm x}N}$. The covariance matrices in \eqref{eq2} are found from
   \begin{equation}
   {\M{R}_{\V{uv}}} = \rm{E} \{(\V{u}-\V{\mu}_{\V{u}})(\V{v}-\V{\mu}_{\V{v}})^{\RM{T}} \},
   \end{equation}
  where ${\V{u}}$ and ${\V{v}} \in \{\V{x},\V{y}\}$. However, the complete SOS depiction involves another matrix,  ${\tilde{\M{R}}_{\V{zz}}}\in \mathbb{C}^{N{\rm x}N}$, named as  \textit{pseudo-covariance matrix} \cite{neeser1993proper,ollila2008circularity}.
  \begin{align} \label{eq3}
 {\tilde{\M{R}}_{\V{zz}}} & = {\rm E} \{(\V{z}-\V{\mu}_{\V{z}})(\V{z}-\V{\mu}_{\V{z}})^{\RM{T}}\}\nonumber \\ & =  {\M{R}_{\V{xx}}} - {\M{R}_{\V{yy}}} + i ({\M{R}_{\V{xy}}^\RM{T}}+{\M{R}_{\V{xy}}}).
 \end{align}
Complete SOS in the sense that ${\tilde{\M{R}}_{\V{zz}}}$ accounts for correlation and unequal power distribution of the quadrature components of a complex \ac{RV} in addition to covariance matrix which assumes equal power and uncorrelated real and imaginary components.
  For non-singular ${\M{R}_{\V{zz}}}$, the following three conditions are
necessary and sufficient for ${\M{R}_{\V{zz}}}$ and ${\tilde{\M{R}}_{\V{zz}}}$ to be the covariance and pseudo-covariance matrices of $\V{z}$ \cite{picinbono1996second}
  \begin{itemize}
 \item ${\M{R}_{\V{zz}}}$ is Hermitian and PSD.
 \item ${\tilde{\M{R}}_{\V{zz}}}$ is symmetric
 \item the Schur complement ${{\M{R}}_{\V{zz}}} - {\tilde{\M{R}}_{\V{zz}}}{{\M{R}}_{\V{zz}}^{ - *}}{{\tilde{\M{R}}_{\V{zz}}}^*}$ is PSD.
  \end{itemize}
SOS are sufficient for the widely known Gaussian RVs as they can completely characterize their distribution, characteristic
function, and higher order moments
 \cite{picinbono1996second}.
 \begin{definition}{1a}\label{def1}
 A multivariate complex Gaussian \ac{RV} $\V{z}$ can be fully characterized using complete SOS properties as $\mathcal{CN} \left( \V{\mu}_{{\V{z}}},{\M{R}_{\V{zz}}},{\tilde{\M{R}}_{\V{zz}}} \right)$ \cite{javed2018multiple}.
   \end{definition}
Consequently, a scalar Gaussian r.v. $z=x+iy$ can be fully described as $\mathcal{CN} \left( {\mu}_z,\sigma_z^2,\tilde{\sigma}_z^2 \right)$ \cite{ollila2008circularity,ollila2011complex}, where ${\mu}_{{{z}}} =  {\rm{E}} \{z\} $,  $\sigma_z^2 = { \rm{E}} \{ |z-{\mu}_{{{z}}}|^2\} $ and  $\tilde{\sigma}_z^2 =  {\rm{E}} \{\left(z-{\mu}_{{{z}}}\right)^2\} $  represent the statistical mean, variance,  and pseudo-variance of $z$, respectively. 
 \begin{definition}{2a}\label{def2}
Two complex RVs $\V{z}_1$ and $\V{z}_2\in\mathbb{C}^{N}$ are uncorrelated iff both  ${{\M{R}}_{\V{z}_1\V{z}_2}}$ and ${\tilde{\M{R}}_{\V{z}_1\V{z}_2}}$ matrices vanish, where cross covariance  ${{\M{R}}_{\V{z}_1\V{z}_2}}= {\rm E} \{(\V{z}_1-\V{\mu}_{\V{z}_1})(\V{z}_2-\V{\mu}_{\V{z}_2})^{\RM{H}}\} $ and cross pseudo-covariance  ${\tilde{\M{R}}_{\V{z}_1\V{z}_2}} = {\rm E} \{(\V{z}_1-\V{\mu}_{\V{z}_1})(\V{z}_2-\V{\mu}_{\V{z}_2})^{\RM{T}}\}$ \cite{neeser1993proper}.
\end{definition}

\paragraph{Real Composite Representation}
The complex \ac{RV} $\V{z}$ can be alternately represented as the real composite \ac{RV} ${\V{u}} = [{\V{x}}^{\RM {T}} \;  {\V{y}}^{\RM{T}}]^{\RM{T}} \in \mathbb{R}^{2N}$. 
The first- and second-order statistical characteristics of this  representation are described by $\V{\mu}_{\V{u}}\in \mathbb{R}^{2N}$ and $ {\M{R}_{\V{uu}}}\in \mathbb{R}^{2N{\rm x}2N}$, respectively~ \cite{schreier2003second,adali2011complex,picinbono1996second}.
\begin{equation}
\V{\mu}_{\V{u}} = {\rm E}\{\V{u}\} =  \left[{\begin{array}{*{20}{c}}
\V{\mu}_{\V{x}}\\
\V{\mu}_{\V{y}}\end{array}} \right], \; {\M{R}_{\V{uu}}} = \left[ {\begin{array}{*{20}{c}}
{\M{R}_{\V{xx}}}&{\M{R}_{\V{xy}}}\\
{\M{R}_{\V{xy}}^\RM{T}}&{\M{R}_{\V{yy}}}.
\end{array}} \right] 
\end{equation} 
\begin{definition}{1b}A complex Gaussian RV $\V{z}$ with alternate representation $\V{u}$ can be fully described as $\mathcal{N} \left(\V{\mu}_{\V{u}},{\M{R}_{\V{uu}}} \right)$
\cite{kariminezhad2016improper}.
\end{definition}
\begin{definition}{2b}
 Two complex RVs  $\V{z}_1$ and  $\V{z}_2$ with real representations  ${\V{u}_1} = [{\V{x}_1^{\RM T}} \;  {\V{y}_1^{\RM T}}]^{\RM{T}}$  and  ${\V{u}_2} = [{\V{x}_2^{\RM T}} \;  {\V{y}_2^{\RM T}}]^{\RM{T}}$  are uncorrelated iff all four cross covariance matrices ${\M{R}_{\V{x}_1\V{x}_2}}$, ${\M{R}_{\V{x}_1\V{y}_2}}$, ${\M{R}_{\V{y}_1\V{x}_2}}$ and ${\M{R}_{\V{y}_1\V{y}_2}}$ vanish \cite{neeser1993proper}.
\end{definition}

\paragraph{Augmented Representation}
The complex \ac{RV} $\V{z}$ is sometimes represented as an augmented complex vector $\underline{\V{z}} = [{\V{z}}^{\RM{T}} \;  {\V{z}}^{\RM {H}}]^{\RM {T}} \in \mathbb{C}^{2N{\rm x}2N}$ for convenience. The presented representations ${\V{u}}$ and $\underline{\V{z}}$ are interchangeable as $\underline{\V{z}} = \sqrt{2} \M{T} {\V{u}}$ and $ {\V{u}} = \frac{1}{\sqrt{2}}\M{T}^{\RM{H}}\underline{\V{z}}$. where $\M{T}$ is unitary transformation matrix defined as \cite{schreier2003second,picinbono1996second,adali2011complex}  
\begin{equation}
\M{T} = \frac{1}{\sqrt{2}} \left[ {\begin{array}{*{20}{c}}
\M{I}_{N}&{i\M{I}_{N}}\\
\M{I}_{N}&{ - i\M{I}_{N}}
\end{array}} \right].
\end{equation} 
Analogous to other representations, the complete first and second order characterization is given by $\V{\mu}_{\underline{\V{z}}}\in \mathbb{C}^{2N}$ and the augmented covariance matrix ${\M{R}_{\underline{\V{zz}}}} \in \mathbb{C}^{2N{\rm x}2N}$, respectively, which are expressed as \cite{picinbono1996second}
 \begin{equation}
\V{\mu}_{\underline{\V{z}}} = {\rm E}\{\underline{\V{z}}\} =  \left[{\begin{array}{*{20}{c}}
\V{\mu}_{\V{x}}+i\V{\mu}_{\V{y}}\\
\V{\mu}_{\V{x}}-i\V{\mu}_{\V{y}}\end{array}} \right], \; {\M{R}_{\underline{\V{zz}}}} = \left[ {\begin{array}{*{20}{c}}
\M{R}_{\V{zz}}&{\tilde{\M{R}}_{\V{zz}}}\\
{\tilde{\M{R}}_{\V{zz}}^\RM{*}}&{\M{R}_{\V{zz}}^*}
\end{array}} \right]. 
\end{equation}
The matrix  ${\M{R}_{\underline{\V{zz}}}}$, which was first used in \cite{van1995multivariate}, has the following features\cite{schreier2010statistical}
 \begin{itemize}
 \item Block pattern structure $\begin{psmallmatrix} \square	 & \blacksquare \\ \blacksquare^* & \square^*	\end{psmallmatrix}$
 \item Hermitian and PSD 
 \end{itemize}
 Owing to the special structure of ${\M{R}_{\underline{\V{zz}}}}$, the matrix factorization like eigen-vlaue decomposition (EVD),
singular value decomposition, or Cholesky factorization work differently to the regular matrices in the sense that all decomposed factors must follow the similar block pattern structure \cite{schreier2003second}. ${\M{R}_{\underline{\V{zz}}}}$ can be connected with  ${\M{R}_{{\V{uu}}}}$ as 
  \begin{equation}
   {\M{R}_{\underline{\V{zz}}}} = {\rm E}\{\underline{\V{z}} \underline{\V{z}}^{\RM H}\} = 2{\rm E}\{\M{T}{\V{u}}{\V{u}}^{\RM H}{\M{T}}^{\RM H}\}=2 \M{T} {\M{R}_{{\V{uu}}}} {\M{T}}^{\RM H}.
  \end{equation}
 \begin{definition}{1c} A complex Gaussian \ac{RV} $\V{z}$ can alternately be fully characterized as  $\mathcal{CN} \left( \V{\mu}_{\underline{\V{z}}}, {\M{R}_{\underline{\V{zz}}}} \right)$. 
 \end{definition}
 \begin{definition}{2c}
 Two complex RVs  $\V{z}_1$ and  $\V{z}_2$ with complex augmented representations  $\underline{\V{z}}_1$  and  $\underline{\V{z}}_2$  are uncorrelated iff  ${\M{R}}_{\underline{\V{z}}_1\underline{\V{z}}_2} = {\rm E} \{(\underline{\V{z}}_1-\V{\mu}_{\underline{\V{z}}_1})(\underline{\V{z}}_2-\V{\mu}_{\underline{\V{z}}_2})^{\RM{H}}\} $ vanishes.
 \end{definition}

\subsubsection{Quaternion Random Vectors}
The algebra of quaternions was invented by Sir W. R. Hamilton in 1844 \cite{hamilton1844theory} while multiplying triplets of real numbers. Fortunately, he failed and defined an elegant way  to multiply quadruplets of numbers by giving up one familiar feature of ordinary multiplication: commutativity. Quaternions are non-commutative extension of complex numbers to hyper-complex numbers \cite{zhang1997quaternions}. Analogous to the case of complex numbers, we present three different representations of the quaternion~RV.
\paragraph{Complex Representation}\label{secIIA2a}
Consider an $N$-dimensional quaternion RV $\V{q}\in \mathbb{H}^{N}$,  $\V{q}=\V{r}_1+ i \V{r}_i + j \V{r}_{j}+{k} \V{r}_{k}$ where $\V{r}_1,\V{r}_i,\V{r}_{j},\V{r}_{k} \in \mathbb{R}^{N}$ and the basis elements $i$,$j$ and $k$ satisfy $i^2=j^2=k^2=ijk=-1$, $ij=-ji=k$, $jk=-kj=i$ and $ki=-ik=j$ (Non-commutative multiplication). 
The quaternion conjugate is given by $\V{q}^* = \Re\left(\V{q} \right) -\Im \left( {\V{q}} \right)=  \V{r}_1 - i \V{r}_i - j \V{r}_{j}-{k} \V{r}_{k} $ and the involution of $\V{q}$ over a pure unit quaternion  $\alpha \in \{i,j,k \}$ is defined as $\V{q}^{(\alpha)} =  -{\alpha}\V{q}{\alpha}$  \cite{coxeter1946quaternions,via2010properness}. 
For instance, the involution $\V{q}^{(i)}= \V{r}_1 + i \V{r}_i - j \V{r}_{j}-{k} \V{r}_{k}$ inverts the sign of $\V{r}_i$ in $\V{q}^*$. Based on these descriptions, the complete SOS characterization of a zero-mean $\V{q}$ requires the covariance matrix $\M{R}_{\V{q}\V{q}} =  {\rm E}\{ {\V{q}} {\V{q}}^{\RM{H}} \}$ and three pseudo-covariance matrices $\M{R}_{\V{q}\V{q}^{(i)}},\M{R}_{\V{q}\V{q}^{(j)}}$ and $\M{R}_{\V{q}\V{q}^{(k)}}$. These pseudo-covariance matrices quantify the correlation between $\V{q}$ and its involutions $\V{q}^{(\alpha)}$ over three pure unit quaternions~\cite{ginzberg2011testing}
\begin{equation}
\M{R}_{\V{q}\V{q}^{(\alpha)}} =  {\rm E}\{ {\V{q}} {\V{q}}^{(\alpha)\RM{H}} \} = -{\rm E}\{ {\V{q}} \alpha {\V{q}}^{\RM{H}} \alpha \}.
\end{equation}
\begin{definition}{3a}
A zero-mean quaternion Gaussian RV $\V{q}$ is completely characterized using SOS of $\V{q}$ and its involutions ${\V{q}^{(i)}},  {\V{q}^{(j)}}$ and ${\V{q}^{(k)}}$ as $\mathcal{QN}\left( \M{R}_{\V{q}\V{q}},\M{R}_{\V{q}\V{q}^{(i)}},\M{R}_{\V{q}\V{q}^{(j)}},\M{R}_{\V{q}\V{q}^{(k)}} \right)$.
\end{definition}
\paragraph{Real Composite Representation}
The zero-mean quaternion \ac{RV} $\V{q}$ can be alternately represented as the quadrivariate real composite \ac{RV} ${\V{v}} \in \mathbb{R}^{4N}$, ${\V{v}}= [{\V{r}_1^{\RM T}} \;  {\V{r}_i^{\RM T}}\;  {\V{r}_j^{\RM T}}\;  {\V{r}_k^{\RM T}}]^{\RM{T}}$. Similar to complex case, the SOS properties of ${\V{v}}$ are given by the $\mathbb{R}^{{4N}{\rm x}{4N}}$ covariance matrix  \cite{ginzberg2011testing}
\begin{equation} \label{Quaternion-RC}
\M{R}_{\V{v}\V{v}} =\left[ {\begin{array}{*{20}{c}}
\M{R}_{\V{r}_1\V{r}_1} &\M{R}_{\V{r}_1\V{r}_i}&\M{R}_{\V{r}_1\V{r}_j}&\M{R}_{\V{r}_1\V{r}_k}\\
\M{R}_{\V{r}_i\V{r}_1}&\M{R}_{\V{r}_i\V{r}_i}&\M{R}_{\V{r}_i\V{r}_j}&\M{R}_{\V{r}_i\V{r}_k}\\
\M{R}_{\V{r}_j\V{r}_1}&\M{R}_{\V{r}_j\V{r}_i}&\M{R}_{\V{r}_j\V{r}_j}&\M{R}_{\V{r}_j\V{r}_k}\\
\M{R}_{\V{r}_k\V{r}_1}&\M{R}_{\V{r}_k\V{r}_i}&\M{R}_{\V{r}_k\V{r}_j}&\M{R}_{\V{r}_k\V{r}_k}
\end{array}} \right],
\end{equation}
where, $\M{R}_{\V{r}_{\gamma}\V{r}_{\zeta}} = {\rm E}\{ {\V{r}_{\gamma}} {\V{r}_{\zeta}^{\RM{T}}} \} $ and $\M{R}_{\V{r}_{\gamma}\V{r}_{\zeta}} = \M{R}_{\V{r}_{\zeta}\V{r}_{\gamma}}^{\RM T} $ with $\gamma, \zeta \in \{1,i,j,k \}$.
 \begin{definition}{3b} A zero-mean quaternion Gaussian RV $\V{q}$ with quadrivariate real composite representation $\V{v}$ is completely characterized using the symmetric correlation matrix $\M{R}_{\V{v}\V{v}}$.
\end{definition}
\paragraph{Augmented Representation}
Let $\underline{\V{q}}$ denotes the  augmented representation of $\V{q}$ and its involutions as
\begin{equation} \label{eq7}
\underline{\V{q}}=\left[ {\begin{array}{*{20}{c}}
{\V{q}}\\
{\V{q}^{(i)}}\\
{\V{q}^{(j)}}\\
{\V{q}^{(k)}}
\end{array}} \right] = \left[ {\begin{array}{*{20}{c}}
\V{r}_1+ i \V{r}_i + j \V{r}_{j}+{k} \V{r}_{k}\\
\V{r}_1+ i \V{r}_i - j \V{r}_{j}-{k} \V{r}_{k}\\
\V{r}_1- i \V{r}_i + j \V{r}_{j}-{k} \V{r}_{k}\\
\V{r}_1- i \V{r}_i - j \V{r}_{j}+{k} \V{r}_{k}
\end{array}} \right].
\end{equation}
Analogous to complex case, the alternate representations of $\V{q}$ are interchangeable using $\underline{\V{q}} = 2\M{A}_N {\V{v}}$ and $ {\V{v}} = \frac{1}{2}\M{A}_N^{\RM{H}}\underline{\V{q}}$, where $\M{A}_N$ is a unitary transformation matrix defined as  \cite{took2011augmented}
\begin{equation}
\M{A}_{N} =  \frac{1}{2}\left[ {\begin{array}{*{20}{c}}
{\M{I}_{N}}&i{\M{I}_{N}}&j{\M{I}_{N}}&k{{\M{I}_{N}}}\\
{\M{I}_{N}}&i{\M{I}_{N}}&-j{\M{I}_{N}}&-k{{\M{I}_{N}}}\\
{\M{I}_{N}}&-i{\M{I}_{N}}&j{\M{I}_{N}}&-k{{\M{I}_{N}}}\\
{\M{I}_{N}}&-i{\M{I}_{N}}&-j{\M{I}_{N}}&k{{\M{I}_{N}}}
\end{array}} \right].
\end{equation}
The SOS characterization of $\underline{\V{q}}$ is given by the following augmented covariance matrix \cite{via2011generalized}
 \begin{equation}
 {\M{R}_{\underline{\V{q}}\underline{\V{q}}}} = \left[ {\begin{array}{*{20}{c}}
\M{R}_{\V{q}\V{q}} &\M{R}_{\V{q}\V{q}^{(i)}}&\M{R}_{\V{q}\V{q}^{(j)}}&\M{R}_{\V{q}\V{q}^{(k)}}\\
\M{R}_{\V{q}\V{q}^{(i)}}^{(i)} &\M{R}_{\V{q}\V{q}}^{(i)}&\M{R}_{\V{q}\V{q}^{(k)}}^{(i)}&\M{R}_{\V{q}\V{q}^{(j)}}^{(i)}\\
\M{R}_{\V{q}\V{q}^{(j)}}^{(j)}&\M{R}_{\V{q}\V{q}^{(k)}}^{(j)}&\M{R}_{\V{q}\V{q}}^{(j)}&\M{R}_{\V{q}\V{q}^{(i)}}^{(j)}\\
\M{R}_{\V{q}\V{q}^{(k)}}^{(k)}&\M{R}_{\V{q}\V{q}^{(j)}}^{(k)}&\M{R}_{\V{q}\V{q}^{(i)}}^{(k)}&\M{R}_{\V{q}\V{q}}^{(k)}
\end{array}} \right],
 \end{equation}
where $\M{R}_{\V{q}\V{q}^{(\alpha)}}^{(\beta)} = -\beta \M{R}_{\V{q}\V{q}^{(\alpha)}} \beta$  with $\beta \in \{i,j,k \}$. The correlation matrix $\M{R}_{\V{q}\V{q}^{(\alpha)}}^{(\alpha)} = \M{R}_{\V{q}\V{q}^{(\alpha)}}^{(\alpha){\RM H}}$ is $i$-Hermitian, $j$-Hermitian and $k$-Hermitian for $\alpha = i, j \, \text{and}\, k$, respectively. Thus, proving a non-trivial extension of augmented complex statistics
to its quaternion counterpart \cite{took2011augmented}. ${\M{R}_{\underline{\V{qq}}}}$ is linked with ${\M{R}_{{\V{vv}}}}$ as 
  \begin{equation}
   {\M{R}_{\underline{\V{qq}}}} \!\!=\! {\rm E}\{\underline{\V{q}} \underline{\V{q}}^{\RM H}\}\!\! = \! 4 {\rm E}\{\M{A}_N{\V{v}}{\V{v}}^{\RM H}{\M{A}_N^{\RM H}}\}\!\!=\!  4\M{A}_N {\M{R}_{{\V{vv}}}} {\M{A}_N^{\RM H}}.
  \end{equation}
\begin{definition}{3c} A zero-mean quaternion Gaussian RV $\V{q}$ with augmented representation $\underline{\V{q}}$ is completely characterized using the augmented correlation matrix $ {\M{R}_{\underline{\V{qq}}}}$.
\end{definition}

The uncorrelation between two quaternions $\V{q}_1$ and $\V{q}_2$ with augmented representations $\underline{\V{q}}_1$ and $\underline{\V{q}}_2$, respectively, require the regular cross covariance $\M{R}_{\V{q}_1\V{q}_2} = {\rm E} \{ \V{q}_1{\V{q}_2^{\RM H}}\}$ and three involutional cross covariance matrices
 $\M{R}_{\V{q}_1\V{q}_2}^{(\alpha)} = -\alpha  \M{R}_{\V{q}_1\V{q}_2} \alpha$ along with all possible combinations of cross pseudo-covariance matrices ${\rm E}\{ {\V{q}_1^{(\gamma)}} {\V{q}_2}^{(\zeta)\RM{H}} \}$ contained in ${\M{R}_{\underline{\V{q}}_1\underline{\V{q}}_2}}= {\rm E}\{{\underline{\V{q}}_1{\underline{\V{q}}_2^{\RM H}}}\}$ to be zero.
\subsubsection{Summary and Insights}
{In a nutshell, this subsection covers three well-known data presentation techniques i.e., complex, real composite, and complex augmented representation for both complex and quaternion RVs. 
However, why do we need multiple equivalent representations of the same phenomenon? To address this concern, consider the tedious analysis of real composite representation which can be significantly simplified using complex representation}  \cite{picinbono1996second}. {Besides this intuitive reasoning, there are other limitations that will be discussed in next subsection. Another curiosity, that arises, is why do we need redundant complex augmented representation of the RVs. The response to this query is three-folds. 1) The uncorrelation of real $\V{x}_1$ and $\V{x}_2$ is completely defined as ${\M{R}_{\V{x}_1\V{x}_2}}=\M{0}$. However, the uncorrelation of the complex $\V{z}_1$ and $\V{z}_2$ is not completely defined as ${\M{R}_{\V{z}_1\V{z}_2}}=\M{0}$. Thus, we require augmented  $\underline{\V{z}}_1$ and $\underline{\V{z}}_2$ to completely characterize the uncorrelation as ${\M{R}}_{\underline{\V{z}}_1\underline{\V{z}}_2}= \M{0}$} \cite{schreier2010statistical}.
{2) Pertaining to the block structure of the ${\M{R}_{\underline{\V{zz}}}}$, interestingly it can be invertible even when $\M{R}_{\V{zz}} $ and $\tilde{\M{R}}_{\V{zz}}$ are not} \cite{schreier2003second}. 
{3) It is a powerful tool for WL transformations (refer to Section \ref{ssec:transformations})} \cite{adali2011complex}.
{On the other hand, vector representations i.e., real composite and augmented formulation allow easier geometrical
interpretations in high-dimensional space relative to the complex representation} \cite{amblard2004properness}.
In conclusion, we prefer complex representation for comprehensive analysis, real composite representation for easy geometrical interpretations and augmented formulation for complete modeling, characterization, and operations (e.g., inversion and transformation) \cite{took2011augmented}.
 \subsection{Propriety versus Circularity}
Properness evaluation i.e., the identification of correlation between the complex valued RVs  and their complex conjugates is a popular subject in signal processing \cite{picinbono1994circularity,schreier2003second,ollila2008circularity}  and information theory \cite{neeser1993proper}. It has also been extended to quaternion valued r.v. \cite{amblard2004properness} and vectors \cite{via2010properness,took2011augmented} for various applications. This subsection illustrates extensive definitions of impropriety and the measures for degree of impropriety~(DoI) for complex as well as quaternion RVs based on different representations.
 \subsubsection{Complex Random Vectors}
  The full statistical characterization of complex RVs involves the analysis of the moments and probability distributions (if exist). A complex RV is designated as proper/improper and circular/non-circular based on these characteristics.
  \paragraph{Propriety}
 Based on the real and complex representation of complex RVs, a \textit{proper complex} RV is described using the following definition 
  \begin{definition}{4}[Proper RV] A proper complex RV is composed of real and imaginary vectors with identical auto covariance matrices ${\M{R}_{\V{xx}}} = {\M{R}_{\V{yy}}}$ and skew-symmetric cross covariance matrix ${\M{R}_{\V{xy}}} = - {\M{R}_{\V{xy}}}^\RM{T}$. Alternately, a proper complex \ac{RV} $\V{z}$ renders zero pseudo-covariance matrix ${\tilde{\M{R}}_{\V{zz}}}$ \cite{neeser1993proper}. Properness may also be seen as the uncorrelation of $z$ with its complex conjugate $z^*$\cite{picinbono1997second,trampitsch2013complex}. These equivalent propriety definitions are sometimes referred as \textit{Strict Propriety} \cite{schreier2003second}. 
 \end{definition}
 This implies that a proper RV $\V{z}$ has ${\M{R}_{\V{xy}}}$ with zero main diagonal elements rendering uncorrelated real and imaginary components of each element $z_n$ in $\V{z}$. However, the  off-diagonal elements can be non-zero yielding correlated $\Re\{z_{k}\}$ and $\Im\{z_{l}\}$ for $k \neq l$ \cite{neeser1993proper}. For a zero-mean scalar complex r.v. $z=x+iy$, $\sigma_{xy} ={\rm{E}} \{xy\}=0$ is necessary for propriety. 
 
 
   With slight abuse  of terminology, we term $\underline{\V{z}}$ proper if $\V{z}$ is proper where proper $\V{z}$ implies vanishing pseudo-variance  $\tilde{\M{R}}_\V{zz}$. However, proper  $\underline{\V{z}}$ does not imply vanishing $\tilde{\M{R}}_{\underline{\V{zz}}}$ rather it 
 implies a block diagonal structure of  $ {\M{R}_{\underline{\V{zz}}}} = \textbf{Block-Diag}\left(\M{R}_{\V{zz}}, {\M{R}_{\V{zz}}^*}\right)$. This demonstrates the equivalence of the propriety statements for $\V{z}$ and $\underline{\V{z}}$ yet different implications. It is evident that for proper $\V{z}$, all eigenvalues of $ {\M{R}_{\underline{\V{zz}}}}$ are real (same as that of Hermitian $\M{R}_{\V{zz}}$) and have even multiplicity (paired because of diagonal blocks $\M{R}_{\V{zz}}$ and  $\M{R}_{\V{zz}}^*$). Thus, strict propriety implies even multiplicity of the eigenvalues, however, it is not a 
sufficient condition. Therefore, Schreier \textit{et al.}  established the following generalization based on the augmented representation.
 \begin{definition}{5}[Generalized Propriety] A complex RV is
termed as generalized proper if all eigenvalues of $\M{R}_{\underline{\V{z}\V{z}}}$ have even multiplicity \cite{schreier2003second}.
 \end{definition}
The EVD of ${\M{R}_{\underline{\V{z}}\underline{\V{z}}}}$ takes on the form
${\M{R}_{\underline{\V{z}}\underline{\V{z}}}} = \M{V} \left( \M{T}\M{\Lambda} \M{T}^{\RM H} \right) \M{V}^{\RM H}$ with diagonal $ \M{T}\M{\Lambda} \M{T}^{\RM H}$ iff $\V{z}$ is generalized proper, otherwise block matrix 
$ \M{T}\M{\Lambda} \M{T}^{\RM H}$ with diagonal blocks \cite{schreier2003second}.
 Moreover, the complex RVs $\V{z}_1$ and $\V{z}_2$ are \textit{cross proper} iff the cross pseudo-covariance matrix ${\tilde{\M{R}}_{\V{{z_1}z_2}}}$ vanishes. 
  Notably, they are \textit{jointly proper} if the composite RV having $\V{z}_1$ and $\V{z}_2$ as subvectors is proper \cite{neeser1993proper} or if they are proper and cross proper \cite{via2010properness}.
 \begin{definition}{6}[Improper RV]\label{Def:Improper}
 A complex RV $\V{z}\!\!=\!\!\V{x}\!\!+\!\!i\V{y}$ is called improper if any of the following statements holds~\cite{schreier2008bounds}
\begin{itemize}
\item Non identically distributed $\V{x}$ and $\V{y}$ i.e.,  ${\M{R}_{\V{xx}}} \neq {\M{R}_{\V{yy}}}$
\item Correlated $\V{x}$ and $\V{y}$ i.e., ${\M{R}_{\V{xy}}} \neq - {\M{R}_{\V{xy}}}^\RM{T}$
\item Non-zero $\tilde{\M{R}}_{{\V{z}\V{z}}}$
\item Correlated $\V{z}$ and $\V{z}^*$
\item Lack of block diagonal structure in ${\M{R}}_{\underline{\V{z}\V{z}}}$
\item Lack of even multiplicity of eigenvalues in ${\M{R}}_{\underline{\V{z}\V{z}}}$ 
\end{itemize}
\end{definition} 
 Absence of properness of a RV is termed as \textit{improperness} and the extent of improperness is argued by the eigenvalue spread of $\M{R}_{\underline{\V{z}\V{z}}}$  \cite{schreier2008bounds}. For instance, a RV $\V{z}_1$ is \textit{less improper} than another RV $\V{z}_2$ if the eigenvalues of $\M{R}_{\underline{\V{z}}_1 \underline{\V{z}}_1}$ are majorized by (less spread out) those of $\M{R}_{\underline{\V{z}}_2  \underline{\V{z}}_2}$ \cite{schreier2005detection}.
 
   \begin{definition}{7}
[Maximally Improper RV] For a given  $\M{R}_{{\V{z}\V{z}}}$, the vector whose  $\M{R}_{\underline{\V{z}\V{z}}}$ has least eigenvalue spread must be proper $\tilde{\M{R}}_{\V{zz}} = \M{0}$ whereas the vector whose $\M{R}_{\underline{\V{z}\V{z}}}$ has maximum possible eigenvalue spread must be maximally improper \cite{schreier2010statistical}.
 \end{definition}
   \paragraph{Circularity} A stronger version of propriety considering the probability distribution of a RV is named \textit{Circularity} and is defined as
 \newline
 \begin{definition}{8}[Circular RV] A complex RV $\V{z}$ is said to be circular (having CSC distribution
about the origin), iff its distribution remains invariant under
multiplication by any (complex) number on the unit complex
circle i.e., 
$\V{z}$ and ${\V{\hat{z}}} = {\V{z}}{e^{j\alpha}}$ have the same distribution for any real $\alpha$ \cite{picinbono1997second,adali2011complex}.
\end{definition}
 For a scalar $z$, is equivalent to spherical symmetry of the corresponding 
 real composite vector ${{u}} = [x \;y]^{\RM{T}}$ \cite{fang1990symmetric}. On the other hand, the magnitude of circularity of a complex RV $\V{z}$ is further classified based on the behavior of the underlying random variables~\cite{picinbono1994circularity,adali2010complex}:
   \begin{figure}[t]
\begin{minipage}[b]{1.0\linewidth}
  \centering
  \centerline{\includegraphics[width=5.5cm]{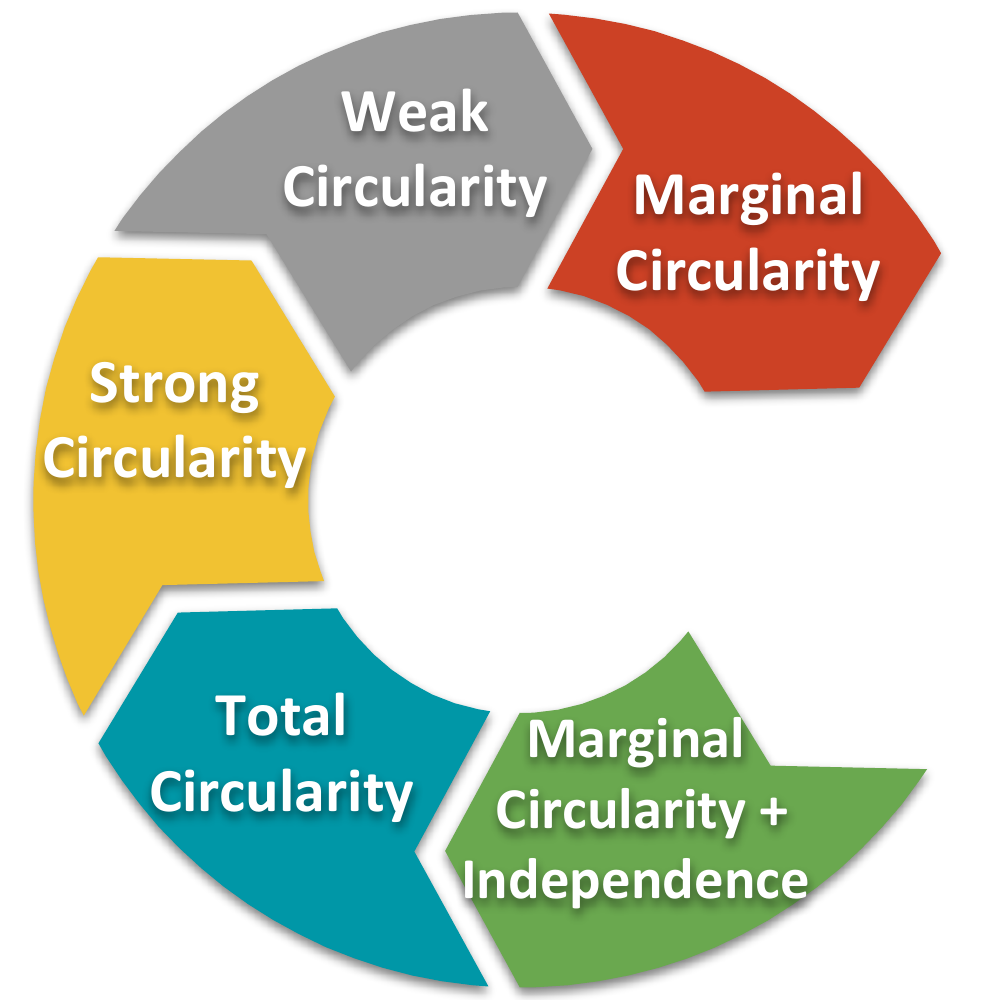}}
\end{minipage}
\caption{Extents of Circularity and their Implications}
\label{fig:CI}
\end{figure}
\begin{itemize}
\item \textit{Marginal Circular:} If its components are complex circular r.v.
\item  \textit{Weakly Circular:} If $\V{z}$ and ${e^{j\alpha}}{\V{z}}$ have the same probability distribution for any real $\alpha$
\item \textit{Strongly Circular:} If $\V{z}$ and ${e^{j\V{a}}} \odot{\V{z}}$ have the same probability distribution for any real vector~$\V{a}$
\item  \textit{Total Circular:} If its components are independent and circular.
The circularity characteristics are related to each other as shown in Fig. \ref{fig:CI}.
\end{itemize}

\paragraph{Relation between Propriety and Circularity}
Often the terms proper and circular are used interchangeably. However, as a matter of fact, they are quite related yet distinct phenomenon.
\begin{corollary} 
{Propriety and circularity are related as:}\\
 Circularity $ \mathbin{\lower.5ex\hbox{$\buildrel\textstyle\Longrightarrow\over
{\smash{\notimplies}\vphantom{_{\vbox to 2.5ex{\vss}}}}$}}$ Propriety, Impropriety $\mathbin{\lower.5ex\hbox{$\buildrel\textstyle\Longrightarrow\over
{\smash{\notimplies}\vphantom{_{\vbox to 2.5ex{\vss}}}}$}}$ Non-circularity
\end{corollary} 
\begin{itemize}
\item For a zero-mean circular RV $\V{z}$, the respective pseudo-covariance matrices ${\tilde{\M{R}}_{\V{zz}}}$ and ${\tilde{\M{R}}_{\V{\hat{z}\hat{z}}}}$ are related as:
 \begin{equation}\label{eq8}
{\tilde{\M{R}}_{\V{\hat{z}\hat{z}}}} = {\rm E}\left[ {\V{\hat{z}}} {\V{\hat{z}}}^{\RM{T}} \right] = {e^{j2\alpha }}{\tilde{\M{R}}_{\V{zz}}}\quad \forall \alpha.
 \end{equation}
Thus, \eqref{eq8} implies ${\tilde{\M{R}}_{\V{zz}}}$ should be a zero-matrix for any given $\alpha$ to satisfy circularly symmetry. Thus, circularity implies properness whereas the converse is not true in general. 
\item Propriety requires the second-order moments to be rotationally invariant. However, circularity requires that the \ac{PDF} and thus all existing moments to be rotationally invariant. Thus, circularity implies zero mean and propriety, but not vice versa \cite{adali2011complex}.
\item In the light of the aforementioned arguments, an improper RV with rotationally variant second moment is essentially NC. However, a NC RV with non-zero mean but zero pseudo-covariance is still proper, proving the fact that non-circularity does not imply improperness in general.
\end{itemize}


\begin{figure}
        \centering
        \begin{subfigure}[b]{0.22\textwidth}
            \centering 
            \includegraphics[width=\textwidth]{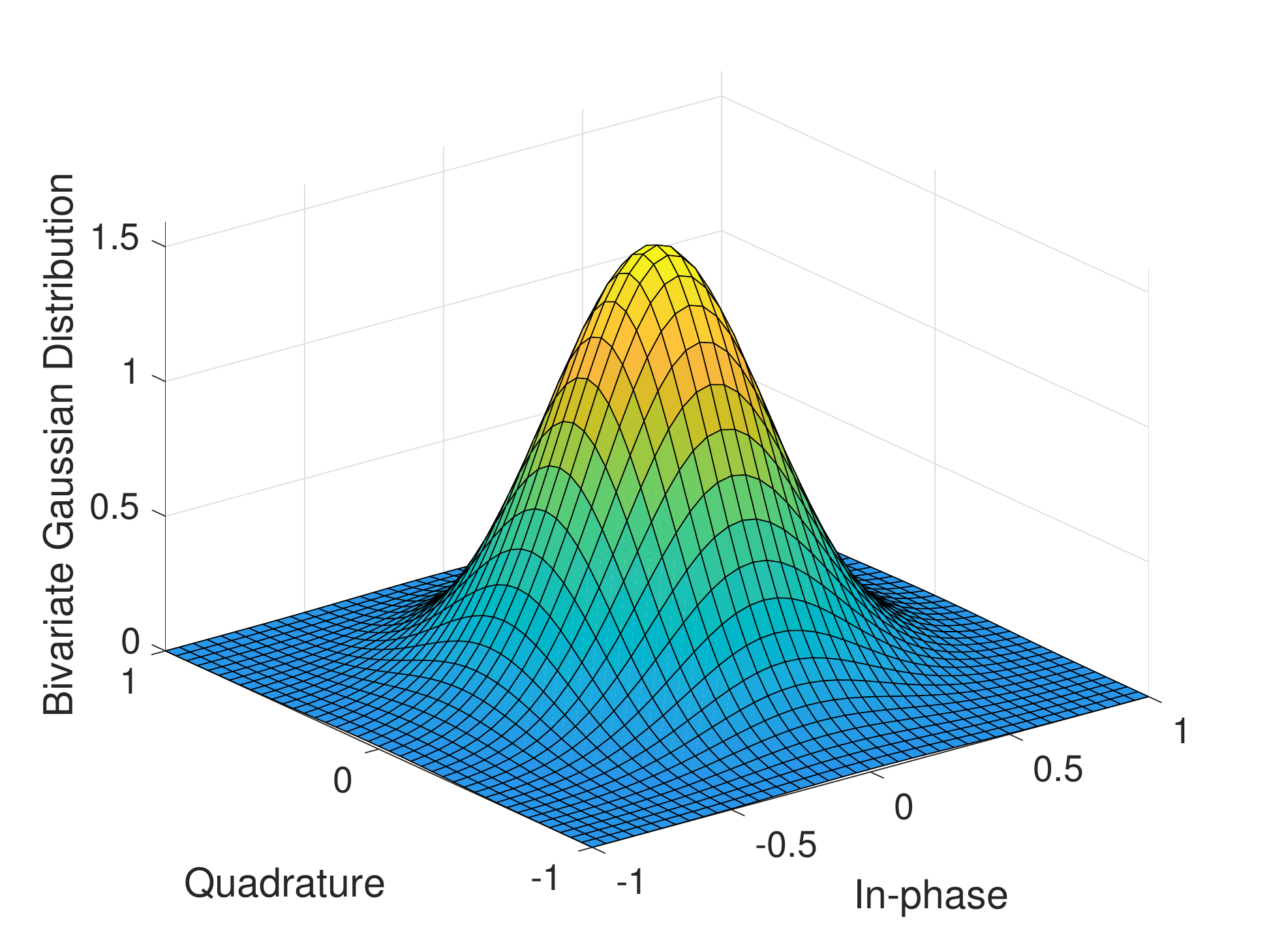}
   \label{fig:3a}
          \end{subfigure}
          \quad
        \begin{subfigure}[b]{0.22\textwidth}  
            \centering 
            \includegraphics[width=\textwidth]{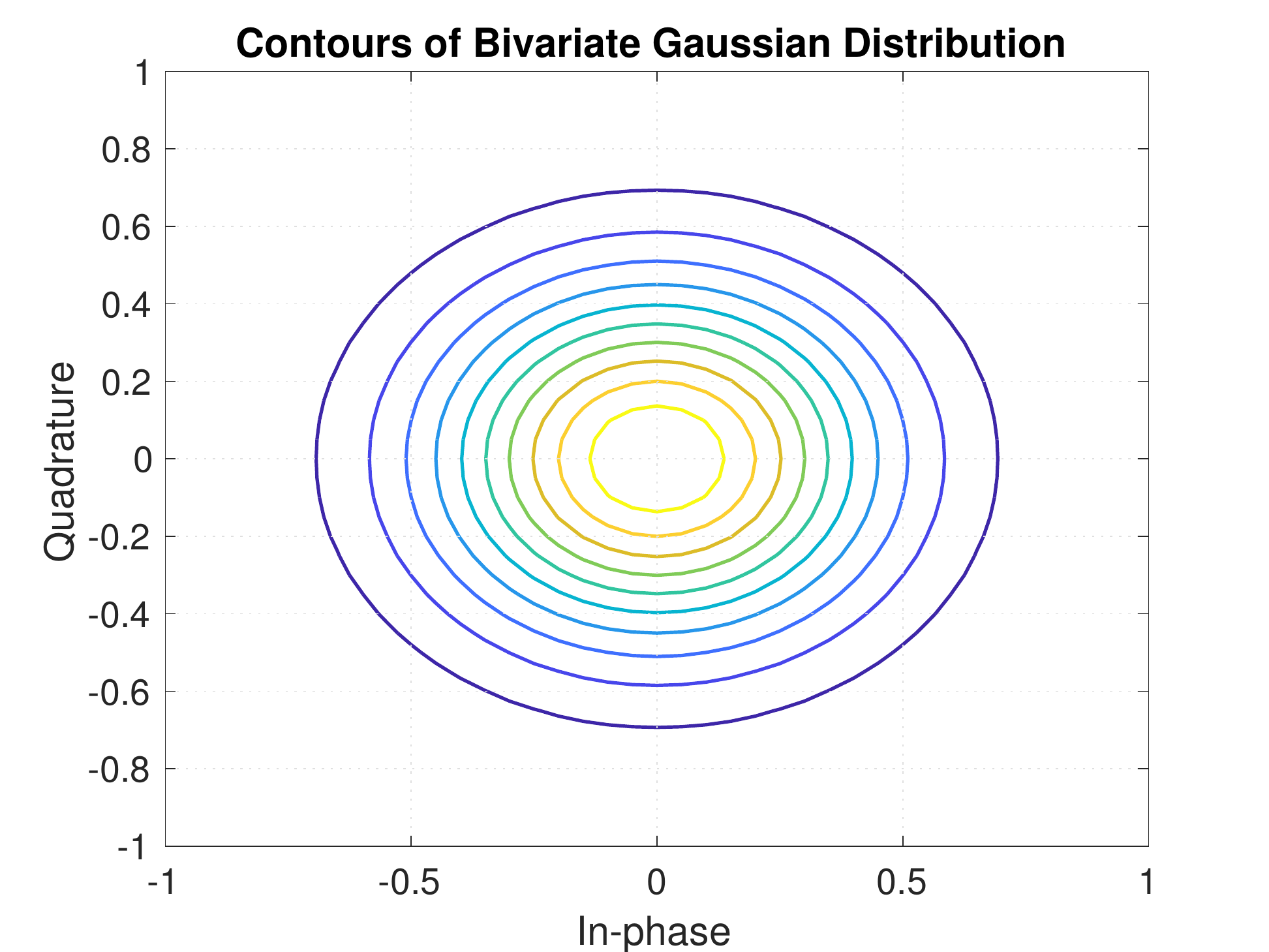}            
                     \end{subfigure}
      \quad   
        \begin{subfigure}[b]{0.22\textwidth}   
            \centering 
            \includegraphics[width=\textwidth]{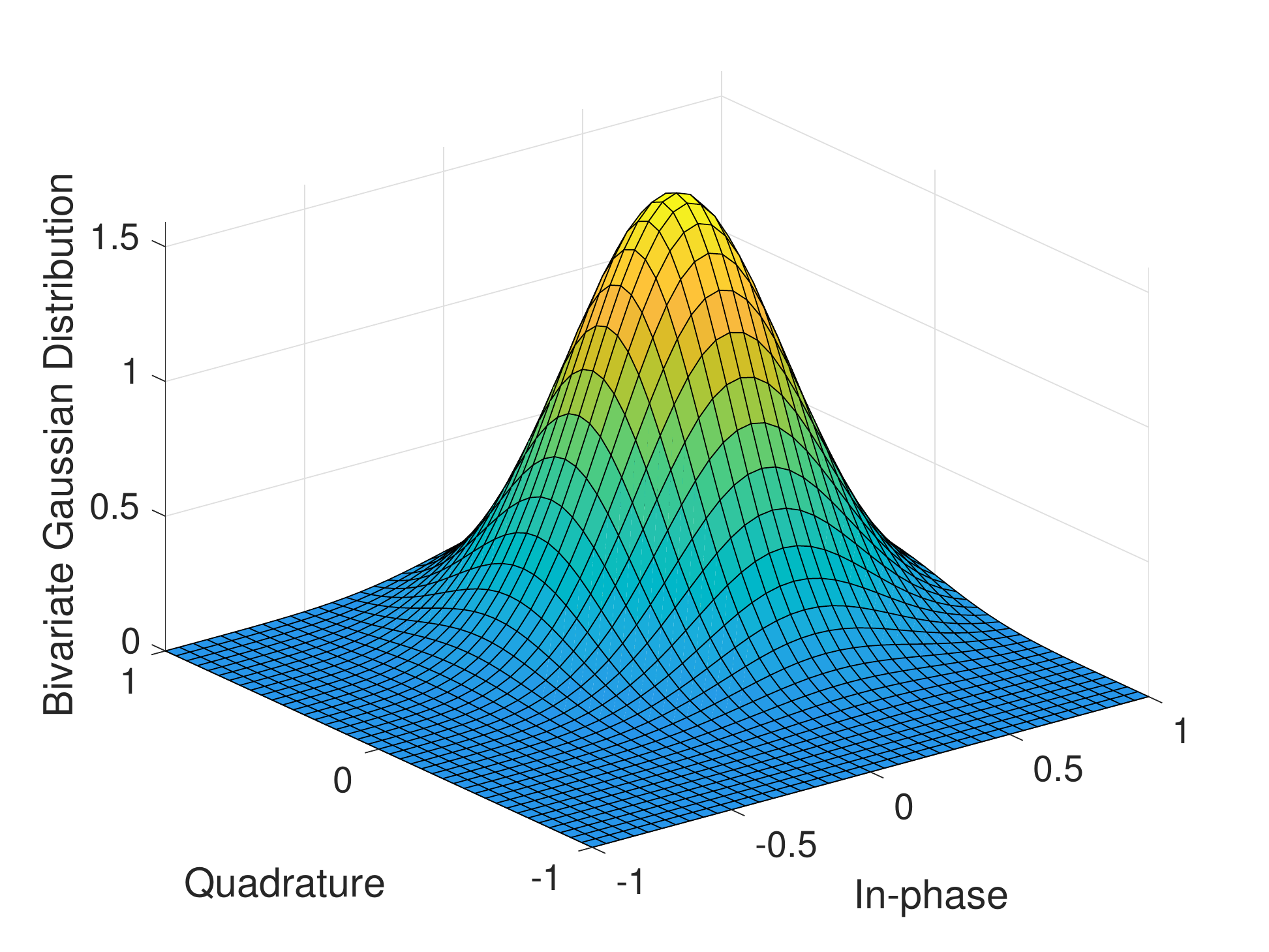}  
            \label{fig:3b}
        \end{subfigure}
        \quad
        \begin{subfigure}[b]{0.22\textwidth}   
            \centering 
            \includegraphics[width=\textwidth]{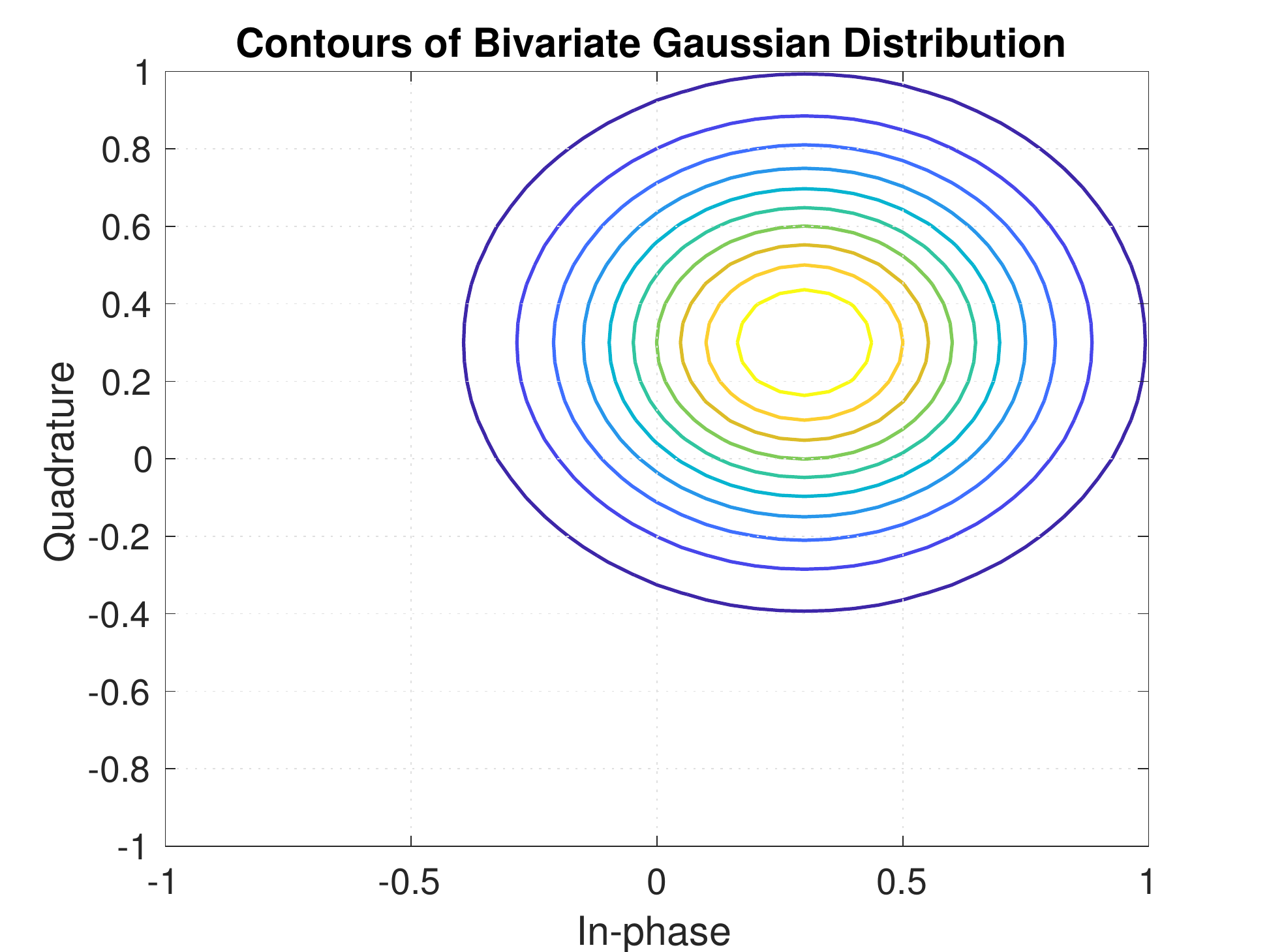}  
        \end{subfigure}
          {{\small  (a) Proper and Circular Gaussian RV (Centered)}} 
          \quad \quad
         {{\small (b) Proper and NC Gaussian RV (Non-Centered)}} 
             \vskip\baselineskip
       \begin{subfigure}[b]{0.22\textwidth}     \label{fig:3c}
            \centering 
            \includegraphics[width=\textwidth]{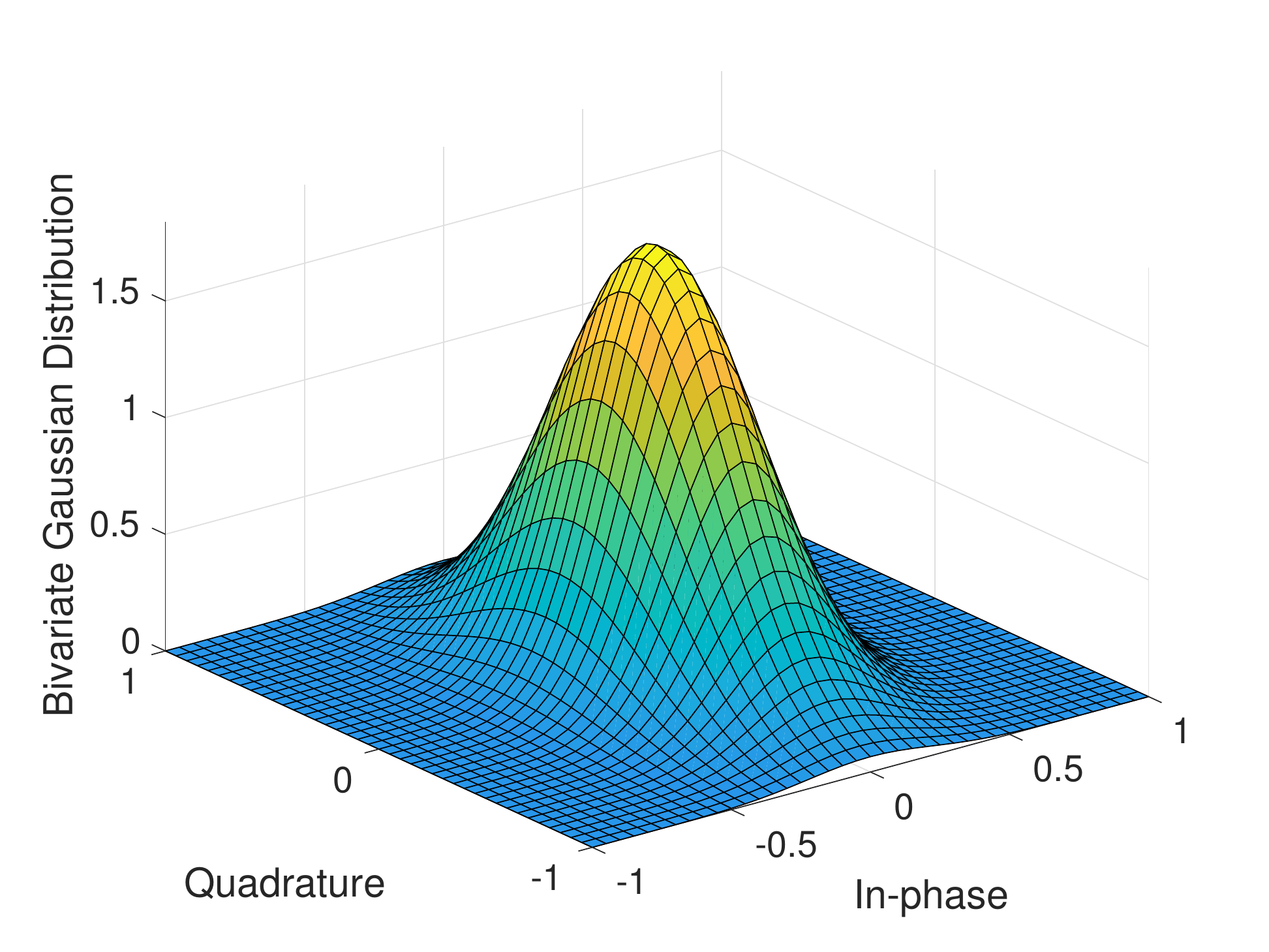}  
        \end{subfigure}
        \quad
        \begin{subfigure}[b]{0.22\textwidth}   
            \centering 
            \includegraphics[width=\textwidth]{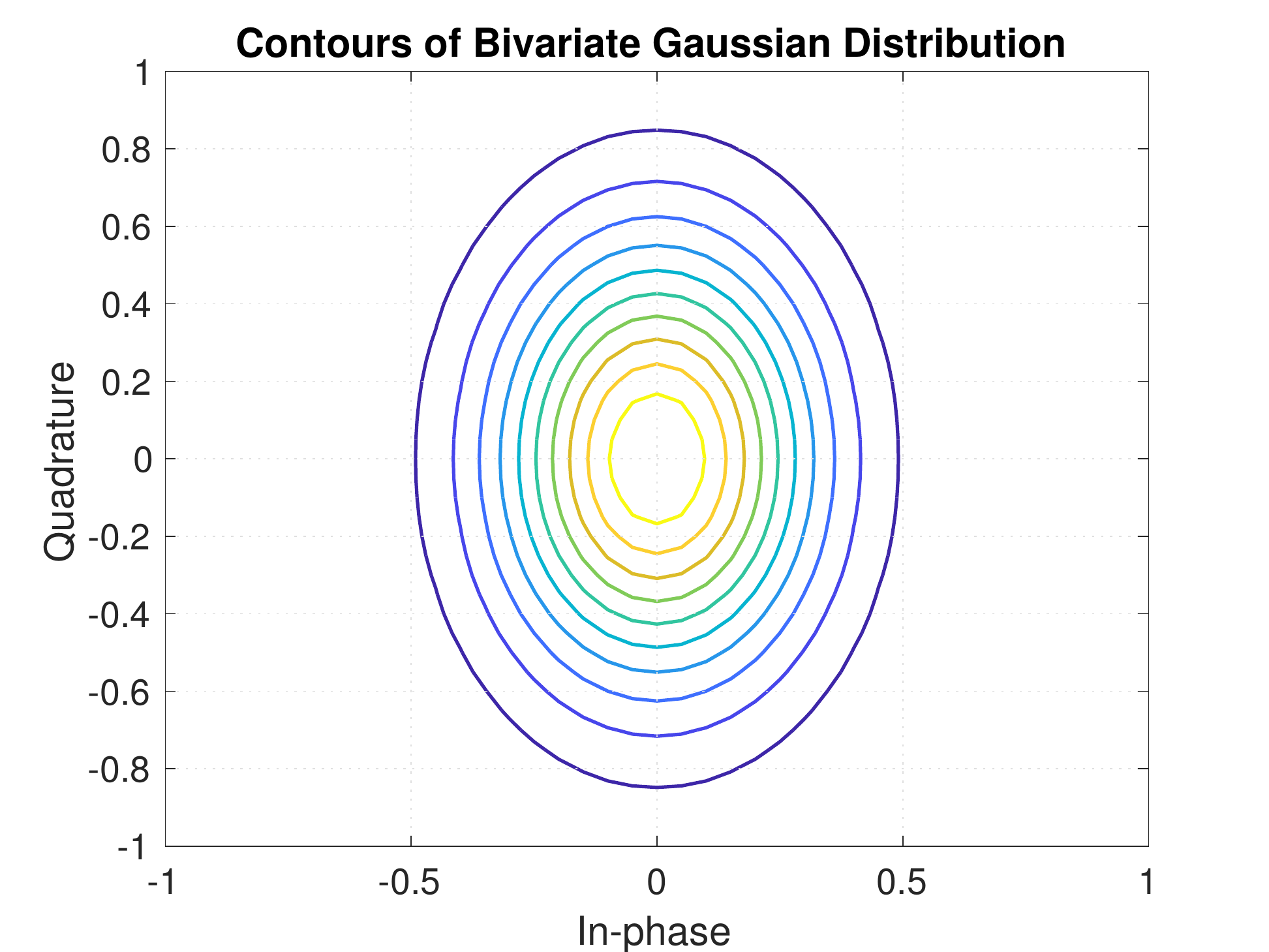}  
        \end{subfigure}
             \quad      
              \begin{subfigure}[b]{0.22\textwidth}     \label{fig:3d}
            \centering 
            \includegraphics[width=\textwidth]{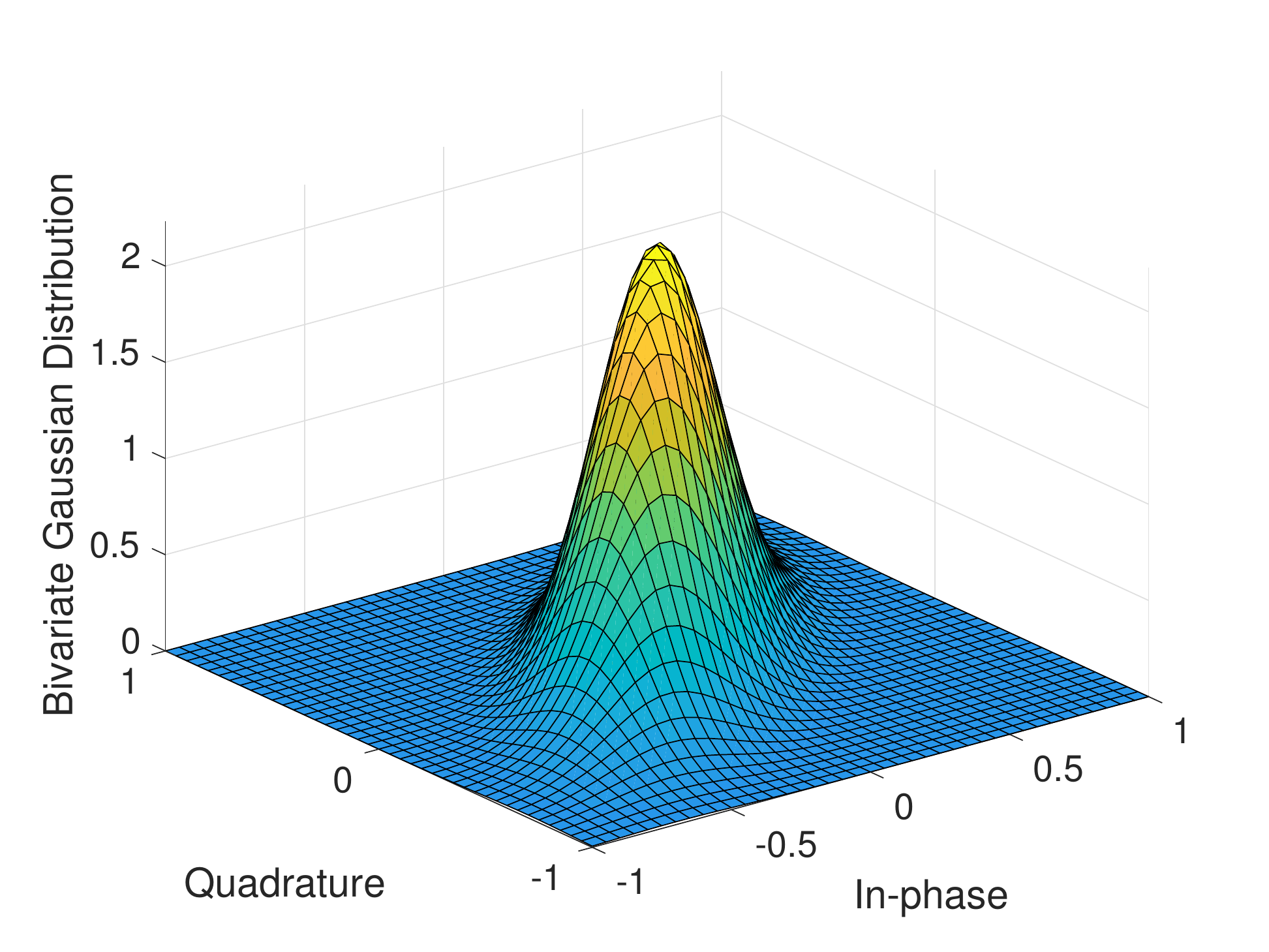}  
        \end{subfigure}
        \quad
        \begin{subfigure}[b]{0.22\textwidth}   
            \centering 
            \includegraphics[width=\textwidth]{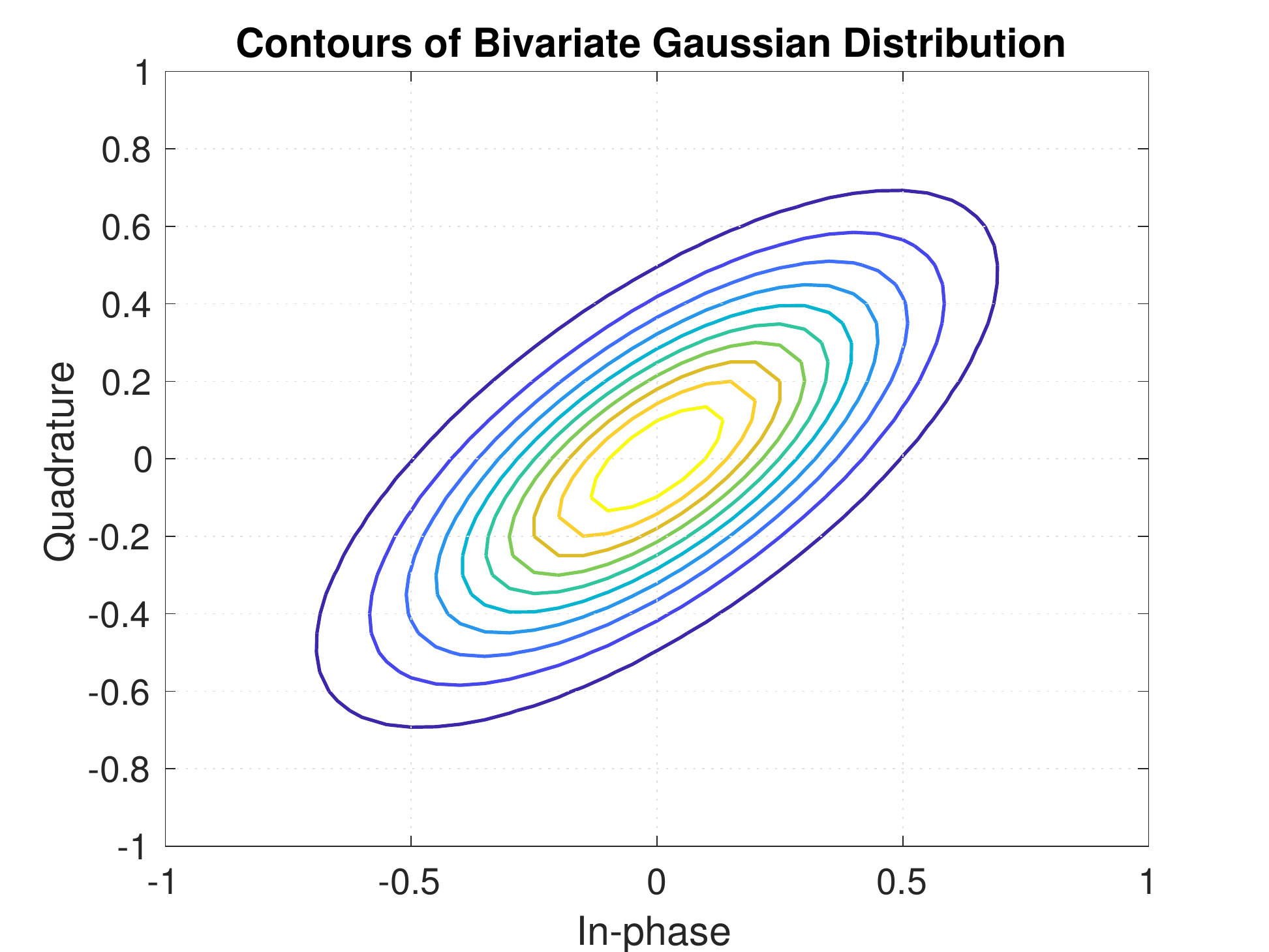}  
     \end{subfigure}
        {{\small  (c) Improper and NC Gaussian RV  (Unequal Power)}} \quad \quad 
       {{\small  (d) Improper and NC Gaussian RV (Correlated) }} 
        \caption[ ]
        {\small The Interplay between Propriety and Circularity} 
        \label{fig:PvC}
    \end{figure}

\textbf{Example.} Propriety and circularity are equivalent for a zero-mean Gaussian r.v. having uncorrelated real and imaginary components with equal power distribution as depicted in Fig. \ref{fig:PvC}a, whereas the same distribution with non-zero mean is proper but not circular in Fig. \ref{fig:PvC}b. Furthermore, the deviation from equal power distribution and uncorrelation result in improper as well as NC complex r.v. as shown in Fig. \ref{fig:PvC}c and Fig. \ref{fig:PvC}d, respectively. The quadrature component has more power/variance than the in-phase component in Fig. \ref{fig:PvC}c and  non-perpendicular distribution contours to x- or y-axis depict the correlation between I/Q components in Fig. \ref{fig:PvC}d.

\begin{figure}[t]
\begin{minipage}[b]{1.0\linewidth}
  \centering
  \centerline{\includegraphics[width=6cm]{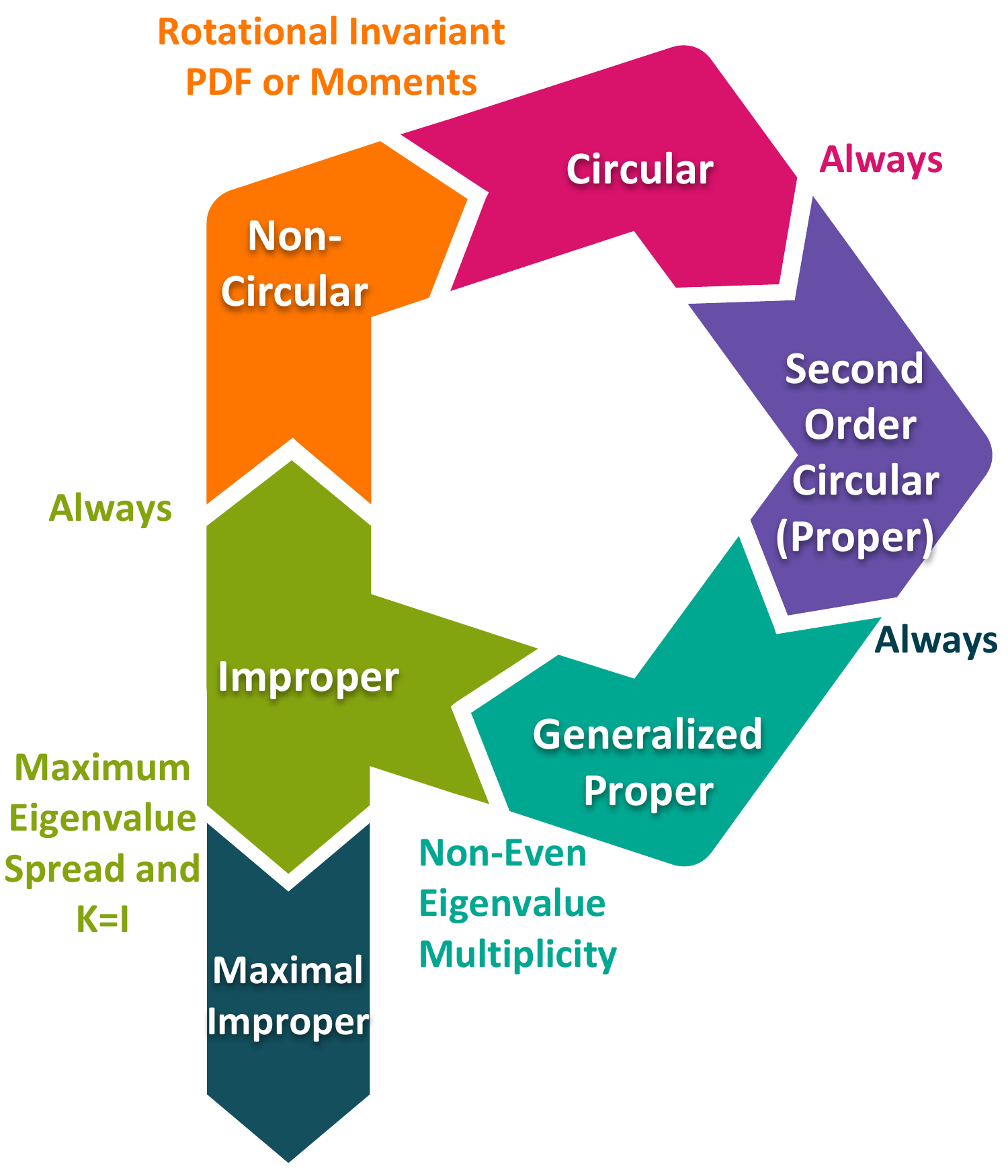}}
\end{minipage}
\caption{Extents of Propriety and their Implications}
\label{fig:PI}
\end{figure}

 For other zero-mean RVs propriety and circularity are related as shown in Fig. \ref{fig:PI}. The rotational invariance of all existing moments of circular RV certainly implies the rotational invariance of second-order moments and thus it is equivalent to designating it as a proper RV. Second-order circular or strictly proper RV demonstrates even multiplicity of eigenvalues owing to the block-diagonal structure of ${\M{R}_{\underline{\V{zz}}}} = \textbf{Block-Diag}\left(\M{R}_{\V{zz}}, {\M{R}_{\V{zz}}^*}\right)$ rendering generalized proper RV. Consequently. a circular RV is always proper/generalized proper. On the other hand, lack of any propriety condition as elaborated in Definition \ref{Def:Improper} leads to improper RV. An improper RV with maximum eigenvalue spread of ${\M{R}_{\underline{\V{zz}}}}$ is termed as maximal improper RV. Moreover, improperness always implies non-circularly as it nullifies the condition of rotational invariant second-moment. 
\paragraph{Degree of Improperness}
Impropriety implies non-circularity, thus the rotational variance of the distribution of a complex entity is characterized by the DoI.
 For a complex scalar r.v. ${z}$, with finite variance $\sigma_{z}^2$ and pseudo-variance $\tilde{\sigma}_{z}^2 $, the measure of correlation between ${z}$ and ${z}^*$ is given by,
 \begin{definition}{9}
 Circularity quotient $\rho_z$ is defined as the  fraction between the  pseudo-variance and the variance 
 \begin{equation}
\rho_z = \frac{\tilde{\sigma}_{z}^2}{\sigma_{z}^2} = k e^{i\phi},
 \end{equation}
where, DoI is measured by the circularity coefficient $k = {\left| \tilde{\sigma}_{z}^2 \right|} / {\sigma_{z}^2}$ and the circularity angle is given by~$\phi$ \cite{ollila2008circularity}.
  \end{definition}
The term \textit{circularity coefficient} for $k$ is originated
from \cite{eriksson2006complex} while the terms \textit{non-circularity rate} and \textit{non-circularity phase} are also used for $k$ and $\phi$, respectively \cite{abeida2006music}. In reality, the circularity coefficient is the \textit{canonical correlation}
between ${z}$ and ${z}^*$ \cite{schreier2006generalized}. The geometric interpretation for the circularity quotient is given as $\rho_z \in \Omega$, where $\Omega  = \left\{ {z \in {\mathbb{C}} :|z| \le 1} \right\}$ is the unit circle \cite{ollila2008circularity}. Thus, the circularity coefficient lies in the range $k \in \left[0,1 \right]$ providing an interesting result $0\leq {\left| \tilde{\sigma}_{z}^2 \right|} \leq \tilde{\sigma}_{z}^2$. The circularity coefficient measures the degree of circularity, as in $k = 0$ means $z$ is second-order circular
whereas $k = 1$  means $z$ is maximally NC (i.e., $x$ or $y$ is constant,
equal to zero, or $x$ is a linear function of $y$) \cite{ollila2011complexw}.
This quantifies DoI for a r.v., next we present DoI of a RV as:
\begin{definition}{10} \label{Def7} For a complex RV $\V{z}$, the coherence matrix $\M{c}={\M{R}_{\V{zz}}^{-1/2}}{\tilde{\M{R}}_{\V{zz}}}{\M{R}_{\V{zz}}^{-\RM{T}/2}}$ with Takagi factorization $\M{c}=\M{f}\M{k}\M{f}^{\RM{T}}$ contains the canonical correlations or circularity coefficients $k_i$ in $\M{K}=\textbf{Diag}(k_1,k_2,\ldots,k_N)$ and $\M{F}$ is a complex unitary matrix. It is important to highlight that $k_i$ is the circularity coefficient of the $i^{\rm th}$ variable in $\V{z}$. The DoI can be defined as a function of canonical
correlations in following ways \cite{schreier2008bounds}
\begin{equation} \label{rho_1}
{\rho _1} = 1 - \prod\limits_{i = 1}^r {\left( {1 - k_i^2} \right)} \mathop  = \limits^{(r = N)}  1- \left| {{\M{R}}_{\underline{\V{zz}}}} \right| \left| {{\M{R}}_{{\V{zz}}}} \right|^{-2},
\end{equation} \end{definition}
\begin{equation}
{\rho _2} = \prod\limits_{i = 1}^r {k_i^2}  \mathop  = \limits^{(r = N)}  \left| {\tilde{\M{R}}_{{\V{zz}}}} {{\M{R}}_{{\V{zz}}}^{-*}} {\tilde{\M{R}}_{{\V{zz}}}}^{*} \right|\left| {{\M{R}}_{{\V{zz}}}} \right|^{-1},
\end{equation}
\begin{equation}
{\rho _3} = \frac{1}{N}\sum\limits_{i = 1}^r {k_i^2}  \mathop  = \limits^{(r = N)}  \frac{1}{N}{\RM{Tr}}\left( {{\M{R}}_{\V{zz}}^{ - 1}}{\tilde{\M{R}}_{\V{zz}}}{{\M{R}}_{\V{zz}}^{ - *}}{{\tilde{\M{R}}_{\V{zz}}}^*} \right),
\end{equation}
where $r =\mathrm{rank}({\M{R}}_{\underline{\V{zz}}})$. The set of canonical correlations ${\{{k_i}\}_{i=1}^N}$ is also referred to as \textit{circularity spectrum} \cite{eriksson2006complex}. DoI must satisfy $0 \leq {\rho_i} \leq 1$, 
ranging from proper signal ($ {\rho_i} = 0$) to maximally improper signal ($ {\rho_i} = 1$) \cite{ollila2008circularity}. The intuitive meaning of the aforementioned measures of improperness is summarized as: 
\begin{itemize}
\item {$\rho_1$ is 1 if at least one $k_i=1$ i.e., any one maximal improper variable in RV $\V{z}$ will result in maximum DoI. Additionally, it is 0 if all $k_i$ are 0. Thus, it helps in identifying if any one element of the RV is maximally improper or if all entries in a RV are proper. However, it fails to discriminate if all or any subset of the entries are maximally improper. Similarly, it is insufficient to identify any subset of proper entries. Nevertheless, $\rho_1$ gives the entropy loss due to improperness of a RV}  \cite{adali2011complex} {(refer to Section \ref{ssec:Entropy}) and it is also used for impropriety likelihood-ratio testing} \cite{schreier2006generalized} {(refer to Section \ref{ssec:Testing}) as it measures the linear dependence between $\V{z}$ and $\V{z}^*$.} 
\item $\rho_2$ is 0 if at least one $k_i=0$ i.e., any one proper variable in RV $\V{z}$ will result in minimum DoI. Moreover, $\rho_2$ is 1 if all $k_i$ are 1. Thus, it helps in identifying if any one element of the RV is proper or if all entries in a RV are improper. However, it fails to apprehend if all or any subset of the entries are proper. Likewise, it cannot assess any subset of maximally improper entries.
\item $\rho_3$ attains maximum value when all entries in $\V{z}$ are maximally improper and attains minimum value when all entries in $\V{z}$ are proper. Nonetheless, it cannot identify subset of proper or maximally improper entries in a RV.
\end{itemize} 
In short,  $\rho_1$ and  $\rho_2$ complement each other and provide the missing information, whereas $\rho_3$  is the preferred choice for joint assessment. Intuitively, $\M{K}=\M{I}$ portrays all elements of the RV $\V{z}$ to be maximal improper. Nevertheless, it is a necessary but not sufficient condition for maximal improperness of a RV On the other hand, maximum eigenvalue spread is the sufficient condition for such maximal impropriety.
\paragraph{Discussion} 
The concepts of propriety and circularity and the absence of these phenomena i.e., impropriety and non-circularity along with the extent of improperness (namely DoI) are discussed at length. Numerous degrees of propriety and circularity from strictly proper to generalized proper and from marginally circular to total circular, respectively, are also distinctly stated.  Eventually, the interplay between propriety and circularity can be concluded as: circularity is a subset of propriety with more restrictions, whereas impropriety is a subset of non-circularity.
\subsubsection{Quaternion Random Vectors}
Impropriety characterization is not limited to the complex domain. For instance, Vakhania studied the concept of ‘properness’ in quaternion domain \cite{vakhania1999random}; however, his definition of $\mathcal{Q}$-properness is
restricted to the invariance of the PDF under some specific rotations around angle of $\pi/2$. Amblard  \textit{et al.} further relaxed the conditions of $\mathcal{Q}$-properness to $\mathcal{C}^\alpha$-properness with an arbitrary axis and angle of rotation $\varphi$, $\V{q} \triangleq e^{\alpha \varphi}\V{q}$  for any pure unit quaternion $\alpha$ \cite{amblard2004properness}. The evolved and refined definitions of quaternions propriety rely on the vanishing properties of the pseudo-covariance matrices as:
\begin{definition}{11}[$\mathcal{R}^\alpha$-properness] \label{def12}
A quaternion RV $\V{q}$ is $\mathcal{R}^\alpha$-proper iff the pseudo-covariance matrix $\M{R}_{\V{q}\V{q}^{(\alpha)}}$ vanishes \cite{via2010properness}.
\end{definition}
\begin{definition}{12a}[$\mathcal{C}^\alpha$-properness] \label{def13} 
A quaternion RV $\V{q}$ is $\mathcal{C}^\alpha$-proper if it is correlated with $ \V{q}^{\left(\alpha\right)}$ and uncorrelated with the rest i.e., ${ \V{q}}^{\left(\bar{\alpha}\right)}$ \cite{took2011augmented}. 
\end{definition}
\begin{table*}[t]	
\renewcommand{\arraystretch}{1.25}
\caption{Consequences of Vanishing Pseudo-Variances}
\begin{center}
  \begin{tabular}{||c||c||c||}
  \toprule
$\M{R}_{\V{q},\V{q}^{(i)}} =  \M{0}$ & $\M{R}_{\V{q},\V{q}^{(j)}} = \M{0}$   & $\M{R}_{\V{q},\V{q}^{(k)}} = \M{0}$  \\
\midrule
$\M{R}_{\V{r}_1\V{r}_1}  + \M{R}_{\V{r}_i\V{r}_i} = \M{R}_{\V{r}_j\V{r}_j}  +\M{R}_{\V{r}_k\V{r}_k}$ 
 & $\M{R}_{\V{r}_1\V{r}_1}  + \M{R}_{\V{r}_j\V{r}_j} = \M{R}_{\V{r}_i\V{r}_i}  +\M{R}_{\V{r}_k\V{r}_k}$  & $\M{R}_{\V{r}_1\V{r}_1}  + \M{R}_{\V{r}_k\V{r}_k} = \M{R}_{\V{r}_i\V{r}_i}  +\M{R}_{\V{r}_j\V{r}_j}$ \\
$\M{R}_{\V{r}_i\V{r}_1}  - \M{R}_{\V{r}_1\V{r}_i} = \M{R}_{\V{r}_k\V{r}_j}  -\M{R}_{\V{r}_j\V{r}_k} $ 
 & $\M{R}_{\V{r}_1\V{r}_i}  + \M{R}_{\V{r}_i\V{r}_1} = - \M{R}_{\V{r}_k\V{r}_j}  -\M{R}_{\V{r}_j\V{r}_k} $ & $\M{R}_{\V{r}_i\V{r}_1}  + \M{R}_{\V{r}_1\V{r}_i} = \M{R}_{\V{r}_k\V{r}_j}  +\M{R}_{\V{r}_j\V{r}_k} $\\
$\M{R}_{\V{r}_1\V{r}_j}  + \M{R}_{\V{r}_j\V{r}_1} = \M{R}_{\V{r}_i\V{r}_k}  +\M{R}_{\V{r}_k\V{r}_i} $ 
 & $\M{R}_{\V{r}_j\V{r}_1} - \M{R}_{\V{r}_1\V{r}_j} = \M{R}_{\V{r}_i\V{r}_k}  -\M{R}_{\V{r}_k\V{r}_i} $  & $\M{R}_{\V{r}_1\V{r}_j}  + \M{R}_{\V{r}_j\V{r}_1} = - \M{R}_{\V{r}_i\V{r}_k}  -\M{R}_{\V{r}_k\V{r}_i} $\\
$\M{R}_{\V{r}_1\V{r}_k}  + \M{R}_{\V{r}_k\V{r}_1} = -\M{R}_{\V{r}_i\V{r}_j}  -\M{R}_{\V{r}_j\V{r}_i} $   & $\M{R}_{\V{r}_1\V{r}_k}  + \M{R}_{\V{r}_k\V{r}_1} = \M{R}_{\V{r}_i\V{r}_j} + \M{R}_{\V{r}_j\V{r}_i} $  & $\M{R}_{\V{r}_1\V{r}_k}  - \M{R}_{\V{r}_k\V{r}_1} = \M{R}_{\V{r}_i\V{r}_j}  -\M{R}_{\V{r}_j\V{r}_i}$ \\   
\bottomrule
  \end{tabular}
\end{center}
\label{tab:Cj}
\end{table*}
\textbf{Example.} A quaternion RV $\V{q}$ is $\mathcal{C}^j$-proper iff the pseudo-covariance matrices $\M{R}_{\V{q},\V{q}^{(i)}}$ and $\M{R}_{\V{q},\V{q}^{(k)}}$ vanish \cite{via2010properness}. Equivalently,  $\mathcal{C}^j$-proper quaternion exhibits the following $\M{R}_{\V{v}\V{v}}$ structure \cite{ginzberg2011testing}
\begin{equation}\label{Cj-proper}
\M{R}_{\V{v}\V{v}} =\left[ {\begin{array}{*{20}{c}}
\M{R}_{\V{r}_1\V{r}_1} &\M{R}_{\V{r}_1\V{r}_i}&-\M{R}_{\V{r}_j\V{r}_1}&-\M{R}_{\V{r}_j\V{r}_i}\\
\M{R}_{\V{r}_i\V{r}_1}&\M{R}_{\V{r}_i\V{r}_i}&-\M{R}_{\V{r}_k\V{r}_1}&-\M{R}_{\V{r}_k\V{r}_i}\\
\M{R}_{\V{r}_j\V{r}_1}&\M{R}_{\V{r}_j\V{r}_i}&\M{R}_{\V{r}_1\V{r}_1}&\M{R}_{\V{r}_1\V{r}_i}\\
\M{R}_{\V{r}_k\V{r}_1}&\M{R}_{\V{r}_k\V{r}_i}&\M{R}_{\V{r}_i\V{r}_1}&\M{R}_{\V{r}_i\V{r}_i}
\end{array}} \right],
\end{equation}
\begin{proof}
Evidently, \eqref{Cj-proper} is obtained from \eqref{Quaternion-RC}
by exploiting the definitions of vanishing $\M{R}_{\V{q},\V{q}^{(i)}}$ and $\M{R}_{\V{q},\V{q}^{(k)}}$, which are expressed as  
\begin{align} 
\!\!\M{R}_{\V{q},\V{q}^{(i)}}\!\!  = & - \!\! {\rm E} \{ \left( \V{r}_1\! +\! i \V{r}_i \!+\! j \V{r}_{j}\!+\!{k} \V{r}_{k} \right)\! \left(\V{r}_1^{\RM T}\! -\! i \V{r}_i ^{\RM T}\! +\! j \V{r}_{j}^{\RM T}\! +\!{k} \V{r}_{k}^{\RM T}  \right) \}. \nonumber \\
\!\!\M{R}_{\V{q},\V{q}^{(k)}}\!\!  = & - \!\! {\rm E} \{ \left( \V{r}_1\!+\! i \V{r}_i \!+\! j \V{r}_{j}\!+\!{k} \V{r}_{k} \right)\!  \left(\V{r}_1^{\RM T} \!+\! i \V{r}_i ^{\RM T}\! +\! j \V{r}_{j}^{\RM T} \!-\!{k} \V{r}_{k}^{\RM T}  \right) \}.
\end{align}
Equating these pseudo-variances to zero render the set of four real-valued equations as shown in Table \ref{tab:Cj}. In case of $\mathcal{C}^j$-proper quaternion, the four set of equations each from $\M{R}_{\V{q},\V{q}^{(i)}}$ and $\M{R}_{\V{q},\V{q}^{(k)}}$ are simultaneously solved. The solution implies 
$\M{R}_{\V{r}_j\V{r}_j} = \M{R}_{\V{r}_1\V{r}_1}$, 
$\M{R}_{\V{r}_k\V{r}_k} = \M{R}_{\V{r}_i\V{r}_i}$, 
$\M{R}_{\V{r}_k\V{r}_j} = \M{R}_{\V{r}_i\V{r}_1}$, 
$\M{R}_{\V{r}_j\V{r}_k} = \M{R}_{\V{r}_1\V{r}_i}$, 
$\M{R}_{\V{r}_1\V{r}_j} = - \M{R}_{\V{r}_j\V{r}_1}$, 
$\M{R}_{\V{r}_i\V{r}_k} = - \M{R}_{\V{r}_k\V{r}_i}$, 
$\M{R}_{\V{r}_1\V{r}_k} = - \M{R}_{\V{r}_j\V{r}_i}$, and 
$\M{R}_{\V{r}_i\V{r}_j} = - \M{R}_{\V{r}_k\V{r}_1}$. Consequently, substituting these relations in \eqref{Quaternion-RC} yields $\M{R}_{\V{v}\V{v}}$ in \eqref{Cj-proper}. 
\end{proof}
\begin{definition}{12b} Alternately, a $\mathcal{C}^\alpha$-proper quaternion is defined to exhibit a distribution that is invariant by \textit{left Clifford translation},
 i.e., $\V{q} \triangleq e^{\alpha \varphi}\V{q} \; \forall \varphi$, for one and only one imaginary unit $\alpha$ \cite{amblard2004properness}.
 \end{definition} 
 \begin{definition}{13}[$\mathcal{Q}$-properness]\label{def14}
  A quaternion RV $\V{q}$ is $\mathcal{Q}$-proper iff all three pseudo-covariance matrices $\M{R}_{\V{q}\V{q}^{(i)}}$, $\M{R}_{\V{q}\V{q}^{(j)}}$ and $\M{R}_{\V{q}\V{q}^{(k)}}$ vanish \cite{via2010properness}. It also implies that $\V{q}$ is uncorrelated with its three vector involutions.  Moreover, the corresponding $\M{R}_{\underline{\V{qq}}}$ is real-valued, positive definite, and symmetric  \cite{took2011augmented}. 
\end{definition}
$\mathcal{Q}$-properness is also referred as $\mathcal{H}$-properness and is equivalently reported as the distribution invariance
of axis $\alpha$ and $\varphi$, i.e., $\V{q} \triangleq e^{\alpha \varphi}\V{q} \; \forall \varphi$, for any imaginary unit $\alpha$ \cite{amblard2004properness}. Analogous to the $\mathcal{C}^j$-proper case, the equivalent  $\M{R}_{\V{v}\V{v}}$ for $\mathcal{Q}$-proper quaternion is obtained by simultaneously solving the 12 set of equations (as given in Table \ref{tab:Cj}) obtained by setting $\M{R}_{\V{q}\V{q}^{(i)}}$, $\M{R}_{\V{q}\V{q}^{(j)}}$ and $\M{R}_{\V{q}\V{q}^{(k)}}$ to zero \cite{ginzberg2011testing}.
\begin{equation}
\M{R}_{\V{v}\V{v}} =\left[ {\begin{array}{*{20}{c}}
\M{R}_{\V{r}_1\V{r}_1} &-\M{R}_{\V{r}_i\V{r}_1}&-\M{R}_{\V{r}_j\V{r}_1}&-\M{R}_{\V{r}_k\V{r}_1}\\
\M{R}_{\V{r}_i\V{r}_1}&\M{R}_{\V{r}_1\V{r}_1}&-\M{R}_{\V{r}_j\V{r}_i}&\M{R}_{\V{r}_i\V{r}_k}\\
\M{R}_{\V{r}_j\V{r}_1}&\M{R}_{\V{r}_j\V{r}_i}&\M{R}_{\V{r}_1\V{r}_1}&-\M{R}_{\V{r}_k\V{r}_j}\\
\M{R}_{\V{r}_k\V{r}_1}&-\M{R}_{\V{r}_i\V{r}_k}&\M{R}_{\V{r}_k\V{r}_j}&\M{R}_{\V{r}_1\V{r}_1}
\end{array}} \right].
\end{equation}
Intuitively, a  $\mathcal{Q}$-proper quaternion is $\mathcal{C}^\alpha$-proper for all pure unit quaternions $\alpha = i, j, \; \text{and} \; k$. Additionally, $\mathcal{R}^\alpha$ and $\mathcal{C}^\alpha$-properness are complementary and together they result in $\mathcal{Q}$-properness.  As a special case, the propriety of scalar quaternion $q = r_1 + i r_i + j r_j + k r_k$, is equivalent to sphericity of $v$, i.e., it is called proper iff $r_1, r_i, r_j \; \text{and} \; r_k$ are independent and identically distributed (i.i.d) \cite{ginzberg2011testing}.

Analogous to the complex case, two quaternion RVs $\V{q}_1$ and $\V{q}_2$ are \textit{cross proper} vectors as
\begin{itemize}
\item Cross $\mathcal{R}^\alpha$-proper iff $\M{R}_{\V{q}_1{\V{q}_2^{(\alpha)}}} = \M{0}$ with $\alpha \in \{i,j,k\} $.
\item Cross $\mathcal{C}^\alpha$-proper iff $\M{R}_{\V{q}_1{\V{q}_2^{(\alpha')}}} = \M{0}$ and $\M{R}_{\V{q}_1{\V{q}_2^{(\alpha'')}}} = \M{0}$ with arbitrary one-to-one mapping between $\{\alpha,\alpha',\alpha''\}$ and $\{i,j,k\}$.
\item Cross $\mathcal{Q}$-proper iff all $\M{R}_{\V{q}_1{\V{q}_2^{(i)}}} \!=\! \M{R}_{\V{q}_1{\V{q}_2^{(j)}}} \!=\!\M{R}_{\V{q}_1{\V{q}_2^{(k)}}}\!=\!\M{0}$.
\end{itemize}
Similarly, two quaternion RVs $\V{q}_1$ and $\V{q}_2$ are \textit{jointly-proper} iff they are $\mathcal{R}^\alpha$-, $\mathcal{C}^\alpha$- or $\mathcal{Q}$- proper and respectively cross proper~\cite{via2010properness}.
\paragraph{Degree of Improperness}
 The degree of ${\Pi}$-properness of a quaternion RV $\V{q}$ with $\Pi \in \left( \mathcal{R}^\alpha,\mathcal{C}^\alpha, \mathcal{Q} \right)$ and augmented covariance matrix ${\M{R}_{\underline{{\V{q}}{\V{q}}}}}$ is evaluated as
 \begin{equation}
\mathcal{P}_{\Pi}  = \mathop {\min }\limits_{{{\hat{\M{R}}_{\underline{{\V{q}}{\V{q}}}}}} \in \mathfrak{R}_{\Pi} } D\left({{\M{R}}_{\underline{{\V{q}}{\V{q}}}}} \parallel {\hat{\M{R}}_{\underline{{\V{q}}{\V{q}}}}}  \right),
 \end{equation}
where $\mathfrak{R}_{\Pi} $ represents the set of proper augmented covariance matrices with ${\Pi}$-properness e.g.,
\begin{equation}\label{SetQ}
\mathfrak{R}_{\mathcal{Q}} =\lbrace {\hat{\M{R}}_{\underline{{\V{q}}{\V{q}}}}} | {\hat{\M{R}}_{\underline{\V{q}}\underline{{\V{q}}}^{(i)}}}={\hat{\M{R}}_{\underline{\V{q}}\underline{\V{q}}^{(j)}}}={\hat{\M{R}}_{\underline{\V{q}}\underline{\V{q}}^{(k)}}}=\M{0}    \rbrace, 
\end{equation}
\begin{equation}\label{SetC_alpha}
\mathfrak{R}_{\mathcal{C}^\alpha} =\lbrace {\hat{\M{R}}_{\underline{{\V{q}}{\V{q}}}}} | {\hat{\M{R}}_{\underline{\V{q}}\underline{\V{q}}^{(\alpha')}}}={\hat{\M{R}}_{\underline{\V{q}}\underline{\V{q}}^{(\alpha'')}}}=\M{0} \rbrace,
\end{equation}
\begin{equation}\label{SetR_alpha}
\mathfrak{R}_{\mathcal{R}^\alpha} =\lbrace {\hat{\M{R}}_{\underline{{\V{q}}{\V{q}}}}} | {\hat{\M{R}}_{\underline{\V{q}}\underline{\V{q}}^{(\alpha)}}}=\M{0} \rbrace.
\end{equation}
Moreover, $D\left({{\M{R}}_{\underline{{\V{q}}{\V{q}}}}} \parallel {\hat{\M{R}}_{\underline{{\V{q}}{\V{q}}}}}  \right)$ is the Kullback-Leibler divergence \cite{cover2006elements} between two zero-mean quaternion Gaussian distributions with ${{\M{R}}_{\underline{{\V{q}}{\V{q}}}}}$ and ${\hat{\M{R}}_{\underline{{\V{q}}{\V{q}}}}}$ \cite[Table III]{via2010properness}. Interestingly, the Pythagorean theorem for exponential families of PDF's render $\mathcal{P}_\mathcal{Q} = \mathcal{P}_{\mathcal{R}^\alpha} + \mathcal{P}_{\mathcal{C}^\alpha}$ with $\mathcal{P}_\mathcal{Q}, \mathcal{P}_{\mathcal{R}^\alpha}, \; \text{and} \; \mathcal{P}_{\mathcal{C}^\alpha}$ as defined in Table \ref{tab:QGLRT} \cite{via2010properness}. 
\paragraph{Discussion}
  Quaternion propriety definitions are restricted to the SOS which completes the analysis for Gaussian quaternions, e.g., the impropriety measure $\mathcal{P}_{\Pi}$ is also the non-circularity measure for a zero-mean Gaussian quaternion RV based on its complete characterization using SOS. Nevertheless, further investigation of the $n^{\RM {th}}$-order properness is required to handle other distributions. As an example, the higher-order discrete rotational invariance analysis is required to tackle 4D constellations in communication systems~\cite{zetterberg1977codes}.  
 \subsubsection{Summary  and Insights} 
The extent of complex and quaternion RV properness is classified in two (strict and generalized proper) and three (${\mathcal{R}^\alpha},{\mathcal{C}^\alpha},{\mathcal{Q}}$-proper) categories, respectively. 
However, the strongest versions of properness in complex and quaternion RVs have different implications i.e.,  a strictly proper complex RV $\V{z}$ may contain correlated $\Re\{z_{k}\}$ and $\Im\{z_{l}\}$ for $k \neq l$ \cite{neeser1993proper} whereas a strictly proper quaternion ($\mathcal{Q}$-Proper) RV $\V{q}$ cannot contain correlated $\Re\{q_{k}\}$ and $\Im\{q_{l}\}$ with ${\rm E} \{\Re \{q_k\} \Im\{q_l\} \}=0 \; \forall  \left(k \neq l\right)$. Furthermore, the correlation between two RVs can be assessed using cross and joint properness in both complex and quaternion RVs. 
Additionally, the respective DoI not only provides the entropy loss due to the improperness but also helps in impropriety testing to identify the underlying type of properness for appropriate transformation and processing \cite{via2011generalized}. {Moreover, the complex domain DoI is also useful in blind source separation}  \cite{schreier2010statistical} {and the quaternion domain DoI can provide the error exponent of Neyman-Pearson detector for binary hypothesis testing. Nonetheless, these perks are only obtained with the usage of complex or augmented representations. The real composite representation can only identify the improperness of a complex RV based on the structure of the real covariance matrix, yet it fails to provide a measure for the DoI.}
 
  \subsection{Transformations and Operations} \label{ssec:transformations}
{Impropriety classification helps to identify the simplified form of processing and transformation in terms of computational complexity} \cite{took2009quaternion}. {This subsection highlights suitable processing models for complex as well as quaternion RVs based on their propriety characterization.}
\subsubsection{Complex Random Vectors} 
{Various transformations and operations of the complex RVs depend on their propriety characterization. The appropriate processing enables us to exploit the additional design freedom offered by the RVs' improperness and extract the information embedded in them.}
   \paragraph{Transformations} {Three forms of transformations,  i.e., \textit{real linear transformation} (RLT) or WL \textit{transformation} (WLT), \textit{complex linear transformation} (CLT) and \textit{widely unitary transformation} (WUT) are reviewed as:}
\begin{itemize}
\item  {RLT  $\M{L} \in \mathbb{R}^{2M{\rm x}2N}$ on the real composite vector $ {\V{u}}\in \mathbb{R}^{2N}$ yields another real composite vector $\V{u}_{\rm RLT}\in \mathbb{R}^{2M}$ as}
   \begin{equation}\label{eq}
\V{u}_{\rm RLT} = \left[\!\! {\begin{array}{*{20}{c}}
\V{x}_{\rm RLT}\\
\V{y}_{\rm RLT}
\end{array}}\!\! \right]= \M{L} \V{u} = \left[\!\! {\begin{array}{*{20}{c}}
{{\M{L}_{11}}}&{{\M{L}_{12}}}\\
{ {\M{L}_{21}}}&{{\M{L}_{22}}}
\end{array}}\!\! \right] \left[ \!\!{\begin{array}{*{20}{c}}
\V{x}\\
\V{y} \end{array}}\!\! \right].
\end{equation}
{Equally, the augmented complex model of the transformation $\V{u}_{\rm RLT} =  \M{L} \V{u} $ is given~as}
 \begin{align}\label{eq}
\underline{\V{w}} = \left[ {\begin{array}{*{20}{c}}
\V{w}\\
{\V{w}^*}
\end{array}} \right]& = \sqrt{2} \M{T}\V{u}_{\rm RLT}  = \sqrt{2} \M{T} \M{L} \V{u} \nonumber \\& =  \{  \M{T} \M{L}  \M{T}^{\RM{H}}\}\{\sqrt{2}\M{T}\V{u}\} = \underline{\M{N}}\underline{\V{z}}, 
\end{align}
where, 
\begin{equation}
 \underline{\M{N}} = \left[ {\begin{array}{*{20}{c}}
{{\M{N}_{1}}}&{{\M{N}_{2}}}\\
{\M{N}_{2}^*}&{\M{N}_{1}^*}
\end{array}} \right].
\end{equation}
{is the augmented description of the WLT or \textit{linear-conjugate-linear transformation} preserving the block pattern structure. Thus, ${\V{w}}=\V{z}_{\rm WLT} = {{\M{N}_{1}}} {\V{z}} + {{\M{N}_{2}}} {\V{z}}^*$ with}
\begin{equation}
{{\M{N}_{1}}} = \frac{1}{2} \left[ {{\M{L}_{11}}} +{{\M{L}_{22}}} + i \left( {{\M{L}_{21}}}-{{\M{L}_{12}}}\right) \right],
\end{equation}
\begin{equation}
{{\M{N}_{2}}} = \frac{1}{2} \left[ {{\M{L}_{11}}} -{{\M{L}_{22}}} + i \left( {{\M{L}_{21}}}+{{\M{L}_{12}}}\right) \right].
\end{equation}
\item {A CLT or \textit{strictly linear} (SL) \textit{transformation} (SLT) is a special case of the WLT when ${\V{z}_{\rm CLT}} = {{\M{N}_{1}}} \V{z}$ with ${{\M{N}_{2}}}=\M{0}$ and the corresponding RLT is given as}
 \begin{equation}\label{eqCLT}
{\V{z}_{\rm CLT}} = \left[ {\begin{array}{*{20}{c}}
{\V{x}_{\rm CLT}}\\
{\V{y}_{\rm CLT}}
\end{array}} \right]= \left[ {\begin{array}{*{20}{c}}
{{\M{L}_{11}}}&{{\M{L}_{12}}}\\
{ - {\M{L}_{12}}}&{{\M{L}_{11}}}
\end{array}} \right] \left[ {\begin{array}{*{20}{c}}
\V{x}\\
\V{y} \end{array}} \right].
\end{equation}
{To summarize, RLT on $\mathbb{R}^{2N}$ are linear on $\mathbb{C}^N$ if they have the matrix structure as in \eqref{eqCLT}, otherwise they are WLT. Surprisingly, WLT in $\mathbb{C}$-domain provides more insight than the RLT in $\mathbb{R}$-domain but offers computationally expensive hardware implementation} \cite{schreier2010statistical}.
\item {Interestingly, the EVD of $\M{R}_{\underline{\V{zz}}}$ is different from EVD of ordinary matrices and takes the unfamiliar form $\M{R}_{\underline{\V{zz}}} = \underline{\M{U}}  \underline{\Lambda} \underline{\M{U}}^{\RM H}$ with WUT $\underline{\M{U}}$, satisfying $\underline{\M{U}}^{\RM H}\underline{\M{U}}=\underline{\M{U}}\underline{\M{U}}^{\RM H}=\M{I}$ and the following structure} \cite{schreier2003second,schreier2008bounds}
 \begin{equation}
 \underline{\M{U}} = \left[ {\begin{array}{*{20}{c}}
{{\M{U}_{1}}}&{{\M{U}_{2}}}\\
{\M{U}_{2}^*}&{\M{U}_{1}^*}
\end{array}} \right].
\end{equation}
{Moreover, the eigen-valued matrix $\underline{\Lambda}$ is given as}
\begin{equation}
\underline{\Lambda} =  \M{T} \M{\Lambda}  \M{T}^{\RM{H}} =  \frac{1}{2} \left[ {\begin{array}{*{20}{c}}
{\Lambda}^{\left(1 \right)} +{\Lambda}^{\left(2 \right)}&{\Lambda}^{\left(1 \right)} -{\Lambda}^{\left(2 \right)}\\
{\Lambda}^{\left(1 \right)} -{\Lambda}^{\left(2 \right)}&{\Lambda}^{\left(1 \right)} +{\Lambda}^{\left(2 \right)}
\end{array}} \right],
\end{equation}
{where, ${\Lambda}^{\left(1 \right)}= \textbf{Diag}\left( \lambda_1, \lambda_3, \ldots , \lambda_{2n-1}\right)$ and ${\Lambda}^{\left(2 \right)}= \textbf{Diag}\left( \lambda_2, \lambda_2, \ldots , \lambda_{2n}\right)$ with $\lambda_1\ge \lambda_2 \geq \ldots \geq \lambda_{2n}$ being the ordered eigenvalues of $\M{R}_{\underline{\V{zz}}}$. This simplifies to a diagonal matrix $\Lambda = \textbf{Diag}\left[ {\Lambda}^{\left(1 \right)},{\Lambda}^{\left(2 \right)}\right]$ for the special case of proper RVs.}
\end{itemize} 
\begin{definition}{14}
Any RV obtained from a complex proper RV $\V{z}$ by a linear or affine transformation, i.e., $\hat{\V{z}} = {\hat{\M{A}}}\V{z} + \hat{\V{b}}$, where $\hat{\M{A}} \in \mathbb{C}^{M{\rm x}N}$ and $\hat{\V{b}} \in \mathbb{C}^{M}$ are constant, is also proper. Thus, properness is preserved under affine transformations \cite{neeser1993proper}.  On contrary, the generalized propriety is also preserved by unitary WLT and arbitrary SLT \cite{schreier2003second}
\end{definition}
\paragraph{Operations}
{Inner products and quadratic forms of complex RVs are distinctly defined for different data representations.}
\begin{itemize}
\item {The inner product of  two $N$-dimensional complex RVs $\V{z}_1,\V{z}_2 \in \mathbb{C}^{N}$ with real composite representation $\V{u}_1,\V{u}_2 \in \mathbb{R}^{2N}$ and augmented representation $\underline{\V{z}}_1,\underline{\V{z}}_2 \in \mathbb{C}_*^{2N}$ is defined as}
  \begin{equation}
{\V{u}_1^{\rm T}}\V{u}_2 =  \frac{1}{2} {\underline{\V{z}}_1^{\RM H}}\M{T}{{\M{T}}^{\RM H}}{\underline{\V{z}}}_2 =\frac{1}{2}{\underline{\V{z}}_1^{\RM H}}\underline{\V{z}}_2 = \Re \{{\V{z}_1^{\RM H}}\V{z}_2 \}.
  \end{equation}
\item {Moreover, the real-valued quadratic operation can be described using different representations as} \cite{adali2011complex,picinbono1996second}. 
 \begin{equation}
 {\V{u}^{\rm T}} \M{L}  \V{u} = \frac{1}{2}{\underline{\V{z}}^{\RM H}}\underline{\M{N}}\underline{\V{z}} = \Re \{{\V{z}^{\RM H}} \M{N}_1 \V{z} + {\V{z}^{\RM H}} \M{N}_2 \V{z}^* \}.
 \end{equation}
\end{itemize}
\paragraph{Discussion}
{Remarkably, the maximal invariants of $\M{R}_{\underline{\V{zz}}}$ depend on the underlying transformations. Meaning thereby, any function of $\M{R}_{\underline{\V{zz}}}$ that is invariant under particular transformation must be a function of these maximal invariants only. For example, the eigenvalues of $\M{R}_{\underline{\V{zz}}}$ constitute a maximal invariant for $\M{R}_{\underline{\V{zz}}}$ under WUT whereas the circularity coefficients constitute a maximal invariant for $\M{R}_{\underline{\V{zz}}}$ under non-singular SLT} \cite{schreier2010statistical}. 

  \subsubsection{Quaternion Random Vectors}
The type of quaternion properness defines several types 
of linear processing \cite{via2011generalized}
\begin{itemize}
\item \textit{Full-widely} linear processing is the general optimal linear processing with the simultaneous operation on the quaternion and its involutions as
\begin{equation}
{\V{q}_{\rm {f}}}=  {\M{F}_1^{\RM{H}}} \V{q}  + {\M{F}_{i}^{\RM{H}}} \V{q}^{(i)} +{\M{F}_{j}^{\RM{H}}}\V{q}^{(j)} +{\M{F}_{k}^{\RM{H}}} \V{q}^{(k)},
\end{equation}
where, ${\V{q}_{\rm {f}}} \in \mathbb{H}^{P{\rm x}1} $ and ${\M{F}_1^{\RM{H}}},{\M{F}_{i}^{\RM{H}}},{\M{F}_{j}^{\RM{H}}},{\M{F}_{k}^{\RM{H}}}\in \mathbb{H}^{P{\rm x}N} $.
Full-WLT for augmented representation takes the form $\underline{\V{q}}_{\RM {f}} = {\M{F}_{\underline{\V{q}}}^{\RM H}}{\underline{\V{q}}} $ with ${\M{F}_{\underline{\V{q}}}} = \left[{\M{F}_1^{\RM{T}}}   \;{\M{F}_{i}^{\RM{T}}}\; {\M{F}_{j}^{\RM{T}}}\;{\M{F}_{k}^{\RM{T}}} \right]^{\RM T}$, whereas for the real composite representation it becomes ${\V{v}_{\RM {f}}} = {\M{F}_{\V{v}}^{\RM{T}}} \V{v}$ with $\M{F}_{\V{v}} = {\M{A}_N^{\RM H}}{\M{F}_{\underline{\V{q}}}} \M{A}_P$.  
\item \textit{Semi-widely} linear processing is for the $\mathcal{C}^\alpha$-proper vectors and takes the form 
\begin{equation}
{\V{q}_{\rm {s}}}=  {\M{F}_1^{\RM{H}}} \V{q}  + {\M{F}_{\alpha}^{\RM{H}}} \V{q}^{(\alpha)}.
\end{equation}
\item The \textit{conventional} linear processing of the $\mathcal{Q}$-proper vector take the simplified form ${\V{q}_{\rm {c}}}=  {\M{F}_1^{\RM{H}}} \V{q}$.
\end{itemize}
Notably, $\mathcal{R}^\alpha$-proper quaternion RVs do not result in a simplified linear processing model \cite{via2011generalized}. 
A $\mathcal{Q}$-proper RV is invariant under affine transformation \cite{took2011augmented} whereas the $\mathcal{C}^\alpha$-properness is invariant under semi-WLT \cite{via2010properness}. 
 \subsubsection{{Summary and Insights}}
{In conclusion, all the transformations and operations can be equivalently carried out in all three data formulations. Of all transformations, the properness is only preserved under affine or SLT i.e., the resultant vector after transformation will carry the same propriety value as that of the original vector. Propriety characterization is especially significant to apply the most simplified form of processing while carrying out transformations and operations. Furthermore, exploiting the WL model of the RV provides vast benefits, e.g., it can provide four times faster convergence of the quaternion least-mean square adaptive filtering relative to its quadrivariate counterpart} \cite{took2009quaternion,xiang2018performance}.

\subsection{Entropy and Probability Density Functions } \label{ssec:Entropy}
Information entropy quantifies the average ambiguity i.e., the amount of information or uncertainty in an event. Thus, impropriety incorporation is crucial to evaluate the information theoretic entropy losses as well as the probability distributions of the generalized Gaussian RVs. This section presents the generalized framework of these analysis for both complex and quaternion RVs. 
  \subsubsection{Complex Random Vectors}
This part contains the definitions of differential entropy and probability distributions for a general complex Gaussian RV.   
\paragraph{Differential entropy} 
The entropy is broadly categorized as discrete and differential entropy to measure the amount of surprise in discrete and continuous RV, respectively. Focusing on the differential entropy of a complex RV $\V{z}$ with composite real representation ${\V{u}} = [{\V{x}}^{\RM {T}} \;  {\V{y}}^{\RM{T}}]^{\RM{T}}$ when $\V{x}$ and $\V{y}$ are both Gaussian distributed, gives \cite{schreier2003second}
\begin{equation} \label{eq:diffentropy}
H(\V{u}) = \frac{1}{2}\log \left[ {{{\left( {2\pi e} \right)}^{2N}}\left| {{\M{R}_{\V{uu}}}} \right|} \right].
\end{equation}
Equivalently, the differential entropy of $\V{z}$ in terms of the augmented covariance matrix $\M{R}_{\underline{\V{zz}}}$ is \cite{adali2011complex}
\begin{equation}
H(\V{z})=\frac{1}{2}\log \left[ {{{\left( {\pi e} \right)}^{2N}}\left| {{\M{R}_{\underline{\V{zz}}}}} \right|} \right].
\end{equation}
\begin{definition}{15}
 The differential entropy of a complex RV $\V{z}$ with a fixed
correlation matrix is maximum, iff the RV is zero-mean Gaussian and proper \cite{neeser1993proper}, where it is given by
\begin{equation}
H_P(\V{z})=\log \left[ {{{\left( {\pi e} \right)}^{N}}\left| {{\M{R}_{{\V{zz}}}}} \right|} \right].
\end{equation}
 \end{definition}
 \textit{Remark}: For a scalar, the differential entropy is
appropriately defined as the joint differential entropy of its real
and imaginary parts \cite{neeser1993proper}. 
The difference is the differential entropies of a proper and improper complex Gaussian RV $\V{z}$ is given by the mutual information $\mathcal{I}(\V{z};\V{z}^*)$ between $\V{z}$ and $\V{z}^*$ as \cite{adali2011complex}
\begin{equation}
H_I(\V{z})= H_P(\V{z}) - \mathcal{I}(\V{z};\V{z}^*),
\end{equation}
where, $ \mathcal{I}(\V{z};\V{z}^*)\!\! = \! - \frac{1}{2}\log \prod\limits_{i = 1}^N {\left( {1 \!-\! k_i^2} \right)}$ is a function of $\rho_1$~\eqref{rho_1}.  
\paragraph{Complex-valued Gaussian Distribution} 
Van Den Bos was the pioneer to formulate a general multivariate Gaussian distribution for improper complex processes. He demonstrated that 
the conventional definition of the complex Gaussian distribution (based on the covariance matrix) is only a special case and thus, applicable
to proper processes only \cite{van1995multivariate}. The \ac{PDF} of a general complex Gaussian RV $\V{z}$ with ${\V{u}} = [{\V{x}}^{\RM {T}} \;  {\V{y}}^{\RM{T}}]^{\RM{T}}$ when $\V{x}$ and $\V{y}$ are both Gaussian distributed, is given as
\begin{equation}
p\left( \V{u} \right) \!= \!\frac{1}{{\sqrt {{{\left( {2\pi } \right)}^{2N}}\left| {{\M{R}_{\V{uu}}}} \right|} }}\exp \left\{ { \!- \frac{1}{2}\left( {\V{u}\! -\! {\V{\mu}_{\V{u}}}} \right)^{\RM{T}}\!R_{\V{u}\V{u}}^{ - 1}\!\left( {\V{u}\! -\! {\V{\mu}_\V{u}}} \right)} \right\}.
\end{equation}
Equivalently, the \ac{PDF} of a complex RV $\V{z}$ with augmented covariance matrix $\M{R}_{\underline{\V{zz}}}$ is written as \cite{van1995multivariate}
\begin{equation} \label{eq19}
p\left( {\V{z}} \right)\! = \!\frac{1}{{\sqrt {{\pi ^{2N}}\left| {{\M{R}_{ \underline{\V{z}}  \underline{\V{z}} }}} \right|} }}\exp \left\{ {\! - \frac{1}{2}\left( { \underline{\V{z}}\!  -\! {\V{\mu} _{ \underline{\V{z}} }}} \right)^{\RM{H}} \!\M{R}_{ \underline{\V{z}}  \underline{\V{z}} }^{ - 1}\!\left( { \underline{\V{z}}\!  - \!{\V{\mu} _{ \underline{\V{z}} }}} \right)} \right\}.
\end{equation}
\subsubsection{Quaternion Random Vector}
 This subsection contains the definitions of differential entropy and probability distributions for a general multivariate quaternion~RV.   
\paragraph{Differential entropy}  
 The differential entropy of a quaternion RV $\V{q}$ with real representation $\V{v}$ and the respective covariance matrix $\M{R}_{\V{vv}}$ is given by \cite{via2010properness}
\begin{equation}
H_{\V{q}}\left(\M{R}_{\V{vv}}\right)= 2N \ln \left( 2 \pi e\right) + \frac{1}{2} \ln |\M{R}_{\V{vv}}|.
\end{equation}
 Equivalently, the augmented representation renders \cite{took2011augmented}
\begin{equation}
H_{\V{q}}\left(\M{R}_{\underline{\V{qq}}}\right) = 2N \ln \left( \frac{\pi e}{2}  \right) + \frac{1}{2} \ln |\M{R}_{\underline{\V{qq}}}|.
\end{equation} 
 \begin{definition}{16}
The differential entropy of any quaternion RV is upper bounded by the differential entropy of a centered $\mathcal{Q}$-proper Gaussian RV with $\M{R}_{\V{vv}} = \sigma^2 \M{I}_{4N}$ and is given by \cite{took2011augmented}
\begin{equation}
H_{\V{q}} \leqslant H_{ \mathcal{Q}-{\rm proper}} =2N \ln \left(2\pi e \sigma^2 \right).
\end{equation}
\end{definition}
 Beyond the mutual information between two complex RVs, its generalization to higher dimensions is termed as \textit{interaction information} \cite{mcgill1954multivariate}. Took \textit{et al.} presents the interaction information between quaternion-valued Gaussian RVs $\V{q}, \V{q}_i, \V{q}_j$ and $\V{q}_k$ in \cite[eq. 43]{took2011augmented}. Unlike mutual information, interaction information can be negative depicting a decrease in the degree of association between the variates in a multivariate quantity, when one variable is dealt as a constant.
The impropriety measure $\mathcal{P}_\mathcal{Q}$ measures the interaction information between $\V{q}$ and its involutions. It represents the entropy loss due to improperness as $ \mathcal{P}_\mathcal{Q} = H_{\V{q}}\left(\M{R}_{\V{qq}}\right) - H_{\V{q}}\left(\M{R}_{\underline{\V{qq}}}\right)$. Similarly, $ \mathcal{P}_{\mathcal{C}^\alpha}$ and $ \mathcal{P}_{\mathcal{R}^\alpha}$ presents the entropy losses due to ${\mathcal{C}^\alpha}$- and ${\mathcal{R}^\alpha}$-improperness of $\V{q}$, respectively \cite{via2010properness}. 
In other words, $\mathcal{P}_{\mathcal{R}^\alpha}$ measures of the entropy loss due to the $\mathcal{Q}$-improperness of the $\mathcal{C}^\alpha$-proper quaternion vector.  
\paragraph{Quaternion-valued Gaussian Distribution} 
The PDF of a zero-mean Gaussian RV $\V{q}$ with real representation $\M{R}_{\V{vv}}$ is given by \cite{via2010properness}
\begin{equation}
p\left({\V{v}}\right)= \frac{\exp \left( -\frac{1}{2} \V{v}^{\RM T} {\M{R}_{\V{vv}}^{-1}} \V{v} \right)}{\left(2 \pi \right)^{2N} |\M{R}_{\V{vv}}|^{1/2}}. 
\end{equation}
 Equivalently, it can be expressed using the augmented representation as \cite{took2011augmented}
  \begin{equation} \label{eq:QGPdf}
p\left({ \underline{\V{q}}}\right)= \frac{\exp \left( -\frac{1}{2} \underline{\V{q}}^{\RM H} {\M{R}_{\underline{\V{qq}}}^{-1}} \underline{\V{q}} \right)}{\left( \pi/2 \right)^{2N} |\M{R}_{\underline{\V{qq}}}|^{1/2}}.
 \end{equation}
In the Gaussian case, the distribution of a $\mathcal{H}$-proper variable is contained in a 4D hypersphere \cite{amblard2004properness}.
\subsubsection{{Summary and Insights}}
{Impropriety i.e., correlation between a RV and it's conjugate results in loss of entropy and this loss can be quantified in terms of mutual information and interaction information in case of complex and quaternion RVs, respectively.  Geometrically, it is the loss in capacity by enclosing the codewords in a multidimensional ellipsoid instead of a hypersphere} \cite{fisher2002precoding}. {The differential entropy is maximum for a zero-mean proper Gaussian RV. Any deviation from this trio i.e., zero-mean, properness, or Gaussianity will result in a loss of entropy. Conventionally, researchers adhered to this trio to maximize entropy in a communication system, however later divergence from this property was exploited to gain benefits in some interference-limited applications.}
\subsection{Testing for Impropriety}\label{ssec:Testing}
Impropriety testing is a procedure to characterize the impropriety features of the random signals. In particular, it involves the identification of a RV as proper or improper RV based on its random samples. Moreover, it sometimes includes the quantification of the extent of non-circularity of the potential improper RV. Impropriety testing is an important consideration in order to exploit the significant performance gains offered by the improper signals through appropriate processing. Impropriety defining pseudo-covariance matrix is practically estimated from the available data. However, the general non-zero estimate does not indicate that the source is actually improper. 
 \begin{table}[t]	
\renewcommand{\arraystretch}{1.25}
\caption{Impropriety Evaluation based on various Tests involving Different Data Representations}
\begin{center}
  \begin{tabular}{||p{2.5cm} ||c||c||l||}
 \toprule
\textbf{ Testing Criteria} &\textbf{ Analysis} & \textbf{ Variables} & \textbf{ Assumptions}  \\
 \midrule
\multirow{9}{*}{\shortstack[2]{GLRT with \\ augmented \\ representation}}& \multirow{4}{*}{Conventional} & \multirow{4}{*}{Complex RV} &  i.i.d samples of Gaussian RV and simulated test threshold \cite{ollila2004generalized} \cite{schreier2006generalized}   \\ \cline{4-4}
  & & & independent but non-identical samples  \cite{delmas2011asymptotic} \\\cline{4-4}
& &  &  i.i.d complex elliptically symmetric distributions  \cite{ollila2011complexw}\\\cline{4-4}
& &  & i.i.d samples of Gaussian RV and analytical test threshold \cite{ramirez2014testing} \\
\cline{2-4}
 & \multirow{3}{*}{Frequency }  & {\shortstack[c]{\\Complex vector \\ time-series}} & stationary Gaussian vector sequence with spectral identification \cite{chandna2017frequency}\\
 \cline{3-4}
        &  & \multirow{2}{*}{{\shortstack[c]{\\Complex vector\\ sequences}}}  & stationary non-Gaussian vector sequence \cite{tugnait2017testing} \\\cline{4-4}
        &   &    & Stationary non-Gaussian signal with improper/colored noise \cite{tugnait2017multisensor}  \\
 \cline{2-4}
& Spectral image & 2D-spectrum & Stationary non-Gaussian image with improper Gaussian noise \cite{matalkah2008generalized} \\
 \cline{2-4}
 & Conventional & Quaternion RV &  i.i.d sample of quaternion Gaussian RV  \cite{via2011generalized} \\
\cline{1-4}
LRT with real representation & \multirow{5}{*}{ Conventional} & Quaternion RV &  i.i.d sample of quaternion Gaussian RV \cite{ginzberg2011testing} \\
\cline{1-1} \cline{3-4}
GLRT with real representation & & \multirow{4}{*}{Complex RV} &  i.i.d samples of Gaussian RV and analytical test threshold \cite{walden2009testing} \\
\cline{1-1}\cline{4-4}
Wald's type detector &  & &  i.i.d CES distributions \cite{ollila2011robust} \cite{ollila2011complex} \\
\cline{1-1}\cline{4-4}
Circularity coefficients estimation& & & asymptotic analysis for arbitrary distribution \cite{delmas2009asymptotic} \\
\cline{1-1}\cline{4-4}
Invariant testing &  &  &  i.i.d samples of Gaussian RV and analytical test threshold \cite{walden2009testing} \\
\bottomrule
  \end{tabular}
\end{center}
\label{tab:ImproprietyTesting}
\end{table}

Therefore, various studies have proposed strategies to test for the impropriety of the signal from its observations based on different assumptions as highlighted in Table \ref{tab:ImproprietyTesting}. They include hypothesis testing based on likelihood ratio tests (LRT) \cite{ollila2004generalized,schreier2006generalized,delmas2011asymptotic,pralon2016probabilistic,ollila2011complexw,ramirez2014testing,chandna2017frequency,tugnait2017testing,tugnait2017multisensor,matalkah2008generalized,via2011generalized,ginzberg2011testing,walden2009testing}, maximum likelihood (ML) estimation of circularity coefficients \cite{delmas2009asymptotic}, Wald's type detector \cite{ollila2011robust,ollila2011complex} and invariant testing \cite{walden2009testing} etc. For instance, Schreier \textit{et al.} \cite{schreier2006generalized} and Olilla  \textit{et al.}\cite{ollila2004generalized} independently proposed a binary hypothesis test for impropriety of the complex Gaussian data based on generalized LRT (GLRT) and simulated test thresholds. Whereas, \cite{walden2009testing} proposed analytical test threshold based on the theoretical analysis of the null asymptotic distribution of the test statistic. 
In \cite{hellings2015impropriety}, a real-valued formulation based on block-skew circulant matrices is considered for GLRT. Interestingly, these studies deal with the i.i.d random samples taken from the Gaussian distribution. The case of independent  but not necessarily identical or non-Gaussian distribution of RVs is discussed in \cite{delmas2011asymptotic}. A robust Wald’s type ML (WTML)-detector for propriety is also presented, robust to deviations from the Gaussianity assumption, based on multiple i.i.d samples in the broad class of complex
elliptically symmetric (CES) distributions~\cite{ollila2011robust,ollila2011complex}.
\newline

Previous parametric impropriety tests are limited to a sequence of independent RV. To address this concern, \cite{chandna2017frequency} and \cite{tugnait2017testing} extended the results to stationary Gaussian and non-Gaussian sequences, respectively, using power spectral representations. Interestingly, \cite{chandna2017frequency} specifies the frequencies causing improperness when propriety is invalid, whereas the approach in \cite{tugnait2017testing} provides only a decision on propriety. The impropriety testing is not only limited to vectors and sequences rather it has also been extended to quaternions \cite{ginzberg2011testing,via2011generalized} and spectral images \cite{matalkah2008generalized}. A compelling practical application is the detection of random transmit waveforms from the Noise Radars (electromagnetic systems that use random signals for detecting and locating reflecting objects) based on the circularity tests \cite{pralon2016probabilistic}. 
\subsubsection{GLRT for Complex Random Vectors}
Of all impropriety tests, hypothesis testing based on GLRT and its variants are the most popular owing to the simple derivation and intuitive interpretation of the detection rules. Although the procedure is not generally optimal in the Neyman-Pearson sense, it is still practical and reliable \cite{walden2009testing,ollila2004generalized,mardia1979multivariate}.
Moreover, GLR is well-known for its invariance characteristics as the hypothesis test (like propriety) must be invariant to LT but not WLT\cite{schreier2006generalized,kay2003invariance}. 

Let $\V{z} \in \mathbb{C}^{N}$ be a complex Gaussian RV with the PDF given by  \eqref{eq19}. Now consider $M$ i.i.d random samples ${{\V{z}} }_1 ,{{\V{z}} }_2  ,\ldots,{{\V{z}} }_M$ taken from the
Gaussian distribution with augmented mean ${\V{\mu} _{\underline{\V{z}} }}$  and augmented covariance ${{\M{R}_{\underline{\V{z}} \underline{\V{z}}}}}$. Let
$\underline{\M{Z}} = [{\underline{\V{z}} }_1 ,{\underline{\V{z}} }_2  ,\ldots,{\underline{\V{z}} }_M ]$ denotes the augmented
sample matrix, where ${\underline{\V{z}} }_m = [\V{z}_m^{\RM{T}} \; \V{z}_m ^{\RM{H}}]^{\RM{T}} $ is the augmented
sample vector. Then, the joint PDF of these samples is given by
\begin{align} \label{eq19}
p\left( {\M{z}} \right)  
& = \pi^{-MN} \left| {{\M{R}_{ \underline{\V{z}}  \underline{\V{z}} }}} \right|^{-M/2}  \rm{x}  \nonumber \\
 &\;\exp \left\{ { - \frac{1}{2}  \sum\limits_{m=1}^{M} {  \left( { \underline{\V{z}}_m  - {\V{\mu} _{ \underline{\V{z}} }}} \right)^{\RM{H}} \M{R}_{ \underline{\V{z}}  \underline{\V{z}} }^{ - 1}\left( { \underline{\V{z}}_m  - {\V{\mu} _{ \underline{\V{z}} }}} \right)  } } \right\}, \\
 &  \! = \!  \pi^{-MN} \left| {{\M{R}_{ \underline{\V{z}}  \underline{\V{z}} }}} \right|^{-M/2}  \! \exp \left\{ \! - \! \frac{M}{2} {\RM{Tr}}({\M{R}}_{\underline{\V{zz}}}^{-1}\hat{{\M{R}}}_{\underline{\V{zz}}} )\right\},
\end{align}
where $\hat{{\M{R}}}_{\underline{\V{zz}}}$ is the sample augmented covariance matrix
\begin{equation}
\hat{{\M{R}}}_{\underline{\V{zz}}} \!=  \!\frac{1}{M} \!\sum\limits_{m=1}^{M} \! {({ \underline{\V{z}}_m  \! - \! {\hat{\V{\mu}} _{ \underline{\V{z}} }}}){( \underline{\V{z}}_m   \!-  \!{\hat{\V{\mu}} _{ \underline{\V{z}} }}})^{\RM{H}}} \!=  \!\left[ {\begin{array}{*{20}{c}}
{\hat{\M{R}}_{\V{zz}}}&{\hat{\tilde{\M{R}}}_{\V{zz}}}\\
{{{\hat{\tilde{\M{R}}}_{\V{zz}}^*}}}&{{\hat{\M{R}}_{\V{zz}}}^*},
\end{array}} \right]
\end{equation}
where ${\hat{\V{\mu}} _{ \underline{\V{z}} }} =  \frac{1}{M}\sum\limits_{m=1}^{M} {\underline{\V{z}}_m}$ is the sample augmented mean vector. The aim is to distinguish signals based on the binary hypothesis tests presented in Table \ref{tab:CGLRT}. GLRT is the ratio of likelihood with ${{\M{R}}}_{\underline{\V{zz}}}$ constrained to have zero off-diagonal blocks i.e., $\tilde{{\M{R}}}_{{\V{zz}}} = \M{0}$, to likelihood with unconstrained ${{\M{R}}}_{\underline{\V{zz}}}$. Thus, GLRT is testing the block-diagonal structure of the augmented covariance matrix as \cite{schreier2006generalized,ollila2004generalized}
\begin{equation}
\lambda = \frac{\max \limits_{{{\M{R}}}_{\underline{\V{zz}}},\tilde{{\M{R}}}_{{\V{zz}}} = \M{0} } p\left( {\M{z}} \right)}{\min \limits_{{{\M{R}}}_{\underline{\V{zz}}}} p\left( {\M{z}} \right)}.
\end{equation}
\begin{table}[t]	
\renewcommand{\arraystretch}{1.25}
\caption{Test Thresholds for ${\mathbb{C}}-$GLR Hypothesis Tests}
\begin{center}
  \begin{tabular}{||p{3cm}||p{3cm}||p{3cm}||c||}
 \toprule
\textbf{Null Hypothesis }  & \textbf{Alternate Hypothesis } &  \textbf{Test Threshold }& \multirow{2}{*}{\textbf{Ref}}  \\
(Proper)& (Improper)& & \\
\midrule
$\mathsf{H}_0: \; \tilde{{\M{R}}}_{{\V{zz}}} = \M{0}$& $\mathsf{H}_1: \; \tilde{{\M{R}}}_{{\V{zz}}} = \tilde{{\M{R}}}_{{\V{zz}}}^\dagger$& Absent & \cite{ollila2004generalized}   \\ 
$\mathsf{H}_0: \; \tilde{{\M{R}}}_{{\V{zz}}} = \M{0}$& $\mathsf{H}_1: \; \tilde{{\M{R}}}_{{\V{zz}}} \neq \M{0}$ & Simulated& \cite{schreier2006generalized}  \\ 
$\mathsf{H}_0: \; \text{all}\; \delta_{n} = {0}$& $\mathsf{H}_1: \; \text{all}\; \delta_{n} \neq {0}$& Approximated & \cite{walden2009testing}  \\ 
   $\mathsf{H}_0: \; \rho = {0}$& $\mathsf{H}_1: \; \rho \neq {0}$ & Asymptotic & \cite{ollila2011complex}   \\ 
    \bottomrule
  \end{tabular}
\end{center}
\label{tab:CGLRT}
\end{table}
 In a GLR, the unknown parameters are replaced by ML estimates  \cite{mardia1979multivariate}. It is well known that the unconstrained and constrained ($\tilde{{\M{R}}}_{{\V{zz}}} = \M{0}$) ML estimate of  ${{\M{R}}}_{\underline{\V{zz}}}$
is the sample covariance matrix ${\hat{\M{R}}}_{\underline{\V{zz}}}$ and  ${\hat{\M{R}}}_\M{0}$, respectively, where ${\hat{\M{R}}}_\M{0}$ is defined as 
\begin{equation}
{\hat{\M{R}}}_\M{0} = \left[ {\begin{array}{*{20}{c}}
{\hat{\M{R}}_{\V{zz}}}&{{\M{0}}}\\
{{\M{0}}}&{{\hat{\M{R}}_{\V{zz}}}}^*
\end{array}} \right].
\end{equation}
 Therefore, GLR can be expressed as
 \begin{align} \label{eq28}
  l  = \lambda^{\frac{2}{M}}&  =  \left| {{\hat{\M{R}}_{\M{0} }}}^{-1} {{\hat{\M{R}}_{\underline{\V{z}}\underline{\V{z}}}}}  \right| \left(  \exp \left\{ - \frac{M}{2} {\RM{Tr}}( {{\hat{\M{R}}_{\M{0} }}}^{-1} {{\hat{\M{R}}_{\underline{\V{z}}\underline{\V{z}}}}} - \M{I} )\right\} \right)^{\frac{2}{M}} \nonumber \\
  & = \frac{\left|{{\hat{\M{R}}_{\underline{\V{z}}\underline{\V{z}}}}}  \right|}{\left| \hat{\M{R}}_{\V{zz}}  \right|^2}=\frac{\left| \hat{\M{R}}_{\V{zz}}-
{\hat{\tilde{\M{R}}}_{\V{zz}}}
{\hat{\M{R}}_{\V{zz}}^{-*}}
  {\hat{\tilde{\M{R}}}_{\V{zz}}^*}   
    \right|}{\left| \hat{\M{R}}_{\V{zz}} \right|}.
 \end{align}
Thus, GLR test statistic is the quotient between 
the determinant of the Schur complement of ${{\hat{\M{R}}_{\underline{\V{z}}\underline{\V{z}}}}}$ and the determinant
of $\hat{\M{R}}_{\V{zz}}$. This test statistic is defined in \cite{ollila2004generalized} and the hypothesis testing involves the test for null $\tilde{\M{R}}_{\V{zz}}$ signifying proper RV $\V{z}$ or known $\tilde{\M{R}}_{\V{zz}}^\dagger$ signifying improper RV $\V{z}$. On the other hand, \cite{schreier2006generalized} presents slightly different hypothesis test as given in Table \ref{tab:CGLRT} along with the simulated thresholds. We can also employ the estimated canonical correlation matrix $\hat{\M{K}}$ using $\hat{\M{R}}_{\V{zz}}$ and $\hat{\tilde{\M{R}}}_{\V{zz}}$ as discussed in Definition \ref{Def7}. We then have 
 \begin{equation}
 l=\left| \M{I} - \hat{\M{K}}\hat{\M{K}}^{\RM{H}}   \right| = \prod\limits_{n=1}^{N} {\left( 1-\hat{k}_n^2 \right)}. 
 \end{equation}
This explains that the GLR is a function of the squared canonical correlations which make up a complete/maximal set of invariants under LT.
Contributions like \cite{delmas2009asymptotic} suggest to directly find the ML estimates of circularity coefficients which are the singular values of the empirical coherence matrix $\hat{\M{C}}$, estimated from $\hat{\M{R}}_{\V{zz}}$ and $\hat{\tilde{\M{R}}}_{\V{zz}}$. Moreover, two invariant tests in \cite{walden2009testing} not only rely on these maximal invariants but also suggest the critical regions to minimize false alarms based on the Box approximation. These two invariant tests based on the real composite representation state null-hypothesis as all $\delta_k = 0$ where Takagi Factorization of $\M{R}_{\V{uu}} = \M{G}\Delta_1\M{G}^{\RM T}$ with $\Delta_1 = \textbf{Diag}\left( \left( \M{I}+ \Delta_2 \right) , \left( \M{I}- \Delta_2 \right)  \right)$ and $\Delta_2 = \textbf{Diag}\left(\delta_1,\delta_2,\ldots,\delta_N \right)$. They accept $H_0$~iff 
\begin{equation}
T_1 \equiv  \prod\limits_{n=1}^{N} {\left( 1-\hat{\delta}_n^2 \right)} \ge c_1 \left( \eta_1,M,N\right),
\end{equation}
\begin{equation}
T_2 \equiv  \sum\limits_{n=1}^{N} {\left( \hat{\delta}_n^2 \right)} \le c_2\left(\eta_1,M,N \right),
\end{equation}
where $c_1\left(\eta_1,M,N\right)$ and $c_2\left(\eta_1, M,N\right)$ are constants based on the probability of false alarm (PFA)  and $\eta_1$ is the specific size of the test \cite{walden2009testing}. Interestingly, $T_2$ is the locally most powerful (LMP) test  as it has as high a power as possible for alternatives $H_1$ i.e., those for which all ${k}_n$ are small. However, no uniformly most powerful (UMP) invariant
test for impropriety exists for the problem $N\ge 2$ \cite{walden2009testing}. Based on the standard ML theory, the statistic $l$ possesses
an asymptotic chi-squared distribution with $F=N(N + 1)$
DoF under the null hypothesis \cite{walden2009testing,ollila2011complexw,delmas2011asymptotic}. Thus, the statistical analysis allows us to accept $H_0$ iff $-(M-N)\log T_1 < \chi_F^2(1-\eta_1)$ where $ \chi_F^2(1-\eta_1)$ is the 100$(1-\eta_1)\%$ point of the chi-square
distribution with $F$ DoF. Additionally, the Wald's type detector \cite{ollila2011complex} for the uni-variate case suggests to reject null hypothesis based on the ML estimate of circularity quotient $\hat{\rho}$ of the underlying CES distribution with finite fourth-order moment~if
\begin{equation}
M | \hat{\rho}|^2/\hat{\varsigma}_0 \ge \chi_{2;1-\eta_2}^2
\end{equation}
where $\chi_{2;1-\eta_2}^2$ denotes $(1-\eta_2)^{\rm th}$ quantile of the chi-square distribution with 2 DoF, $\eta_2$ PFA and $\hat{\varsigma}_0 $ as given by \cite[(20)]{ollila2011complex}.
 An interesting extension is to look for the extent of non-circularity when the null-hypothesis is rejected e.g., \cite{ollila2011complexw} tests the hypothesis that a particular number of circularity coefficients vanish. 

To conclude, the impropriety testing can be implemented with great computational efficiency using the block-skew circulant matrices for the composite real formulation of the GLRT  \cite{hellings2015impropriety}. One of the proposed efficient implementation involves the following test statistic
\begin{equation}
 r = \frac{2^{2N} \left| \hat{\M{R}}_{\V{uu}} \right|}{{\left| \hat{\M{R}}_{\V{zz}}  \right|^2}},
 \end{equation} 
where $\hat{\M{R}}_{\V{uu}}  = \frac{1}{M}\sum\limits_{m=1}^{M} {({ {\V{u}}_m  - {\hat{\V{\mu}} _{ {\V{u}} }}}){( {\V{u}}_m  - {\hat{\V{\mu}} _{ {\V{u}} }}})^{\RM{T}}}$ is the sample covariance matrix based on the real representation in Definition \ref{def1}. Hellings \textit{et al.} claim that this implementation reduces the computational complexity roughly by a factor of four when compared to the implementation of the augmented complex
version in \eqref{eq28}, owing to its inherent redundancy.
 \begin{table*}[t]	
\renewcommand{\arraystretch}{1.25}
\caption{Quaternion GLR Hypothesis Tests and Decision Criterion}
\begin{center}
  \begin{tabular}{||c||c|c|c||}
    \toprule
& $\mathcal{Q}$\textbf{-properness} & $\mathcal{C}^\alpha$\textbf{-properness} & $\mathcal{R}^\alpha$\textbf{-improperness}  \\
\midrule
Null Hypothesis & $\mathsf{H}_\mathcal{Q}$: $ {\M{R}_{\underline{\V{q}}\underline{\V{q}}}} \in \mathfrak{R}_\mathcal{Q}$   &  $\mathsf{H}_{\mathcal{C}^\alpha}$: $ {\M{R}_{\underline{\V{q}}\underline{\V{q}}}} \in \mathfrak{R}_{\mathcal{C}^\alpha}$  &  $\mathsf{H}_\mathcal{Q}$: $ {\M{R}_{\underline{\V{q}}\underline{\V{q}}}} \in \mathfrak{R}_\mathcal{Q}$  \\ 
Alternative Hypothesis & $\mathsf{H}_\mathcal{I}$: $ {\M{R}_{\underline{\V{q}}\underline{\V{q}}}}  \notin \mathfrak{R}_\mathcal{Q}$   &  $\mathsf{H}_\mathcal{I}$: $ {\M{R}_{\underline{\V{q}}\underline{\V{q}}}}  \notin \mathfrak{R}_{\mathcal{C}^\alpha}$ &  $\mathsf{H}_{\mathcal{C}^\alpha}$: $ {\M{R}_{\underline{\V{q}}\underline{\V{q}}}} \in \mathfrak{R}_{\mathcal{C}^\alpha}$  \\ 
    \hline
GLRT Statistic & $\mathcal{P}_\mathcal{Q} \!\!= \!\!-\frac{1}{2} \ln \!\left| {\hat{\M{D}}_\mathcal{Q}^{-1/2}}   {\hat{\M{R}}_{\underline{\V{q}}\underline{\V{q}}}}   {\hat{\M{D}}_\mathcal{Q}^{-1/2}}\right|$ & $\mathcal{P}_{\mathcal{C}^\alpha}\!\! = \!\!-\frac{1}{2} \ln\! \left| {\hat{\M{D}}_{\mathcal{C}^\alpha}^{-1/2}}   {\hat{\M{R}}_{\underline{\V{q}}\underline{\V{q}}}}   {\hat{\M{D}}_{\mathcal{C}^\alpha}^{-1/2}}\right|$ &  $\mathcal{P}_{\mathcal{R}^\alpha} \!\!= \!\!-\frac{1}{2} \ln \! \left| {\hat{\M{D}}_{\mathcal{Q}}^{-1/2}}   {\hat{\M{D}}_{\mathcal{C}^\alpha}}   {\hat{\M{D}}_{\mathcal{Q}}^{-1/2}}\right|$ \\   
\hline 
GLRT Comparison & $\mathcal{P}_\mathcal{Q} \mathop \lessgtr \limits_{\mathsf{H}_\mathcal{Q}}^{\mathsf{H}_\mathcal{I}}  \gamma_\mathcal{Q} $ &  $\mathcal{P}_{\mathcal{C}^\alpha} \mathop \lessgtr \limits_{\mathsf{H}_{\mathcal{C}^\alpha}}^{\mathsf{H}_\mathcal{I}} \gamma_{\mathcal{C}^\alpha}$ &   $\mathcal{P}_{\mathcal{R}^\alpha} \mathop \lessgtr \limits_{\mathsf{H}_\mathcal{Q}}^{\mathsf{H}_{\mathcal{C}^\alpha}}  \gamma_{\mathcal{R}^\alpha}$ \\
\bottomrule
  \end{tabular}
\end{center}
\label{tab:QGLRT}
\end{table*}
\subsubsection{GLRT for Quaternion Random Vectors}
Like complex RV, propriety testing has also been expanded for the quaternion RV based on its real composite \cite{ginzberg2011testing} as well as augmented representation \cite{via2011generalized}. For instance, Via \textit{et al.} suggest three GLRTs to identify two main kinds of quaternion properness. Based on the ML estimates of ${\M{R}_{\underline{\V{q}}\underline{\V{q}}}}$ under three distinct hypotheses namely $\mathcal{Q}$-proper, $\mathcal{C}^\alpha$-proper and possibly improper vectors, the GLRTs binary hypothesis tests are given in Table \ref{tab:QGLRT}. The hypothesis testing is based on the test whether ${\M{R}_{\underline{\V{q}}\underline{\V{q}}}}$ belongs to the convex set of  $\mathcal{Q}$-proper and $\mathcal{C}^\alpha$-proper augmented covariance matrices, given as $\mathfrak{R}_\mathcal{Q}$ \eqref{SetQ} and $\mathfrak{R}_{\mathcal{C}^\alpha}$ \eqref{SetC_alpha}, respectively, or not. 
Moreover, the GLRT test statistics are based on the ML estimate of ${\M{R}_{\underline{\V{q}}\underline{\V{q}}}}$ i.e., ${\hat{\M{R}}_{\underline{\V{q}}\underline{\V{q}}}}$ and its constraint formulations ${\hat{\M{D}}_\mathcal{Q}}$ and ${\hat{\M{D}}_{\mathcal{C}^{\alpha}}}$ under the hypothesis $\mathsf{H}_\mathcal{Q}$ and $\mathsf{H}_{\mathcal{C}^{\alpha}}$, respectively \cite{via2011generalized}
\begin{equation}
\hat{\M{D}}_\mathcal{Q} = \textbf{Block-Diag} \left( 
\hat{\M{R}}_{\V{q}\V{q}},\hat{\M{R}}_{\V{q}\V{q}}^{(i)},\hat{\M{R}}_{\V{q}\V{q}}^{(j)},\hat{\M{R}}_{\V{q}\V{q}}^{(k)} \right),
\end{equation}
\begin{equation}
\hat{\M{D}}_{\mathcal{C}^{\alpha}}= \left[ {\begin{array}{*{20}{c}}
\hat{\M{R}}_{\V{q}\V{q}} &\hat{\M{R}}_{\V{q}{\V{q}^{(i)}}} &\M{0}_N&\M{0}_N\\
\hat{\M{R}}_{\V{q}\V{q}^{(i)}}^{(i)}&\hat{\M{R}}_{\V{q}\V{q}}^{(i)}&\M{0}_N&\M{0}_N\\
\M{0}_N&\M{0}_N&\hat{\M{R}}_{\V{q}\V{q}}^{(j)}&\hat{\M{R}}_{\V{q}\V{q}^{(i)}}^{(j)}\\
\M{0}_N&\M{0}_N&\hat{\M{R}}_{\V{q}\V{q}^{(i)}}^{(k)}&\hat{\M{R}}_{\V{q}\V{q}}^{(k)}
\end{array}}  \right].
 \end{equation}
 GLRT comparisons for $\mathcal{Q}$-properness, $\mathcal{C}^\alpha$-properness and improperness are established with predefined fixed thresholds $\gamma_{\mathcal{Q}}, \gamma_{\mathcal{C}^\alpha}$ and $\gamma_{\mathcal{R}^\alpha}$, respectively.
 \subsubsection{Summary and Insights}
{Various studies present different test statistic and threshold levels for the GLR hypothesis tests. Nonetheless, no invariant test for impropriety is uniformly most powerful, they all are inclined and thus locally robust tests} \cite{walden2009testing}. {Most of the contributions describe the GLRT of circularity assuming complex normal data} \cite{ollila2004generalized,schreier2006generalized}, {which is further adjusted to accommodate a broad class of CES distributions. If the propriety test is invalid, it is then useful 
to detect the number of latent NC signals in a complex Gaussian RV based on multiple hypothesis tests}~\cite{novey2011testing}. {Similarly, GLRT for quaternions is focused on the identification of $\mathcal{Q}$-proper, $\mathcal{C}^\alpha$-proper or possibly improper RV.}
 \section{Impropriety Sources and Consequences}
\begin{figure*}[t]
\begin{minipage}[b]{1.0\linewidth}
  \centering
  \centerline{\includegraphics[width=16cm]{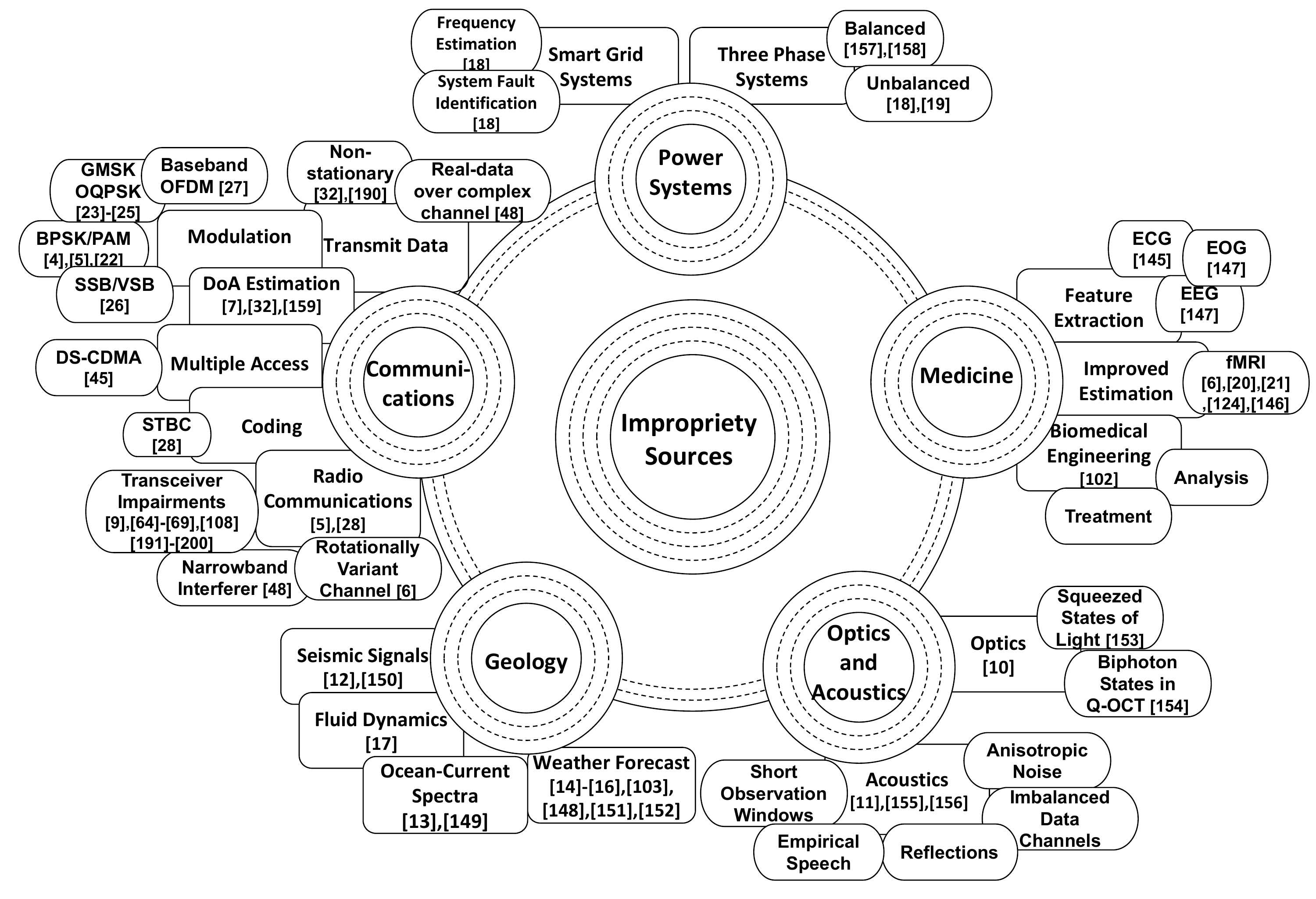}}
\end{minipage}
\caption{{Interdisciplinary Sources of Improperness}}
\label{fig:SI}
\end{figure*}

The most common speculation in the complex analysis assumes propriety and/or circularity of the r.v. under investigation. However, this  phenomenon assumes uncorrelated real and imaginary components of the complex entities with/without identical variance which is generally not true. The two main sources of impropriety include asymmetric signals and asymmetric data.  Asymmetric signals may occur naturally or result from some transforming phenomenon whereas empirical data is generally asymmetric. The improper/asymmetric signatures in numerous fields are highlighted in various technical contributions as shown in Fig. \ref{fig:SI}.
\subsection{Communications} 
A necessary, yet insufficient condition for impropriety
is that the real-valued random process be at least nonstationary \cite{schreier2003stochastic}. This implies that the complex baseband representations of wide-sense stationary thermal noise and nonstationary transmitted data signals are appropriately modelled as proper and improper (may reduce to proper) \cite{schreier2005detection}. Thus, the potentially improper nature of signals must be taken into account when designing detection algorithms.

Improperness may originate from various modulation, coding or access schemes. Important examples of such digital modulation schemes include BPSK \cite{schreier2003second,buzzi2001new}, PAM \cite{ollila2008circularity}, GMSK \cite{hellings2015block,chevalier2006new}, OQPSK or staggered QPSK \cite{mirbagheri2006enhanced}, SSB, VSB \cite{yeo2011optimal} and baseband (but not passband) OFDM \cite{taubock2007complex}. On the other hand, coding schemes like STBC also result in improper signals \cite{alamouti1998simple}.  Interestingly, the real-valued
data transmission over a complex-valued channel also results in non-zero pseudo-covariance of the received observations or non-zero cross pseudo-covariance between the received observations and the desired variable, thus resulting in improper received signal. Such a scenario arises in the GSM, binary CPM, offset QAM \cite{gerstacker2003receivers} and general simplex signals. In many sensor applications, DoA estimation with electronically steerable antenna arrays is improved by employing NC signal constellations \cite{schreier2005detection,adali2011complex,roemer2006efficient}.
In addition, multiple access scheme like direct-sequence code-division multi-access (DS-CDMA) \cite{yoon1997maximizing} with narrowband interferer results in NC interference \cite{gerstacker2003receivers}. Consequently, such modulation, coding and multiple access procedures can 
corrupt  sensor observations with NC complex noise. This necessitates appropriate observation models and distributed complex filters to tackle them~\cite{mohammadi2015distributed}.

In areas such as radio communications, beamforming and
spectral sensing, the amplitude and phase imbalances between its I/Q components may also introduce non-circularity in the received signals \cite{xia2017augmented}. This can be caused by transceiver imperfections \cite{javed2018multiple,anttila2008circularity,hakkarainen2013widely,li2017noncircular}, communication channels that are not rotationally invariant \cite{schreier2010statistical} or NC interference from other sources  \cite{ollila2008circularity}. 
Therefore, WL detection of the multiple-input multiple-output (MIMO) signals is emphasized by arguing the non-circularity of the data signal arising inherently within an iterative receiver~\cite{alamouti1998simple}.
\subsection{Power Systems}
Accurate real-time estimation of system frequency is a major technical challenge for future smart grids with dynamically updating generation and loading topology. Recently, the complex signal retrieved after the Clarke's ($\alpha\beta$) transformation of unbalanced three-phase voltages is shown to be second-order NC \cite{xia2012widely,dini2013widely}.
Therefore, the augmented statistics and the corresponding WL models are exploited to manipulate varying degrees of non-circularity relative to different frequency variation sources. Hence supporting the estimation of instantaneous frequency in a three-phase system under both balanced and unbalanced conditions \cite{xia2012adaptive,arablouei2013adaptive}. The complete second-order information offers next-generation solutions for accurate, adaptive and robust frequency estimation as well as system fault identification \cite{xia2012widely}.
\subsection{Medicine}
In medical data analysis, non-circularity is exploited for features extraction and improved estimations in electrocardiograms (ECG) \cite{zarzoso2010robust} and functional magnetic resonance imaging (fMRI)  \cite{adali2014optimization,li2011application}, respectively.
The fMRI measured neural activities in the brain or spinal cord are recently recognized as improper signals \cite{schreier2010statistical,rowe2005modeling, rodriguez2012noising}. 
Thus, the NC probability distributions of the sources enable real-time extraction of eye muscle activity: electrooculogram (EOG) from the electroencephalogram (EEG) recordings using some blind source extraction algorithms \cite{javidi2010complex}. Rhythms of brain waves are non-stationary signals \cite{clark2012existence} and thus potentially improper because the temporal information arises from correlativity in the frequency domain.
Hence, NC characteristics are a useful resource for suitable analysis and treatment in biomedical engineering.
\subsection{Optics and Acoustics}
Analogous to stochastic complex fields, arbitrary second moments of the complex envelopes of scalar optical fields are also completely characterized by its phase-insensitive and phase-sensitive correlation functions \cite{erkmen2006optical}. Phase-insensitive correlation function is analogous to the covariance matrix whereas phase-sensitive correlation function is equivalent to the pseudo-covariance matrix in the optics. Thus, the non-zero phase-sensitive correlation function marks improper optical field. Conventional light sources like sunlight, LED and lasers have trivial phase-sensitive correlation whereas advanced non-linear optical sources like squeezed states of light possess phase sensitive correlation \cite{yuen1976two}. The biphoton states in quantum
optical coherence tomography (Q-OCT) depict entanglement properties and offer various benefits owing to the phase-sensitive correlation between its two photons \cite{erkmen2006phase}. Optical coherence theory should accommodate phase-sensitive fluctuations for complete second order characterization pertaining to the significant propagation differences between phase-insensitive and phase-sensitive optical sources.

It is quite natural to assume real-world acoustic signals as NC owing to the anisotropic noises, unequal powers of data channels and reflections, and short observation windows. Thus, augmented statistical framework is required to accurately model the acoustic sources for speech recognition \cite{looney2011augmented,layton2006augmented}.
It is based on the fact that speech can be empirically improper in the frequency domain \cite{rivet2007log}.
\subsection{Oceanography and Geophysics}
Complex trace analysis of seismic signals \cite{taner1979complex}, interpretation of ocean-current spectra  \cite{calman1978interpretation} and the analysis of wind fields \cite{khalili2014collaborative,knight2018long,burt1974mesoscale} exploit the improperness of their respective complex representations. The impropriety concept is well-known and appreciated by meteorologists and oceanographers since the early 1970s \cite{gonella1972rotary,mooers1973technique}.
Recently, the improper nature of the real-time wind data along with the tracking of degree of circularity is identified for renewable energy applications \cite{jelfs2012adaptive,adali2010complex}. Similarly, adaptive filtering of the real-world wind signals is based on its NC and nonstationary observations \cite{xia2011augmented}. Others employ impropriety concepts to further physical applications with asymmetric/anisotropic data structures e.g., in fluid dynamics \cite{sykulski2017frequency} and seismic data signals \cite{sykulski2016widely} to overcome modeling challenges.
Most of the above mentioned contributions are backed by \cite{mandic2009complex},  which claims the improper nature of signals involved in various real problems.
     \begin{figure}
        \centering
        \begin{subfigure}[b]{0.23\textwidth}
            \centering
            \includegraphics[width=\textwidth]{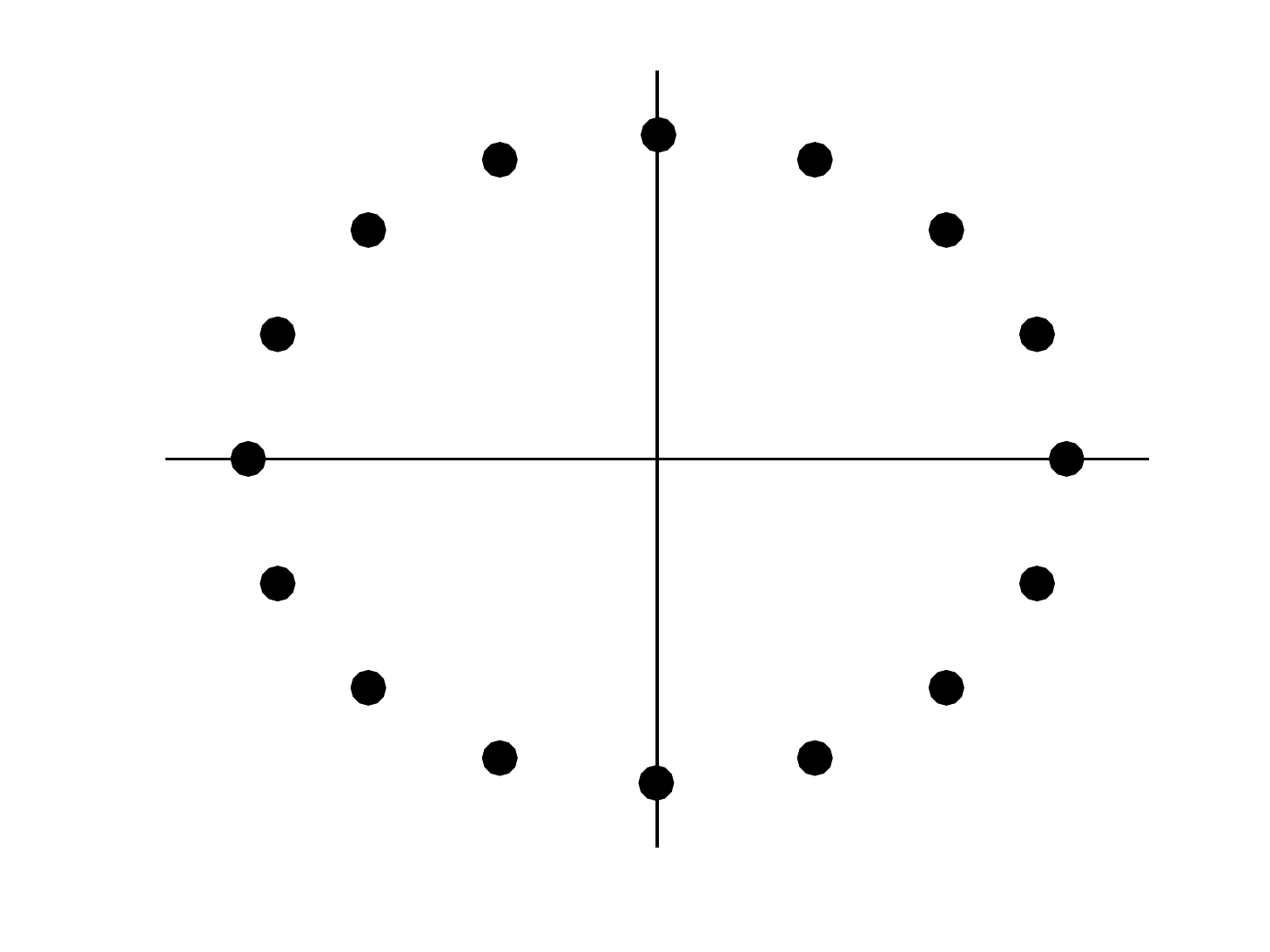}
            \caption[]%
            {{\small Symmetric 16-PSK}}    
            \label{fig:a}
        \end{subfigure}
      \quad
        \begin{subfigure}[b]{0.23\textwidth}  
            \centering 
            \includegraphics[width=\textwidth]{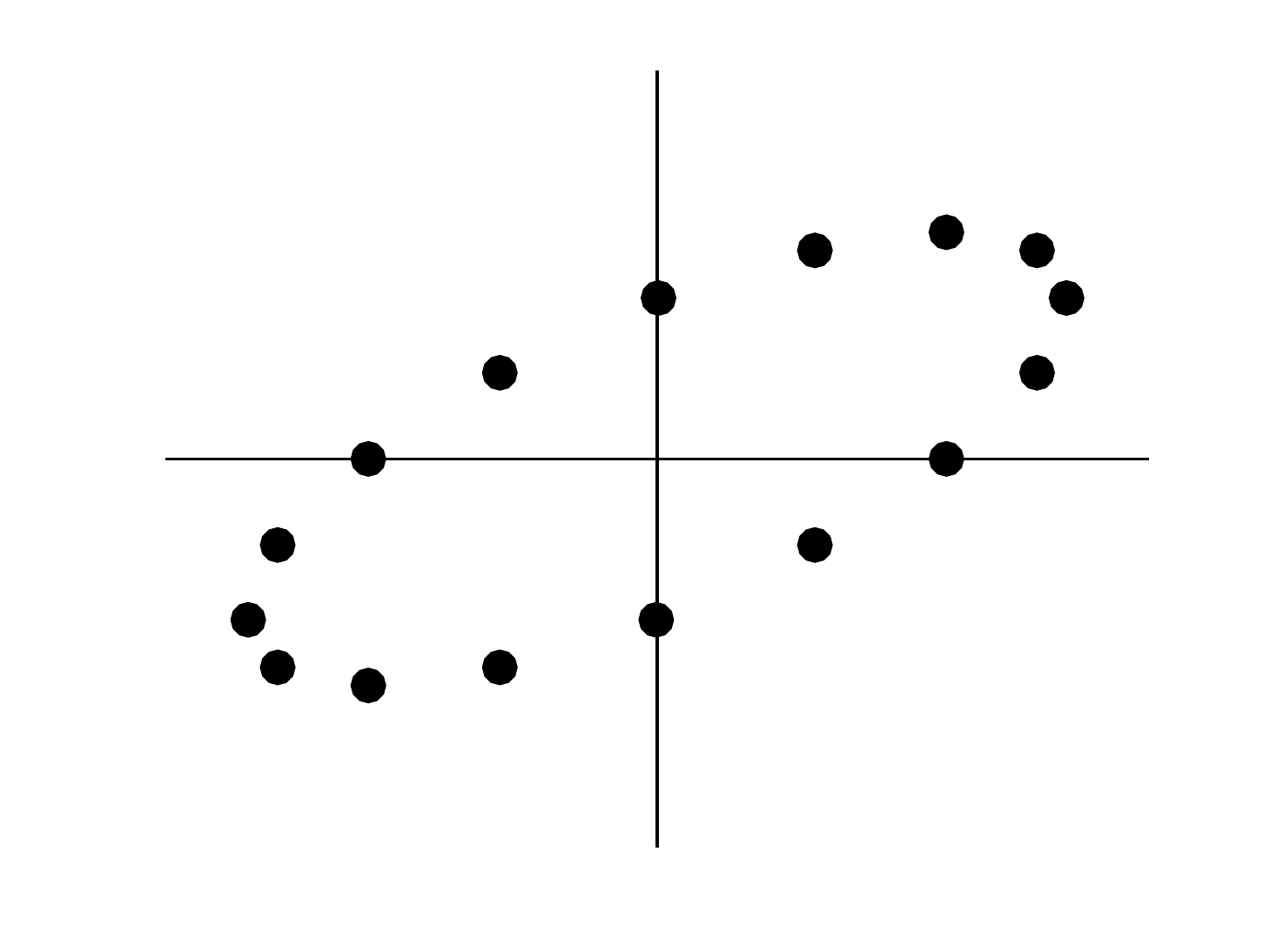}
            \caption[]%
            {{\small Asymmetric 16-PSK}}    
            \label{fig:b}
        \end{subfigure}
        \vskip\baselineskip
        \begin{subfigure}[b]{0.23\textwidth}   
            \centering 
            \includegraphics[width=\textwidth]{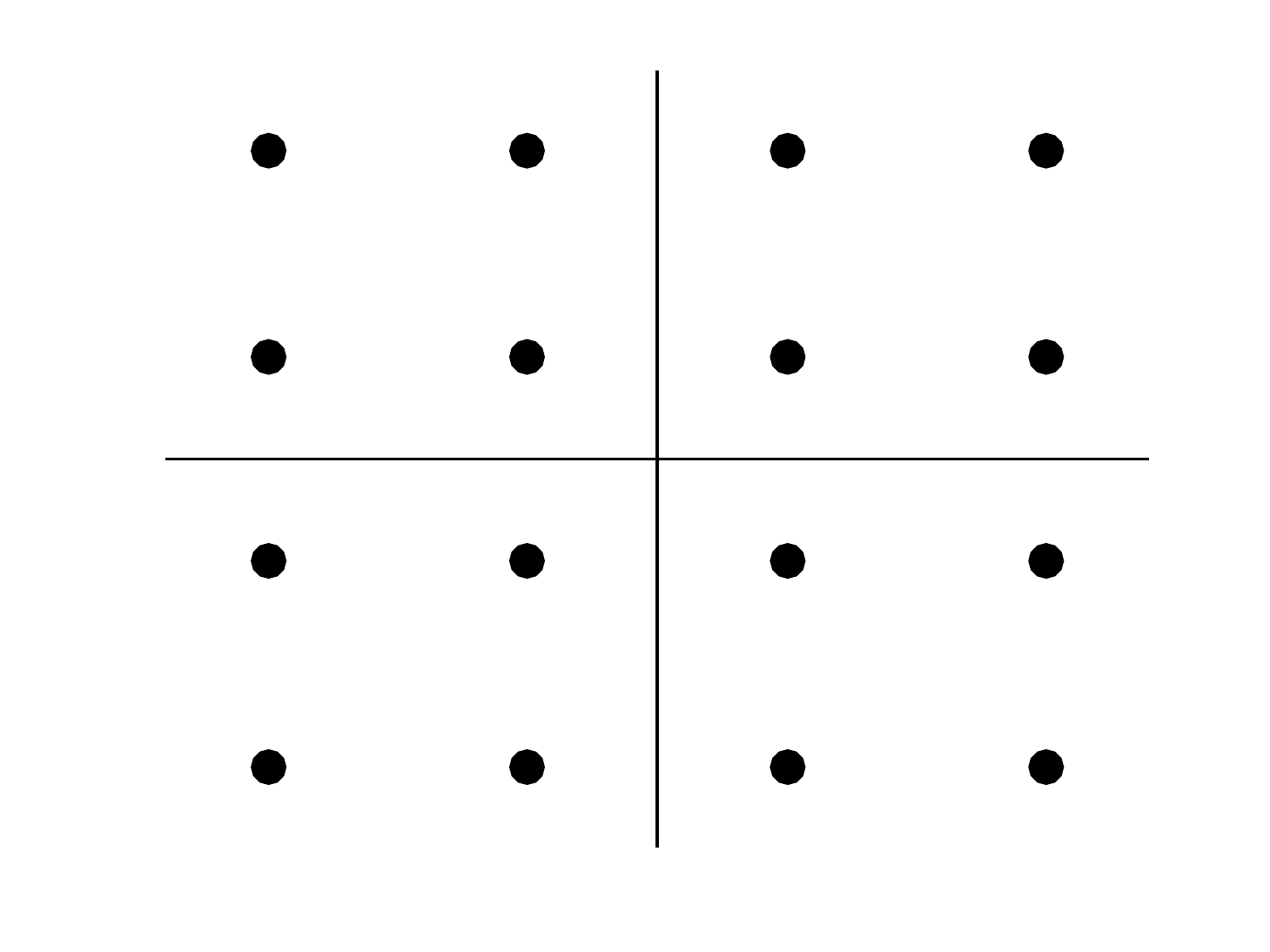}
            \caption[]%
            {{\small Symmetric 16-QAM }}    
            \label{fig:c}
        \end{subfigure}
        \quad
        \begin{subfigure}[b]{0.23\textwidth}   
            \centering 
            \includegraphics[width=\textwidth]{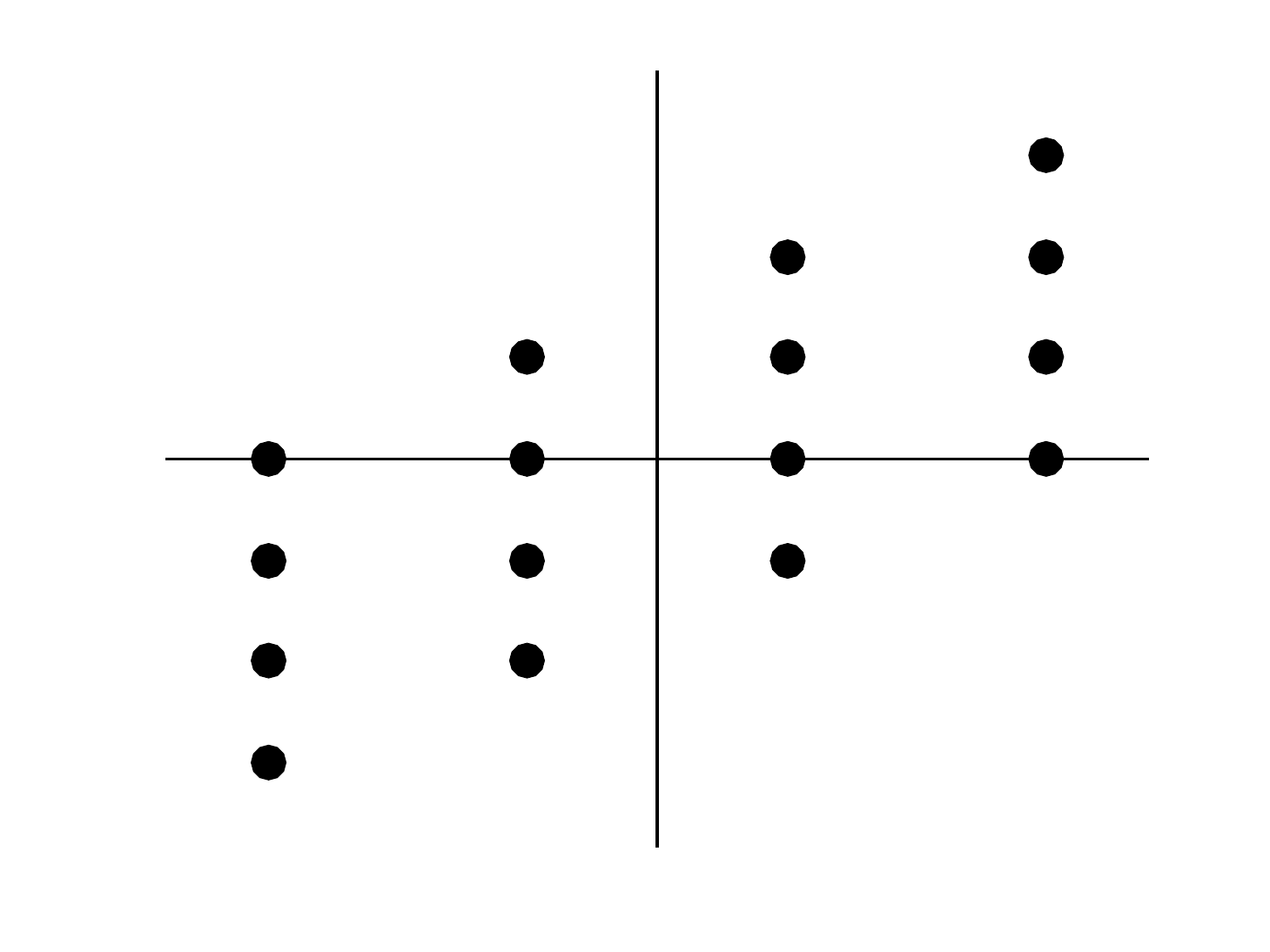}
            \caption[]%
            {{\small Asymmetric 16-QAM}}    
            \label{fig:d}
        \end{subfigure}
        \caption[ ]
        {\small Widely Linear Transformed Signal Constellations }
        \label{fig:WLTconstellations}
    \end{figure}
\subsection{Hardware Imperfections}
Hardware impairments (HWIs) like non-linearity, IQI and phase noise rise in various radio frequency (RF) and baseband blocks such as analog-to-digital convertors, high power and low noise amplifiers, low-pass and band-pass filters etc. \cite{krishnan2015impact, schenk2008rf}.
These impairments accumulate at both the transmitter (TX) and receiver (RX) RF chains and yield undesired effects. For instance, non-linearities result in additive Gaussian distortion noise  \cite{bjornson2013new,xia2015hardware,duy2015proactive}, 
 whereas IQI not only induces phase and amplitude errors but also mixes the desired  and image signals \cite{hakkarainen2013widely,zhang2017widely,schenk2008rf,korpi2014widely}. Considering the information bearing transmitted signal $x_{0} \sim \mathcal{  CN}(0,  \sigma_{x}^2 , 0 )$, the noiseless down-converted IQI signal is modeled as WLT~\cite{boulogeorgos2016q}  
\begin{equation} \label{WLT}
x_{\mathrm{IQI}} = \alpha_1 x_0 + \alpha_2 x_0^*, 
\end{equation} 
 where,  $\alpha_1$ and $ \alpha_2$ are complex scalars which capture the amplitude and phase errors. The WL modeling of IQI is thoroughly investigated and analyzed in the literature
 \cite{madero2015robustness}. However, only few works have acknowledged the asymmetric characteristics induced by this WL structure of IQI \cite{zarei2016q,anttila2008circularity,javed2018multiple,madero2015robustness}. The WLT of IQI transforms PGS to IGS, which can be verified by finding non-zero pseudo-variance of the resulting signal $\tilde{\sigma}_{x_{\rm{IQI}}}^2 =  2 \alpha_1 \alpha_2 \sigma_{x}^2$ \cite{javed2017impact}. Similarly, the RX-IQI is 
 mainly responsible for the improper Gaussian additive noise  \cite{mokhtar2013ofdm,li2014q}. 
 Therefore, transceiver impairments specifically IQI play a vital role in transforming proper transmitted signal to improper received signal as well as transforming proper Gaussian noise to improper Gaussian noise \cite{javed2018improper,soleymani2019improper,alsmadi2018imperfect,li2017noncircular,javed2018multiple}. An example of baseband equivalence of WLT of discrete 16-PSK and 16-QAM owing to TX-IQI is depicted in Fig. \ref{fig:WLTconstellations}. 
 Besides wireless communications, other noise models such as underwater propeller noise from maritime ship also demonstrate an improper nature \cite{clark2012existence}.
  \begin{figure*}[t]
\begin{minipage}[b]{1.0\linewidth}
  \centering
  \centerline{\includegraphics[width=17cm]{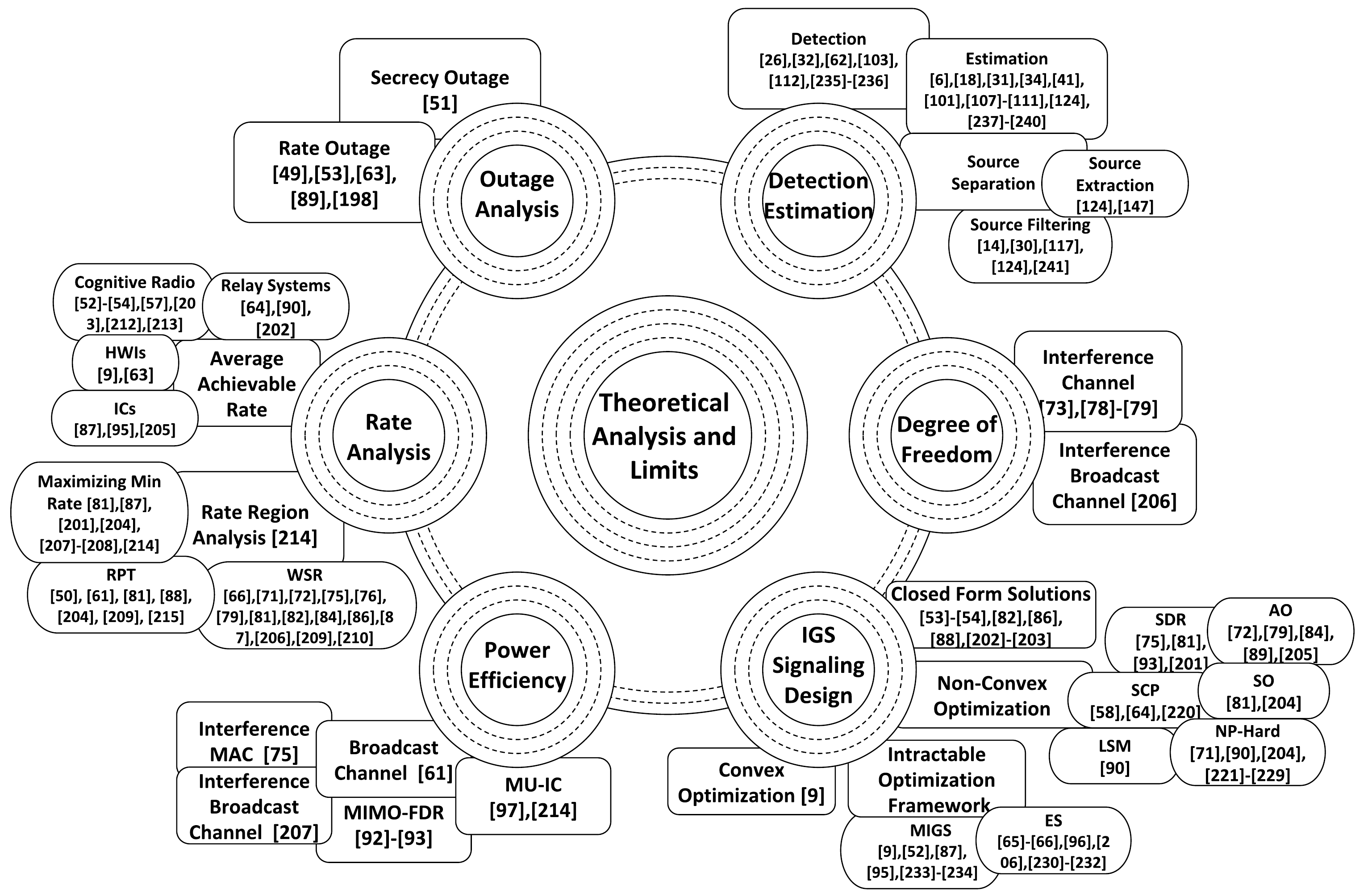}}
\end{minipage}
\caption{{Overview and Configuration of Theoretical Analysis}}
\label{fig:TAPL}
\end{figure*}
\section{Theoretical Analysis and Performance Limits} \label{sec:TAPL}
General analysis of a wireless communication system with improper transmission, interference, or noise should at least accommodate the complete SOS for valid theoretical analysis and performance limits. Improper Gaussian noise or interference are inevitable, however, the employment of proper or improper Gaussian signaling is questionable. Conventional information theoretic studies advocated the capacity achieving PGS in the point-to-point (P2P), BC and MAC \cite{cover2006elements}.  However, recent studies have demonstrated some scenarios and conditions when PGS gains are inferior to that obtained from the IGS owing to the additional design freedom \cite{taubock2012complex}. The performance gains are theoretically evaluated by metrics of interest like achievable rate, outage probability, power efficiency, and DoF. This section highlights the performance limits obtained by both conventional PGS and appealing IGS in various system settings. Followed by the design guidelines to fine tune the IGS transmission parameters in order to exploit the possible performance gains as detailed in Fig. \ref{fig:TAPL}.
\subsection{Achievable Rate}
The performance of a wireless communication system can be theoretically quantified as achievable rate. A rate $R$ is achievable if there exists a sequence of codes such that the maximal probability of error tends
to 0 for sufficiently large block lengths.
Recent contributions have highlighted a more general framework to quantify the achievable rate in order to accommodate the improper nature of the participating signals arising from IGS \cite{lameiro2015benefits,lameiro2018improper,zeng2013improper,
amin2017overlay,lameiro2017rate,javed2018improper,gaafar2016alternate,
hedhly2017interweave,javed2018multiple,kurniawan2015improper,
zeng2013optimized,ho2012improper,hedhly2020benefits}, WL precoding of PGS \cite{zhang2017widely,lagen2014decentralized,lameiro2013degrees,
lagen2016coexisting,lameiro2018performance,lagen2016superiority,
bai2018optimal,lin2018multi,zeng2013transmit}, or complex beamforming resulting in dependent real and imaginary parts \cite{shin2012new,kim2012asymmetric}.  
As an illustrative example, consider the most general case of $K$-user MIMO-IC, where each user intends to communicate with its own RX but results in an interference to other $K-1$ RXs. Each user transmits general IGS $\V{x}_k \in \mathbb{C}^{M}$ with the distribution $\mathcal{CN}\left( \V{0},\M{R}_{\V{x}_k \V{x}_k }, \tilde{\M{R}}_{\V{x}_k \V{x}_k } \right) \forall k=1,\ldots,K$ from $M$ transmitting antennas. Thus, the accumulated received signal vector at user $k$ with $N$ receive antennas is
\begin{equation}
\V{y}_k = \M{H}_{kk} \V{x}_k + \sum\limits_{\ j \neq k, j = 1}^{K} { \M{H}_{kj} \V{x}_j  } + \V{n}_k,
\end{equation}
where, $ \V{n}_k \sim \mathcal{CN}\left( \V{0},\sigma^2 \M{I}_N, \M{0} \right)$ is the CSC Gaussian thermal noise and ${ \M{H}_{kj}} \in \mathbb{C}^{N{\rm x} M}$ is the quasi-static fading channel from TX-$j$ to RX-$k$.
Based on the augmented representations and Schur complement, the differential entropy of $\V{y}_k \in \mathbb{C}^{N} $ and interference plus noise term  $\V{w}_k =  \sum{ \M{H}_{kj} \V{x}_j + \V{n}_k } $ are given as $H(\V{z})$ for $\V{z}=\V{y}_k$ and  $\V{z}=\V{w}_k$, respectively \cite{schreier2010statistical}.
\begin{equation}\nonumber
H(\V{z})\!=\!\log\left( {{{\left( {\pi e} \right)}^{N}}\left| {{\M{R}_{{\V{zz}}}}} \right|} \right)\!+\! \frac{1}{2}\log \left( \left|\M{I}\! -\! {{\M{R}}_{\V{zz}}^{ - 1}}{\tilde{\M{R}}_{\V{zz}}}{{\M{R}}_{\V{zz}}^{\! -\! \RM{T}}}{{\tilde{\M{R}}_{\V{zz}}}^{\RM{H}}} \right|\right).
\end{equation}
Shannon’s capacity formula assumes additive Gaussian noise and  coded transmission with codewords drawn from a Gaussian codebook. Therefore, the instantaneous achievable rate per unit bandwidth (spectral
efficiency) at RX-$k$ with IGS can be obtained as \cite{zeng2013transmit}
\begin{align} \label{eqR}
R_k = &  \mathcal{I} \left( \V{y}_k;\V{x}_k  \right)=H( \V{y}_k)-H( \V{y}_k | \V{x}_k)=H( \V{y}_k)-H( \V{w}_k) \nonumber \\
& = \frac{1}{2}\log \frac{{\M{R}}_{{\underline{\V{y}}_k}{\underline{\V{y}}_k}}}{{\M{R}}_{{\underline{\V{w}}_k}{\underline{\V{w}}_k}}}= \underbrace{
\log \frac{\left|\sigma^2 \M{I}+   \sum\limits_{\ j = 1}^{K} { \M{H}_{kj}   \M{R}_{\V{x}_j \V{x}_j }{\M{H}_{kj}^{\RM H}} }  \right|}{\left|\sigma^2 \M{I}+   \sum\limits_{\ j \neq k, j = 1}^{K} { \M{H}_{kj}   \M{R}_{\V{x}_j \V{x}_j }{\M{H}_{kj}^{\RM H}} } \right|}}_{\triangleq  R_{k,\text{proper}} \left(  \M{R}_{\V{x}_j \V{x}_j } \right)   }  \nonumber \\
& \quad + \frac{1}{2} \log \frac{ \left|\M{I} - {{\M{R}}_{\V{y}_k \V{y}_k }^{ - 1}}{\tilde{\M{R}}_{\V{y}_k \V{y}_k }}{{\M{R}}_{\V{y}_k \V{y}_k }^{ - \RM{T}}}{{\tilde{\M{R}}_{\V{y}_k \V{y}_k }}^{\RM{H}}} \right|}{ \left|\M{I} - {{\M{R}}_{\V{w}_k \V{w}_k }^{ - 1}}{\tilde{\M{R}}_{\V{w}_k \V{w}_k }}{{\M{R}}_{\V{w}_k \V{w}_k }^{ - \RM{T}}}{{\tilde{\M{R}}_{\V{w}_k \V{w}_k }}^{\RM{H}}} \right|}. 
\end{align}
The achievable rate $R_k $ at RX-$k$ has two components, the first term $R_{k,\text{proper}} \left(  \M{R}_{\V{x}_j \V{x}_j } \right) $ is the conventional achievable rate attainable by traditional PGS scheme as it only depends on covariance $  \M{R}_{\V{x}_j \V{x}_j } $. Whereas, the additional term is the consequence of employing IGS 
as it is a function of both $  \M{R}_{\V{x}_j \V{x}_j } $ and $ \tilde{ \M{R}}_{\V{x}_j \V{x}_j } $. Therefore, IGS yields another privilege to appropriately choose/optimize  $ \tilde{ \M{R}}_{\V{x}_j \V{x}_j } $ under the power constraints to yield the second term in \eqref{eqR} strictly positive. Interestingly, this property cannot be exploited in the equivalent real composite domain analysis \cite{zeng2013transmit}. 

The rate expression obtained for multiuser (MU) MIMO-IC can be easily reduced to MU single-input single-output (SISO)-IC \cite{zeng2013improper}, SISO Z-IC  \cite{lameiro2017rate}, single-input multiple-output (SIMO) systems with hardware distortions (HWDs) \cite{javed2017asymmetric}, MU multiple-input single-output (MISO)-IC \cite{zeng2013optimized}, MU MIMO-IC \cite{lagen2016coexisting}, MU MIMO-BC \cite{bai2018optimal}, MIMO-IBC \cite{lin2018multi}, cognitive radio settings \cite{lameiro2015benefits,lameiro2018improper,amin2017overlay,
hedhly2017interweave}, point-to-point (P2P) systems with asymmetric HWIs \cite{javed2018multiple}, relaying systems \cite{gaafar2016alternate}, full-duplex systems \cite{javed2018improper} and other similar interference-limited scenarios where interference can be treated as Gaussian noise. Various studies argue that IGS can offer significant achievable rate improvement relative to the PGS. 
Optimizing the covariance and pseudo-covariance matrices allow us to optimize the system performance to maximize the average achievable rate or achievable sum rate, minimize achievable rate, or to decide achievable Pareto rate region boundaries~\cite{zeng2013transmit}. 
\subsubsection{Average Achievable Rate Limits}
Average achievable rate or ergodic rate is an insightful performance metric to quantify the capability of a network to transmit the number of bits per second per Hertz. This section summarizes the average achievable rate performance gains of IGS over PGS in various interference limited scenarios arising in cognitive radio schemes \cite{lameiro2015benefits,amin2017overlay,hedhly2017interweave}, multi- antenna or MU systems \cite{lagen2014decentralized,lameiro2018performance,lagen2016superiority,javed2018multiple},  and relay systems \cite{kim2012asymmetric,javed2018improper,gaafar2016alternate} etc. 
 
 In a \textbf{cognitive radio} setting for dynamic spectrum access, the unlicensed secondary user (SU) coexists with a licensed primary user (PU) and opportunistically utilize its spectrum resources to improve the overall spectral efficiency \cite{yucek2009survey,yeo2011optimal}. However, this approach renders undesirable interference on the legit primary network. It is noteworthy that the primary TX sticks to PGS as it is using its own spectrum being unaware of the secondary IC \cite{amin2016underlay,lameiro2015benefits}. Interestingly, the least entropy loss due to IGS makes it a suitable transmission scheme for SU as the improper interference on PU is far less detrimental compared to proper interference. Thus, IGS is the preferred choice for SU to maximize its achievable rate while maintaining PU rate or quality-of-service (QoS) constraints. The average achievable rate performance gain of SU with IGS over PGS are analyzed in three different cognitive setups i.e., underlay, overlay and interweave.  
\begin{itemize}
\item In an underlay cognitive system, IGS is only beneficial
when the ratio of the squared modulus between the SU-PU
interference link and the SU direct link exceeds a given threshold. Upon meeting this criteria, the SU adopts IGS as optimal signaling and demonstrates magnified relative performance gains up to 256\% over PGS especially when PU is not heavily loaded~\cite{lameiro2015benefits,lameiro2016maximally,gaafar2015spectrum}.
\item In an overlay cognitive system, where SU broadcasts a mixture of PGS and IGS to aid the primary message transmission and minimize the interference effect of SU on PU respectively. 
 The optimized IGS offers 33.33\% performance gain for 30dB SNR with partial CSI, while meeting PU quality of service (QoS). This gain is conditional and improves with decreasing primary network direct-link gain~\cite{amin2017overlay,amin2017impact}. 
\item In an interweave cognitive system, the achievable rates of both the PU and the SU depend on the activity of the PU and the detection ability of the SU. Employed IGS yield percentage increase up to 8.26\% over PGS with 50\% probability of detection (PoD) and 10\% PFA. The gain is significant especially at low sensing and detection capabilities of the SU, lower PU direct link and higher SU interference on the PU side \cite{hedhly2017interweave}.
\end{itemize}

Inspired by the perks of IGS in interference limited environments,
various contributions have reaped IGS benefits in full duplex (FD)/half-duplex (HD) multi-hop decode-and-forward (DF) \textbf{relay systems} by effectively compensating residual self-interference (RSI), inter-relay interference (IRI), and/or HWDs \cite{kim2012asymmetric,javed2018improper,gaafar2016alternate}. 
For instance, IGS is proposed in single antenna DF-FD relay channels to eliminate the self-interference (SI) and increase the throughput. This scheme with SI-elimination provides significant improvement of 107.14\% over the conventional symmetric
signaling without SI-elimination \cite{kim2012asymmetric}.
 IGS is further employed to compensate not only RSI but also improper 
HWDs in multi-hop DF-FDR system. IGS parameters can be centrally optimized at one node or distributively optimized at multiple nodes to maximize the end-to-end (E2E) achievable rate. Centralized and distributed optimization offer relative performance gains of 166.67\% and 80\%, respectively. Gain can rise up to 355.56\% with centralized approach for higher interference-to-noise ratio with increased communication overhead and system complexity \cite{javed2018improper}. 
Moreover, the potential benefits of IGS are also exploited in two-hop alternate relaying system to relieve IRI and maximize total achievable rate in the absence of CSI at source. Two detection schemes are employed based on the IRI level relative to the desired signal level i.e., 1) for low IRI: single-user decoding (SUD) treating IRI as noise and 2) for strong IRI: successive decoding (SD) that first decodes IRI and subtracts it from received signal before detection. Average rate percentage improvement up to 161.54\% and 66.67\% is achieved with SUD and adaptive scheme that switches between SUD and SD, respectively. This gain is especially significant when the source-relay channel is a bottleneck \cite{gaafar2016alternate}.

IGS superiority is also proven in multi-antenna setup to mitigate various \textbf{HWIs} \cite{javed2017asymmetric,javed2018multiple}. Interestingly, aggregate impairments specially IQI transform symmetric noise and transmitted signal to asymmetric noise and received signal, respectively, motivating the employment of IGS to combat them. Javed \textit{et al.} proposed optimal IGS and maximal IGS schemes for tractable and intractable optimization of SIMO and MIMO systems, respectively. Percentage average achievable rate improvement of 7.76\% and 3\% is depicted in highly impaired SIMO systems with optimal IGS and adaptive maximal IGS-PGS, respectively. Similarly, the employment of maximal IGS yields up to 10\% increase in average achievable rate of MIMO relative to PGS even in low SNR regime \cite{javed2018multiple}. 

Other contributions highlight the perks of IGS in single user \cite{lagen2016superiority} and \textbf{MU MIMO-ICs} \cite{lameiro2018performance,lagen2014decentralized}. 
Firstly, IGS transmission with uniform- (UPA) or optimal- (OPA) power allocation strategies is exploited in MIMO P2P channel with interference (P2P-I) to improve the achievable rate. Relative rate gains of OPA-UPA of IGS over PGS are 102-115\% for 1$\rm x$1, 46-37\% for
2 $\rm x$2, 24-20\% for 4$\rm x$4, and 17-13\% for 8$\rm x$8 in MIMO P2P-I \cite{lagen2016superiority}. Secondly, for 2-user (2U) MIMO-IC, transmit covariance matrix is designed based on two maximal IGS schemes i.e., Improper-LB that provides the SU with the minimal rate, and Improper-UB that provides the SU with the maximum rate \cite{lameiro2018performance}.
Percentage relative gain of user-1 and user-2 achievable rate is 195.45\% increase and 8.7\% decrease with Improper-LB relative to PGS. However, the percentage relative gain of user-1 and user-2 achievable rate is  10.53\% decrease and 209.52\% increase with Improper-UB relative to PGS. Interestingly, in an interference limited scenario, if the IGS scheme is beneficial for interferer, it will be detrimental for the sufferer and vice versa.
Lastly, for MU MIMO-IC transmission rates increase with IGS and transmit coordination and IGS instead of conventional interference management (IM) with PGS. Lagen \textit{et al.} claims IGS performance gains over PGS in terms of mean user throughput are 10.64\%, 13.95\% and 22.92\% with centralized-IM, decentralized-IM, and no-IM, respectively \cite{lagen2014decentralized}.

{
\subsubsection{Rate Region Analysis}
Pareto boundary of the achievable rate region comprises of all the Pareto optimal points, which are defined as:
\begin{definition}{17}[Pareto-Optimal]
The rate pair (${R}_1,{R}_2$) is called Pareto-optimal if (${\bar{R}_1},{{R}_2}$) and (${{R}_1},{\bar{R}_2}$ ), with ${\bar{R}_1}> {{R}_1}$ and ${\bar{R}_2}>{{R}_2}$, are not achievable \cite{lameiro2017rate}. 
\end{definition}
Interestingly, IGS and/or WL processing significantly improves the achievable rate region in MU interference setup. Various studies evaluated such improved rate regions whereas many others focused on the boundary of an achievable rate region, called the Pareto
boundary, based on sum rate analysis, rate profile technique (RPT), or minimum weighted rate maximization \cite{soleymani2020performance}. 
  \begin{table}[htbp]
\renewcommand{\arraystretch}{1.25}
\caption{Achievable Rate Improvement by IGS over PGS in Different Settings}
\begin{center}
  \begin{tabular}{||c|p{1.1cm}|p{1.1cm}|p{1.1cm}|c|c|c|c||}
    \toprule
 \textbf{System} & \textbf{Tech} & \textbf{Desc} & \textbf{IGS-TX} &
\textbf{Metric} & \textbf{Procedure} & \textbf{Improvement}   & \textbf{Ref} \\
      \midrule
   \multirow{4}{*}{\shortstack[c]{\\Cognitive \\ Radio}} & \multirow{2}{*}{Underlay}&P2P&SU&{\shortstack[c]{\\Max SU-Rate with \\ PU-Rate Constraint}}& Maximal IGS & up to 256\%& \cite{lameiro2015benefits}    \\ 
\cline{3-8}
 & &S-MAC& {\shortstack[c]{\\S-MAC\\ Users}}  & {\shortstack[c]{\\Rate Region Boundary \\ Maximize Sum Rate}}  &   {\shortstack[c]{\\Algorithm: Closed- \\Form Expressions}} &  {\shortstack[c]{\\R1:133\%, R2:117\%\\ up to 200\%}}& \cite{lameiro2018improper} \\ 
\cline{2-8}
& Overlay& P2P &SU& \multirow{2}{*}{{\shortstack[c]{\\Max SU-Rate with\\PU-QoS Constraint}} }  & \multirow{2}{*}{{\shortstack[c]{\\Piecewise Closed-\\Form Solution}} }& up to 191.7\%& \cite{amin2017overlay} \\ 
\cline{2-4}
\cline{7-8}
 & {\shortstack[c]{\\Interweave}}&P2P&{\shortstack[c]{SU}} &&& up to 8.3\%&\cite{hedhly2017interweave}\\ 
   \hline
    \hline
\multirow{9}{*}{MIMO}   & Massive &MU DL & BS & {\shortstack[c]{\\Maximize \\Sum Rate }} & {\shortstack[c]{\\Numerical \\ Evaluation}} & {\shortstack[c]{\\ ZF/MMSE:75\% \\ BD:97\%}} & \cite{zhang2017widely} \\
       
          \cline{2-8}

      & \multirow{3}{*}{X-IC} & 2U & SU & Maximize SU-Rate &  Maximal IGS & {\shortstack[c]{\\ R1:195.5\% increase  \\ R2:8.7\% decrease}} & \cite{lameiro2018performance} \\
   
   \cline{3-8}
      & &  MU & All & Minimize MSE &  AO & {\shortstack[c]{\\ cent-IM:10.6\%  \\ decent-IM:13.9\% \\ no-IM:22.9\%}} & \cite{lagen2014decentralized} \\
   \cline{3-8}
   &  & {\shortstack[c]{\\LP/WLP \\Users }} & {\shortstack[c]{\\Only WL \\Users }} &{\shortstack[c]{\\Maximize WSR \\($ \equiv$ Minimize WMSE)}} &  BCD with AO & {\shortstack[c]{\\ WLP:133\%, \\HetTX(1):84.2\%  \\ HetTX(K/2):28.9\%}} & \cite{lagen2016coexisting} \\

    \cline{2-8}
      & Z-IC & OPA & Interferer & {\shortstack[c]{\\Average Sum Rate \\ Min-Rate Performance}} & Maximal IGS  &{\shortstack[c]{\\Sum rate:55.2\% \\ Min rate:640\% }}  & \cite{lagen2016superiority} \\
    \cline{2-8}
         & BC & {\shortstack[c]{Multicell\\N/w}} & BS & \multirow{2}{*}{Maximize WSR }&BCD with AO & up to 11.8\% & \cite{bai2018optimal} \\
   \cline{2-4}
    \cline{6-8}
            & \multirow{2}{*}{IBC} & HCRAN &{\shortstack[c]{\\Femto BSs \\in all cells }} &  &{\shortstack[c]{\\WMMSE and \\ ADMM for AO }}  & {\shortstack[c]{\\WMMSE-IGS:12.5\% \\ WMMSE-PGS:21.7\% }} & \cite{lin2018multi} \\
                     \cline{3-8}
     & & { MU }& {BS} &  {Maximize min rate}&  {Path following Algo} & 30dBm: 37.93\% & \cite{nasir2019improper} \\
            \cline{2-8}
               & P2P-I & UPA/OPA & MIMO TX & {\shortstack[c]{\\Maximize Rate  \\ Minimize MSE}}  & {\shortstack[c]{\\Majorization\\ Theory Tools}} & {\shortstack[c]{\\Rate Gap:1.55\\ MSE GAP:0.2 \\ Rate Gain:115\% }}  &  \cite{lagen2016superiority}\\
    \hline
    \hline 
 \multirow{4}{*}{  MISO }& IC & MU P2P  &  Users & {\shortstack[c]{\\Pareto Rate Region\\ Max-Min Rate}} &  {\shortstack[c]{\\SDR with SOCP \\ and GR}}&  {\shortstack[c]{\\R1:70\%, R2:25\% \\ up to 42.8\%}}  & \cite{zeng2013optimized} \\
\cline{2-8}
   & \multirow{2}{*}{ BC} & 2U & BS & Pareto Rate Region &  {\shortstack[c]{\\Separate Optimization \\ with Bisection Search}} & R1:150\%, R2:100\%  & \cite{zeng2013optimized} \\
       \cline{3-8}
     & & 3U & BS & Pareto Rate Region &  {\shortstack[c]{\\Gradient based \\ Rate Balancing}} & {\shortstack[c]{\\R1:100\%,\\ R2/R3:42.86\%}}  & \cite{hellings2017reduced} \\
            \cline{2-8}
     &  NOMA &  {\shortstack[c]{Multicell\\N/w}}&  {BS} & {Max-min Fairness} &   {\shortstack[c]{\\LMI based Path\\ following Algorithm}} &  {\shortstack[c]{NOMA: 87.50\% \\ OMA:121.05\%}} &  \cite{tuan2019non} \\
    \hline
    \hline
   SIMO & P2P &  HWI &  TX & {\shortstack[c]{\\ Maximize \\  Achievable Rate }}  &  {\shortstack[c]{\\Convex QCQP \\  using IPM}} & up to 10\%  & \cite{javed2018multiple} \\
   \hline
    \hline  
 \multirow{6}{*}{  SISO }&  \multirow{6}{*}{X-IC} &   \multirow{4}{*}{2U} & {1U} & {Rate Region} & {Closed-form} & {R1:133.7\%,R2:86.6\%}  & \cite{soleymani2019robust}  \\
 \cline{4-8}
  &&&   \multirow{7}{*}{All users} & {\shortstack[c]{\\Max-Min Fairness\\ Proportional Fairness}} &{\shortstack[c]{\\Closed-Form\\ Solution }}& {\shortstack[c]{\\ up to 83\% \\ up to 633\%}}  & \cite{ho2012improper} \\
  \cline{5-8}
 & & & & Max-Min Rate & SDR & up to 228.57\%  & \cite{zeng2013optimized} \\
  \cline{5-8}
&&& &{\shortstack[c]{\\ Pareto Rate Region\\ Max-Min Rate \\ Sum Rate}}  & SDR for QCQP &  {\shortstack[c]{\\ R1:357\%, R2:389\% \\ JO:58.6\%, SO:54.8\%  \\JO:21.5\%, SO:18.8\%}} & \cite{zeng2013transmit} \\
\cline{3-3}
\cline{5-8}
& &   4U&  & {\shortstack[c]{\\ Maximize \\ Sum Rate}}  & {\shortstack[c]{\\ Alternating \\ Minimization }} &  up to 35.2\% & \cite{lameiro2013degrees} \\
\cline{3-3}
\cline{5-8}
 &  &  MU &  & {\shortstack[c]{\\ Max-Min \\ Weighted Rate }}  & SDR and GR & {\shortstack[c]{\\2U:221.4\%,\\ 3U:304.3\%}} & \cite{zeng2013improper} \\
   \bottomrule
  \end{tabular}
\end{center}
\label{tab:Rate}
\end{table}

  \begin{table}[htbp]
\renewcommand{\arraystretch}{1.25}
\begin{center}
  \begin{tabular}{||c|p{1.1cm}|p{1.3cm}|p{1.3cm}|c|c|c|c||}
    \toprule
\multirow{5}{*}{SISO}  & \multirow{3}{*}{Z-IC} &\multirow{2}{*}{P2P}  &  &  {\shortstack[c]{\\ Maximize\\ Sum Rate }}  & \multirow{3}{*}{ {\shortstack[c]{\\ Closed-Form\\ Solution }} }& up to 30.8\%  & \cite{kurniawan2015improper} \\
\cline{5-5}
\cline{7-8}
  & & &&  \multirow{2}{*}{ {\shortstack[c]{\\ Pareto \\Rate Region }} } &&  R1:83\%, R2:150\%   &  \cite{lameiro2017rate} \\
    \cline{7-8}
  & & &&  && {R1:30.9\%,R2:36.2\%}&{\cite{soleymani2019robust}}   \\
   \cline{2-8}
&  {BC-NOMA} &  { 2U} &  {BS }&  {Max Sum Rate} & {KKT conditions} &  $\Delta P= 0.4{\rm P_t}$: 23.60\%& \cite{mahady2019sum}  \\   \cline{2-8}
  & IBC & {\shortstack[c]{\\MU in \\ 3-Cells }}  & {\shortstack[c]{\\Each BS \\ in all Cells }}   & {\shortstack[c]{\\ Maximize \\ Sum Rate }}  &{\shortstack[c]{\\ Exhaustive \\ Search }}   & {\shortstack[c]{\\ $K$=5: 13.7\%\\ $K$=10:17.6\% \\$K$=20:20.6\% }}   & \cite{shin2012new} \\
 \hline
  \hline
\multirow{3}{*}{{ \shortstack[c]{\\ DF \\ Relays }} }& \multirow{2}{*}{ FD} & {\shortstack[c]{\\Multi-hop \\ with HWIs }}  &\multirow{2}{*}{{\shortstack[c]{\\Source \\ and Relays }}}  &{\shortstack[c]{\\Maximize \\ E2E Rate }}   & SCP for QCLP   & {\shortstack[c]{\\Centralized:\\ up to 355.6\% \\ Distributed:\\ up to 80\%}}    & \cite{javed2018improper} \\
  \cline{3-3}
 \cline{5-8}
 &   & Dual-hop&  & Throughput  &  {\shortstack[c]{\\Line Search \\ Method}}   &    up to 107\%& \cite{kim2012asymmetric} \\

   \cline{2-8}
  &  HD & {\shortstack[c]{\\Alternate\\  Relaying }}  &Relays only & {\shortstack[c]{\\Maximize Total \\  Achievable Rate }}  &   {\shortstack[c]{\\Piecewise Closed-\\Form Solution  }} &  {\shortstack[c]{\\SUD:161.5\%
 \\ SUD-SD:66.7\% }}   & \cite{gaafar2016alternate} \\
    \bottomrule
  \end{tabular}
\end{center}
\end{table}
\paragraph{Sum Rate Analysis}
Pareto boundary of the rate region can be acquired by maximizing the sum rate when dealing with multiple nodes. Impropriety characterization renders substantial increase in the pareto boundaries as observed in interference MAC \cite{kariminezhad2016improper,kariminezhad2017interference},  
Z-IC \cite{kurniawan2015improper}, and X-IC \cite{ho2012improper}. Apart from the complete boundary characterization, some studies focused on the weighted sum-rate (WSR) maximization achieved by IGS in various system settings like cognitive radio \cite{lameiro2018improper}, massive MIMO with IQI \cite{zhang2017widely}, X-IC \cite{soleymani2019robust,zeng2013transmit,lameiro2013degrees,lagen2016coexisting}, Z-IC \cite{lagen2016superiority,kurniawan2015improper}, BC  \cite{bai2018optimal,mahady2019sum}, and IBC \cite{shin2012new,lin2018multi}.}

{In \textbf{underlay cognitive radio} scheme, IGS has emerged as a promising candidate to improve the sum rates of the MAC users. Firstly, for \textbf{primary-MAC} (P-MAC) interfered by a P2P channel, IGS and symbol extensions can improve the achievable rates up to three times based on the interferer strength~\cite{kariminezhad2016improper,kariminezhad2017interference}. 
Secondly, underlay 2U \textbf{secondary-MAC} (S-MAC) exploited IGS to improve average sum rates up to 200\% \cite{lameiro2018improper}. Moving on to the \textbf{large-scale MIMO} systems with TX-IQI,  Zhang \textit{et al.} analyzed WLP algorithms based on ZF (WL-ZF), matched filter (WL-MF), WL-MMSE, and block-diagonalization (WL-BD) for the downlink (DL) scenario \cite{zhang2017widely}. They argued the achievability of same multiplexing gains and WSR with WL-ZF and WL-BD in the presence of IQI as their counterparts ZF and BD in the absence of IQI. However, this performance is attained at the expense of minor power loss owing to the increased system scale. Interestingly, the WSR analysis of IQI system with WL-ZF/WL-MMSE and WL-BD depicted percentage increase up to 97.67\% and 75\% over ZF/MMSE and BD for single- and multi-antenna users,  respectively \cite{zhang2017widely}. }
\newline

{Rate region improvement of \textbf{Z-IC} has also been studied extensively \cite{kurniawan2015improper}. This contribution employs real-composite representation for easy optimization and characterizes only one point of rate region \cite{kurniawan2015improper}.
Improved sum rates offered by IGS with effective IM in Z-IC are also widely investigated in 2U SISO Z-IC. PGS is preferred for weak interference whereas optimal IGS can provide WSR improvements up to 30.8\% in strong interference regime \cite{kurniawan2015improper}.}

{Sum rate analysis is also extended to \textbf{X-IC} in various multi-user and multi-antenna setups. For instance, Ho \textit{et al.} explore the Pareto region for the 2U SISO-IC with cooperative (IGS) as well as non-cooperative (PGS) transmission strategy. They prioritize improper rank one signals over full-rank signals because of their simplicity, easy implementation and close to optimal sum rate \cite{ho2012improper}. The study focuses on improving the system efficiency in terms of max-min fairness and proportional fairness while carrying out the rate region analysis. Evidently, at the max-min fairness, both users share the same maximum possible rate in Pareto region however at proportional fairness, the aim is to maximize the product of improvement over the Nash equilibrium. IGS provides remarkable percentage max-min fairness and proportional fairness improvements up to 83.33\% and  633.33\%, respectively, in the medium SNR regime. The rate region improvement is more substantial for decreasing SNR and asymmetric channel i.e., one IC is stronger than the other \cite{ho2012improper}. On the other hand, the Pareto region attained with minimum mean square error (MMSE) scheme is larger than and contains the corresponding region with ZF scheme. On the other hand, for 2U SISO-IC, Zeng \textit{et al.} propose a joint (JO) and separate (SO) IGS optimization framework which achieves 21.52\% and 18.85\% WSR improvement, respectively, relative to PGS scheme \cite{zeng2013transmit}. Similarly, Soleymani \textit{et al.} report average sum rate increase up to 150\% in  2U SISO-IC even with imperfect CSI \cite{soleymani2019robust}. Moreover, Lameiro \textit{et al.} illustrate WSR improvement up to 35.14\% at 60dB SNR with IGS and linear interference alignment (IA) for 4U SISO-IC 
\cite{lameiro2013degrees}. The WSR analysis is also extended from MU SISO-IC to MU MIMO-IC in a transitional heterogeneous (HetTX) setting where some legacy linear transceivers i.e., linear precoding and linear estimation (LP-LE) coexist with other WL transceivers i.e., WL precoding and WL estimation (WLP-WLE) \cite{lagen2016coexisting}. This work addresses WLT filter design to maximize WSR and presents iterative procedure to solve equivalent minimum weighted-mean square error (W-MSE) problem. Transition from LP-LE to WLP-WLE, HetTX(1)-WLE (1 LP,$K-1$ WLP), and HetTX($K/2$)-WLE ($K/2$ LP,$K/2$ WLP) achieve the percentage improvements up to 133.33\%, 84.21\%, and 28.95\%, respectively. Interestingly, this performance gain increases with increasing number of users, increasing aggregate interference levels or decreasing number of antennas. Surprisingly, the WL transceivers with no interference coordination performed worse than the linear transceivers with full coordination among users \cite{lagen2016coexisting}.}


{Apart from IC, WSR maximization problem is also extensively studied for \textbf{BC} and \textbf{IBC} \cite{bai2018optimal,shin2012new,lin2018multi,mahady2019sum}.  
IGS offers significant sum-rate maximization for the downlink non-orthogonal multiple access i.e., non-OMA (NOMA) with imperfect successive interference cancellation (SIC). In a 2U SISO-BC NOMA setup, IGS optimization based on Karush-Kuhn-Tucker (KKT) conditions renders 
23.60\% and 18.71\% sum-rate improvement when the power difference between two users is 40\% and 20\% of the total power (${\rm P_t}$), respectively \cite{mahady2019sum}.
Similarly, increasing number of users, receiver antennas and transmission power of the base-station (BS) offer considerable WSR gain of 11.84\% with WL design in MU MIMO-BC \cite{bai2018optimal}.
In contrast to the IC and BC, few studies focused on analyzing the potential benefits of IGS in the combined IBC \cite{shin2012new,lin2018multi}.
For instance, Shin \textit{et al.}  propose a new IA strategy based on IGS and MU diversity (MUD) for 3-cell SISO-IBC where each BS covers $K$ users per cell. The percentage WSR improvement of 13.68\%, 17.65\%, and 20.56\% are achieved with the proposed strategy relative to conventional IA strategies for $K=5$, $K=10$ and $K=20$ users, respectively, at 20dB SNR \cite{shin2012new}. 
WSR maximization problem in MIMO-IBC (e.g., heterogeneous cloud radio access network (H-CRAN)) is a non-trivial extension of MIMO-IC \cite{lagen2016coexisting} and MIMO-BC \cite{hellings2015iterative}.
Thus, Lin \textit{et al.} propose a distributed beamforming algorithm for separate optimization. This algorithm outperforms existing WMMSE with IGS and PGS in terms of WSR by 12.5\% and 21.74\%, respectively~\cite{lin2018multi}.}
  
{
\paragraph{Rate Profile Technique}
Rate region boundary can also be established using RPT instead of maximizing the sum rate. This technique also advocates impropriety incorporation to render improved pareto boundaries in numerous interference-limited setups like 
underlay MAC \cite{lameiro2018improper},
Z-IC \cite{lameiro2017rate,soleymani2019robust},
X-IC  \cite{soleymani2019robust,zeng2013transmit,
 soleymani2019ergodic,zeng2013optimized}, and BC 
\cite{zeng2013optimized,hellings2017reduced}.}

{Lameiro \textit{et al.} extended their work in underlay P2P cognitive system \cite{lameiro2015benefits} to \textbf{underlay MAC} setup in order to study the improved rate region by IGS. The IGS transmitting unlicensed S-MAC coexists with PGS transmitting licensed primary link. The numerical results for 2U S-MAC with zero-forcing (ZF) decoding present the rate improvements up to 117\% and 133\% for SU-2 and SU-1, respectively. IGS guarantees rate improvement if the sum of IC gains is above a certain threshold and surprisingly, the relative gain increases with increasing number of users \cite{lameiro2018improper}.}

{Rate region improvement of \textbf{Z-IC} has also been studied extensively \cite{lameiro2017rate}. The contribution \cite{lameiro2017rate} employs augmented complex representation for more insightful analysis and characterizes entire rate region boundary.  Lameiro \textit{et al.} emphasize the conditional optimality of IGS in SISO Z-IC that attains 83.33\% and 150\% percentage increase in the rates of user-1 (R1) and user-2 (R2), respectively, at the maximum sum rate point on Pareto boundary \cite{lameiro2017rate}.  Likewise,  Soleymani \textit{et al.} claim 30.95\% and 36.25\% increase in R1 and R2, respectively, in a 2U SISO Z-IC with IGS transmission under imperfect CSI~\cite{soleymani2019robust}.}

{The perks of IGS are not only limited to Z-IC but also extend to \textbf{X-IC}. By far, 2U SISO-IC is the mostly studied X-IC for the employment of IGS with some substantial results \cite{soleymani2019robust,soleymani2019ergodic,
zeng2013transmit,zeng2013optimized}. For instance, Soleymani \textit{et al.} claim 86.67\% and 133.77\% increase in R2 and R1, respectively, with IGS in a 2U SISO X-IC under imperfect CSI of the interfering links \cite{soleymani2019robust}. Moreover, they also propose a practical IGS scheme i.e., maximal IGS (which does not require any optimization) for 2U SISO-IC with Rayleigh fading. They again present substantial increase in Ergodic rate region as well as sum-rate with IGS under strong interference \cite{soleymani2019ergodic}. Similarly,  Zeng \textit{et al.} present significant increase in the Pareto rate region of 2U SISO-IC as IGS improves up to 388.89\% and 357.14\% rates for user-2 and user-1, respectively \cite{zeng2013transmit}.
Additionally, for a given the optimal rank-1 transmit
covariance matrices, rank-1 pseudo-covariance matrices are proven optimal for
achievable rate region in MU MISO-IC. Percentage improvements of R1:25\% and R2:70\% are attained for 2U MISO-IC \cite{zeng2013optimized}.}

{IGS can render effective interference suppression in \textbf{BCs} when interference is treated as noise.
For instance, 2U MISO-BC can achieve up to R1:100\% and R2:150\% rate improvement \cite{zeng2013optimized}. Similarly, for a 3U MISO-BC, IGS offers R2/R3:42.86\% and R1:100\% increase relative to PGS~\cite{hellings2017reduced}.}

{
\paragraph{Minimum Achievable Rate}
%
%
IGS transmission can also be optimized in order to maximize the minimum achievable rate of the system. Such contributions analyzed MU SISO-IC \cite{zeng2013transmit,zeng2013improper}, MU MISO-IBC \cite{tuan2019non}, MU MISO-IC  \cite{zeng2013optimized}, MU MIMO Z-IC  \cite{lagen2016superiority}, MU MIMO-IC \cite{soleymani2020performance}, and MU MIMO-IBC~\cite{nasir2019improper}.} 

{Firstly, average max-min rate improvement of 228.57\% and 42.86\% is reported in 2U \textbf{SISO-IC} with joint optimal and 3U \textbf{MISO-IC} with suboptimal optimization, respectively \cite{zeng2013optimized}. Similarly, percentage improvements
up to 58.62\% and 54.83\% in average max-min rate based on the JO and SO, respectively, are observed in 2U SISO-IC \cite{zeng2013transmit}. Furthermore, numerical results revealed up to 221.43\% and 304.26\% improvement in 2U and 3U SISO-IC, respectively. The performance gains further improved with increasing SNR and number of users~\cite{zeng2013improper}.}

{Secondly, IGS under OMA or NOMA is reported to exhibit almost two-fold gain in users’ minimum throughput, combating both intra-cell and inter-cell interference, in a MU multi-cell broadcast network. The proposed IGS design algorithms based on linear
matrix inequality (LMI) optimization result upto 87.50\% and 121.05\% improvement in worst user rate relative to PGS in 4U, \textbf{3-cell MISO-IBC} under NOMA and OMA schemes, respectively~\cite{tuan2019non}.}

{Lastly, the IGS benefits to maximize the minimum user rate are also reaped in MU MIMO ICs. For instance, IGS offers up to 640\% improvement in min-rate performance of 2U \textbf{MIMO Z-IC} \cite{lagen2016superiority}. Moreover, a hardware impaired MU MIMO-IC depicts more than 80\% improvement in fairness rate with IGS over PGS in a 10U setup \cite{soleymani2020performance}. 
 Similarly, a path following IGS optimization algorithm yields up to 37.93\% max-min rate improvement at 30dBm in a \textbf{MIMO-IBC} with 3 cells and 6U/cell \cite{nasir2019improper}.}

{In a nutshell, various contributions presented rigorous analysis and substantial results to demonstrate significant achievable rate improvements for different interference limited scenarios as summarized in Table \ref{tab:Rate}. }

\subsection{Outage Probability}
Several researchers resort to
more conceptual analysis such as outage
probability, i.e., the probability of the event when the system performance falls below a pre-defined threshold. Such abstracted analysis sacrifices
the model depth for simplicity leading to simple expressions
that characterize high-level network behavior, highlight general
trade-offs, and facilitate network design. Outage probability can be quantified in terms of rate outage, SNR outage, secrecy outage and error outage etc. Of all these, rate and secrecy outage are focused as they suitably reflect the IGS system operation quality. 
\subsubsection{Rate Outage Probability (ROP)}
ROP is generally defined as the probability of the event $\Pr \{ \text{Rate} < \text{Threshold} \}$. Various contributions have demonstrated substantial decrease in the ROP when IGS transmission is adopted as opposed to PGS. These studies analyzed various systems including but not limited to cognitive radio \cite{amin2016underlay,gaafar2017underlay}, multi-antenna \cite{javed2017asymmetric} and relay systems \cite{gaafar2018full} suffering from external interference or internal  HWIs. 

The utility of IGS at SU in underlay \textbf{cognitive radio} setting is supported by the claim that it decreases SU ROP by 
77.5\% while meeting PU QoS and SU power constraints. The performance gain of IGS increases as the interference-to-noise ratio (INR) of the SU to the PU increase for a certain SU target rate  \cite{amin2016underlay}. In another spectrum sharing scenario with coexisting FD PU and HD SU, IGS offers up to 91.43\% reduction in the ROP when the allowable INR at the PUs exceed a certain threshold and the SU follows a maximum allowable target rate \cite{gaafar2017underlay}. Gaafar \textit{et al.} extended their work to mitigate RSI in \textbf{FDR} using IGS under Nakagami-$m$ fading.
The observed percentage decrease in ROP of IGS relative to PGS with maximal power allocation (MPGS) and PGS with optimal power allocation (OPGS) is 90\% and 68.18\%, respectively, with 5W power budget at relay. Unlike PGS, IGS maintained a fixed performance even with increasing RSI \cite{gaafar2018full}.
 \begin{table}[t]	
\renewcommand{\arraystretch}{1.25}
\caption{Outage Probability Reduction with IGS Relative to PGS}
\begin{center}
  \begin{tabular}{||c|c|c|c|c|c|c|c||}
    \toprule
 \textbf{System} & \textbf{Technology} & \textbf{Description} & \textbf{Improper TX} &
\textbf{Metric} & \textbf{Procedure} & \textbf{Improvement}   & \textbf{Ref} \\
      \midrule
    { \shortstack[c]{\\Multiple \\Antennas }} &SIMO&HWDs&Single TX &{\shortstack[c]{\\Minimize \\Rate Outage}}& {\shortstack[c]{\\Closed-Form \\Solution}}&{\shortstack[c]{\\SISO:85\% \\ SIMO:100\%}}& \cite{javed2017asymmetric}    \\
        \hline
        { \shortstack[c]{\\DF\\Relays}} & FDR &Dual-hop&Relays &{\shortstack[c]{\\Minimize \\ E2E Outage}}& {\shortstack[c]{\\Coordinate \\Descent }}& {\shortstack[c]{\\MPGS: 90\% \\OPGS: 68.18\%}}& \cite{gaafar2018full}    \\
              \hline
   \multirow{3}{*}{\shortstack[c]{ Cognitive \\ Radio \\ Setting }} & \multirow{3}{*}{Underlay}&HD-PU&  \multirow{3}{*}{SU only} & \multirow{2}{*}{{\shortstack[c]{\\Minimize \\Rate Outage}}}&  {\shortstack[c]{\\Closed-Form\\Solution}}& up to 77.5\%& \cite{amin2016underlay}    \\ 
\cline{3-3}
\cline{6-8}
  & &FD-PU& & & {\shortstack[c]{\\Algorithm\\Design}}& up to 91.43\%& \cite{gaafar2017underlay}    \\ 
\cline{3-3}
\cline{5-8}
  & &Eavesdropper& &{\shortstack[c]{\\Minimize \\Secrecy Outage}} & Numerically&up to 95\%& \cite{oliveira2018physical}    \\ 
 \bottomrule
  \end{tabular}
\end{center}
\label{tab:Outage}
\end{table}

Furthermore, Javed \textit{et al.} demonstrated the effectiveness of IGS transmission to efficiently combat the drastic effects of asymmetric HWDs on \textbf{SIMO} systems. IGS parameters were optimized to maximize instantaneous achievable rate and consequently reduce ROP up to 100\% and 85\%
for the adopted SIMO \cite{javed2017asymmetric} and SISO \cite{javed2017impact} system, respectively. The ROP gain in imperfect hardware system is especially significant with increased distortion levels and more receiver streams $N_{\rm R}$  for any given threshold rate.
\subsubsection{Secrecy Outage Probability (SOP)}
A secrecy outage is the probability of event when the mutual information
of the desired  link (A$\rightarrow$B) is lower/equal to that of the undesired link (A$\rightarrow$E) i.e., $\Pr \{ \mathcal{I}_{\text{AB}} -  \mathcal{I}_{\text{AE}} < \text{Threshold} \}$. Consider the \textbf{underlay cognitive} setup with primary network link source-destination (S$\rightarrow$D) and a secondary link Alice-Bob (A$\rightarrow$B) in the presence of an eavesdropper Eve (E). Alice adopts IGS scheme in order to reduce the SOP. It is evident from the fact that IGS demonstrate lower differential entropy than PGS. Thus, an improper interference will be less harmful on the achievable rates relative to a regular proper interference when interference is treated as noise. Therefore, this allows Alice to transmit with higher power and reap higher achievable rates without violating S$\rightarrow$D licensed link. Nonetheless, SOP reduction is a trade-off between impropriety and power of the transmitted signals.
Oliveira \textit{et al.} claim to be the first to establish IGS superiority in physical layer security of cognitive setup. They present closed formulation for the SOP and revealed up to 95\% lower SU-SOP with IGS  \cite{oliveira2018physical}. 

\begin{figure}[t]
\begin{minipage}[b]{1.0\linewidth}
  \centering
  \centerline{\includegraphics[width=8.5cm]{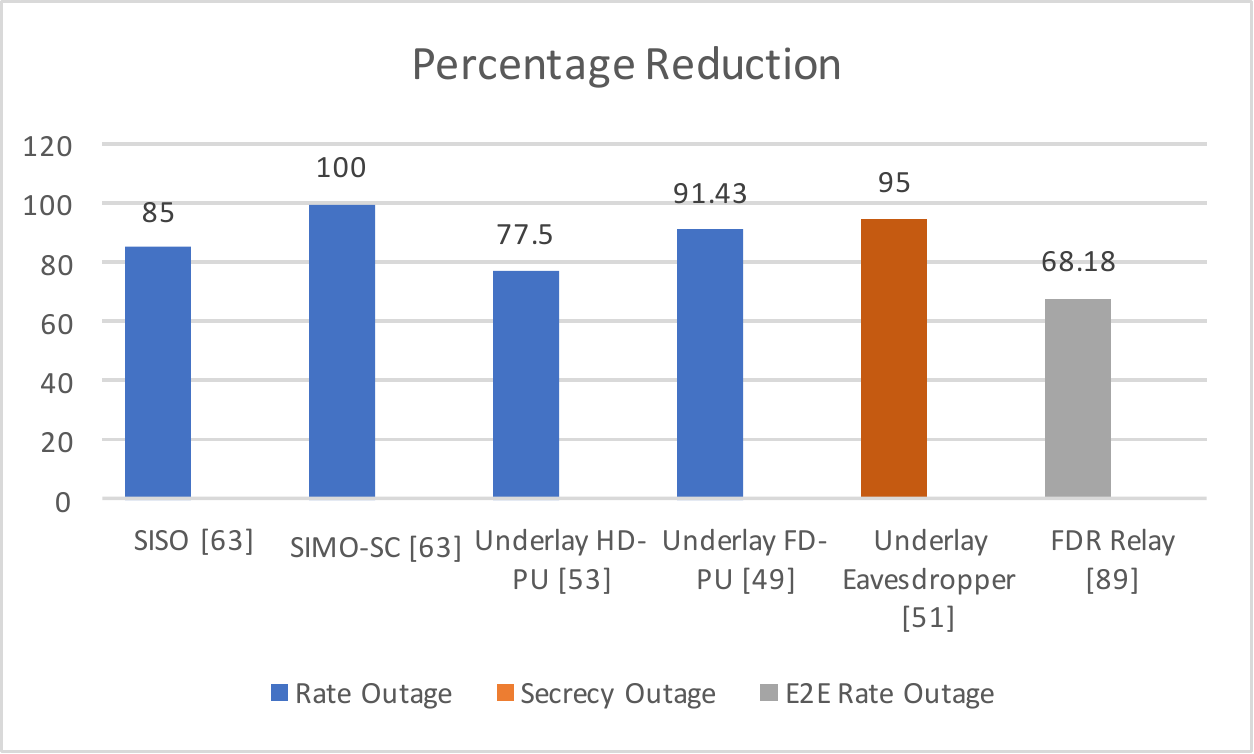}}
\end{minipage}
\caption{Percentage Reduction in Outage Probability with IGS }
\label{fig:Outage}
\end{figure}

All of these contribution support the advantageous trend of IGS in reducing outage probability of the adopted systems suffering from interference or improper noise as highlighted in Table \ref{tab:Outage} and Fig. \ref{fig:Outage}.
\subsection{Power Efficiency}
Interestingly, IGS allows achieving the desired QoS while spending less power at the TXs owing to the additional tunable parameter. This leads to power efficient solutions for various MU, multi-antenna, and relay setups. IGS transmission with and without symbol extensions for \textbf{MAC} interfered by a P2P channel renders up to 32\% and 40\% power saving ratios, respectively \cite{kariminezhad2016improper}.
Likewise, IGS optimization offers up to 19.38\%, 8.47\% and 3.5\% energy efficiency with 4, 5 and 6 transmitting BS antennas  in a \textbf{MU MIMO-IBC} with 3 cells and 6Us per cell \cite{nasir2019improper}. 

IGS is further incorporated for joint rate-energy optimization in multi-antenna heterogeneous two-tier networks where users are subject to TX noise, MU interference and RSI. Although these impairments are detrimental for achievable rate, they are beneficial for energy harvesting because they carry RF energy \cite{kariminezhad2016heterogeneous}.
Energy harvesting of 4$\mu$watts/channel-use is achieved with 0.5bits/channel-use achievable rate in FD D2D 2U pairs coexisting with HD \textbf{MU-BC} in a cellular network. Similarly, for a single-carrier \textbf{MISO BC} the decrease in transmit power requirements by IGS over optimal PGS is 1dB, 3dB and 12dB for 2,4 and 6 users, respectively \cite{hellings2017reduced}.

IGS is beneficial to improve the power efficiency of 
one-way DF-FD MIMO relay \cite{kariminezhad2017power} and two-way AF-FD MIMO relay \cite{zhang2013widely} for MU interference networks. MU network with \textbf{MIMO-FDR} is not only prone to inter-user interference and transceiver noise but also self-interference due to MIMO-FDR relay operation. Optimal IGS at the sources and relay requires 0.95dB minimum sum-power requirement relative to 1.2dB, 1.6dB and 1.75dB for OPGS, PGS-ZF and PGS-MRC, respectively, under QoS demand of 1 bit/channel-use by two-pair MIMO relay network. Power efficiency attained by IGS scheme improves with the increasing interference levels and TX noise for the entire range of rate demands. Interestingly, only IGS scheme is capable of meeting high rate demands by single real streaming whereas PGS fails to do so even with infinite transmission powers \cite{kariminezhad2017power}. Additionally, AF FD MIMO relay system enhances power efficiency through SNR balancing or transmit power minimization using WL transmit strategies. By fine-tuning non-circularity of WL complex transmitted signals significant performance gains up to 150\% percentage increase in average minimum SINR and 5dBW less average minimum  relay transmit power relative to linear precoding schemes \cite{zhang2013widely}.
The WL gain increases with increasing number of pairs but decreases with increasing number of relay antennas.
{Although IGS offers significant power/energy efficiency gains over PGS, the relative energy efficiency benefits are less than rate benefits \cite{soleymani2020performance}.}

\subsection{Degrees of Freedom}
Apart from the discussed metrics, IGS is also beneficial to improve the achievable 
DoF in a given interference scenario including but not limited to X-IC \cite{lameiro2013degrees,yang2014interference}, BC \cite{hellings2013qos}and IBC \cite{shin2012new}. For instance, Lameiro \textit{et al.} illustrated improved DoF $\approx 4/3$ (requiring minimal symbol extensions) with IGS and linear IA for 4U \textbf{SISO-IC} \cite{lameiro2013degrees}. Similarly, Yang \textit{et al.} demonstrated achievable DoF of 0.5 more than the outer bound of DoF of \textbf{MIMO X-IC} when spatial IA and ZF framework are employed along with the IGS  \cite{yang2014interference}. Moreover, in \textbf{MIMO-BC} system with WL transceivers IGS per-user transmit signals outperform PGS counterpart under QoS constraints without time-sharing \cite{hellings2013qos}. IGS guarantees faster convergence of  
 effective DoF to its upper bound relative to PGS for a given rate requirement \cite{hellings2013qos}. Extension to \textbf{3-cell SISO-IBC} new IA strategy based on IGS and MUD revealed 1.5 DoF (1 by IGS + 0.5 by MUD) whereas higher cells required symbol extensions to achieve proportional DoFs \cite{shin2012new}. 
\subsection{IGS Signaling Design}
Theoretical analysis of the IGS revealed tremendous payoffs in various MU scenarios such as the X-IC
\cite{nguyen2015improper,lagen2016coexisting,zeng2013transmit,ho2012improper}, Z-IC
\cite{lagen2016superiority,lagen2014improper,kurniawan2015improper,lameiro2017rate}, BC
\cite{zeng2013miso,hellings2013performance}, cognitive radio networks
\cite{hedhly2017interweave,lameiro2015benefits,amin2016underlay,amin2017overlay}, HWI systems
\cite{javed2018asymmetric,xia2015hardware,soleymani2019improper,javed2017asymmetric}
and relay channels
\cite{gaafar2018full,javed2018improper,gaafar2016alternate}. The vast majority of them assumed
Gaussian codebooks for more efficient IM \cite{santamaria2018information}. IGS scheme requires optimization of the following parameters based on the underlying presentation.
\begin{itemize}
\item Transmit covariance and pseudo-covariance matrices with complex representation which reduces to power and circularity coefficient for small-scale systems.
\item Augmented covariance matrix with complex augmented representation
\item Composite covariance matrix with real-composite representation
\end{itemize}
These parameters are optimized in order to maximize some performance metric like ergodic rate, minimum rate, sum rate, or Pareto rate region. They can also be fine-tuned to minimize ROP, SOP, or MSE etc. This section includes the guidelines and insights of the popular design problems with celebrated optimization techniques.
\subsubsection{Closed Form Solutions}
The convexity of the IGS parameter optimization problem depends on the objective function as well as constraints. Few objective functions like complete characterization of Pareto boundary in \textbf{SISO Z-IC} and \textbf{SISO-IC} (by restriction to rank-1 transmit
covariance matrices) \cite{lameiro2017rate,ho2012improper} and maximizing sum-rate in SISO Z-IC \cite{kurniawan2015improper} yield closed form solutions. Alternately, other optimization problems produce piece-wise closed form solutions. Consider an \textbf{overlay cognitive radio} \cite{amin2017overlay} or \textbf{interweave cognitive radio}  \cite{hedhly2017interweave} setting which involve licensed primary and unlicensed secondary link. The conventional PU is transmitting PGS whereas the SU employs optimized IGS to maximize its achievable rate $R_s$ while meeting PU QoS requirements  $R_p\left( {p_s,\mathcal{C}_x} \right)  \geq R_{\text{min}}$ ($R_p$ is the achievable rate of the PU). The optimization problem is similar to $\textbf{P1}$ and tunes IGS transmission parameters of SU i.e., transmit power $p_s$ and circularity coefficient $ \mathcal{C}_x $.
\begin{align}
 \textbf{P1}:\;   \mathop {\max }\limits_{p_s,\mathcal{C}_x} \quad & R_s\left( {p_s,\mathcal{C}_x} \right) \nonumber \\ 
 {\rm{s. \;t.}} \quad & R_p\left( {p_s,\mathcal{C}_x} \right)  \geq R_{\text{min}}, \nonumber \\
  &0 \leq p_s \leq p_{s,\text{max}}, \nonumber \\
    &0 \leq  \mathcal{C}_x    \leq 1. \nonumber
\end{align}
SU achievable rate is maximized under PU QoS constraint such that the interference from SU can be limited. This ensures a minimum achievable rate for the PU to maintain QoS. Moreover, $p_s$ is constrained under the transmission power budget $p_{s,\text{max}}$ and $ \mathcal{C}_x $ range decides the transmission to be anywhere between proper and maximally improper. Likewise, maximizing total achievable rate in alternate \textbf{HD-DF relay} system \cite{gaafar2016alternate} is equivalent to $\textbf{P1}$ with maximum power transmission eliminating any primary rate constraint and can also be solved as piece-wise closed form solution.

Another form of $\textbf{P1}$ is the ROP minimization problem $\textbf{P2}$ in \textbf{cognitive radio} setup. The SU optimizes IGS transmission parameters $p_s $ and $\mathcal{C}_x $ to minimize its ROP i.e., $P_{\text{out},s}\left( {p_s,\mathcal{C}_x} \right)=\text{Pr}\{  R_s\left( {p_s,\mathcal{C}_x} \right) \le R_{\text{min}}\}$  while  maintaining PU QoS $P_{\text{out},p}\left( {p_s,\mathcal{C}_x} \right)  \leq P_{\text{out,th}}$ \cite{amin2016underlay}.
\begin{align}
 \textbf{P2}:\;   \mathop {\min }\limits_{p_s,\mathcal{C}_x} \quad & P_{\text{out},s}\left( {p_s,\mathcal{C}_x} \right) \nonumber \\ 
 {\rm{s. \;t.}} \quad & P_{\text{out},p}\left( {p_s,\mathcal{C}_x} \right)  \leq P_{\text{out,th}}, \nonumber \\
  &0 \leq p_s \leq p_{s,\text{max}}, \nonumber \\
    &0 \leq  \mathcal{C}_x    \leq 1. \nonumber
\end{align}
Interestingly, problems similar to  $\textbf{P2}$ can also be solved in closed form by investigating the monotonic trend of the objective function with respect to the optimization variables.
\subsubsection{Convex Optimization}
The closed-form solutions of the convex optimization problems are attractive but not always achievable. For instance, the achievable rate maximization problem $\textbf{P3}$ of \textbf{SIMO P2P} system with transceiver HWIs reformulates as a quadratic-constraint quadratic programming (QCQP) problem. The IGS transmission parameters enclosed in $\V{s} = {\left[  \Re \{ \tilde \sigma _x^2\}  \;\Im \{ \tilde \sigma _x^2\} \; \sigma _x^2 \; \right]^T}$ are efficiently optimized using interior point method (IPM) pertaining to the convex quadratic constraints \cite{javed2018multiple}.
\begin{align}
 \textbf{P3}:\;  \mathop {\max }\limits_{\V{s}}\quad & {{\text{R}}_{{\text{SIMO-IGS}}}}\left( \V{s} \right)     \nonumber \\
\rm{\;s.\; t.} \quad & {\M{A}_1}\V{s} \leqslant {\bf b}, \nonumber \\
 & {\V{s}^{\rm T}}{\M{A}_2}\V{s} \leqslant {\bf 0}, \nonumber
\end{align}
where  ${\M{A}}_1 = {\text{diag}} \left[ 0 \; 0 \; 1 \right]^{\rm T}$ and ${\bf{b}}=  \left[ 0 \; 0 \; \rm{P}_{\rm T} \right]^{\rm T}$ signify the transmission power constraint whereas ${\M{A}}_2 = {\text{diag}}\left[ 1 \; 1 \; -1 \right]^{\rm T}$ tracks the magnitude of transmit pseudo-covariance. Thankfully, the complexity of convex optimization problems is polynomial in the problem dimension. However, such fancy convex optimization problems which yield elegant solutions are occasional. 
{\subsubsection{Non-Convex Optimization}
Most of the IGS design problems are non-convex in nature and require exponential efforts. Interestingly, all non-convex problems are not hard but lack convexity owing to their inappropriate formulation. As a matter of fact, many non-convex optimization problems admit a convex reformulation using relaxation approaches like semi-definite relaxation (SDR) and sequential convex programming (SCP). Moreover, for separable problems, alternate optimization is preferred over separate optimization if the underlying sub-problems are convex. Furthermore, line-search methods with gradient descent or Newton method converge to a local solution for unconstrained optimization. However,  the NP-hard class of non-convex problems requires a different treatment. 
\paragraph{Semi-Definite Programming}
The joint optimization of covariance ${\M{R}}_{{x}_k{x}_k}$ and pseudo-covariance $\tilde{\M{R}}_{{x}_k{x}_k}$ matrices to achieve the Pareto-optimal rates emerges is a non-convex problem like $ \textbf{P4}$ in case of \textbf{2U Gaussian SISO-IC} \cite{zeng2013transmit}.
\begin{align}
 \textbf{P4}:\;   \mathop {\max }\limits_{{\M{R}}_{{x}_k{x}_k},\tilde{\M{R}}_{{x}_k{x}_k},R} \quad & R \nonumber \\ 
 {\rm{s. \;t.}} \quad & {R_k}  \geq \alpha_k R,  \; \forall k \nonumber \\
  &0 \leq  \text{Tr} \left({\M{R}}_{{x}_k{x}_k} \right) \leq  P_k,  \; \forall k \nonumber \\
    &  {\M{R}}_{\underline{x}_k \underline{x}_k}  \succeq  \M{0}, \; \text{or}   \;
    0 \leq  |\tilde {\mathcal{C}}_{x_k}|^2    \leq {\mathcal{C}_{x_k}^2}  \; \forall k \nonumber
\end{align}
where $\alpha_k$ is the target ratio between user $k$'s achievable rate ${R_k}$ and the sum-rate of all users $R$. Zeng \textit{et al.} suggests SDR to transform non-convex QCQP (as in $\textbf{P4}$) to quasi-convex semi-definite programming problem which can either be solved using bisection search or Gaussian Randomization (GR) procedure based on the achievable rank-1 constrained solution \cite{zeng2013transmit}. The extension of this problem to MU SISO-IC is a sequence of non-convex minimum weighted rate maximization problems which cannot be solved optimally. Thus, SDR technique along with GR provides an efficient approximation to jointly optimize the transmission parameters~\cite{zeng2013improper}.} 

{The non-convex Pareto rate-region boundary characterization problem of \textbf{interfering MAC} setup can also be transformed from quadratic to linear form using SDP. Moreover, SDR of rank constraints convexifies the problem at hand which can then be solved using IPM. This problem is equivalent to $\textbf{P4}$ except that the objective is to maximize $R_\Sigma$ in place of auxiliary variable $R$ and the first constraint is replaced by the individual link capacity constraint i.e., $\alpha_q R_\Sigma \leq L_q \forall q $. The solution of the relaxed problem can then be projected into the feasible set of the original problem using GR \cite{kariminezhad2016improper}.}

{Another important problem is the sum-power minimization in \textbf{MIMO FDR} for MU interference networks with QoS demands of the communicating pairs. IGS design to tackle RSI and IQI  with minimal power requirement leads to an SDP optimization problem with non-convex constraint set.
Therefore, linearization of the second concave function in difference of concave (DC) programming problem 
using Fenchel’s inequality \cite{boyd2004convex} is suggested \cite{kariminezhad2017power}.
\paragraph{Sequential Convex Programming}
Transmit parameters of the source and participating relays can be optimized in a \textbf{Multi-hop DF-FDR} system to improve E2E achievable rate by efficiently mitigating RSI and HWDs. Javed \textit{et al.} propose maximum allowable power transmission $\sigma _m^2 = P_t$ along with the optimized pseudo-variances  $\tilde \sigma _m^2 $ for all transmitting nodes to promote fairness \cite{javed2018improper}. Considering the achievable link rate between nodes $m$ and $n$ as ${{R_{nm}}\left( {\tilde \sigma _m^2,\tilde \sigma _n^2} \right)}$,  ${\textbf{P5}}$ aims at maximizing the minimum link rate, i.e.,
\begin{align}
{{\textbf{P5}}:\quad }&{\mathop {\rm{maximize} }  \quad \mathop {\min }\limits_n \left\{ {{R_{nm}}\left( {\tilde \sigma _m^2,\tilde \sigma _n^2} \right)} \right\}} \nonumber \\
{}&{{\rm{subject \; to}} \quad 0 \le \left| {\tilde \sigma _m ^2} \right| \le \sigma _m ^2,\;\forall m.} \nonumber
\end{align}
The joint optimization ${\textbf{P5}}$ turned out to be a max-min fractional programming problem which can be solved in two different ways.
\begin{enumerate}
\item Generalized Dinkelbach algorithm (GDA) \cite{zappone2015energy} can transform non-linear fractional programming to non-linear parametric programming and SCP can transform the resultant DC problem to QCQP convex optimization problem \cite{javed2017full}
\item Alternately, the logarithmic properties can transform fractional programming problem to DC problem and eliminate GDA step. However, SCP is inevitable to solve further \cite{javed2018improper}.
\end{enumerate}
SCP for DC programming approximates second concave to its first-order Taylor series expansion and solves the resultant convex problem iteratively.
\paragraph{Alternate Optimization (AO)}
Joint optimization of transmission parameters is not always manageable therefore some researchers suggest AO method to iteratively improve the approximate solution of maximizing WSR problems. For instance, Lagen \textit{et al.} propose block coordinate descent (BCD) algorithm for AO in \textbf{MU MIMO-IC} with heterogeneous (some LT and WLT) deployment \cite{lagen2016coexisting} and decentralized processing  \cite{lagen2014decentralized}. They emphasize WLP of the information symbols $\V{b}_k$ from $k^{th}$-user i.e., $\V{x}_k = \M{T}_{1,k} \V{b}_k + \M{T}_{2,k}{\V{b}_k^*}$ and WLE from the received vector $\V{y}_k$ at $k^{th}$-user i.e., $\hat{\V{b}}_k = {\M{R}_{1,k}^{\RM  H}} \V{y}_k + {\M{R}_{2,k}^{\RM H}}{\V{y}_k^*} $. The goal is to design these precoding ${\M{T}_{1,k}},{\M{T}_{2,k}}$ and estimation matrices ${\M{R}_{1,k}},{\M{R}_{2,k}}$ for all users to maximize WSR by equivalently minimizing W-MSE with weights $\mu_k$, i.e.,
\begin{align}
{{\textbf{P6}}:\quad }&{\mathop {\min }\limits_{\scriptstyle{\M{T}_{1,k}},{\M{T}_{2,k}}\hfill\atop
\scriptstyle{\M{R}_{1,k}},{\M{R}_{2,k}}\hfill} \sum\limits_{k \in \mathcal{K}} \frac{\mu_k}{2}\log_2 \left|\M{E}_k {\M{F}_k^*} \right|  } \nonumber \\
{}&{{\rm{s.t.}} \quad \text{Tr} \left({\M{T}_{1,k}}{\M{T}_{1,k}^{\RM H}} + {\M{T}_{2,k}}{\M{T}_{2,k}^{\RM H}} \right) \leq  P_k  \;\forall k}, \nonumber
\end{align}
where, $\M{E}_k $ is the MSE matrix and $\M{F}_k =  \M{E}_k  -  \tilde{\M{E}}_k { \M{E}_k^{-*}}  {\tilde{\M{E}}_k ^*}$.
Similarly, WLP and WLE design to maximize WSR for the \textbf{MU MIMO-BC} was attained by equivalently minimizing W-MSE ${\textbf{P6}}$ with BCD and AO \cite{bai2018optimal}. Intuitively, this approach renders suboptimal stationary point solutions of precoding and estimation matrices through iterative computation. Additionally, the transmit characteristics of \textbf{DF-FDR} are also optimized using coordinate descent algorithm with AO owing to the monotonic objective function in the individual optimization variables \cite{gaafar2018full}. Furthermore, Lameiro \textit{et al.} also relied on alternating minimization algorithm to design IA precoders and decoders with IGS to provide achievable DoF bounds in \textbf{4U SISO-IC} \cite{lameiro2013degrees}.
\paragraph{Separate Optimization (SO)}
Alternating optimization is the preferred choice with iterative convergence  especially if the sub-problems are convex for the subset of optimization variables by treating the remaining variables as constants. Otherwise, we resort to SO. The separate tuning of transmit covariances ${\M{R}}_{{x}_k{x}_k}$ and pseudo-covariances $\tilde{\M{R}}_{{x}_k{x}_k}$  for $k=1,2$ can be carried out in two ways \cite{zeng2013transmit}.
\begin{enumerate}
\item Exclusive optimization
\item Optimizing ${\M{R}}_{{x}_k{x}_k}$ assuming zero $\tilde{\M{R}}_{{x}_k{x}_k}$ and then obtaining  $\tilde{\M{R}}_{{x}_k{x}_k}$ with given ${\M{R}}_{{x}_k{x}_k}$.
\end{enumerate}
As an illustration, $ \textbf{P4}$ for \textbf{2U SISO-IC} can also be dealt using SO. The covariance optimization problem emerges as a linear feasibility problem necessitating bisection algorithm for its efficient solution. However, pseudo-covariance optimization with fixed ${\M{R}}_{{x}_k{x}_k}$ is a set of feasibility problems and thus can be solved as a finite number of second-order cone programming (SOCP) problems \cite{zeng2013transmit}.}

{Similarly, the joint optimization of non-convex $\textbf{P4}$ for MU Gaussian \textbf{MISO-IC} and \textbf{MISO-BC} does not yield a global optimal solution. Therefore, Zeng \textit{et al.} propose SO of ${\M{R}}_{{x}_k{x}_k}$ and $\tilde{\M{R}}_{{x}_k{x}_k}$. ${\M{R}}_{{x}_k{x}_k}$ by solving feasibility problem with $\tilde{\M{R}}_{{x}_k{x}_k} = \M{0}$ using bisection algorithm. Whereas, $\tilde{\M{R}}_{{x}_k{x}_k}$ is obtained with fixed ${\M{R}}_{{x}_k{x}_k}$ by solving equivalent minimum weighted sum-rate maximization (WSR-Max) \cite{zeng2013optimized}.
\paragraph{Line Search Methods (LSM)}
\textbf{Dualhop DF-FDR} systems can adopt improper signaling by finding the optimum weights to maximize the minimum SNR between the two hops under the perfect SI nulling constraint. The parameterization of the adopted problem renders one-dimensional optimization problem which can be efficiently solved using LSM \cite{kim2012asymmetric}. 
\paragraph{Algorithms for NP-Hard Optimization}
Unfortunately, all non-convex problems cannot be relaxed or convexified, rendering a class of NP-hard optimization problems. 
In computational complexity theory, these problems are informally "at least as hard as the hardest problems in NP". Generally, WSRMax problems are proven to be NP-hard \cite{liu2011coordinated}. Surprisingly, some subclasses of
the general NP-hard problem can still be solved in polynomial
time \cite{luo2008dynamic}. Whereas, others are solved using suboptimal/approximation algorithms e.g., game-theory based algorithm \cite{scutari2009mimo},  interference pricing
based algorithm \cite{huang2006distributed},  gradient descent algorithm \cite{ye2003optimized} with line search methods for unconstrained optimization \cite{kim2012asymmetric}, iterative weighted MMSE based algorithm \cite{shi2011iteratively}, monotonic optimization frameworks \cite{utschick2012monotonic,qian2009mapel}, graph theory for combinatorial optimization \cite{gross2004handbook} and SDR for solving non-convex QCQPs with GR by restricting to rank-1 solutions \cite{zeng2013optimized}.  To summarize, some NP-hard problems can be efficiently solved by combining multiple techniques with certain restrictions or by breaking the problem into sub-problems and then employing suitable optimization technique to solve each sub-problem.}

{Here we present an example of such a scenario employing separate optimization and then WMMSE algorithm and alternating direction method of multipliers (ADMM) are used to solve two subproblems, respectively. 
The non-convex NP-hard WSR maximization in \textbf{MIMO-IBC} poses a huge challenge and cannot be straightforwardly solved using any of the aforementioned techniques. Such problem can be dealt by the separate optimization of transmission parameters where covariance matrices are designed (assuming zero pseudo-covariance) using WMMSE algorithm. Next, target is to design the pseudo-covariance matrices using the pre-designed covariance matrices. This non-convex quadratic programming problem can neither be solved using SDR nor with SCP (as they do not warrant a unique and globally optimal solution). Thus, ADMM-based multi-agent distributed algorithm is suggested to solve an AO sub-problem \cite{lin2018multi}. However, the global optimal solution of AO sub-problem only guarantees the convergence to a stationary solution of the overall problem. This problem is a classical example of employing both separate and alternate optimization to solve a NP-hard problem.
\subsubsection{Intractable Optimization Framework}
Unfortunately, some complicated system configurations result in the intractable optimization problems with no definite framework. Therefore, we have to resort to the brute-force attack with exhaustive search. Such exhaustive search comes at the cost of factorial time complexity and may not be desirable. Thus, a fairly simple but suboptimal procedure is to adopt Maximal IGS conditional to IGS superiority.     
\paragraph{Exhaustive Search (ES)}
IGS parameters can be fine tuned using ES in the feasible domain. For example, the parameters of WLP in downlink \cite{zhang2017widely} and WL RX for uplink \cite{zarei2016q} in \textbf{multi-cell massive MIMO} systems can be chosen using ES. Similarly, exhaustive user scheduling algorithms combined with IGS in \textbf{SISO IBC} can improve its sum-rate performance \cite{shin2012new}. Moreover, the Pareto-optimal transmit covariance matrices for SINR balancing to improve the worst-user rate in the \textbf{2U SISO-IC} case are also obtained exhaustively \cite{park2013sinr}. Although, ES leads to a near-optimal solution but it comes at the expense of factorial time complexity  \cite{pendharkar2010exhaustive,dijkman2009graph,lee2019resource}. Therefore, a rule of thumb is to employ ES when nothing else works.
\paragraph{Maximal IGS (MIGS)}
MIGS is usually adopted either to ease optimization overhead or to overcome intractable optimization issues. For instance,  Javed \textit{et al.} propose adaptive scheme which switches between PGS and maximal IGS in \textbf{multi-antenna} systems under HWIs based on some switching criterion \cite{javed2018multiple}. 
However, MIGS is rarely the optimal improper signaling choice e.g., Lameiro \textit{et al.} argues that maximal IGS is the optimal signaling for SU transmission while operating in IGS favorable domain
in an \textbf{underlay cognitive radio} network \cite{lameiro2015benefits}. They further advocated MIGS in underlay MIMO cognitive radio networks in order to restrict the SU interference 
to protect interference temperature constraint of the PU \cite{lameiro2019maximally}. MIGS dominance over conventional PGS can be evaluated using majorization theory tools (MTT). 
Extension of \textbf{MIMO P2P-I} to two-tier HCN (with multiple MIMO Z-ICs) \cite{lagen2016superiority} and 
multi-antenna systems (specifically 2U MIMO-IC) \cite{lameiro2018performance} exploit MTT to demonstrate the superiority of MIGS \cite{ando1994majorizations}.
Furthermore, MTT also help to demonstrate that the eigenvalue spread of augmented covariance matrix is greater for improper signals and becomes maximum for maximal improper signals~\cite{adali2011complex}.}
\subsubsection{{Summary and Insights}}
The applicability and effectiveness of the most popular optimization frameworks is presented in this subsection. However, the theoretical analysis of IGS in more complicated systems and scenarios opens the research areas for other efficient, optimal, and fast-converging optimization techniques.
\subsection{IGS Detection and Estimation}
\begin{figure}[t]
\begin{minipage}[b]{1.0\linewidth}
  \centering
  \centerline{\includegraphics[width=12cm]{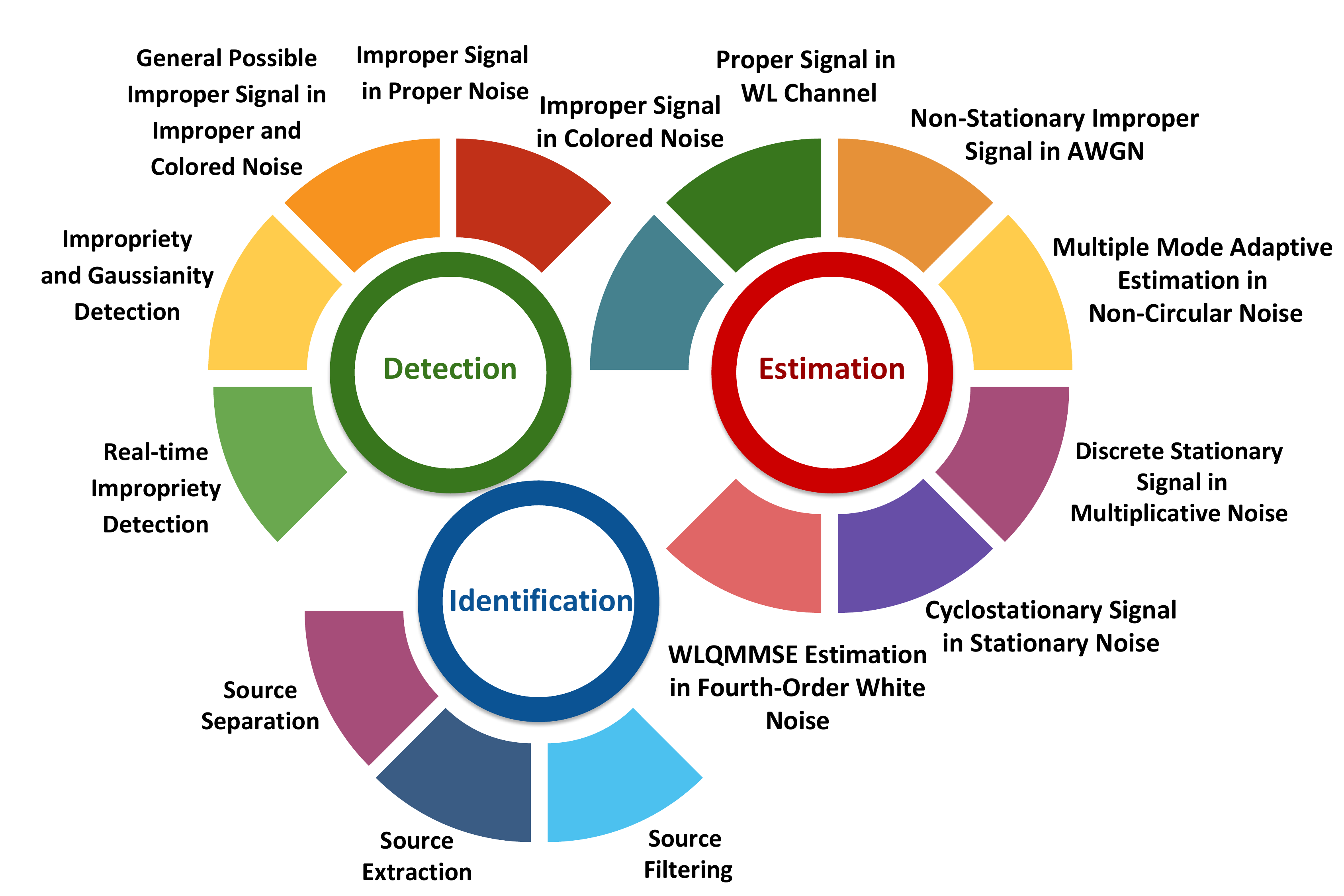}}
\end{minipage}
\caption{Conceptual Classification of IGS Detection, Estimation and Identification}
\label{fig:EC}
\end{figure}

In the engineering sciences, the three main branches of statistical signal processing are estimation, detection, and signal analysis \cite{schreier2010statistical}. Therefore, various contributions have addressed these issues related to IS as enumerated in Fig. \ref{fig:EC}.
\subsubsection{Detection}
 The problem of detecting the presence of improper complex random signal $s(t)$ from the observed complex signal $r(t)$ under additive noise $n(t)$ is carried out by a simple hypothesis test 
\begin{equation}
\mathsf{H}_0:  \; r(t) = n(t), \qquad \mathsf{H}_1: \;  r(t) = s(t) + n(t)  \nonumber 
\end{equation}
\paragraph{Improper Signal in Proper Noise}
The detection of an improper signal is based on a finite-dimensional log-likelihood ratio which can be designated as a cascade of an estimator and a correlator. For a zero-mean complex Gaussian signal $s(t)$, Schreier \textit{et al.} propose the detection based on improper version of Karhunen-Lo\`eve (K-L) expansion \cite{schreier2005detection}. The performance metric in terms of deflection yields double performance gain when pseudo-covariance is taken into account, generalizing the 3-dB gain of coherent processing over non-coherent processing \cite{schreier2005detection}. On the other hand, for improper complex second-order cyclostationary random signal $s(t)$, Yeo \textit{et al.} suggest properizing frequency shift  vectorizer to exploit periodic and symmetric correlations of the complex envelope in the frequency domain. The probability of miss is significantly reduced by the joint utilization of cyclostationarity and impropriety  \cite{yeo2011optimal}.

\paragraph{General Possible Improper Signal in Improper and Colored Noise}
Another interesting scenario is the detection of possible improper complex-valued signal
 common among two or more sensors, in the presence of possible improper and colored noise. 
Tugnait \textit{et al.} propose GLRT using asymptotic distribution of a frequency-domain sufficient statistic, based on the discrete Fourier transform of an augmented measurement sequence.
Interestingly, they present 133.33\% and 70.91\% increase in the PoD of improper signals in improper noise relative to that in proper noise at -10dB and -7.5dB SNR, respectively, while achieving 0.1 PFA \cite{tugnait2017multisensor}.
\paragraph{Impropriety and Gaussianity Detection}
Let us now consider the case when the presence of complex random signal is known but we would like to evaluate if it is improper and/or Gaussian signal or not. Novey \textit{et al.} address this problem of detecting possible improper Gaussian signal $z=x+iy$ using GLRT based on complex generalized Gaussian distribution (CGGD) \cite{novey2009complex} i.e.,
\begin{equation}
p\left(\underline{\V{z}},c, \M{R}_{\underline{\V{z}}\underline{\V{z}}} \right) = \frac{\beta (c)}{{\sqrt {\left| \M{R}_{\underline{\V{z}}\underline{\V{z}}} \right|} }} \exp \left\{-\alpha\left( c \right) \left( \V{z}^{\RM H} {\M{R}_{\underline{\V{z}}\underline{\V{z}}}^{-1}} \V{z} \right) \right\}^{c},
\end{equation}
where, $\beta \left(c \right) \!\!=  \!\!\Gamma\left(2/c \right)/\pi \Gamma\left( 1/c\right)^2$ and $\alpha \left(c \right) \!\!= \!\!c \Gamma\left(2/c \right)/2 \Gamma\left( 1/c\right)$ with Gamma function $\Gamma\left(. \right)$ and shape parameter $c$. 
This problem involves two detection mechanisms, 1) Non-circularity  2) Non-Gaussianity detection based on the following two binary hypothesis tests \cite{novey2009circularity}
\begin{align}
\text{Non-circularity }  & \qquad \qquad\text{Non-Gaussianity}  \nonumber  \\
 \mathsf{H}_0: \M{R}_{\underline{\V{z}}\underline{\V{z}}} = \sigma_z^2 \M{I},  &    \qquad \qquad \mathsf{H}_0: c=1, \M{R}_{\underline{\V{z}}\underline{\V{z}}} = \sigma_z^2 \M{I},   \nonumber \\
\mathsf{H}_1: \M{R}_{\underline{\V{z}}\underline{\V{z}}} \neq \sigma_z^2 \M{I}.  &\qquad \qquad \mathsf{H}_1: c \neq 1,  \M{R}_{\underline{\V{z}}\underline{\V{z}}} \neq \sigma_z^2 \M{I}.  \nonumber
\end{align}
The adjusted-GLRT detector \cite{ollila2009adjusting} performs fairly good as CGGD \cite{novey2009circularity} for Gaussian data ($c=1$) but inferior to CGGD for sub-Gaussian ($c=1.5$) and super-Gaussian data ($c=0.25$). For instance,  
PoD of 0.75 with CGGD is reported relative to 0.3 with adjusted-GLRT for super-Gaussian data~\cite{novey2009circularity}.
\paragraph{Real-time Impropriety Detection}
Sometimes the static detection of improper signals is inadequate as the underlying applications may require real-time identification of improperness. Thus, Jelfs \textit{et al.} propose collaborative adaptive filters trained by the complex least mean square (LMS) algorithms to detect and track improperness in real-time unlike competing static detectors \cite{jelfs2012adaptive}. 
\paragraph{Discussion}
The detection process of improper signals varies under the presence of proper, improper, or colored noise. Once detected, the interest may reside in the evaluation of DoI as well as the underlying distribution (dictated by the shape parameter). Apart from this static detection, contributions have successfully dealt with the problem of impropriety detection in real-time applications.
\subsubsection{Estimation}
While detection problems merely identify the presence or absence of improper complex signal, estimation problems include estimating the value of a parameter, or vector of parameters, from a sequence of measurements. Several engineering applications require complex-valued estimations, such as training neural networks \cite{mandic2009complex}, passive radar tracking \cite{dini2012kalman}, target tracking \cite{mazor1998interacting}, power systems frequency estimation \cite{xia2012widely} and fault diagnosis \cite{zhang1998detection} etc. 
From communication theory perspective, extraction of the transmitted information signal $\V{x}$ from the received observations $\V{y}$ after undergoing a  system, namely channel, is sometimes carried out using estimation based on the likelihood $p\left( \V{y}/\V{x} \right)$. This conditional probability relies on the prior probabilities as per \textit{Baysian} approach whereas other estimation techniques may not require priors e.g., \textit{Frequentists} approach treats $\V{x}$ as a vector of unknown constants \cite{schreier2010statistical}. 

 Considering the problem of estimating $y$ from complex observations vector $\V{x}$, linear estimation aims to design $\V{u}$ such that ${\hat y}=\V{u}^{\RM H}\V{x}$ minimizes linear MSE. Alternately, WLE aims to design $\V{v}$ and $\V{w}$ such that ${\hat y}=\V{v}^{\RM H}\V{x}+\V{w}^{\RM H}\V{x}^*$ minimizes WL-MSE \cite{adali2014optimization}. MSE of the real data can be accurately carried out by linear estimators whereas  WL estimators are generally optimum for complex data. Few relaxations from this rule, which prefer one estimator over the other, are highlighted for a broader picture: 
\begin{itemize}
\item For jointly circular observations and trivial correlation between observations and estimandum, WL estimators reduce to SL estimators \cite{schreier2010statistical}
\item For joint circularity between observations however correlated observations and estimandum, WL-MSE still offers better estimates than linear MSE \cite{picinbono1995widely}
\item For NC observations, it is possible to design $\V{v}$ and $\V{w}$ such that $y$ is uncorrelated with $\V{x}$. This implies zero estimation with the SL procedure and perfect estimation with the WL procedure even when the MSE is zero \cite{picinbono1995widely}
 \item For maximally improper i.e., $\V{x}=\varphi \V{x}^*$ with probability 1, WLE is unnecessary as $\V{x}$ and $\V{x}^*$ carry the same information regarding proper/improper $y$ \cite{schreier2010statistical}
 \item The WL-MMSE estimate of a real signal from a complex signal is always real whereas the LMMSE estimate is generally complex \cite{lang2017classical}
\end{itemize}
This section further characterizes estimation problems of different signals with different types of noise using SL and WL estimators based on MMSE criterion.

\paragraph{Discrete Stationary Signal in Multiplicative Noise}
Estimation problem of discrete second-order stationary signals can be efficiently solved using WL recursive algorithms. Interestingly, WL predictor proposed by Navarro \textit{et al.}  offers significant performance gains. These benefits increase with the increasing improperness of observations and stabilize at a certain value~\cite{navarro2009widely}. 

\paragraph{Cyclostationary Signal in Stationary Noise}
Blind estimation is a key concept to facilitate spectral efficiency by eliminating the need of pilot transmission. Napolitano \textit{et al.} propose a blind algorithm to estimate amplitude, phase,
relative time delay, and frequency shift of each user transmitting NC signals in a multiple access system. The presented algorithm, based on the cyclostationarity properties,
not only provides mean-square consistent estimates of the unknown parameters but is also robust to stationary noise and non-stationary narrowband interference 
\cite{napolitano2004doppler}.

\paragraph{Non-stationary Improper Signal in additive white Gaussian noise (AWGN)}
Next is the estimation of non-stationary improper complex
zero-mean random signal in AWGN. Schreier \textit{et al.} suggest 
WL-MMSE estimator using improper version of K-L expansion to address this problem.  Interestingly, this procedure yields perfect estimates rendering arbitrarily large performance gain over SL estimator in the presence of improper noise. Moreover, WL-MMSE estimation of an improper complex signal in uncorrelated noise can render twice performance advantage over
LMMSE estimation at diminishing noise levels~\cite{schreier2010statistical}. 

\paragraph{Multiple Mode Adaptive Estimation in NC Noise}
The improper stochastic hybrid system with discrete and continuous states can be estimated using WL multiple model adaptive estimation (MMAE) algorithms. These algorithms are based on augmented Kalman filters which are matched to different modes of the hybrid system. WL-MMAE utilizing pseudo-covariance not only converges faster but also offers up to 30\% less MSE than their counterparts \cite{mohammadi2015improper}. Mohammadi \textit{et al.} also extended their work to distributed estimation using diffusion strategies, when a system is observed distributively using an agent/sensor network \cite{mohammadi2015distributed}. 

\paragraph{Improper Signal in Colored Noise}
Estimation of a random improper signal in the presence of colored noise having an additive white part is carried out with Hilbert space theory yielding
 10\% less MSE with WL estimator as compared to SL estimator \cite{navarro2009estimation}.

\paragraph{Proper Signal in WL Channel}
Underlying channel also impacts the performance of the MSE estimators, irrespective of the correlations between data. For instance, 
Trampitsch demonstrates the superiority of WL-MMSE over LMMSE for white Gaussian data with complex AWGN. This lead is observed either due to the WL characteristics of the underlying channel even in the absence of correlations between the data or for highly correlated data in a SL channel \cite{trampitsch2013complex}.

\paragraph{WL Quadratic Estimation in Fourth-order White Noise}
For complex NC case, a scalar complex $y$ can also be estimated from a measurement $\V{x}$ using the WL quadratic estimator as ${\hat y} = c + \V{g}^{\RM H} \underline{\V{x}} + \underline{\V{x}}^{\RM H}  \M{H}  \underline{  \V{x}} $, where $c$ is chosen as $-{\rm Tr} \left( \M{R}_{\underline{yy}} \M{H} \right)$ to ensure zero-mean $\hat y$ if the observations are also zero-mean. 
Interestingly, the WL part of WLQ-MMSE  is not the same as WL-MMSE. Moreover, the better estimation obtained by WLQ-MMSE relies on the complete statistical information up to fourth order \cite{schreier2010statistical}.

\paragraph{Discussion} The performance of various estimators can be distinguished using complementary MSE analysis which quantifies the DoI of the SL and WL estimation errors \cite{xia2017augmented}. In a nutshell, the main difficulty in the state estimation comes from structural uncertainty which arises from the lack of knowledge of the true behavior of observations and noise in the underlying system. Therefore, a generalized approach in terms of WL estimators is preferred to accommodate all possibilities and uncertainties. However, this may come at the cost of over-fitting and 
slower convergence owing to the increased dimensions \cite{adali2014optimization}.
\subsubsection{Source Separation}
Separating one source from a mixture of noisy sources can be carried out using source extraction or source filtering.
\paragraph{Source Extraction}
Source extraction
aims to recover the original sources from their linear (or non-linear) mixtures in both noisy and noise-free environments. Moreover, blind source separation (BSS) does so with neither explicit
knowledge of the sources nor the linear mixing process. Such source extraction is crucial in diverse areas like biomedical engineering, communications, radar, and sonar etc. 

Only few contributions cater for the signals with NC PDFs in the source extraction process.
For instance, Javidi \textit{et al.} propose
second-order complex domain blind source extraction algorithms permitting normalized MSE prediction, at the output of a WL predictor, to be the extraction criterion. Interestingly, the presented framework is suitable for both circular or NC sources with possible improper noise. An important application of BSS is the removal of useful artifacts from the corrupted EEG signals \cite{javidi2010complex}. Another extension of BSS i.e., independent component analysis (ICA), which assumes the statistical independence of the underlying unknown source signals, is the most popular way to separate sources.
Some important separability/identification results from the complex ICA based on the circularity coefficients include \cite{adali2014optimization}:
\begin{itemize}
\item For the real-valued case, identification is only possible in the absence of two Gaussian sources having proportional covariance matrices in the mixture. 
Moreover, the knowledge of sample correlation also allows the segregation  of Gaussian sources.
\item For the complex case, a mixture of improper Gaussian sources with distinct circularity coefficients can be separated using the strong uncorrelating transform even without sample correlation.
\item Two Gaussian sources are non-identifiable in a mixture if both of their covariance and pseudo-covariance matrices are proportional to each other.
\item A unique maximally improper source in the mixture of sources can be perfectly separated. 
\end{itemize}
\paragraph{Source Filtering}
As opposed to extraction, another important phenomenon is the filtering of NC complex signals which can be achieved using various adaptive algorithms based on the WL modeling. Few of these algorithms are:
\begin{itemize}
\item Augmented affine projection algorithm \cite{xia2010augmented}
\item WL least mean squares to minimize MSE \cite{adali2014optimization} 
\item WL least stochastic entropy algorithm \cite{adali2014optimization} 
\item Complex augmented least-mean kurtosis algorithm \cite{mengucc2018augmented}
\item  Incremental augmented complex LMS \cite{khalili2014collaborative}
\item Complex-valued Gaussian sum filter \cite{mohammadi2015complex}
\end{itemize}
\paragraph{Discussion}
Source extraction, especially BSS, is mainly dependent on the structure/distribution of the mixing sources. For instance, two Gaussian sources with proportional augmented covariance matrices or two improper Gaussian sources with similar circularity coefficients cannot be separated. Coming over to the filtering process, the performance of source filters can be quantified using different metrics such as mean square deviation, mean square
error, prediction gain, and convergence rate, etc. which mainly depends on their adaptive or batch-wise
implementation.
\section{Practical Realization and Implementation} \label{sec:PRI}
\begin{figure*}[t]
\begin{minipage}[b]{1.0\linewidth}
  \centering
  \centerline{\includegraphics[width=13cm]{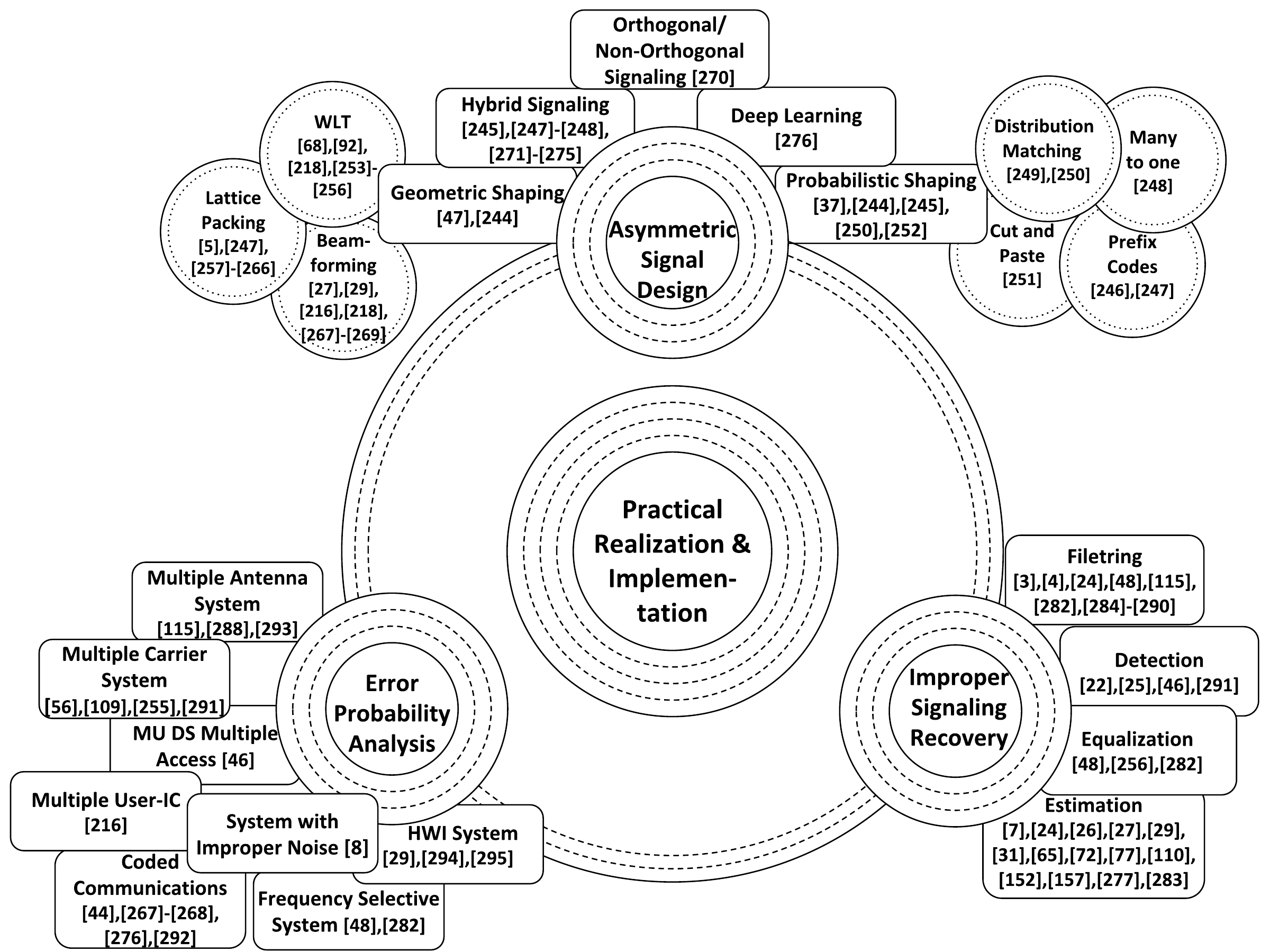}}
\end{minipage}
\caption{{Practical Implementation of IS: From Realization to Recovery and Analysis}}
\label{fig:PRI}
\end{figure*}

Overwhelming theoretical performance advantages of IGS in
various interference-limited system configurations motivated the researchers to propose the practical improper discrete realization i.e., asymmetric signaling. 
Finite discrete constellations are the preferred choice for implementation over Gaussian signals owing to their robustness, reduced detection complexity and bounded peak-to-average power ratio \cite{santamaria2018information}.
 Therefore, the design of new family of asymmetric constellations is imperative for practical realization which is the counterpart of the standard proper discrete constellations i.e., symmetric signaling.
Such asymmetric signaling along with the appropriate signal recovery mechanism can significantly reduce the error probability as discussed in the studies presented in Fig. \ref{fig:PRI}.
\subsection{Asymmetric Signal Design}
Apart from the inherently asymmetric signaling schemes like $M$-PAM, OQPSK and GMSK, what are the possible ways to induce asymmetry in a symmetric discrete constellation like $M$-QAM and $M$-PSK?  What should be the design objective?
How to optimize the design parameters such as transmit power and circularity quotient to meet our objectives? How to generate optimal asymmetric signaling given a certain power constraint and circularity coefficient?
Will the generated signaling be capable of achieving superior performance as demonstrated by the theoretical bounds? This section intends to address all of these concerns and highlights the major contributions in this regard. 
The asymmetric signaling is designed to achieve improper Gaussian capacity except for the 1.53dB shaping loss between IGS capacity and the envelop of capacity curves with sufficiently large $M$-ary constellations \cite{santamaria2018information}. Moreover, they are 
 optimized to reduce the EP of the constrained systems.
  Asymmetric signaling may arise from conventional symmetric signaling in the following ways:
\begin{enumerate}
\item Probabilistic Shaping (PS): Non-uniform probability distribution of the alphabets/symbols  
\item Geometric Shaping (GS): Equally or unequally spaced (due to correlated and/or unequal power distribution between quadrature components of the signals) symbol constellation in a distinct geometric envelop 
\item Orthogonal/Non-Orthogonal Sharing: Assigning unequal orthogonal/non-orthogonal resources to users in a MU environment 
\item Hybrid Signaling (HS)
\item Deep Learning
\end{enumerate}
Asymmetry provides new tools for the constellation design and can be considered as choices much as symbol separation, the number of bits transmitted per symbol, or power. Interestingly, the introduction of asymmetry into the signal set offers another design freedom which neither affects the bandwidth nor the power requirements of the system \cite{divsalar1987trellis}. Surprisingly, the digital modulation schemes yield cyclostationary signals with periodic mean, auto covariance and auto complementary-covariance functions. Thus, improved performance can be attained by utilizing this cyclostationary property besides impropriety \cite{han2012capacity,yeo2015joint,lameiro2017rate}. In the following,  we highlight the famous shaping techniques and procedures to design appropriate asymmetric signaling for a given application. 

\subsubsection{Probabilistic Shaping} 
Given a fixed number of
symbols and the symbol locations,
an asymmetric constellation can be obtained by adjusting
the symbol probabilities \cite{thaiupathump2000asymmetric}. Therefore, PS maps equally distributed input bits into constellation symbols with non-uniform distribution \cite{batshon2017coded}.
 Intuitively, manipulating symbol probabilities and deviation from uniform distribution will result not only in some entropy loss but also 
added complexity in the encoding/decoding process. Despite this implementation penalty, the attained performance gains are totally worth it. So, what should be the ideal non-uniform probability distribution and how can we attain it? Intriguingly, the Gaussian probability distribution is the ultimate goal to approach the channel capacity bounds but this comes with a number of practical problems. Therefore, multiple transformations are presented to tackle this issue including prefix codes \cite{kschischang1993optimal,forney1984efficient}, many-to-one mappings combined with a turbo code \cite{yankov2014constellation}, distribution matching \cite{buchali2016rate,bocherer2015bandwidth} and cut-and-paste method \cite{cho2016low}. 
Coded modulation scheme with PS aims to remove the shaping gap and coarse mode granularity problems \cite{bocherer2015bandwidth}.
Interested reader can read \cite{dunbridge1967asymmetric} for the 
design guidelines of asymmetric signaling in the
coherent Gaussian channel with equal signal energies and unequal a priori probabilities. Probabilistic amplitude shaping is another concept that can only be used for square QAM, which greatly limits its application \cite{pan2016probabilistic}.

\subsubsection{Geometric  Shaping}
Geometric shaping can be characterized in two distinct ways. First, uniformly spaced symbols within distinct geometric envelop. Second, non-uniformly spaced symbols pertaining to either non i.i.d quadrature components or intentional asymmetric placement in rectilinear constellations. GS requires unconventional partitioning without any loss in entropy. Isaka \textit{et al.} emphasize unconventional signal set partitioning for asymmetric constellations to achieve unequal error protection capability with multilevel coding and multistage decoding \cite{isaka2000multilevel}. Nonetheless, moving some symbols close to each other may result in more erroneous symbol decisions depending on the underlying application \cite{thaiupathump2000asymmetric}. GS can be realized using the following well-known methodologies.

\paragraph{Widely Linear Transformation}
WLT is the most popular way of transforming proper signaling to improper one in order to exploit the additional freedom offered by complementary covariance matrix \cite{mohammadi2016improper}. The extension of WLP from Gaussian code books to discrete constellations is paving the way for its practical utilization in different applications. The simplest design of asymmetric constellations with complex symbols $v=v_{\rm I} + i v_{\rm Q}$  from standard symmetric discrete constellation with symbols $x=x_{\rm I} + i x_{\rm Q}$ for a given circularity coefficient $\kappa$  and circularity angle $\phi$ can be attained by the following WLT
\begin{equation}
v = \sqrt{\frac{1}{2}\left(1+\alpha\right)}x + \sqrt{\frac{1}{2}\left(1-\alpha\right)}\exp^{i\phi}  x^*, 
\end{equation}
where, $\alpha = \sqrt{\left(1-\kappa^2\right)}$ and $\phi \in [0,\pi/2]$.  The optimal $\kappa$ and $\phi$ can transform a given $M$-ary symmetric constellation to $M$-ary asymmetric constellation \cite{santamaria2018information}. This transformation has been proven helpful in various applications e.g., WL digital beamforming is employed in
massive antenna arrays suffering from IQI in RF chain. Therefore, WL extension of the well-established minimum variance distortionless response beamformer for $M$-QAM modulation is a promising candidate for complex Gaussian interference, unwanted mirror beam and RF imperfections suppression \cite{hakkarainen2013widely}. Furthermore, adaptive algorithms by combining WL processing and set-membership filtering techniques are proven to improve sensor array processing when the signals under study are second-order NC such as BPSK modulated signals \cite{zelenovsky2012set}. Likewise, WLP in spatially multiplexed MIMO-FBMC system with OQAM provides lower BER as compared to  LP at high SINR \cite{caus2013comparison}. MU two-way MIMO-AF relay system exploits the NC transmitted signals like BPSK using WLP to achieve improved system performance with minimal relay power \cite{zhang2013widely}.
Furthermore, single user and MU MIMO communications systems employing asymmetric modulation like $M$-ary ASK and OQPSK depict superior performance with improved ZF and MMSE precoders without loss of spectrum efficiency \cite{xiao2010improved}. 


\begin{figure}
        \centering
        \begin{subfigure}[b]{0.4\textwidth}
            \centering
            \includegraphics[width=\textwidth]{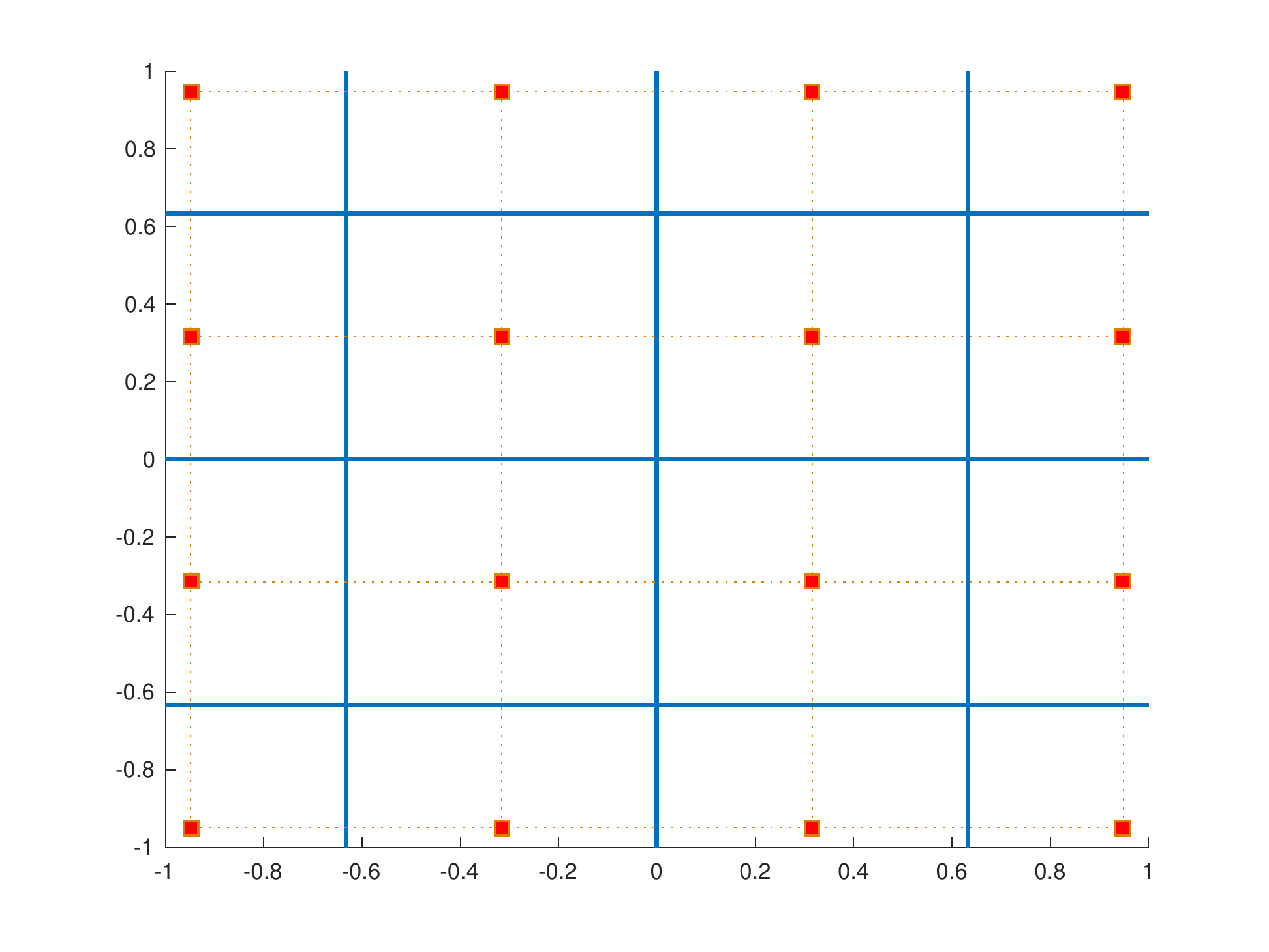}
            \caption[]%
            {{\small Square Lattice}}    
            \label{fig:a}
        \end{subfigure}
      \quad
        \begin{subfigure}[b]{0.4\textwidth}  
            \centering 
            \includegraphics[width=\textwidth]{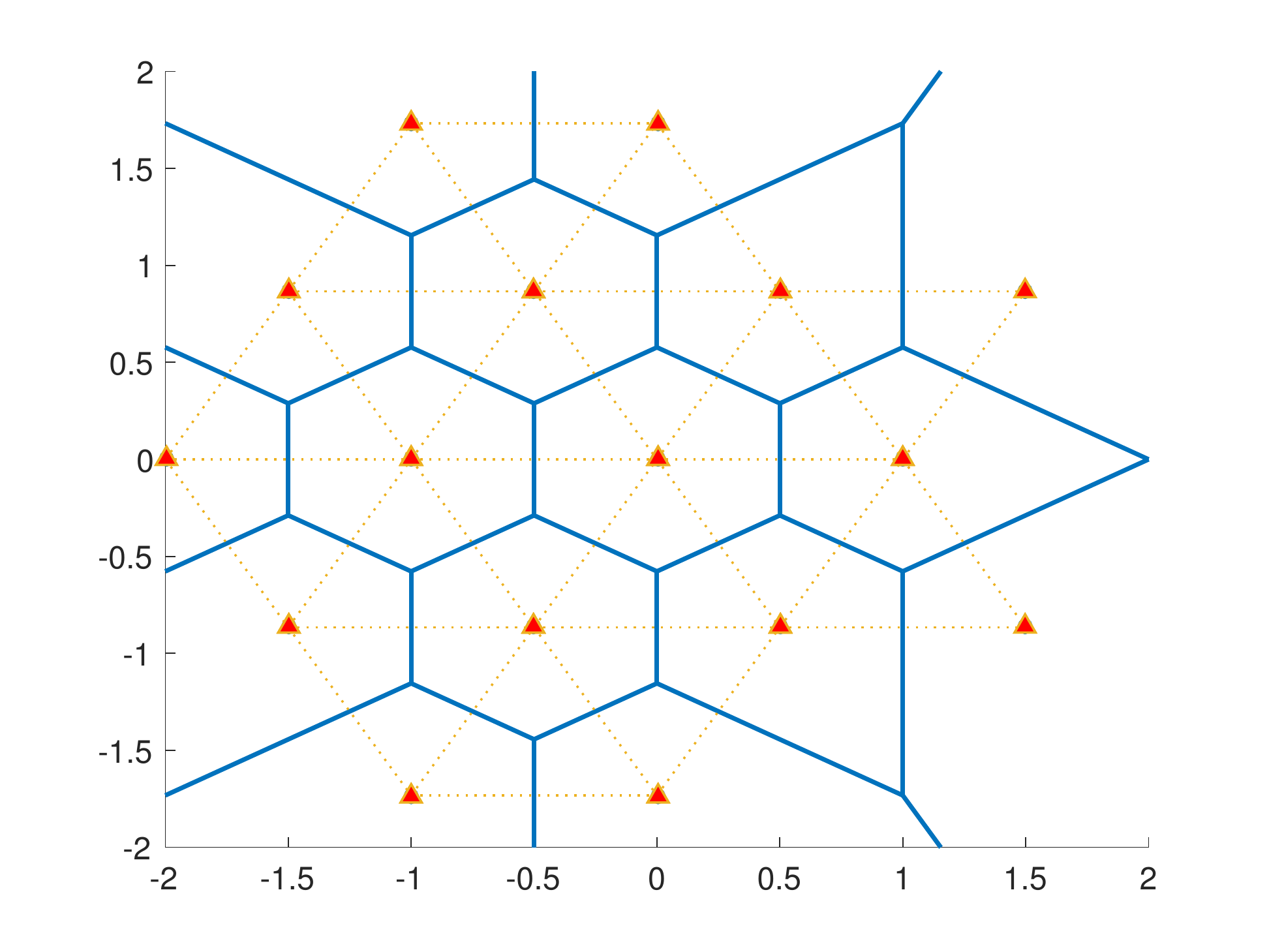}
            \caption[]%
            {{\small Triangular/Hexagonal Lattice}}    
            \label{fig:b}
        \end{subfigure}
        \caption[ ]
        {\small Celebrated 16-QAM Lattice Structures}
        \label{fig:Lattice}
    \end{figure}

\paragraph{Lattice Packing}
Numerous studies have supported the concept of efficient modulation technique by packing a particular lattice structure in some geometric shape \cite{forney1984efficient,fernandez2019design}. Lattice packing is a two-fold procedure. Firstly, lattice structure is chosen from the square, rhombic, triangular or hexagonal lattices \cite{wubben2011lattice,rugini2016symbol}.
A square and rhombic lattice is the periodic arrangement of discrete constellation points/symbols at the corners of the square (as depicted in Fig. \ref{fig:Lattice}a) and rhombus shape \cite{wubben2011lattice}. Unlike these quadrilaterals, triangular lattice has constellation points at the vertexes of contiguous equilateral triangles. This lattice is sometimes referred as hexagonal lattice owing to the hexagonal Voronoi decision region around internal lattice points as shown in Fig. \ref{fig:Lattice}b \cite{rugini2016symbol}. 
 Square QAM (SQAM) is preferred for the simple ML detection mechanism whereas triangular QAM (TQAM) is preferred for power efficiency \cite{park2012performance,rugini2016symbol}. Park \textit{et al.} report the asymptotic power gain of 0.5799dB with TQAM over SQAM, identical peak-to-average-power ratio for significantly large constellation size $M$ and a significant reduction in EP with tolerable detection complexity in an AWGN channel \cite{park2012performance}. Therefore, hexagonal QAM (HQAM) is preferred for various applications, including advanced
channel coding \cite{tanahashi2009multilevel}, multi-antenna systems \cite{srinath2013fast},
multicarrier systems \cite{han2008use}, physical-layer network coding \cite{hekrdla2014hexagonal}, and optical communications \cite{doerr201128}.

\begin{figure}
        \centering
        \begin{subfigure}[b]{0.23\textwidth}
            \centering
            \includegraphics[width=\textwidth]{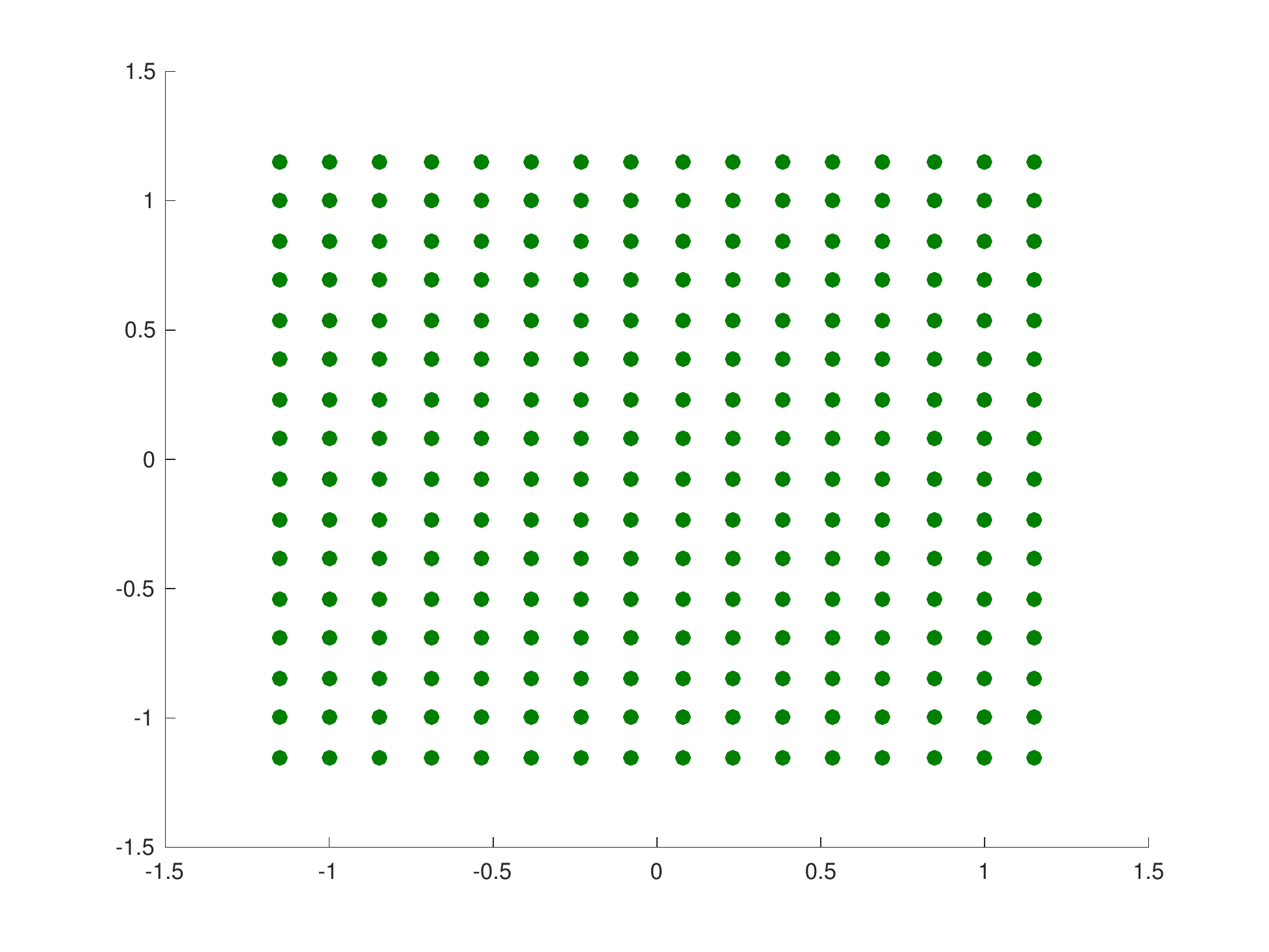}
            \caption[]%
            {{\small Square Packing}}    
            \label{fig:a}
        \end{subfigure}
      \quad
        \begin{subfigure}[b]{0.23\textwidth}  
            \centering 
            \includegraphics[width=\textwidth]{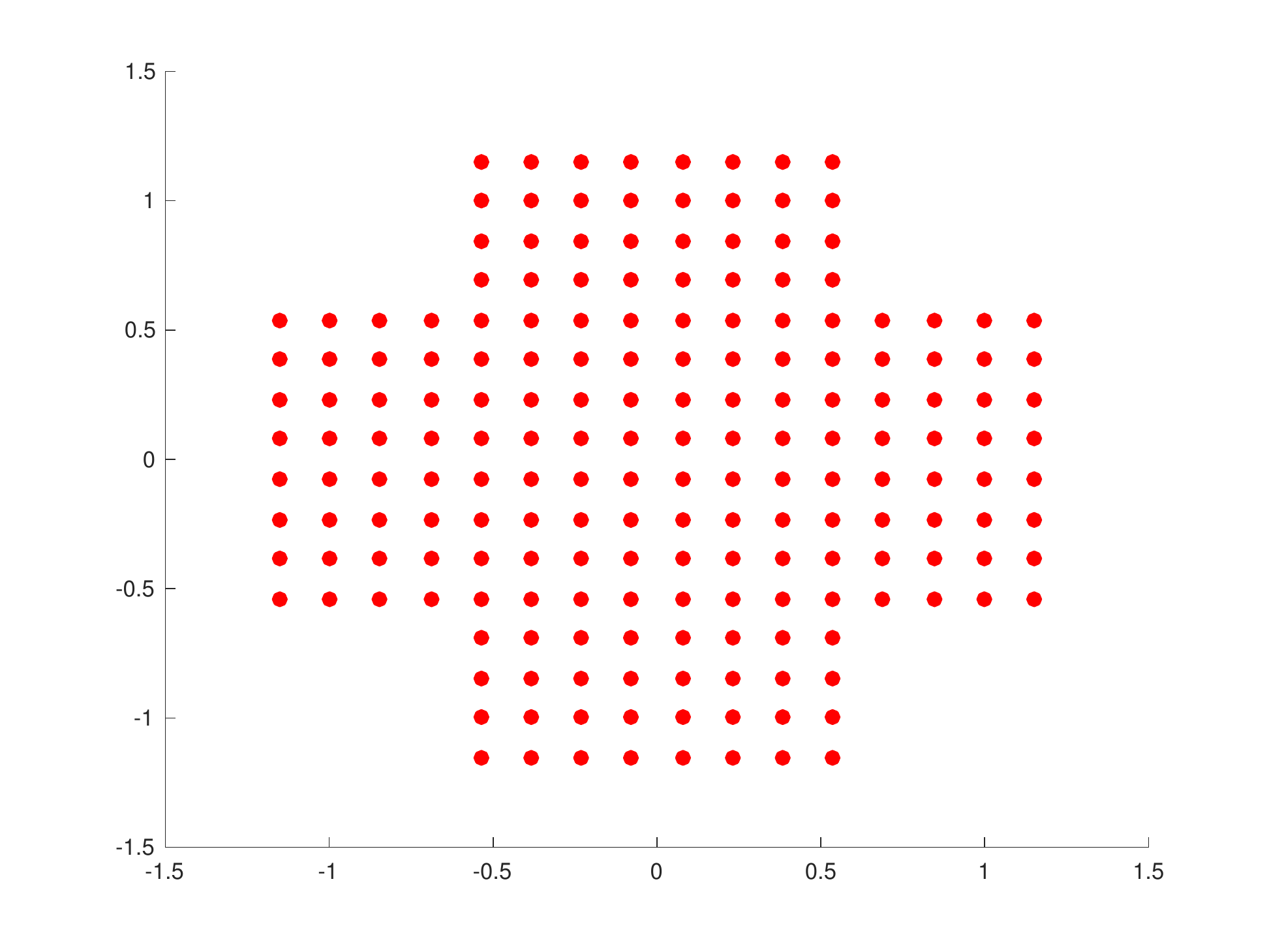}
            \caption[]%
            {{\small Cross Packing}}    
            \label{fig:b}
        \end{subfigure}
     \quad   
        \begin{subfigure}[b]{0.23\textwidth}   
            \centering 
            \includegraphics[width=\textwidth]{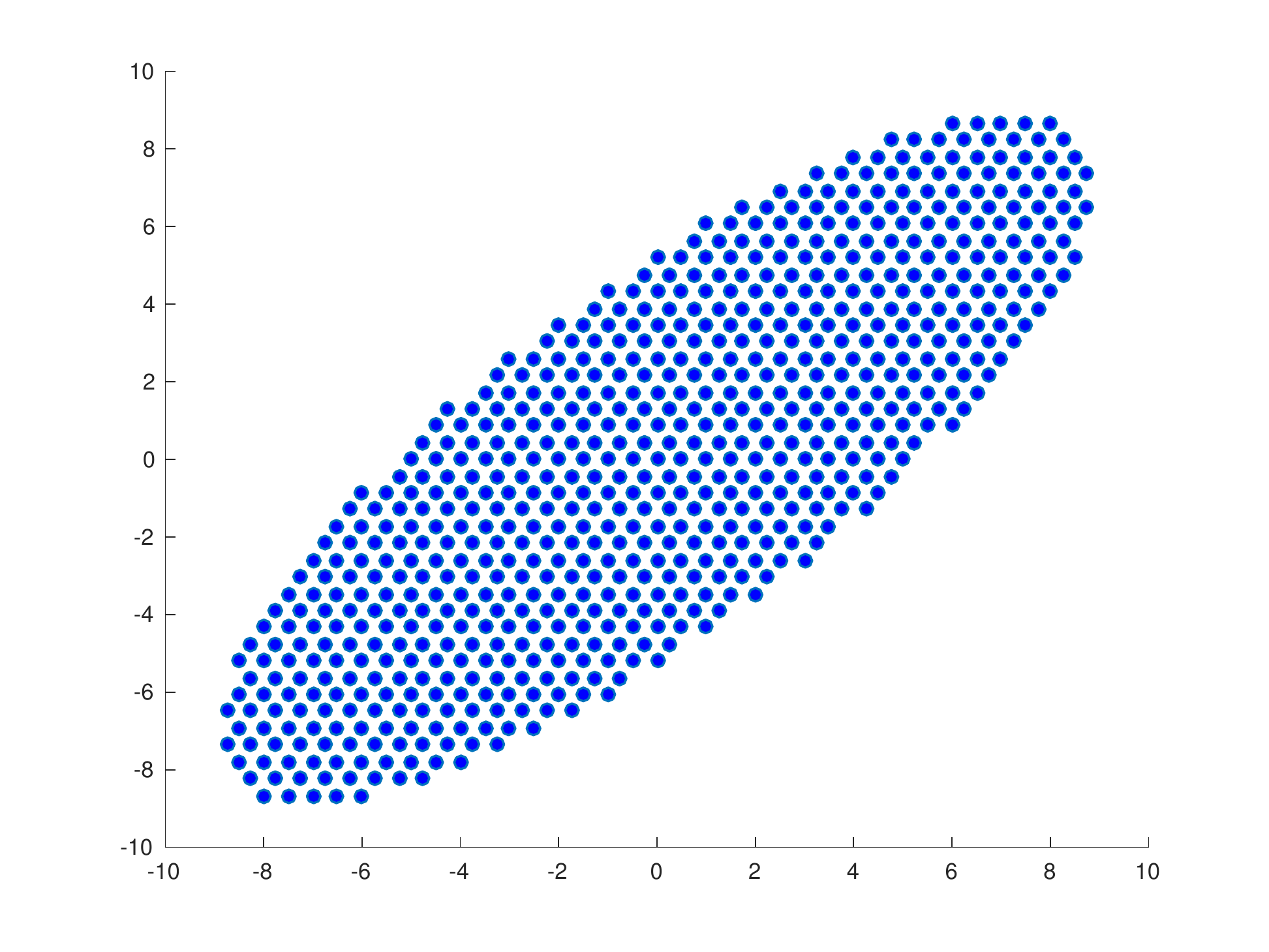}
            \caption[]%
            {{\small Elliptical Packing }}    
            \label{fig:c}
        \end{subfigure}
          \vskip\baselineskip
        \begin{subfigure}[b]{0.23\textwidth}   
            \centering 
            \includegraphics[width=\textwidth]{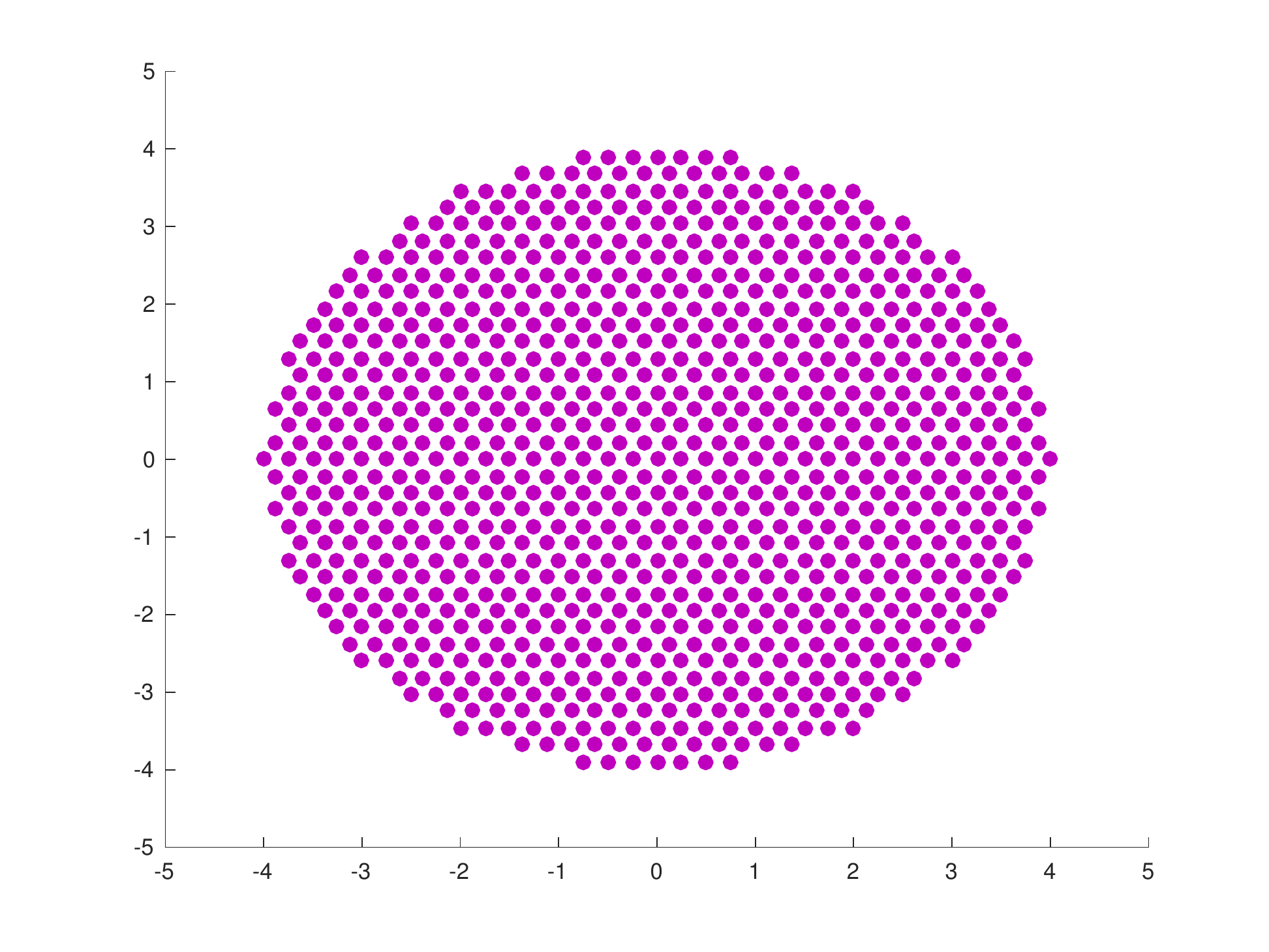}
            \caption[]%
            {{\small Circular Packing}}    
            \label{fig:d}
        \end{subfigure}
          \quad
        \begin{subfigure}[b]{0.23\textwidth}   
            \centering 
            \includegraphics[width=\textwidth]{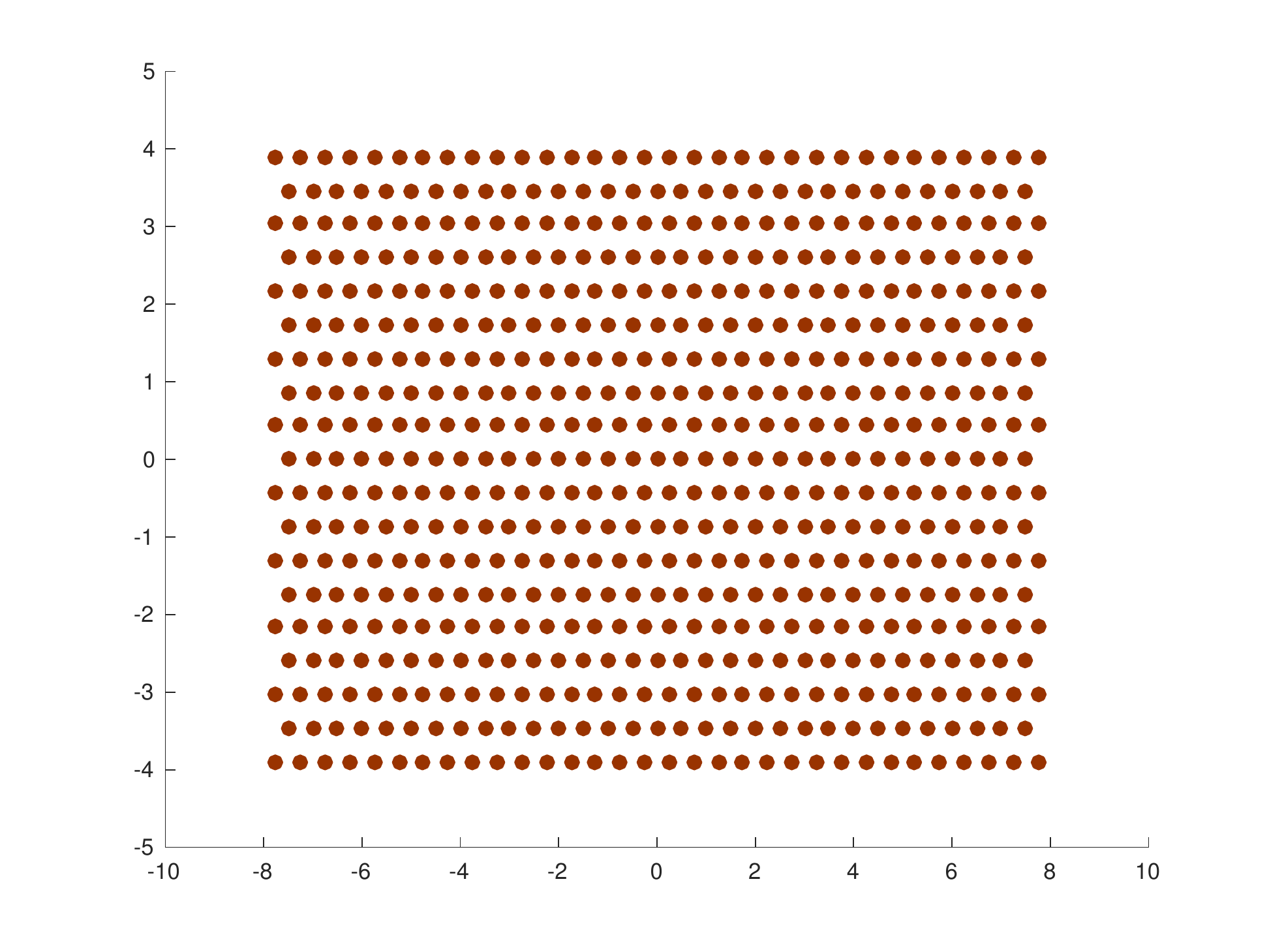}
            \caption[]%
            {{\small Rectangular Packing }}    
            \label{fig:c}
        \end{subfigure}
        \quad
        \begin{subfigure}[b]{0.23\textwidth}   
            \centering 
            \includegraphics[width=\textwidth]{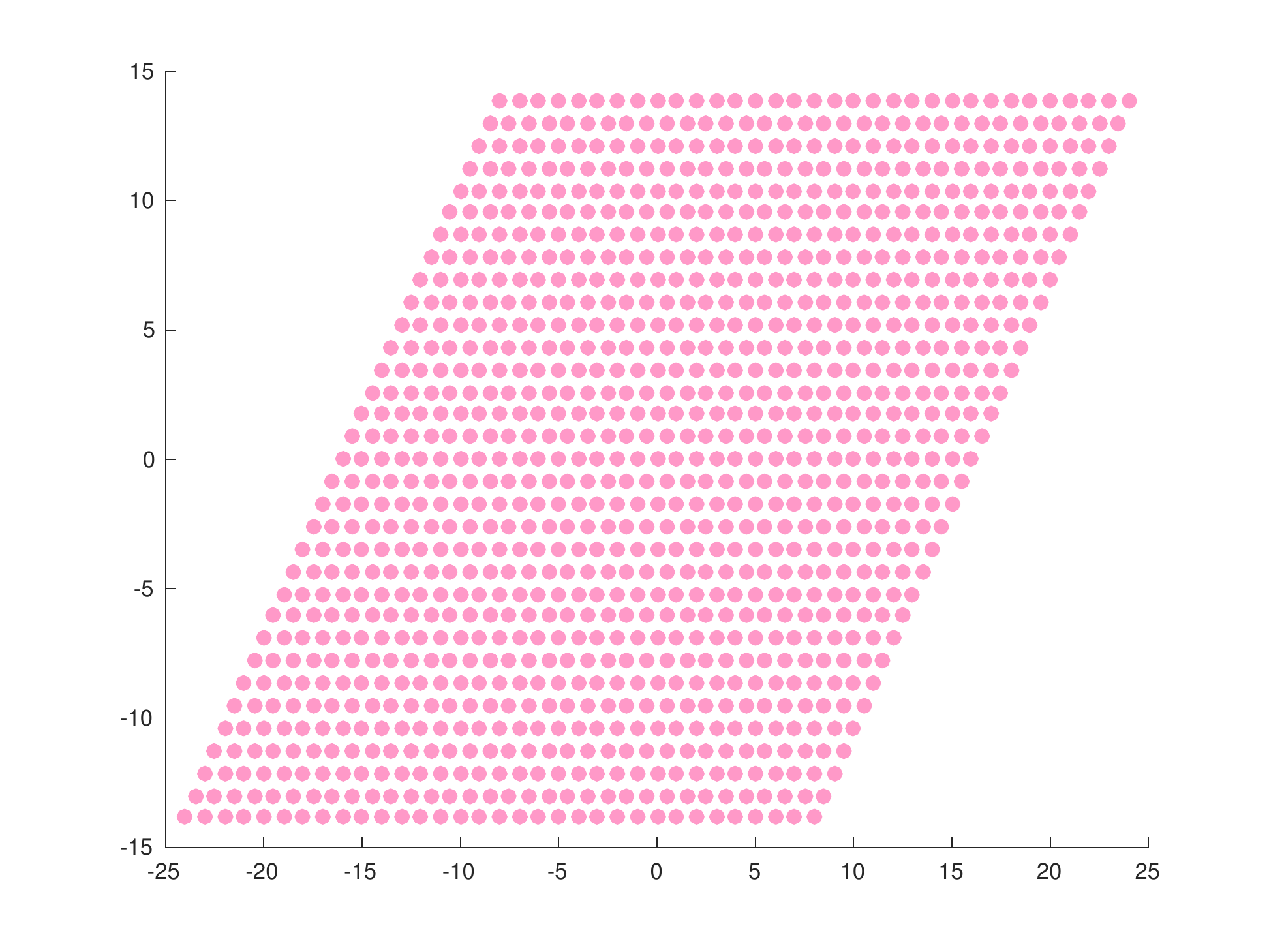}
            \caption[]%
            {{\small Parallelogram Packing}}    
            \label{fig:d}
        \end{subfigure}
        \caption[ ]
        {\small Geometrical Packing/Mask Structures}
        \label{fig:Packing}
    \end{figure}

Next step is the distribution of lattice around origin, which can be packed in square, cross, elliptical, circular, rectangular, or parallelogram envelops
 \cite{fernandez2019design,forney1984efficient,abdelaziz2017triangular}. For instance, Fig. \ref{fig:Packing}a and \ref{fig:Packing}b depict square and cross packing of the square lattice, respectively. However, Fig. \ref{fig:Packing}c-\ref{fig:Packing}f pack hexagonal lattice in elliptical, circular, rectangular, and parallelogram envelops, respectively. Asymmetric constellations formulated by the elliptical packing are based on the geometric interpretation of the circularity quotient \cite{ollila2008circularity}.  Accordingly, the modulus and phase of its principal square-root depict the eccentricity and angle of orientation of the ellipse defined by the covariance matrix. Thus, vanishing eccentricity implies zero circularity quotient (symmetric case) whereas maximum eccentricity implies that the circularity quotient exists on the unit complex circle \cite{ollila2008circularity}.

Recently, Fern{\'a}ndez \textit{et al.} suggest elliptical packing of the hexagonal lattice to realize optimal asymmetric constellations for a certain circularity coefficient \cite{fernandez2019design}. 
The optimal constellation is the overlap between the translated hexagonal lattice and the rotated ellipse (as per circularity quotient). The idea is to capture exactly $M$-constellation points in the intersection area and then apply WLT to transform it to quasi-hexagonal constellation rendering the required circularity coefficient. They are optimal as they yield lowest EP under average power constraint in an AWGN channel especially at high SNR and large constellation size.
Authors claim significant SNR gains resulting from this design scheme as compared to WL transformation for two reasons: \textit{shaping loss} and \textit{packing loss} \cite{fernandez2019design}. Shaping loss of WL transformed constellations is around 1.53dB with respect to the IGS which is equivalent to proper square $M$-QAM constellation. This limitation is addressed by the hexagonal lattice which offers 0.6dB gain over a rectangular one with the same boundary being the densest lattice in 2D \cite{forney1984efficient}. On the other hand, 
 packing loss is pertinent to the parallelogram envelope of WLT asymmetric constellation in place of the optimal elliptical envelope.

\paragraph{Beamforming/Precoding}
Numerous other transformations are based on the fine tuning of the basic constellation parameters to achieve minimum EP with a given power budget. For instance, Zhang \textit{et al.} propose asymmetric 4-PSK constellation design with TCM by calculating the optimum angle 
$\alpha \in [0,\pi]$ (with $\alpha = \pi/2$ yielding the symmetric 4 PSK) 
that minimizes the EP bound \cite{zhang2009optimum}.
Similarly, Subramaniam \textit{et al.} introduce asymmetry in 8-PSK to increase the minimum product distance reducing error events in TCM  \cite{subramaniam2000performance}. For colored noise compensation, Taubock prefers optimally rotated rectangular symbol constellations over common quadratic QAM in order to minimize capacity loss and SEP~\cite{taubock2007complex}. 

Unlike the contributions which propose modulation specific designs, \cite{javed2018asymmetric} and \cite{nguyen2015improper} propose asymmetric constellation design based on the minimum Euclidean distance aiming at minimizing the maximum PEP or SER irrespective of the underlying modulation scheme. 
Nguyen \textit{et al.} optimize the precoding matrices of all users in order to minimize the maximum PEP or SER in MU SISO-IC. 
Alternately, precoding matrices can be designed in order to minimize the sum MSE, maximum MSE, minimize interference leakage or maximize SINR. However, the schemes based on minmax-MSE perform inferior to minmax PEP/SER especially for higher modulation transmission with higher interference \cite{nguyen2015improper}.
Moreover, Javed \textit{et al.} design asymmetric modulation for HWI transceiver systems by separately optimizing the rotation and translation matrices which jointly formulate the transmit precoding matrix \cite{javed2018asymmetric}. Likewise, pairwise optimization algorithm transforms conventional constellations like 8PSK star-8QAM, set-partitioning-8QAM and circular-8QAM to asymmetric optimal constellation in order to minimize the BER \cite{zhang2017generalized}. 

In general, researchers mainly rely on maximizing the minimum Euclidean distance of the constellation ($d_{\rm min}$) as it is the building block of various performance metrics like mutual information, MMSE and SEP. They all have asymptotic behavior which is proportional to the Gaussian Q-function $Q  \left( \sqrt{{\rm SNR}} d_{\rm min}/2 \right)$ \cite{santamaria2018information}. 

\subsubsection{Orthogonal/Non-Orthogonal Sharing}
Another form of asymmetry, which can be induced to attain added benefits, is through the non-uniform allocation of orthogonal resources like time slots and frequency bands. For instance, asymmetric time sharing is assumed to be a potential candidate in future generation of wireless communications for enhanced performance in some interference limited scenarios. On the other hand, some NOMA schemes like sparse carrier multiple access that is going to be an essential part of 5G communications for many excellent properties i.e., shaping and diversity gain by sparse codebooks, resilient to inter-user interference, and robust to codebook collision can also be considered as a form of asymmetric signaling \cite{dai2015non}.

\subsubsection{Hybrid Signaling}
We characterize the joint employment of any two or more types of asymmetric signaling schemes as hybrid signaling. There is an ongoing debate on the superiority of one type of asymmetric scheme over the other depending on the underlying application and employed modulation schemes. Both PS and GS were successfully employed in optical communications \cite{yankov2014constellation}. 
From one perspective, PS outperforms circular based GS in terms of shaping gain for the same number of constellation points \cite{batshon2017coded}. However, another point of view is the superior performance of GS 16-QAM constellations over its PS counterpart \cite{qu2017geometrically}.
A striking way out of this debate is to employ hybrid probabilistic and 
geometric shaping (PS/GS) concept  to bridge the gap to the Shannon limit \cite{batshon2010iterative}. Albeit hybrid PS/GS is popular in optical communications but this concept is yet to find its standings in the wireless communication systems. In contrast to the PS, where redundancy improves power efficiency, Hybrid PS/GS capitalize on redundancy by introducing a transmitted signal structure that improves Euclidean distance and reduces SER \cite{batshon2017coded}. Hybrid PS/GS of any two-dimensional signal constellation outperform the probabilistically shaped as well as regular constellations with universal distribution matchers for asymmetric $M$-QAM \cite{qu2018universal} and multi-dimensional coded modulation format using amplitude-phase shift keying for single-stage \cite{batshon2017coded} and multi-stage \cite{cai201770} nonlinearity compensation. Other optimal strategies involving the combined PS/GS to shape circular 64-QAM constellation attain 1dB sensitivity gain and 28\% gain in transmission reach over compared to conventional 64-QAM  in both linear and nonlinear regime of wavelength division multiplexed systems \cite{jardel2017experimental}. 

To conclude, dense hexagonal packing with optimal circular boundaries 
yields around 1dB improvement over PAM in a band-limited channel \cite{forney1984efficient}. Moreover, source coding with non-uniform probabilities provide more
than 1dB gain which is alternately achievable using higher dimensional modulation in an uncoded system. Furthermore, channel coding with simple block/trellis coding can render coding gains of the order of 3dB
\cite{forney1984efficient}. 

\subsubsection{Deep Learning}
Although communications field has matured over time, however there are scenarios when the accurate mathematical modeling and rigorous analysis seems unattainable or intractable. Such scenario may arise 
while modeling and equalizing various types of channels and hardware imperfections. Equivalently, the optimal signaling design and detection schemes is another major challenge. To address this concern, machine learning especially its branch deep learning has demonstrated some promising results. For instance,  a communication system can be trained as an auto-encoder treating TX, channel and RX as one deep neural network. The auto-encoder learns appropriate symbol representations of information messages to tackle channel imperfections i.e., noise, distortion, and fading, etc. in order to attain small error PoD. Interestingly, the performance enhancement is achieved 
when the auto-encoder learns asymmetrical constellation in 2U IC setting~\cite{o2017introduction}.

\subsection{Asymmetric Signal Recovery}
Communication systems with asymmetric transmission and/or improper noise require appropriate treatment in terms of equalization, estimation, filtering and detection to account for the induced improperness. Therefore, WL processing models are incorporated to design several estimation \cite{schreier2010statistical}, filtering \cite{mandic2009complex}, and detection \cite{aghaei2010widely} algorithms. WL processing is utilized in communication systems that apply asymmetric constellations e.g., 
\cite{gerstacker2003receivers,xiao2010improved,steinwandt2011widely,mattera2006widely,darsena2007universal}
, and/or encounter improper noise, e.g., \cite{yoon1997maximizing}. Additionally, it is exploited in linear-dispersion STBC,
e.g., \cite{aghaei2010widely,hassibi2002high,gerstacker2004equalization}, and, recently, in IC and BC \cite{hellings2013qos,ho2012improper}.
\subsubsection{Equalization}
In band-limited, high data rate digital communication systems, equalizers flatten the channel frequency response in order to minimize channel distortion on the transmitted information symbols. Asymmetric transmission, improper noise/interference, and/or frequency-selective channel necessitate the employment of WL-equalizers in place of SL-equalizers. For example, real-valued data transmission over complex-valued frequency-selective channels producing ISI necessitates MMSE based WL equalizers for RXs with or without decision feedback \cite{gerstacker2003receivers}. Besides this,  STBC  transmission with conventional equalizers and decoders require equal receiver and transmitter antennas for detection. However, this kills the main purpose of STBC to achieve  pure transmit diversity. Employment of MMSE  based WL equalizers with/without decision feedback can overcome this restriction to further enhance the achievable data rates \cite{gerstacker2004equalization}.  
Multicarrier systems with finite impulse response (FIR) linear precoders and asymmetric constellations offer an intrinsic source of redundancy, which aids in efficient design of WL-ZF universal equalizers for immaculate symbol recovery in FIR-channel with narrowband interference \cite{xiao2010improved}.
\subsubsection{Estimation}
Estimation involves approximating/estimating the value of an entity from a sequence of observations or measurements. Conventional linear estimation aims at designing $\V{u}$ to approximate $y$ from a set of observations stacked in $\V{x}$ using the linearly combination ${\hat y}=\V{u}^{\RM H}\V{x}$. Still it fails to exploit the information hidden in the correlation between  the observations and their complex conjugate. 
Thus, WLE designs $\V{v}$ and $\V{w}$ to better approximate $y$ as ${\hat y}=\V{v}^{\RM H}\V{x}+\V{w}^{\RM H}\V{x}^*$.
WLE is advantageous in various applications in signal processing \cite{adali2011complex,xia2017full}, communications \cite{chevalier2006new}, power systems \cite{xia2012adaptive}, biomedical engineering \cite{xia2011augmented} and renewable energy \cite{mandic2009complex}. WLE can be carried out using various variants like WL-MMSE estimation \cite{picinbono1995widely}, WL minimum
variance distortionless response (WLMVDR) estimation [26],
[27], [29], [77], and the WL-LMS
algorithm. The practical application of these estimators in modern digital communication systems with asymmetric constellations and/or improper noise/interference are overwhelming.
The important class of space-time block-codes for MIMO
channels i.e., linear dispersion codes is constructed from linear combination of input symbols and their complex conjugates. 
Linear dispersion codes like orthogonal, quasi-orthogonal
and V-BLAST codes utilize WL-MMSE estimates of transmitted
symbols as the sufficient statistics for ML detection of these symbols \cite{aghaei2010widely}. 
IQI aware WL-MMSE RX with channel estimation and data detection capability outperforms its linear counterpart in uplink multicell massive MIMO systems. It jointly suppresses MU interference, pilot contamination, and IQI while performing closely to the MMSE RX in a perfect system without IQI \cite{zarei2016q}. 
A linear RX cannot reap the maximum benefit from WL precoding in a MU MIMO-BC. Therefore, WL estimator is required to maximize the weighted sum-rate with limited power budget of the participating base station \cite{bai2018optimal}.
\subsubsection{Filtering}
Filtering separates any entity or group of entities from a mixture/amalgam based on their distinct characteristics. Analogous to other signal recovery schemes, optimal WL filtering is superior to SL filtering in the NC context. It is widely used in applications such as detection \cite{picinbono1996extensions}, prediction \cite{picinbono1997second}, modeling \cite{schreier2003second}, interference cancellations like co-channel interference (CCI) \cite{latouche1998mmse} and narrow-band interference (NBI) 
and equalization for SISO \cite{gerstacker2003receivers,chevalier2006new},
MIMO \cite{gerstacker2004equalization},  and DS-CDMA systems 
\cite{gelli1998cyclostationarity,buzzi2002generalized}.
 Asymmetric complex nature of BPSK signals has motivated researchers to apply WL filtering at the RX for improved detection \cite{schreier2003second}. Moreover, co-antenna interference in the generalized MIMO systems (transmitting complex conjugates along with actual data) can be effectively attenuated using an iterative RX with WL filters. In such linear space time mappings, WL filter also accounts for the non-circularity arising inherently within an iterative RX \cite{witzke2005linear}. 
Additionally, multiple CCI cancellation in PAM/QAM modulated SIMO is achieved using WL filtering for demodulation. 
Surprisingly, interference cancellation ability of WL-ML RXs is only dependent on interferer modulation type and RX antennas but irrespective of modulation scheme of the desired signal \cite{kuchi2009performance}.  Similarly, fully WL-MMSE filter opens new perspectives for intra-network and external interferences management for 4th generation radio communication cellular networks using the Alamouti scheme \cite{chevalier2011widely}. 
WL array RXs  are also considered optimal for the demodulation of BPSK, MSK, and GMSK Signals under improper interferences. Effective single antenna interference cancellation can be attained for these modulation schemes in GSM cellular networks using WL filtering \cite{chevalier2006new}. 
Furthermore, Hellings \textit{et al.} suggest block-Hankel-skew-circulant structured matrices for appropriate processing of WL filters and asymmetric signaling in MIMO IC~\cite{hellings2015two}.

\subsubsection{Detection}
Unlike the IGS and impropriety detection, asymmetric detection involves the recovery of the transmitted signals inheriting asymmetric structure from asymmetric transmission and/or asymmetric noise/interference. The gains of such WL detection are two folds: 1) It accounts for the asymmetric characteristics for improved detection. 2) It efficiently suppresses the effects of asymmetric noise/interference. These perks cannot be attained by the conventional detectors neglecting SOS. For example, WL structures for blind MUD in synchronous DS/CDMA systems can effectively suppress both narrowband and wideband multiple-access interference \cite{gelli2000blind}. 
Similarly, WL RX for OQPSK modulated DS/CDMA system  employing least-mean square algorithm efficiently mitigates both symmetric/asymmetric interference \cite{mirbagheri2006enhanced}.
Furthermore, asynchronous DS/CDMA systems employing BPSK requires a new family of MMSE detectors to jointly suppress multi-access and external interference by fully exploiting the SOS through poly periodical processing \cite{buzzi2001new}.
Direct application of WL processing may not be suitable in some scenarios e.g., MIMO FBMC/OQAM. Therefore, a two step RX; first with linear and later WL processing may help in removing the intrinsic interference which keeps us from taking the full advantage of WL-MMSE RX \cite{cheng2013widely}.
\paragraph{Discussion}
 WL extension of various RXs may or may not be helpful in attaining optimum performance depending on the type of
application, employed modulation/coding schemes and degrading noise/interference.  For instance, WL matched filter solution does not take account noise and is still linear solution \cite{adali2011complex}. Alternately, optimum ML detection and WL-MVDR estimation only requires adjustments for improper noise/interference and is irrespective of the symmetric/asymmetric transmission. Whereas, other detectors like WL-MMSE, WL-ZF, and WL-LMS filtering may require the propriety characterization of the transmitted signals as well as received noise.
Importantly, the performance gains of WL model diminish when the underlying system is linear \cite{adali2011complex}.

\subsection{Error Probability Analysis}
EP is a tangible measure used to fairly judge
the performance of communication systems. EP captures prime system details (e.g., modulation scheme, RX
type, symbol constellation, etc.) and is considered the most
revealing metric about the communication system performance \cite{simon2005digital}. It is the probability of receiving the erroneous information and can be studied through pairwise EP (PEP), bit EP (BEP), symbol EP (SEP), and frame/block EP (FEP) etc. In the context of wireless networks, EP has mainly been studied
and conducted for symmetric transmission and additive white proper Gaussian noise. This section summarizes the contributions that tackle EP analysis for asymmetric constellations with/without improper noise/interference in various system configurations.

\subsubsection{Multiuser Direct Sequence Multiple Access  Systems}
Various trellis-coded systems with asymmetric PSK transmission outperform their counterparts with symmetric PSK signaling. For example, the performance of the DS spread-spectrum multiple access (SSMA) system employing trellis coded modulation (TCM) with asymmetric signal constellations outperforms its competitor symmetric signal constellations by 58.02\% for 3U DS/SSMA system and offers 
1dB SNR reduction at 1e-10 BER.  Similarly, 
WL MUD offers perks over SL MUD in a DS/CDMA system with QPSK modulation suffering from BPSK NBI. BER reduction of 100\% and 96\% is observed by WL-MUD RX w.r.t conventional matched filter and L-MUD, respectively, for the first user out of 6 CDMA users. Alternately it offers EbNo gains of 2dB at 1e-5 BER w.r.t L-MUD RX \cite{gelli2000blind}.

\subsubsection{Coded Communication Systems}
Interestingly, deep learning for the physical layer suggests the joint optimization of TX and RX components by modeling 
a communications system as an auto-encoder and training it using stochastic gradient descent. This approach to minimize block error rate (BLER) outperforms all the existing well-known schemes by yielding the asymmetric modulation as the optimum one.
Surprisingly, for a communications system employing BPSK modulation and a Hamming code, the trained auto-encoder outperforms the uncoded and coded scheme with hard decision (HD) by 98.64\% and
96.25\%, respectively. It further offers EbNo gain of 2dB at 8e-4BLER and 1.5dB at 3e-4BLER over uncoded and coded scheme with HD, respectively. The gains further increase with increasing the batch
size while reducing the learning rate during training. Unanticipated asymmetric learned constellations of auto-encoder performs  
equally good as the coded scheme with ML detector without any prior knowledge \cite{o2017introduction}. Similarly, for a 2U IC, the auto-encoder and time-sharing have identical BLERs for (1, 1) and (2, 2), the former yields significant gains of around
0.7dB for (4, 4) and 1dB for (4, 8) at a BLER of 1e-3. 
O'Shea \textit{et al.} further claim that auto-encoder with RTN can outperform differential BPSK with ML estimation and Hamming (7,4) code in a multipath fading environment. 

Asymmetric TCM (ATCM) provides better SNR gains over the traditional symmetric TCM (STCM) for very low data rate systems especially with small number of users in high SNR and less drastic fading circumstance \cite{park1998ds}. Similarly, asymmetric 8-PSK signal sets in 4-state and 8-state rate 2/3 TCM outperforms symmetric 8-PSK TCM owing to the increase in the minimum product distance.  ATCM reduces BER up to 20\%, 76.67\% and 50\% relative to STCM for Rayleigh, Rician and light shadowed Rician channels, respectively. Moreover, an improvement of about 0.3-0.4dB around BER 1e-5 is observed over the Rayleigh and shadowed Rician channels. Whereas, it reduces to 0.2-0.3dB for the Rician channels. Subramaniam \textit{et al.} claim that the improvement of 0.2-0.4dB signifies 5-10\% savings in power, owing to the enlarged minimum product distance and free Euclidean distance by the ATCM \cite{subramaniam2000performance}. Another approach to design ATCM is to minimize BER $\left( \text{or BER bound} \right)$ instead of maximizing the free Euclidean distance. Zhang \textit{et al.} present 75\% and 55\% reduction in BER with 2-state and 4-state ATCM relative to STCM. Furthermore, the optimum ATCM achieves 0.5dB SNR gain at 1e-5 BER and 0.1dB SNR gain at 1e-7 BER with 2-state code and 4-state codes, respectively \cite{zhang2009optimum}. Moreover, asymmetric signals with non-uniform spacing outperform symmetric signals with uniform spacing in the trellis-coded systems. For instance, two-state trellis-coded optimum asymmetric 4PSK offers 99.83\% less BER and 0.5dB EbNo gain at 1e-5 BER as compared to its symmetric counterpart. Additionally, two-state trellis-coded  asymmetric 4-AM 98.33\% less BER and 1dB EbNo gain at 1e-5 BER as compared to its symmetric counterpart \cite{divsalar1987trellis}. Divsalar \textit{et al.} emphasize that the asymmetric signaling does not affect power or bandwidth requirements of the system. Thus, rendering BER performance gains at little or no cost.

\subsubsection{Multiple Antenna Systems}
Conventional iterative RXs which are optimal for the symmetric modulations, such as $M$-QAM and $M$-PSK, are suboptimal for the asymmetric modulations, such as  $M$-ary ASK, OQPSK (for which ${\rm{E}}\left[ {{\V{ s}}{\V{s}^{\rm{T}}}} \right] \ne \M{0}$) in uncoded MIMO systems. Therefore, Xiao \textit{et al.} proposed a novel iterative RX with various decoding strategies like ZF, MMSE and SIC etc.  Accommodating the asymmetric behavior of the 4-ASK and OQPSK signal constellations offer BER percentage decrease up to 
 99.25\% and 99.58\% with ZF RX and 
97.78\% and 97.50\% with MMSE RX, respectively.  In other terms, 
the proposed scheme SNR gains of 
8dB at 2e-3 and 12dB at 1.5e-3 with ZF RX and  6dB at 2e-4
and 8.5dB at 3.5e-5 with MMSE RX, respectively, with  4-ASK and OQPSK modulations in 4x4 uncoded MIMO system. The percentage decrease in BER with 4-ASK modulation employing MMSE based novel iterative RX in 4x3, 4x4 and 4x5 uncoded MIMO system is 98.9\%, 98\% and 94\%, respectively. Interestingly, accounting for asymmetry not only achieves superior performance and faster convergence than the conventional systems but renders equivalent effect of increasing receive diversity order \cite{xiao2009iterative}. 

\begin{table*}[htbp]
\renewcommand{\arraystretch}{1.25}
\caption{Error Probability Reduction with Appropriate Modeling (Signaling, Filtering, Estimation or Detection)}
\begin{center}
  \begin{tabular}{||c|c|p{1cm}|c|p{2.5cm}|c|c||}
    \toprule
 \textbf{System} & \textbf{Transmitter} & \textbf{Detector} & \textbf{Competitor} &
\textbf{Percentage Reduction} & \textbf{SNR Gain}  & \textbf{Ref} \\
      \midrule
{{\shortstack[c]{Generalized\\ MIMO }}}&
{{\shortstack[c]{16-QAM \\with V-Blast }}}&
{{\shortstack[c]{WLD \\ iterative \\ RX}}}&
{{\shortstack[c]{WL filter \\versus SL filter}}} & 30\% & 0.3dB at 0.02 FER
& \cite{witzke2005linear} \\
      \hline
{{\shortstack[c]{Multicarrier\\ Transmission \\ (UW-OFDM)}}}&
{{\shortstack[c]{$M$-ASK \\ modulation \\ with UW-OFDM}}} &
WL-MMSE &
{{\shortstack[c]{WL-MMSE versus \\LMMSE}}}&
{{\shortstack[c]{AWGN: 97.4\%  \\ IEEE indoor \\channel: 94.5\%}}} &
{{\shortstack[c]{1dB at 2e-5 BER\\
3dB at 1e-6 BER}}}
 &
\cite{trampitsch2013complex} \\
      \hline

{{\shortstack[c]{2-16 states \\ trellis \\coding }}}&
{{\shortstack[c]{Asymmetric  \\$M$-PSK,$M$-AM \\ modulation}}}&
{{\shortstack[c]{Viterbi \\ decoder}}}&
{{\shortstack[c]{Asymmetric \\versus symmetric \\ signaling}}}&
{{\shortstack[c]{4PSK:99.83\% \\4AM:98.33\%}}}&
{{\shortstack[c]{0.5dB at 1e-5 BER \\
1dB at 1e-5 BER}}}&
\cite{divsalar1987trellis} \\
  \hline
  
{{\shortstack[c]{DS/SSMA using \\ trellis coding}}}&
{{\shortstack[c]{Asymmetric \\ $M$-PSK}}} &
{{\shortstack[c]{  Viterbi\\ decoder}}} &
{{\shortstack[c]{Asymmetric signaling \\ versus \\Symmetric signaling }}}&
3U: 58.02\%  &
1dB at 1e-10 BER &
\cite{park1998ds} \\
   \hline
   
 {{\shortstack[c]{4- and 8-state \\TCM schemes}}}&
{{\shortstack[c]{Asymmetric \\ 8-PSK }}}&
{{\shortstack[c]{Viterbi \\decoder}}} &
{{\shortstack[c]{ATCM versus STCM \\ to minimize 
 \\Euclidean distance }}} &
{{\shortstack[c]{Rayleigh: 20\% \\
Rician: 76.67\% \\
Shadowed Rician: 50\% }}} &
{{\shortstack[c]{0.3dB at 1e-5 BER \\
0.4dB at 1e-5 BER \\
0.2dB at 1e-5 BER  }}}&
\cite{subramaniam2000performance} \\
   \hline
 
 {{\shortstack[c]{ 2- and 4-state \\TCM schemes}}} &
 {{\shortstack[c]{Asymmetric \\4-PSK}}} &
 {{\shortstack[c]{Viterbi\\ decoder}}} &
 {{\shortstack[c]{ATCM versus STCM \\for BER bound \\
  minimization}}}&
 {{\shortstack[c]{2 states: 75\% \\
4 states: 55\% }}}&
 {{\shortstack[c]{0.5dB at 1e-5 BER \\
0.1dB at 1e-7 BER }}} &
\cite{zhang2009optimum} \\
  \hline
  
 {{\shortstack[c]{ MU IC}}}&
 {{\shortstack[c]{ Asymmetric \\ $M$-PSK \\or $M$-QAM }}}&
 {{\shortstack[c]{ Symbol-by-\\Symbol\\ detection}}} &
 {{\shortstack[c]{ Symmetric versus \\asymmetric for max \\ PEP/SER 
Minimization}}} &
 {{\shortstack[c]{ AWGN: 100\% \\ Cellular: 99\% }}}&
 {{\shortstack[c]{  -\\ 5dB at 8e-2 max(BER) }}}&
\cite{nguyen2015improper} \\
  \hline

 {{\shortstack[c]{Deep learning \\ for commun.\\
  system }}}&
 {{\shortstack[c]{Learned coded \\ modulation by\\ auto-encoder}}} &
 {{\shortstack[c]{Neural \\ networks }}}&
 {{\shortstack[c]{Auto-encoder versus \\ conventional 
 modulation/\\coding to 
Min BLER}}} &
 {{\shortstack[c]{Uncoded: 98.64\%  \\Coded:96.25\% }}}&
{{\shortstack[c]{0.7dB at 1e-3 BLER \\ 
 1dB at 1e-3 BLER }}}&
\cite{o2017introduction} \\
  \hline
  
 {{\shortstack[c]{Phase \\ estimation \\without pilot }}} &
 {{\shortstack[c]{Asymmetric \\8-PSK}}} &
 {{\shortstack[c]{Coherent \\detection }}}&
 {{\shortstack[c]{Asymmetric 8-PSK \\versus\\ symmetric
  8-DPSK}}} &
97.50\% &
3dB at 1e-2 SEP &
\cite{thaiupathump2000asymmetric} \\
  \hline
  
\multirow{2}{*}{{{\shortstack[c]{Uncoded \\ MIMO \\ systems}}}}&
\multirow{2}{*}{{{\shortstack[c]{$M$-ASK, \\ QPSK/OQPSK }}}}&
ZF RX &
\multirow{2}{*}{{{\shortstack[c]{Iterative RX with \\asymmetric versus \\ symmetric modulation}}}}&
{{\shortstack[c]{ASK: 99.25\% \\ OQPSK: 99.58\% }}}&
{{\shortstack[c]{8dB at 2e-3 BER \\ 12dB at 1.5e-3 BER  }}}&
\cite{xiao2009iterative} \\
  \cline{3-3}
    \cline{5-7}
& & MMSE RX & &
{{\shortstack[c]{ASK: 97.78\% \\ OQPSK: 97.50\%}}}&
{{\shortstack[c]{6dB at 2e-4 BER \\ 8.5dB at 3.5e-5 BER }}}&
\cite{xiao2009iterative} \\
\hline

{\shortstack[c]{HWI AF \\ Relaying}} &
{{\shortstack[c]{QAM with \\ Optimal power \\ allocation}}}&
\multirow{3}{*}{{{\shortstack[c]{MED \\ and \\  MLD}}}}&
{{\shortstack[c]{Optimal versus \\ suboptimal \\ detectors}}}&
{{\shortstack[c]{ RX only IQI: 37.50\% \\
TX only IQI: 95.48\% \\
TX\& RX IQI: 55.56\% }}} &
{{\shortstack[c]{ 2dB at 1e-2ASEP\\
3.5dB at 1e-2ASEP\\
3dB at 1e-2ASEP}}} &
{\cite{canbilen2019impact}} \\
\cline{1-2}
\cline{4-7}
\multirow{2}{*}{{{\shortstack[c]{HWI system}}}}&
{{\shortstack[c]{Asymmetric \\ GPSK, QAM }}}&  &
{{\shortstack[c]{Asymmetric versus\\ symmetric modulation to \\ minimize max PEP/SEP}}}&
{{\shortstack[c]{10dB SNR: 84\% \\ 20dB SNR: 97.8\% }}}&
10dB at 8e-2 SER &
\cite{javed2018asymmetric} \\
\cline{2-2}
\cline{4-7}
 & {{\shortstack[c]{Grey coded \\ $M$-QAM \\ modulation}}}& &
\multirow{2}{*}{{{\shortstack[c]{Optimal MLD with \\ IGN versus traditional \\ MLD with PGN}}}}&
 {{\shortstack[c]{$\sigma_{\rm t}^2$ = 0.001: 8.33\% \\ $\sigma_{\rm t}^2$ = 0.010: 38.64\% \\ $\sigma_{\rm t}^2$ = 0.100: 52.78\%}}}&
 {{\shortstack[c]{10dB at 1.2e-1 BER \\ 9dB at 4.4e-2 BER \\ 4dB at 1.7e-3 BER}}}&
\cite{javed2017optimal} \\
\cline{1-3}
\cline{5-7}

{{\shortstack[c]{MIMO system \\under IGN}}}&
{{\shortstack[c]{SSK \\ modulation}}} &
{{\shortstack[c]{ ML \\ RX}}} & &
{{\shortstack[c]{PGN: 0\% \\
n.id IGN: 20\% \\
correlated IGN:32\% \\
correlated and \\ n.id  IGN: 44\% }}}&
{{\shortstack[c]{0dB at 5e-3 EP \\
1dB at 5e-3 EP \\
 2dB at 5e-3 EP \\
 3dB at 5e-3 EP }}}&
\cite{alsmadi2018ssk} \\
\hline

\multirow{2}{*}{{{\shortstack[c]{MIMO with \\ data like\\ CCI and ISI \\ in FS channel}}}}&
\multirow{2}{*}{{{\shortstack[c]{SC block \\ transmission \\ of  OQPSK}}}}&\multirow{2}{*}{
{\shortstack[c]{WL- \\ MMSE \\ with  FF, \\ NP-FB}}}&
\multirow{2}{*}{{{\shortstack[c]{Linear versus \\ WL FF and NP-FB \\ filters}}}}&
{{\shortstack[c]{w/ CCI: 98.50\% \\ w/o CCI: 99.67\% }}}&
{{\shortstack[c]{20dB at 6e-2 BER \\ 8dB at 1e-5 BER }}}&
\cite{kim2016asymptotically} \\
\cline{5-7}
& & & &
{{\shortstack[c]{w/ CCI: 97.10\% \\  w/o CCI: 99.75\% }}}&
{{\shortstack[c]{20dB at 6.2e-2 BER \\ 8dB at 1e-5 BER}}}&
\cite{kim2016asymptotically} \\
  \bottomrule
  \end{tabular}
\end{center}
\label{tab:Error}
\end{table*}

\begin{table*}[htbp]
\renewcommand{\arraystretch}{1.25}
\begin{center}
  \begin{tabular}{||c|c|p{1cm}|c|p{3cm}|c|c||}
    \toprule
{{\shortstack[c]{SIMO with \\ multiple CCIs}}}&
{{\shortstack[c]{PSK data \\ modulation and \\PAM/QAM CCI}}}&
{{\shortstack[c]{WL \\ML \\RX}}}&
{{\shortstack[c]{WL RX \\versus \\conventional \\  RX}}}&
{{\shortstack[c]{Desired – CCI \\
QPSK-BPSK: 83.53\%\\
BPSK-BPSK: 98.15\%\\
BPSK-PAM+QAM: 80\%\\
BPSK-2BPSK: 42\%}}}&
{{\shortstack[c]{7dB at 8e-4 SER\\
12dB at 1.3e-1 SER\\
4dB at 1.4e-3 SER\\
3dB at 1e-3 SER }}}&
\cite{kuchi2009performance} \\
\hline

\multirow{2}{*}{ {{\shortstack[c]{MIMO-FBMC\\  systems}}}}&
\multirow{2}{*}{ {{\shortstack[c]{FBMC \\ modulation \\based on \\offset
QAM}}}} &
MSE RX &
{{\shortstack[c]{SL versus WL \\ processing}}}&
{{\shortstack[c]{ITU:PAG: 91.25\% \\
ITU:VAG: 91.9\% }}}&
{{\shortstack[c]{2.5dB at 8e-6 BER \\
5dB at 5e-4 BER }}}&
\cite{caus2013comparison} \\

 \cline{3-7}
& & {{\shortstack[c]{ LP and\\ WLP }}}&
{{\shortstack[c]{MMSE versus\\ WL-MMSE }}}&
 {{\shortstack[c]{Partial IC:  97.6\% \\
Full IC: 99.6\%}}}&
{{\shortstack[c]{9dB at 2.5e-3 BER \\
10.5dB at 2.5e-3 BER }}}&
\cite{cheng2013widely} \\
\hline
 
{{\shortstack[c]{DS/CDMA \\ systems}}}&
{{\shortstack[c]{ QPSK \\ modulation}}}&
{{\shortstack[c]{MMSE \\MUD}}}&
{{\shortstack[c]{L-MUD versus \\ WL-MUD}}}&
{{\shortstack[c]{MF: 100\% \\ L-MUD: 96\% }}}&
{{\shortstack[c]{ - \\ 2dB at 1e-5BER }}}&
\cite{gelli2000blind} \\
 \hline
 
 \multirow{2}{*}{{{\shortstack[c]{ Frequency\\ selective\\ channels }}}}&
{{\shortstack[c]{ASK, OQAM \\ or BMSK}}}&
 \multirow{2}{*}{ {{\shortstack[c]{WL-\\MMSE}}}}&
\multirow{2}{*}{{{\shortstack[c]{ LE versus WLE \\ with or without \\decision feedback}}}}&
{{\shortstack[c]{MMSE-LE: 96.07\% \\ MMSE-DFE: 56\%}}}&
{{\shortstack[c]{5dB at 1e-4BER \\ 0.7dB at 1e-4BER}}}&
\cite{gerstacker2003receivers} \\
\cline{5-7}
\cline{2-2}
 
&  {{\shortstack[c]{STBC \\ with $M$-PSK\\ modulation}}} & & &
{{\shortstack[c]{ RA SISO-LE: 93.5\% \\
HT SISO-DFE: 72.5\% \\
TU SIMO-DFE: 74.29\% }}}& 
{{\shortstack[c]{7dB at 2e-3 BER \\
2dB at 1e-3 BER\\
2.3dB at 7e-4 BER}}}&
\cite{gerstacker2004equalization} \\
    \bottomrule
  \end{tabular}
\end{center}
\end{table*}

Co-antenna interference suppression in generalized MIMO systems with linear dispersion codes like V-BLAST achieve up to 30\% less frame error rate with WL detector (WLD) as compared to SL detector with iterative RX. Furthermore, it renders 0.3dB SNR gain at 2e-2 FER \cite{witzke2005linear}. 

WL filtering is beneficial for effective demodulation of PAM/QAM modulated SIMO systems suffering from multiple data-like CCIs.  
Kuchi \textit{et al.} argue the tradeoff between diversity advantage and interference cancellation in a flat Rayleigh fading channel. Interestingly, WLF offers SER reduction up to 83.53\%, 98.15\%, 80\% and 42\% for desired-CCI combinations of 1x1 QPSK-BPSK, 1x1 BPSK-BPSK, 1x1 BPSK-PAM+QAM and 1x2 BPSK-2BPSK, respectively. In other words, the respective SNR gain of these four examples are given by 7dB at 8e-4 SER, 12dB at 1.3e-1 SER, 4dB at 1.4e-3 SER and 3dB at 1e-3 SER, respectively. Compelling WL RX with $N$-antennas is capable of rejecting any combination of $M_1$ PAM and $M_2$ QAM interferers satisfying $M_1 + 2M_2 < 2N$, whereas SL RX can only reject up to $M_1 + M_2$ interferers with $M_1 + M_2 < N$ \cite{kuchi2009performance}. In a nutshell, WL RX can handle more CCIs relative to SL RX while maintaining a certain SER. 

\subsubsection{Multi-carrier and Single-carrier Systems}
WL processing with asymmetric transmission is advantageous in both multi-carrier and single-carrier systems. For example,  WL-MMSE is beneficial over LMMSE for unique word (UW)-OFDM systems with ASK modulation which introduces asymmetry in the system. Interestingly, WL-MMSE yields up to 97.4\% and 94.5\% less BER relative to LMMSE for 8-ASK modulation over AWGN and IEEE indoor channels, respectively. Additionally, it offers SNR reduction of  1dB at 2e-5 BER for 16-ASK modulation and 3dB at 1e-6 BER for 4-ASK modulation over AWGN and IEEE indoor channels, respectively \cite{trampitsch2013complex}. 
Filter bank multicarrier (FBMC) scheme to combat frequency selective fading in MIMO systems employing OQAM modulation require WL processing to account for NC transmitted signals. Caus \textit{et al.} present two conditions i.e., number of streams $\left(S \right)\leq N_{\rm R}\leq3, N_{\rm T}\geq	N_{\rm R}$ and Coherence BW $ >>$ Subcarrier Spacing, 
when linear processing is superior to WL processing. Otherwise, linear processing gives an error floor and thus, WL processing is the preferred choice in terms of BER, especially in low noise. Intuitively, for a $N_{\rm R} = 2$, $N_{\rm T} = 3$ and $S=2$ MIMO system, WL processing to minimize the sum MSE outperforms it's counterpart by 91.25\% and 99.9\% when the underlying channel follows international telecommunication union (ITU): Pedestrian A guidelines (PAG) and Vehicular A guidelines (VAG), respectively.  Moreover, the respective SNR gains are obtained as 2.5dB at 8e-6 BER and 5dB at 5e-4 BER, respectively, at high SNR regime \cite{caus2013comparison}. Alternately, Cheng \textit{et al.} suggest two step RX based on linear and WL processing to reap the benefits of both domains. First step cancels intrinsic interference that prevents to reap maximum benefit from WL processing and second step employs WL-MMSE RX to cater for RSI. This technique renders 97.6\% and 99.6\% BER reduction with partial  and complete interference cancellation in first step, respectively, in 4x4 MIMO FBMC/OQAM system. Equivalently, the respective SNR gains are 9dB and 10.5dB at 2.5e-3 BER \cite{cheng2013widely}. Similarly, MIMO single carrier (SC) block transmission suffering from inter-symbol interference (ISI) and data-like CCI requires WL filters with feed forward (FF) and noise prediction feedback (NP-FB) attributes to outperform its linear counterparts. Both suboptimal and optimal SC frequency domain equalizers with WL filters are proposed for OQPSK modulated data and CCI. Surprisingly, WL filters outperform SL filters by 98.50\% and 99.67\% with optimal equalizer and by 97.10\% and 99.75\% with low-complexity suboptimal equalizer in the presence and absence of CCI, respectively. Moreover, both equalizers offer EbNo gains of around 20dB at 6e-2 and 8dB at 1e-5 with and without CCI, respectively, for a 2x2 MIMO SC block transmission with a cyclic prefix~\cite{kim2016asymptotically}.

\subsubsection{Systems with Frequency Selective Channels}
equalization schemes based on WL processing outperform their linear counterparts in frequency selective channels with underlying ASK, OQAM, BMSK type modulation schemes.  Gerstacker \textit{et al.} propose FIR filters for WL-MMSE equalization
without and with decision feedback , termed as
MMSE-WLE and MMSE-WDFE, respectively. The novel MMSE-WLE and MMSE-WDFE RXs provide BER reduction up to 96.07\%
and 56\% at 20dB EbNo relative to their linear counterparts MMSE-LE and MMSE-DFE, respectively. Alternately, they offer respective EbNo gains of 5dB and 0.7dB at 1e-4 BER for real-valued transmission over complex ISI channel \cite{gerstacker2003receivers}. On the other hand, if the zeros of the channel are far from unit circle then WLP can only offer small-to-moderate gains.
Gerstacker \textit{et al.} further extended their findings of WLE to STBC over frequency selective channels. Numerical examples show BER reduction of  93.5\%,  72.5\% and 74.29\%  with 2x1 STBC WLE employing 8PSK, 2x1 STBC WDFE employing BPSK and 2x2 STBC WDFE employing 8PSK relative to SISO-LE, SISO-DFE and 1x2 SIMO-DFE over rural area (RA), hilly terrain (HT) and typical urban area (TU), respectively \cite{gerstacker2004equalization}.

\subsubsection{Systems with Improper Noise}
Similarly for general improper Gaussian noise (IGN),  Alsmadi \textit{et al.} present an optimal detector for MIMO system with space shift keying (SSK) RX. The accommodation of improper nature of the additive noise in optimal detection results in the average EP percentage reduction up to 0\%, 20\%, 32\% and 
44\% for RXs affected by proper Gaussian noise, non-identical
uncorrelated IGN , identical
correlated IGN and non-identical correlated IGN, respectively, relative to the traditional ML detector considering PGN.
Moreover, respective SNR gains are given by 0dB, 1dB, 2dB and 3dB at 5e-3, respectively \cite{alsmadi2018ssk}.

\subsubsection{Hardware Impaired Systems}
Evidently, HWI systems under IQI and additive distortions alter the symmetry of the signals under study. Thus, accounting for the induced asymmetry in the detection process presents benefits in terms of average BER reduction up to 8.33\%, 38.64\% and 52.78\% for low, moderate and high transmit distortion levels, respectively. Alternately, it renders SNR gain up to 10dB at 1.2e-1, 9dB at 4.4e-2
and 4dB at 1.7e-3, respectively, for the three aforementioned transmit distortion cases \cite{javed2017optimal}.
Asymmetric signaling is favorable in suppressing improper accumulated noise and self-interference resulting from the hardware imperfections like IQI and non-linear distortions. Thus, Javed \textit{et al.} suggest to employ asymmetric QAM to minimize maximum PEP or SEP. The proposed asymmetric transmission renders up to 84\% and 97.8\% reduced average SER with respect to symmetric transmission with ML detector at 10dB and 20dB SNR, respectively. Additionally, asymmetric transmission attains the same SER performance (8e-3) with 10dB SNR as that of symmetric modulation with 20dB SNR \cite{javed2018asymmetric}. 
Moreover, Canbilen \textit{et al.} advocate the effectiveness of optimal MLD which incorporates the asymmetric characteristics over suboptimal MLDs in a dual-hop AF relay system suffering from IQI. They claim  37.50\%, 95.48\% and 55.56\% reduction in average SEP at 30dB SNR when the system is subject to RX only, TX only and both TX-RX IQI. This reduction exhibits 2dB, 3.5dB and 3dB SNR gain at 1e-2 average SEP for RX only, TX only and both TX-RX IQI systems \cite{canbilen2019impact}.
%
\subsubsection{Multiuser Interference Channel}
Asymmetric signaling is also beneficial in MU IC when interference is treated as noise. Thus, Nguyen \textit{et al.} optimize the precoding matrices in MU SISO ICs to minimize maximum pairwise
EP (PEP) and symbol error rate (SER) \cite{nguyen2015improper}. They present the advantageous asymmetric signaling under both AWGN and cellular networks with or without channel coding.
Asymmetric signaling offers 100\% and 99\% decrease in SER with minmax-PEP/minmax-SER objectives relative to symmetric signaling with power control for 3U SISO IC in AWGN and cellular setup, respectively. Furthermore, it renders 5dB SNR gain at 8e-2 BER in a cellular setup with 3-edge users employing low-density-parity-check (LDPC) coding with QPSK modulation \cite{nguyen2015improper}.  

In a nutshell, the generalized approach to account for system asymmetry instead of an unrealistic symmetry assumption equip system designers with appropriate tools and additional design freedom to achieve lower EP, depending on the considered application, as emphasized in Table~\ref{tab:Error}.
 \begin{figure}[t]
\begin{minipage}[b]{1.0\linewidth}
  \centering
  \centerline{\includegraphics[width=14cm]{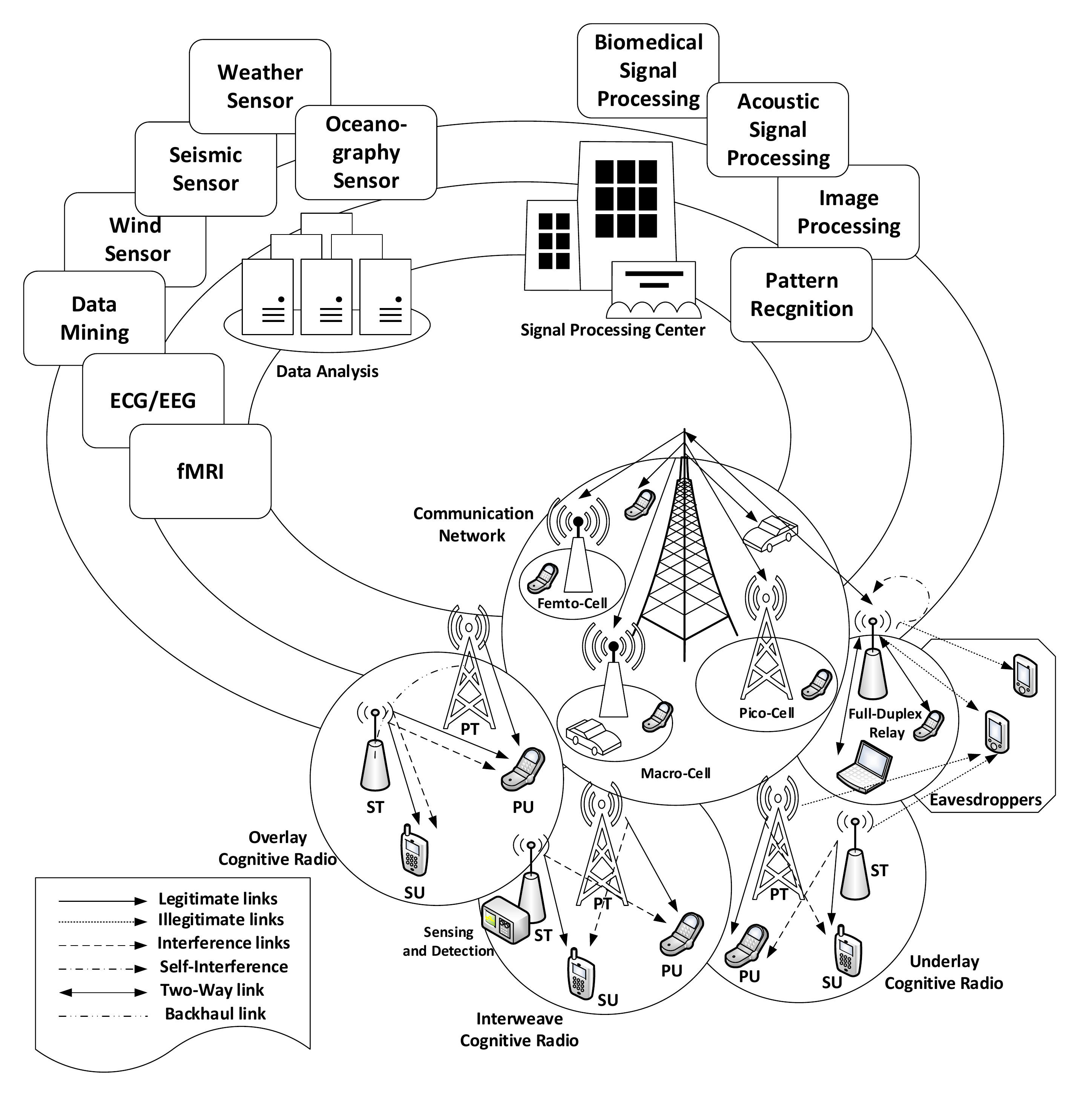}}
\end{minipage}
\caption{Applications of Impropriety in Data Analysis, Signal Processing and Communication Networks}
\label{fig:App}
\end{figure}
\section{Applications} 
Previously, we discussed various scenarios and sources which arise impropriety in the system. This impropriety can be exploited to achieve performance gain in numerous diverse fields including but not limited to medicine, communication, geology and computer vision. The existing impropriety exploitation and/or introduced impropriety utilization have wide applications
in data analysis, signal processing and communications \cite{taubock2012complex,adali2011complex,schreier2010statistical,ho2012improper}.
Intuitively, asymmetric signaling and WL processing are the concepts that go hand in hand \cite{hellings2015block} and are beneficial in the various settings as demonstrated in Fig. \ref{fig:App}.
\subsection{Data Analysis}
{Impropriety incorporation and appropriate processing provide tremendous advantages in data analysis and characterization. Key data analysis tools such as ICA and principal component analysis (PCA) demonstrate enhanced performance owing to NC characterization} \cite{hyvarinen2001independent,eriksson2006complex,li2011noncircular}.
{Other miscellaneous data analysis techniques with impropriety incorporation successfully categorize seismic, oceanographic and weather data sets} \cite{samson1983pure,adali2011complex,wang2015framework}. 
\subsubsection{{Independent Component Analysis}}
{ICA is a relatively new statistical and computational technique for revealing hidden factors that underlie sets of random variables, measurements, or signals. It particularly aims at the blind recovery of the source signal from the observations} \cite{eriksson2006complex}. {Many complex ICA algorithms either assume that the underlying sources are circular or rely on a magnitude-only model. However, this greatly limits the performance of ICA} \cite{rowe2005modeling}. {Thus, NC characterization enabled ICA has found real-world applications in various diverse fields such as medicine, economics, and data mining}
\cite{li2011application,adali2007complex,hyvarinen2001independent,wang2006data}.
\paragraph{{Biomedical Data Analysis}}
{Human brain functioning can be thoroughly examined using complex-valued fMRI data which measures the electrical and magnetic activity of the human brain} \cite{hyvarinen2001independent}. {However, fMRI poses tremendous
challenges for data analysis techniques, including the design of robust yet flexible framework to capture the richness of human brain activities}
\cite{adali2007complex}. 
{Adaptive noncircular ICA algorithms can effectively address these challenges rendering vast applications} \cite{eriksson2006complex,li2011application}.  {For instance, feature extraction in electrocardiograms and fMRI data analysis lead to the improved neural activity estimation} \cite{schreier2010statistical,rowe2005modeling, rodriguez2012noising}. 
{Moreover, real-time brain-computer interfacing relies on the extraction of eye muscle activity: electrooculogram (EOG) from electroencephalogram (EEG) recordings} \cite{javidi2010complex}. {EEG records electrical
potentials at various locations on the scalp and can render immense WL predictability by using blind source extraction algorithms like~ICA.} 
{Furthermore, a robust ICA technique is also proposed to extract atrial activity in atrial fibrillation electrocardiograms (ECGs) \cite{zarzoso2009robust}.}

\paragraph{{Econometric and Data Mining}}
{One major concern in econometric is the identification of underlying independent causes of a phenomenon, e.g., economic indicators, interest rates, and psychological factors of the exchange rates.
These causes are quite insightful and can be identified by the decomposition of the financial time series analysis using appropriate ICA algorithms} \cite{hyvarinen2001independent}. {Another diverse application of such data analysis tool is in data mining, such as latent variable decomposition, multivariate time series analysis and prediction, extracting hidden signals in satellite images, text document data analysis, and weather data mining}~\cite{wang2006data,eriksson2006complex}.

{
\subsubsection{{Principal Component Analysis}}
PCA transforms correlated observed variables into a subset of uncorrelated variables, that account for total variance. PCA identifies patterns in data based on the correlation between features and thus it is less stringent than ICA \cite{hyvarinen2001independent}. PCA is a classical technique in statistical data analysis for pattern recognition \cite{zhang2001nonlinear,kwak2008principal,zhao2006novel,martis2012application}, feature extraction
\cite{xiao2010principal,suganthy2012principal}, data compression \cite{Jolliffe2002pca,huang2001application}, data reduction \cite{deluzio2007biomechanical,crescenzi2001main,duforet2016detecting}, data visualization \cite{gabriel1971biplot}, noise reduction \cite{does2019evaluation,eldeep}, factor analysis \cite{abdi2010principal,guimet2004application}, model selection \cite{li2011noncircular}, rank reduction \cite{schreier2003second}, dimensionality reduction \cite{hellings2015composite}, etc. 
Classical PCA for real-valued systems relies on SOS i.e., variance maximization. A staright forward extension to the proper complex-valued systems i.e., circular PCA (cPCA) relies on Hermitian covariance matrix \cite{picinbono1994circularity}. However, a more general extension to the complex systems which can be circular/NC is based on:
\begin{itemize}
\item Both covariance and pseudo-covariance estimates to maximize likelihood function in complex representation rendering NC PCA (ncPCA) \cite{li2010principal}. 
\item  The augmented covariance matrices yielding WL PCA (wlPCA) for direct extension from the cPCA \cite{schreier2003karhunen}. 
\item The real-composite covariance matrices resulting composite real PCA (crPCA) \cite{hellings2015composite}.
\end{itemize}
Aforementioned forms of general complex PCA exhibit their own merits and demerits based on the underlying applications and their linear/non-linear models.
\paragraph{Rank, Dimensionality and Data Reduction}
Rank reduction finds balance between model bias and model variance to reconstruct signals from noisy observations. It can be achieved using PCA with eigen analysis of complex vectors. WL transformations take full advantage of SOS when compared to SL transformation. The wlPCA offers concentrated signal variance in first few principal components relative to ncPCA for rank reduction \cite{schreier2003second}.
Interestingly, crPCA demonstrates its superiority in dimensionality reduction owing to the finer granularity and lower computational complexity than that of wlPCA~\cite{hellings2015composite}.
PCA is widely used as a preliminary step for data reduction in various biomedical applications e.g., to extract and differentiate biomechanical features of gait waveform data related to knee osteoarthritis \cite{deluzio2007biomechanical}, in order selection for complex NC fMRI data \cite{adali2007complex}, in data analysis for complementary DNA microarray experiments \cite{crescenzi2001main} and for genome data analysis in bioinformatics \cite{duforet2016detecting}. PCA offers numerous other applications including data visualization for biplot graphic display of matrices \cite{gabriel1971biplot} and data compression of meteorological parameters obtained from high-resolution infrared spectra \cite{huang2001application}. 
\paragraph{Feature Extraction and Pattern Recognition}
Feature selection selects a subset of the original features, whereas feature extraction constructs a new feature subspace based on the feature set information. Feature extraction of image sequences by PCA can be later used for classification and recognition process in quality control applications \cite{xiao2010principal} and fast iris recognition \cite{suganthy2012principal}, respectively. Moreover, PCA is inevitably useful in pattern recognition like handwritten digits recognition \cite{zhang2001nonlinear}, face recognition \cite{zhao2006novel} and automated diagnosis of cardiac health using principal components of segmented ECG beats \cite{martis2012application}.
\paragraph{Model Selection, Digital Filtering and Data Denoising}
Some complex mixing models i.e., DoA estimation, BSS and NC signal detection cannot employ wlPCA owing to their linear models. Thus, ncPCA is particularly advantageous for applications like model selection which aim to determine the subspace order and the number of noncircular signals. Furthermore, it is capable of detecting circular and noncircular signals  and estimating signal subspace \cite{li2011noncircular}. ncPCA is preferred for scenarios with high SNR, large number of samples and high degree of noncircularity. On the other hand, non-linear component analysis requires WL PCA with complex kernel (wlPCA-ck) \cite{scholkopf1997kernel} for the  design of digital filters and regression frameworks
\cite{papaioannou2013principal}. Furthermore, statistical PCA ia also employed for data denoising especially for image denosing on multi-exponential MRI relaxometry \cite{does2019evaluation} and signal denoising in stock market trading \cite{eldeep}.} 

{Conclusively, non-circularity exploitation in complex PCA is particularly significant when the underlying entities are improper. 
\subsubsection{Complex Least Mean Square Analysis}
LMS extenstion to the complex domain i.e., complex-LMS (CLMS) with non-circularity incorporation leads to variants like augmented CLMS (ACLMS) and complementary CLMS. These analysis techniques can be employed for adaptive estimation with numerous applications in system identification \cite{xia2017full}, real-time impropriety detection \cite{jelfs2012adaptive}, 
communications \cite{schober2004data}, signal processing \cite{douglas2010performance}, renewable energy \cite{mandic2009complexv}, power systems \cite{mandic2015mean,xia2012widely}, and medicine  \cite{li2012modelling}.}

{
CLMS algorithm for SL estimation with general second-order noncircular (improper) Gaussian input is found useful in identifying system coefficients which formed a strictly linear FIR channel \cite{xia2017full}. Moreover,  collaborative adaptive filters trained by the CLMS can detect and track improperness in real-time unlike competing static detectors \cite{jelfs2012adaptive}. Additionally, multiple access interference in DS-CDMA systems can be efficiently supressed using WL LMS algorithms \cite{schober2004data}. Interestingly, ACLMS algorithm renders lower steady-state mean-squared error than conventional CLMS in adaptive beamforming for multi-port antenna arrays \cite{douglas2010performance,krishna2013hybrid}. Furthermore, ACLMS usefulness in wind modelling and forecasting is unprecedented in renewable energy domain \cite{mandic2009complexv,aali2019adaptive}. Similarly, the approximate uncorrelating transform improved adaptive frequency estimations using ACLMS in three-phase power grid systems \cite{mandic2015mean}.
 Last but not the least, a hybrid filter with standard CLMS and ACLMS algorithms can discriminate between discrete states of brain consciousness i.e., coma and quasi-brain-death using nonlinear features in EEG \cite{li2012modelling}.}

\subsubsection{{Miscellaneous}}
{Numerous other data analysis techniques which incorporate complete SOS of the complex improper observation data lead to the improved estimations of seismic traces} \cite{samson1983pure}, {wind measurements} \cite{khalili2014collaborative,adali2011complex}, {and oceanographic velocity measurements} \cite{lilly2006wavelet}.

{WL complex auto-regressive processing of the seismic signals helps to capture essential data characteristics like elliptical oscillations} \cite{sykulski2016widely}. {Other climate and seismology applications with improper complex-valued stochastic models can be efficiently simulated using circulant embedding} \cite{sykulski2016exact}. {Compressive sensing of weather sensor network application can effectively exploit the asymmetrical features for energy estimation to cool down a given structure}~\cite{wang2015framework}.  
\subsection{Signal Processing}
{Augmented and WL processing have demonstrated remarkable performance gains in different signal processing domains, e.g., signal estimation} \cite{schreier2010statistical,xia2017complementary}, 
{filtering} \cite{mandic2009complex}, {and detection} \cite{aghaei2010widely}. {Therefore, it is a leading competitor rendering vast applications in neuro-science, image processing, pattern recognition, and computer vision} \cite{hyvarinen2001independent,mohammadi2016improper}.
\subsubsection{{Array Processing}}
{The merits of impropriety adaptation in array processing algorithms are widely studied in}
\cite{mcwhorter2003widely,roemer2009multidimensional,
delmas2004asymptotically,chevalier2007widely,charge2001non}. 
{For example, coherent
processing (incorporating complementary covariance) for detection and estimation enjoys a 3dB gain over non-coherent processing} \cite{schreier2005detection}. {Similarly, estimation accuracy is substantially enhanced by employing NC signal constellations
in 1-D and 2-D DoA estimation} \cite{roemer2006efficient}. {Furthermore, such enhanced DoA estimation and identification methods for mixed circular and NC sources also improve the resolution capacity}~\cite{xian2008direction}. 
\subsubsection{{System Identification and Feature Extraction}}
{Superior system identification can be achieved using WL adaptive estimation of general IGS using augmented complementary least mean square analysis} \cite{xia2017full,xia2017complementary}. {Similarly, impropriety incorporation is crucial in blind source identification or accurate estimation and then separation or equalization in NC mixing arrangements like acoustic sources and fault diagnosis} \cite{looney2011augmented,de2002blind,eriksson2006complex,adali2014optimization}. 
{Additionally, noncircular ICA has demonstrated huge feature extracting potential in neuro-science, image processing, and vision research where we aim to find features that are as independent as possible}~\cite{hyvarinen2001independent}. 
\subsubsection{{Pattern Recognition and Image Processing}}
{Major signal processing tasks such as  compression, denoising, classification, feature extraction, image processing, and pattern recognition require sophisticated generalized models}\cite{hyvarinen2001independent}.  {These problems can be efficiently solved using a statistical generative model based on NC ICA or
require statistical measures like Bhattacharyya coefficient/distance and Kullback-Leibler divergence for appropriate modeling. These measures are well defined for the real signals or proper complex signals. However, their extension to more generalized scenarios of improper complex-valued Gaussian densities has enabled superior and reliable performance in the aforementioned applications} \cite{mohammadi2016improper}.
{Similarly, target detection in multi-band spectral images suffering from improper Gaussian noise is only possible with the advancement in impropriety literature}~\cite{matalkah2008generalized}.
\subsection{Communication Systems}
Communication systems can reap tremendous benefits from the asymmetric signaling in various interference limited scenarios. In this way, enhanced system performance in terms of improved achievable data rates and more reliable communication with lower EP can be achieved without exhausting the already saturated resources. 

\subsubsection{Cellular Networks}
Improper/asymmetric signaling techniques are implemented
in GSM \cite{otterseten2005receiver,mostafa2003single} and 3GPP networks \cite{MUROS12009,MUROS22009}.
Similarly, in mobile MU communications, non-circularity characterization can render better tradeoff between power consumption and spectral efficiency \cite{adali2011complex}

\subsubsection{Cognitive Radios}
To address the sparse temporal and spatial utilization of spectrum bands, cognitive radio settings permit a secondary network to opportunistically utilize the spectrum resources of a licensed primary network \cite{yeo2011optimal}. Secondary network senses the network availability and transmits under primary network QoS constraints. 
Cognitive radio settings are broadly categorized as underlay, overlay and interweave. Underlay cognitive system limits the transmission power of secondary network to maintain licensed users' QoS  \cite{amin2016underlay}. Overlay cognitive enjoys part of spectrum resources for its transmission while assisting primary network transmission. This scheme maintains primary network QoS through assistance with minimal interference and its own QoS by effectively canceling primary interference at its RX \cite{amin2017overlay}. Lastly, interweave setting utilizes unused spectrum holes for its transmission as long as it is available \cite{hedhly2017interweave}.
Interestingly, efficient interference management by IGS permits secondary network to effectively utilize spectrum resources and enhance their system efficiency while maintaining primary network QoS with PGS. It is evident from the fact that improper interference from secondary network to primary network is way less deteriorating than its counterpart proper interference. Surprisingly, the gains reaped by secondary network with IGS transmission over PGS transmission are conditional. For instance, SU rate improvement with IGS is only feasible if the fraction of the squared modulus between the SU-PU
interference link and the SU direct link surpass a threshold in underlay \cite{lameiro2015benefits} and overlay
cognitive radio paradigm \cite{amin2017overlay}.
Moreover, for underlay MAC, IGS is optimal if the accumulative IC gains exceed a certain threshold \cite{lameiro2018improper}. On the other hand, 
with adequate detection capabilities SU can employ IGS with maximum power to unconditionally improve its rate while satisfying PU QoS \cite{hedhly2017interweave}. Apart from the traditional role of IGS to enhance data rate or reduce EP in communication systems, it can also be employed to enhance the secrecy performance. Consequently, unlicensed user achieves lower SOP with IGS employment in cognitive radio setup \cite{oliveira2018physical}.

\subsubsection{Full Duplex Systems}
The performance of full duplex systems with simultaneous transmission and reception is invariably limited by the inherent self-interference (SI).  This can be efficiently mitigated with the optimal asymmetric signaling transmission for in-band full-duplex capable transceivers with \cite{gaafar2017underlay} or without  \cite{sornalatha2017modeling} spectrum sharing, SISO \cite{gaafar2018full} and MIMO \cite{kariminezhad2017power,liu2016optimal} full duplex relaying. 
as well as heterogeneous multi-tier network involving cellular and D2D full duplex communications \cite{kariminezhad2016heterogeneous}. 
Improper transmission is even more beneficial for the joint compensation of SI and hardware imperfections in full duplex HWI systems. For instance, asymmetric transmission is capable of suppressing SI along with transmitter power amplifier nonlinear distortion and transceiver IQI  \cite{korpi2014widely} and asymmetric hardware distortions \cite{javed2018improper}. 
Besides full-duplex relaying, WL processing is also rewarding for other relay networks including two-way AF-MIMO relayed MU systems \cite{zhang2013widely} and multi-layered relay systems \cite{ho2013optimal}. Surprisingly, IGS is also favorable for the alternate relaying systems which mimic as full-duplex systems \cite{gaafar2016alternate}.

\subsubsection{Hardware Impairments Mitigation}
Improper/asymmetric signaling is a promising candidate for the compensation of various hardware imperfections including asymmetric hardware distortions in receive diversity systems \cite{javed2017asymmetric} 
and IQI in space-time coded transmit diversity systems \cite{zou2008digital}. Severe performance losses caused by the IQI (which leads to improper received signals) can be efficiently compensated by WL RX for uplink multi-cell massive MIMO \cite{zarei2016q}, WL precoding for Large scale MIMO \cite{zhang2017widely}, WL beamforming of linear antenna arrays \cite{hakkarainen2013rf}, and massive antenna arrays \cite{hakkarainen2013widely}, and
circularity based compensators for wideband direct-conversion RXs \cite{anttila2008circularity} and OFDM based
WLAN transmitters \cite{li2017noncircular}.
Improper transmission is also recognized to jointly compensate multiple HWIs such as additive hardware distortions and transceiver IQI
in single antenna \cite{javed2018asymmetric} and multi-antenna systems \cite{javed2018multiple}. Interestingly, impairments in I/Q modulators are also accurately modelled using widely non-linear model using compressed sensing \cite{madero2015robustness}. Moreover, the expectation maximization based ML channel estimation in multicarrier scenarios under phase distortion namely, phase noise and carrier frequency offset holds true for both proper and improper signaling \cite{carvajal2013based}.
\subsubsection{Phase Estimation}
In the absence of pilot training sequence, conventional symmetric signaling employ differential coding scheme for phase estimation. On contrary, asymmetric signaling can do the needful without differential coding hence saving around 3dB loss in SNR at 1e-2 SER. However, the performance gain up to 97.50\% reduced SEP along with the absolute phase estimation comes at some cost. It is very small reduction
in entropy and/or minimum distance owing to the unequally probable symbols and/or unequal symbol spacing, respectively, introducing asymmetry in the constellation \cite{thaiupathump2000asymmetric}.

\subsubsection{Interference Channels}
An interesting question is the suitability of improper signaling and WL filters when the system under study is proper (i.e., information and noise signals are proper). Among many others, Cadambe \textit{et al.} in their pioneering work demonstrated the superiority of IGS in interference alignment scenarios \cite{cadambe2010interference,cadambe2008interference} counter intuitive to the PGS and linear filtering optimality in P2P communications \cite{telatar1999capacity}. Performance gains are also reported for 
a variety of interference-limited settings spanning 
SISO Z-IC \cite{kurniawan2015improper,lameiro2017rate}, 
MIMO Z-IC \cite{lagen2016superiority},
MIMO P2P \cite{yang2014interference,lagen2014improper},
MU SISO X-IC\cite{nguyen2015improper,zeng2013improper,kim2013potential,park2013sinr,ho2012improper,lameiro2013degrees},
MU MISO IC \cite{zeng2013optimized,zeng2013miso},
MU MIMO IC \cite{zeng2013transmit,lagen2016coexisting},
MIMO-BC \cite{zeng2013optimized,hellings2013performance,bai2018optimal,hellings2013qos}, 
MIMO-IBC \cite{shin2011achievable,lin2018multi},
cognitive MAC with primary P2P \cite{lameiro2018improper},
cognitive P2P with primary MAC \cite{kariminezhad2016improper},
MU diversity systems \cite{shin2012new}, multicarrier systems \cite{fusco2004ml,darsena2007universal,cheng2013widely}, multi-antenna systems \cite{lameiro2018performance}, and single-/multi- antenna NC interference cancellation \cite{chevalier2011widely,chevalier2006new}.
Other forms of interference i.e., co-channel interference \cite{kuchi2009performance,kim2016asymptotically}, intra- and intersystem multiple access interference in radio navigation satellite services \cite{enneking2018gaussian}, and
wideband multiple access and narrow band interference in CDMA systems \cite{buzzi2001new,gelli2000blind} can also be suppressed using improper characteristics. The application of the analysis to HCN multi-tier deployment with one macro eNode BS and multiple small eNode BS render large gains with guaranteed rate improvement for all SeNBs \cite{lagen2016superiority}. Additionally, IGS with symbol extension can outperform PGS for interference alignment within the context of linear precoding schemes where all interference is treated as noise \cite{cadambe2010interference,jafar2011interference}.

\subsubsection{Noisy Channels}
 For MU Gaussian MIMO P2P, MAC, BC and IC with proper Gaussian per-user input signals, proper noise is the worst case for the rates under any constraint on the noise covariance matrices \cite{hellings2017worst}.
However, there are instances when noisy channels render improper/asymmetric signatures. The treatment of improper complex noise is carried out in various systems including CDMA \cite{yoon1995matched, yoon1997maximizing}, discrete multitoned systems \cite{taubock2007complex} and spectral image target detection \cite{matalkah2008generalized}. Asymmetric noise characterization is necessary for appropriate estimation \cite{lang2017classical}, detection \cite{tugnait2017multisensor,aghaei2008maximum,kassam1982robust}, filtering \cite{mohammadi2015complex}, processing \cite{taubock2012complex,han2012capacity}, and compensation \cite{alsmadi2018ssk}. 
 
\subsubsection{Trelli's Coding}
A common perception is the optimality of the symmetric discrete signal constellations for both coded and uncoded communication systems. Although this holds true for the uncoded transmission, but may stand false for the coded systems \cite{divsalar1987trellis}. Various contributions have supported this statement by designing asymmetric signal constellations to obtain a performance gain with trellis coding. 
Trellis-based detection is essential to close the gap
between suboptimum DFE and optimum but computational complex MLSE. Trellis-based detection with WL preprocessing enables better suppression of noise and ISI for improper transmission in frequency selective channels \cite{gerstacker2003receivers}.
Moreover, Divsalar \textit{et al.} proposed joint design of $n/(n + 1)$ trellis codes and $2^{n+1}$-point asymmetric signaling, with same bandwidth requirement as an uncoded  $2^{n}$-point symmetric signal constellation. The joint treatment depicts significant improvement in minimum free Euclidean distance of the TCM which corresponds to maximum reduction in EbNo for a given BER \cite{divsalar1987trellis}. Similar studies aim to design TCM parameters not only to increase the effective length but also
the minimum product distance of the code with
Rayleigh or Rician fading channels. Interestingly, the later asymmetric TCM scheme offers gains
without additional bandwidth or power requirements\cite{subramaniam2000performance}. 
Other contributions optimize asymmetric constellations to target minimal BER of TCM systems over Gaussian channels unlike the conventional criterion of maximizing the free Euclidean distance \cite{zhang2009optimum}.
Equivalently, asymmetric signal constellation with Trellis coding outperforms conventional symmetric signaling in terms of BER in DS/SSMA systems \cite{park1998ds}. 
\section{Lessons Learned and Future Research Directions}
\begin{figure*}[t]
\begin{minipage}[b]{1.0\linewidth}
  \centering
  \centerline{\includegraphics[width=12cm]{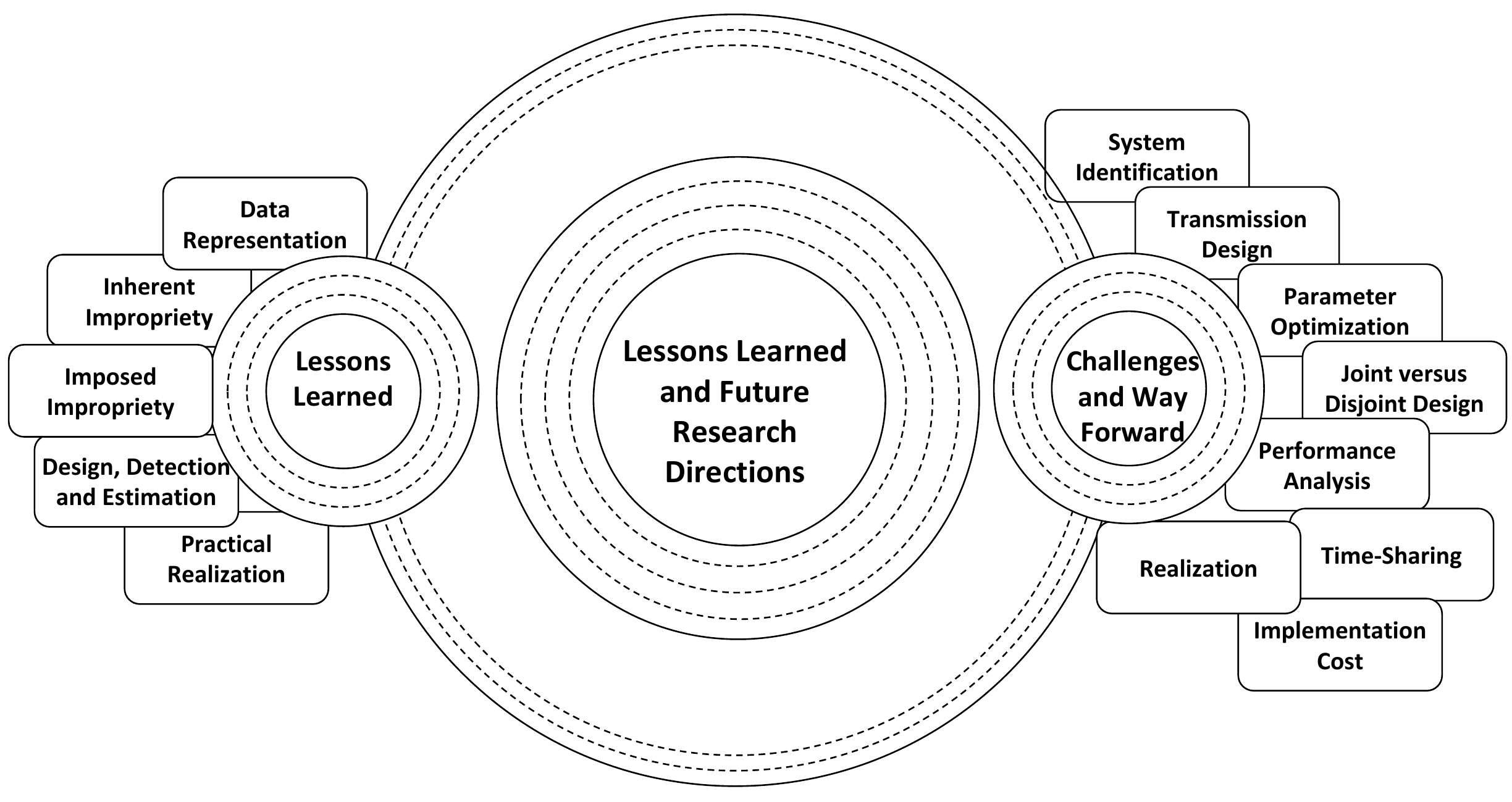}}
\end{minipage}
\caption{Lessons Learned and Way Forward}
\label{fig:Lessons}
\end{figure*}

This section summarizes the valuable lessons learned throughout this survey. It also highlights the challenges and limitations in getting the maximum benefit from impropriety and asymmetry concepts. These limitations open new research directions in numerous fields, especially communication theory as categorized in Fig. \ref{fig:Lessons}.
\subsection{Lessons Learned}
{The importance of complete SOS cannot be advocated more, but this is just the beginning. It completes the analysis for Gaussian distribution; however, others may require higher-order statistics and rotational invariance characteristics for their complete description. It may bring added benefits at the expense of increased computational cost in terms of resources such as time and energy.}
\subsubsection{Data Representations}
 {To summarize, three different yet interchangeable data representations with their own merits are presented. Studies advocate complex representation for easy and compact analysis, real composite representation for straightforward geometrical interpretations, and augmented representation for complete characterization, transformation, and analysis. Moreover, real-composite can identify improperness but fails to provide the DoI, unlike the rest of the two representations. DoI is used to identify the extent of impropriety and quantify the entropy loss. Notably, circularity implies propriety whereas impropriety implies non-circularity but not vice versa. Propriety can vary from generalized to strict whereas circularity of a RV may range from marginal, weak, strong to total circularity. It is important to identify the extent of properness using propriety tests in order to apply the simplified form of processing as properness is only preserved under SL or affine transformation. Impropriety incorporation requires extended definitions of differential entropy and joint distribution for a RV e.g., Gaussian RV \eqref{eq:diffentropy}-\eqref{eq:QGPdf}.}
\subsubsection{Inherent Impropriety}
{Exploiting the inherent impropriety in various configurations such as non-circular modulation, linear time-space coding, iterative receiver, improper interference and hardware impairments in communication and improper empirical data in fluid dynamics etc. can reap numerous benefits. For instance, it can pay significant rewards in the diverse fields like system fault identification in power systems, feature extraction and enhance estimation of EOG, EEG, ECG, and fMRI in biomedical engineering, quantum OCT in optics, speech recognition in acoustics, seismic and wind fields estimation in geophysics, and ocean-current spectra in oceanography etc. Impropriety incorporation is essential for appropriate modeling, analysis and optimum performance. Considering wireless communication systems, augmented representation is crucial for accurate SNR, achievable rate, and outage analysis.} 
\subsubsection{Imposed Impropriety}
{The substantial performance improvement can be achieved by imposing impropriety in various interference-limited communication scenarios. To summarize the comparative study carried out in Table IV, underlay cognitive setup reaps the maximum benefit of IGS when compared to overlay and interweave cognitive setups. Similarly, IGS is proven to be more beneficial, with a substantial increase in achievable rate, in IC relative to BC in a MIMO setup. SISO and SIMO follow the same trend for achievable rate whereas MISO reverses this trend by offering more rate-region improvement for BC and IBC relative to the IC. Furthermore, IGS has an added advantage of suppressing RSI in FD relaying and thus rendering higher achievable rates when compared to the HD relay mode. As per the IGS merits in outage analysis, multiple antenna systems depict lower outage than single-antenna systems. Moreover, percentage improvement relative to maximal PGS is far more than the optimal PGS in dual-hop DF-FDR. Also, the FD underlay cognitive mode renders lower rate outage than its HD counterpart whereas significant secrecy outage improvement is observed in the presence of an eavesdropper.}
\subsubsection{Design, Detection, and Estimation}
{System performance gains can only be attained by the appropriate design of IGS as per the underlying application. The enumerated design guidelines and tools signify the importance of problem identification. For instance, simple convex problems can either be solved in closed form or using  algorithms like IPM. Alternately, for non-convex problems, semi-definite relaxation or sequential convex programming can help to convexify and find the approximate solution. Other times, it is difficult to solve a joint optimization problem. It is then recommended to employ alternate optimization like ADMM if the problem is convex for the subgroups of variables. If this condition fails, then separate optimization is a better alternate with suboptimal solution. For minimal dimensions, line search or even exhaustive search can give promising results whereas other algorithms are needed for NP-hard problems. Eventually, resorting to maximal IGS in place of conventional PGS can also be beneficial for other intractable optimization problems. As for the detection, the presence or DoI of an improper signal can be identified in the presence of proper, improper, or colored noise. Estimation is required to approximate the value of such improper signal at any instance. Furthermore, the separation, filtering and feature extraction of such sources can also be of great interest which can be efficiently carried out using BSS, ICA and PCA etc.}
\subsubsection{Practical Realization}
{Out of the theoretical discussion, the practical realization is of utmost interest. The crux of the matter is that evaluation of the superiority of one form of asymmetric signaling over the other is critical yet tricky. Thus, hybrid signaling can reap tremendous payoffs in terms of improved achievable rates, energy efficiency, DoF and reduced outage and error probabilities at the expense of increased computation complexity in the modulation and detection phase. Error probability analysis depicts significant performance enhancement by asymmetric signaling in trellis coded systems especially with DS/SSMA scheme. Apart from the coded systems, uncoded MIMO with iterative receiver or with CCI and ISI in frequency selective channels reap the maximum benefit with asymmetric signaling. WL extension of various asymmetric signal recovery methods like equalization, estimation, filtering,  and detection outperform their SL counterparts when dealing with asymmetric systems. Interestingly, asymmetric signaling can open new dimensions for user separation based on their asymmetric characteristics.}

In a nutshell, the impropriety characterization in data analysis and signal processing renders numerous applications in medicine, economy, geology, oceanography, data mining, data denoising, data compression, dimensionality reduction, array processing, feature extraction, pattern recognition, and image processing etc. Focusing on the communication systems, the improper signaling has played a vital role in improving the system performance in cellular networks, cognitive radio setups, full-duplex communication, multi-antenna, multi-user and multi-cell setups. Additionally, it can effectively compensate the drastic effects of hardware impairments, interference and noisy channels. Although the journey from improper signaling to asymmetric signaling is quite appealing, it comes with few challenges and limitations as discussed next.
\subsection{Challenges and Way Forward} 
{Throughout this article, we have discussed the advantages of exploiting or incorporating non-circularity and impropriety. However, this performance comes with few challenges in terms of applicability, suitability, and practicality, etc which need further investigations. In this subsection, we discuss these limitations and suggest some way forward.}
\subsubsection{System Identification}
{Improper/asymmetric signaling techniques are already implemented in GSM and 3GPP networks} \cite{ho2012improper}. {Now we need to systematically scrutinize all the applications/systems/scenarios in numerous diverse fields where impropriety characterization can beat the traditional processing.  It is noteworthy that not all the systems benefit from improper/asymmetric signaling and even if they do, the advantages may be conditional.
In wireless communications, the thorough examination to determine the superiority conditions of improper signaling or WL processing over proper signaling or SL processing is as crucial as system scrutinizing itself. 
 As an example,  MIMO-BC with proper Gaussian noise can achieve sum rate capacity under a sum power constraint with dirty paper
coding and PGS in place of IGS}  \cite{vishwanath2003duality,hellings2015block}. {Similarly, IGS in combination
with WL transceivers is beneficial in MIMO-BC under specific scenarios which are not yet understood}  \cite{hellings2019high}. {Moreover, the perks of IGS over PGS for SU transmission are conditional in underlay and overlay cognitive setups} \cite{lameiro2015benefits,amin2017overlay}. 
{On the other hand, IGS is the all time favorite in other interference limited, hardware impaired or improper/non-circular noise based systems}\cite{javed2017asymmetric,zarei2016q,zeng2013transmit,hellings2017reduced} 
{Therefore, a major challenge is to assess the usefulness of improper transmission in the underlying system which is not straightforward. Nonetheless, we would like to highlight a broader guideline as proper/symmetric signaling is the preferred choice in the noise limited regime whereas improper/asymmetric signaling is favorable in the interference limited regime} \cite{ho2012improper}. {Similarly, circular models are favored with small number of samples, low signal-to-noise ratio, or minimal degree of non-circularity} \cite{adali2014optimization}.
\subsubsection{Transmission Design}
{Considering the favorable scenarios, when impropriety can reap benefits, the majority of the studies advertise the employment of IGS transmission which is practically not feasible. Thus, we need to resort to discrete asymmetric transmission, which poses new challenges. 
The main challenge is choosing the optimal asymmetric signaling scheme based on the underlying system. For example, probabilistic shaping is widely applied in optical communications, whereas geometrical shaping is recently introduced for wireless communication systems. Performance superiority of one over the other in a particular application is yet to be investigated. Therefore, we advocate the employment of hybrid signaling in order to meet the upcoming demands of the communication systems for the internet-of-things era. Conclusively, hybrid geometrical and probabilistic asymmetric constellation designs can return significant performance merits while closely approaching Shannon limits.} 
\subsubsection{Parameter Optimization}
{The next challenge is the optimization of the opted asymmetric signaling to fine tune transmission parameters, e.g., prior probabilities for probabilistic shaping, optimal rotation/translation, lattice and envelope for geometric shaping, non-uniform allocation of orthogonal/non-orthogonal resources, or some/all of these for hybrid signaling. The intricate search for an optimal solution is especially complicated by the significant number of feasible transmission strategies, mainly
for a large number of participating users and/or antennas} \cite{kurniawan2015improper}. {For instance, joint optimization of transmission parameters is doable for SISO-IC but we have to resort to suboptimal solutions when it comes to MISO- and MIMO-IC} \cite{zeng2013transmit,zeng2013optimized}. 
{Similarly, maximal IGS is adopted owing to the intractable optimization in MIMO hardware impaired systems} \cite{javed2018multiple}.
{In fact, either there is a lack of optimization tools for non-convex or NP-hard structural problems, or the existing algorithms render suboptimal solutions with excessive computational surcharge. Thus, low complexity algorithms with the near-optimal performance are required to fill the gap opened by the lack of optimal solutions in the complex systems} \cite{kurniawan2015improper}. {Consequently, the search for least-complex near-to-optimal optimization strategy is an open research area.}
\subsubsection{Joint and Disjoint Design}
 {Another conflict is the choice of cooperative or non-cooperative signaling in cases like multi-antenna, multi-user and multi-cell configurations. Non-cooperative signaling may render suboptimal performance whereas cooperative scheme requires the global knowledge of system parameters to yield optimal performance. This may lead to excessive communication overhead besides increased computational cost. For example, the performance comparison of multi-hop DF FDR communication under HWIs with distributed optimization framework reveals enhanced performance gains with increasing cluster size. The maximum gain is achieved with joint optimization of all nodes, however, this performance comes at the expense of increased complexity, communication overhead and processing delays. 
As a general guideline, distributed optimization approach is the favorable choice for large systems whereas joint optimization is preferred for relatively small systems. Another concern is the unavailability or inaccurate estimation of few system parameters on IGS performance. For example, can IGS gains surpass PGS  in large scale MU and/or multi-antenna systems in the absence of instantaneous CSIT?} \cite{soleymani2019ergodic}. {Future research may address the limitation arising from the imperfection or lack of instantaneous/average system parameters while tuning improper transmission parameters.}
\subsubsection{Performance Analysis}
{Accurate analysis is the key to design appropriate system parameters
which can attain the expected system performance. Most of the studies employ complex representation relying on the covariance matrices for SNR analysis. Adopting such representation in the analysis ignores the correlation between the entities and their respective conjugates. Therefore, we suggest the employment of complex augmented covariance matrices to evaluate SNR and subsequent SNR outage performance. For instance, the accurate rate analysis and rate outage with augmented representations is advocated for multi-antenna ICs} \cite{zeng2013transmit}.   
{Similarly, error probability analysis should exploit the improper noise characteristics which is particularly emphasized in a hardware imapired system configuration} \cite{javed2018asymmetric}. 
{Nevertheless, another limitation is the lack of numerical tools for exact performance analysis. The design of improper/asymmetric signaling parameters highly depends on the objective function which can be maximizing achievable rates like average achievable rate, achievable sum rate, minimum achievable rate or achievable rate region, minimizing outage probabilities like rate outage, SNR outage, or secrecy outage, or minimizing error probabilities. Thus, the accurate analysis of these performance metrics will dictate the optimality of improper/asymmetric transmission. 
 For instance, the asymmetric system design is mostly based on the derived bounds instead of exact EP analysis, which yields loosely fitted model parameters} \cite{zhang2009optimum}. {Thus, tools are inevitably required to derive exact EP based on accurately estimated parameters.}
\subsubsection{Time-Sharing (TS)}
{Most of the studies focused on the perks of improper/asymmetric signaling in the absence of TS. Comparison study between PGS and IGS in the MIMO-BC when TIN at high SNR reveals different trends with or without TS. In the absence of TS, gains due to IGS occurs both in systems with enough antennas at the base station and in overloaded systems. Whereas, if TS is allowed then IGS cannot bring any gains in a system with enough antennas at the BS as opposed to overloaded system where it is still advantageous. Similarly, IGS with TS is yet to find it's standings in a MU multicell MIMO IBC} \cite{nasir2019improper}.
{Likewise, the superiority of IGS over PGS to enlarge rate region in SISO Z-IC with TS is subject to the underlying assumptions. IGS is only beneficial under short-term average power constraints, whereas it cannot bring any gains under the long-term average power constraints} \cite{hellings2017improper}. {Although improper rate TS can outperform proper rate TS for SISO Z-IC, the investigation of this trend in a general SISO-IC is an open research problem. Nevertheless, situation is altogether different with three or more users where IGS is bound to bring the benefits even with TS owing to the added DoF} \cite{hellings2018proper}. Therefore, the superiority examination of IGS in a TS context under certain assumptions is an open research area in various interference-limited systems~\cite{hellings2019high,nasir2019improper}.
\subsubsection{Realization}
{The realization of the optimal asymmetric complex signaling along with the appropriate detection mechanism (to cater for the induced asymmetry) is also one the challenges the obstruct the journey from IGS to asymmetric signals. Undoubtedly, the asymmetric modulation does not work in isolation in modern communication setups. Interestingly, the survey of asymmetric discrete constellation with Trellis coding clarifies that the coding and modulation schemes are sometimes interrelated and hence cannot be treated independently} \cite{divsalar1987trellis}. {Moreover, the optimal detection should exploit the non-i.i.d. noise components e.g., maximum a posterior detection in a HWI system outperforms regular ML or minimum Euclidean distance detection} \cite{javed2018asymmetric}. {Similarly, 
asymmetric model of the aggregate HWIs in a wireless communication system will help in accurate system analysis and design}
\cite{javed2018multiple}. {Such appropriate modeling will also dictate the requisite resources to tackle and deal with the interference challenges. For example, asymmetric HWIs generate more errors requiring special buffer management approaches in wireless networks 
to maintain QoS} \cite{showail2016buffer,shihada2014buffer}. {Efficient buffering and queue management schemes are inevitably required to tackle the latency issues in delay-sensitive applications} \cite{bouacida2018practical}.  
{Consequently, the appropriate realization of the system containing asymmetric signatures is immensely important to achieve the target performance.}
\subsubsection{Implementation Cost}
{The existing infrastructure employs SL transceivers, and thus the up-gradation to WL transceivers is a sequential process} \cite{lagen2016coexisting}. {Last but not least is the evaluation of the tradeoff between performance gains and computational/implementation complexity.
Performance gains in terms of increased capacity, reduced outage, and  minimized EP are attained at the cost of added communication overhead to transfer system parameters, computational complexity to find a near-to-optimal solution, and implementation complexity to practically realize/detect the asymmetric transmission. Another intriguing concern is the power saving affair as emphasized in} \cite{divsalar1987trellis}, {i.e., introducing asymmetry does not affect the power or bandwidth needs of the systems. 
Nevertheless, a fair comparison is required to inspect whether the power saved by exploiting additional design freedom offered by asymmetric constellation is greater than the computational power spent on its fine tuning or not?}

These are few challenges and limitations which need dedicated efforts for comprehensive treatment and effective realization in order to attain the maximum benefit from the rising asymmetry concept.
\section{Conclusion}
The journey from proper signaling to improper signaling and then from improper Gaussian to asymmetric discrete constellation is captured and summarized in this article. Various complex data representations, their complete SOS characterization and appropriate processing models are presented for comprehensive illustration. Furthermore, some intrinsic sources of impropriety as well as the vast applications of asymmetric signaling in various diverse fields i.e., medicine, communication, geology and computer vision are elaborated. 
This review article takes readers from the theoretical achievable bounds to practical realization of impropriety concepts.
One of the notable contribution of this work is the performance comparison of improper signaling versus traditional proper signaling in terms of achievable rate, system outage and EP in numerous system configurations. The comparison captures all the necessary details including maximum achievable percentage improvement, transceiver types, design metric, employed strategies and optimization procedure etc. We believe that this survey along with the presented challenges and future research directions will not only compel readers to incorporate propriety concepts but also increase the activity in this critical realm.
\bibliographystyle{IEEEtran}

\bibliography{IEEEabrv,Survey_Ref}

\vspace{-1 cm}
\begin{IEEEbiography}[{\includegraphics[width=1in,height=1.25in,clip,keepaspectratio]{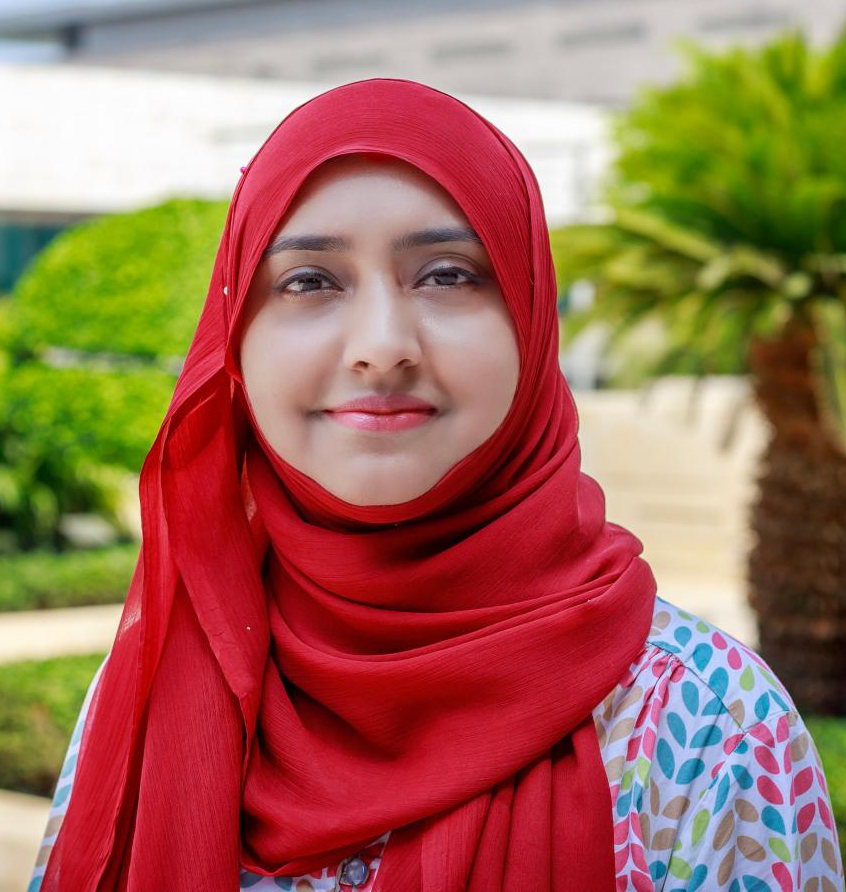}}]{Sidrah Javed} (S'16)
received the B.E. degree in Electrical (Telecommunication) Engineering from National University of Science and Technology (NUST), Islamabad, Pakistan, in 2012. From 2012 to 2015, she has worked as a Research Engineer in National Radio and Telecommunication Corporation, Pakistan. She is currently pursuing M.S./Ph.D. degree at the Computer, Electrical and Mathematical Sciences \& Engineering (CEMSE) Division, King Abdullah University of Science and Technology (KAUST). Her research interests include hardware impairment and interference management using asymmetric signaling, constellation shaping, spatial modulation, and cooperative communications in wireless communications, vehicular tracking/monitoring and solar energy harvesting. 
\end{IEEEbiography}
\vspace{-1 cm}
\begin{IEEEbiography} [{\includegraphics[width=1in,height=1.25in,clip,keepaspectratio]{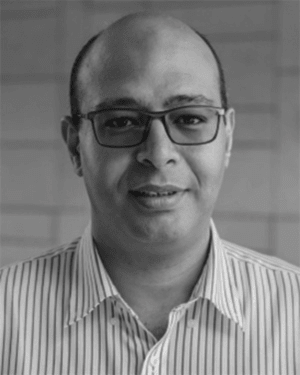}}]{Osama Amin} (S'07, M'11, SM'15) received the B.Sc. degree in electrical and electronics engineering from Aswan University, Aswan, Egypt, in 2000,  the M.Sc. degree in electrical engineering from Assiut University, Assiut, Egypt, in 2004, and the Ph.D. degree in electrical and computer engineering from the University of Waterloo, Canada, in 2010. In 2012, he joined the Electrical and Electronics  
Engineering Department, Assiut University, as an Assistant Professor. He is currently a Research Scientist with the King Abdullah University of Science and Technology, Thuwal, Saudi Arabia. His general research interests lie in communications systems and signal processing for communications with special emphasis on wireless applications, optical wireless communication, molecular communications, terahertz communications, green communications, and cognitive radio.  

Dr. Amin has served as a technical program committee (TPC) member  
for ICC, GLOBECOM, IEEE VTC, CROWNCOM, PIMRC, and ISSPIT conferences. He  
has served also as a Co-Organizer and a Co-Chair of the Next Generation  
Green ICT and 5G Networking 2015 at the IEEE International Conference  
on Ubiquitous Wireless Broadband in Montreal, Canada. In 2018, he served as the chair of the IoT, M2M, Sensor Networks, and Ad-Hoc Networking track in the VTC2018-fall held in Chicago, USA.  He serves as an Associate Editor for the IEEE COMMUNICATIONS LETTERS.
\end{IEEEbiography}
\vspace{-1 cm}
\begin{IEEEbiography}[{\includegraphics[width=1in,height=1.25in,clip,keepaspectratio]{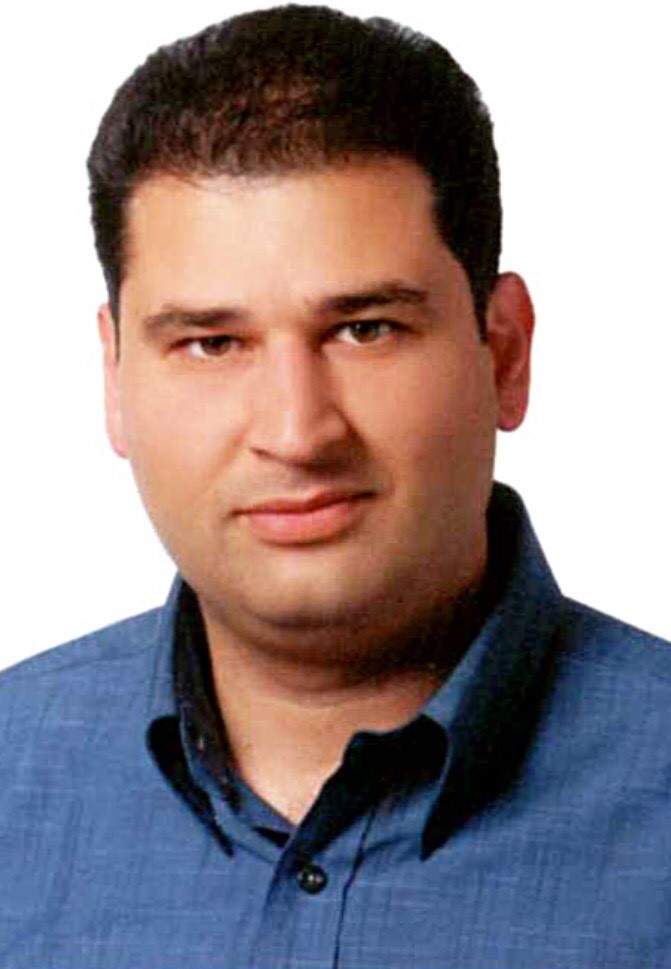}}] {Basem Shihada} (SM'12) is an associate \& founding professor in the Computer, Electrical and Mathematical Sciences \& Engineering (CEMSE) Division at King Abdullah University of Science and Technology (KAUST). He obtained his PhD in Computer Science from University of Waterloo.

In 2009, he was appointed as visiting faculty in the Department of Computer Science, Stanford University. In 2012, he was elevated to the rank of Senior Member of IEEE.
His current research covers a range of topics in energy and resource allocation in wired and wireless networks, software defined networking, internet of things, data networks, network security, and cloud/fog computing. 
\end{IEEEbiography}
\vspace{-1 cm}
\begin{IEEEbiography}[{\includegraphics[width=1in,height=1.25in,clip,keepaspectratio]{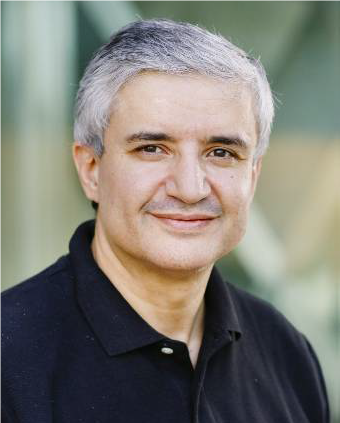}}] {Mohamed-Slim Alouini} (S'94, M'98, SM'03, F'09) 
was born in Tunis, Tunisia. He received the Ph.D. degree in Electrical Engineering from the California Institute of Technology (Caltech), Pasadena, CA, USA, in 1998. He served as a faculty member in the University of Minnesota, Minneapolis, MN, USA, then in the Texas A\&M University at Qatar, Education City, Doha, Qatar before joining King Abdullah University of Science and Technology (KAUST), Thuwal, Makkah Province, Saudi Arabia as a Professor of Electrical Engineering in 2009. His current research interests include the modeling, design, and performance analysis of wireless communication systems.
\end{IEEEbiography}

\end{document}